\documentclass[a4paper,11pt]{article}
\pdfoutput=1

\usepackage{jheppub}

\usepackage{subfig}
\usepackage{xspace}
\usepackage[countmax]{subfloat}
\usepackage{slashed}
\usepackage{hyperref}

\usepackage{bbm}

\usepackage{color}
\definecolor{darkblue}{rgb}{0,0,0.5}
\definecolor{darkred}{rgb}{0.5,0,0}
\definecolor{darkgreen}{rgb}{0,0.5,0}

\newcommand{\GeV}{\text{GeV}}

\def\zcut{z_{\text{cut}}}

\newcommand{\ecf}[2]{{e_{#1}^{(#2)}}} 
\newcommand{\ecfnobeta}[1]{e_{#1}} 

\newcommand{\ecfvar}[3]{{_{#1}e_{#2}^{(#3)}}} 
\newcommand{\ecfvarnobeta}[2]{{_{#1}e_{#2}}} 

\newcommand{\Dobs}[2]{D_{#1}^{(#2)}} 
\newcommand{\Dobsnobeta}[1]{D_{#1}} 

\newcommand{\Uobs}[2]{U_{#1}^{(#2)}} 
\newcommand{\Uobsnobeta}[1]{U_{#1}} 

\newcommand{\Nobs}[2]{N_{#1}^{(#2)}} 
\newcommand{\Nobsnobeta}[1]{N_{#1}} 

\newcommand{\Mobs}[2]{M_{#1}^{(#2)}} 
\newcommand{\Mobsnobeta}[1]{M_{#1}}

\newcommand{\Cobs}[2]{C_{#1}^{(#2)}} 
\newcommand{\Cobsnobeta}[1]{C_{#1}}

\def\nn{{\nonumber}}

\DeclareRobustCommand{\Sec}[1]{Sec.~\ref{#1}}
\DeclareRobustCommand{\Secs}[2]{Secs.~\ref{#1} and \ref{#2}}
\DeclareRobustCommand{\Secss}[3]{Secs.~\ref{#1}, \ref{#2}, and \ref{#3}}
\DeclareRobustCommand{\App}[1]{App.~\ref{#1}}
\DeclareRobustCommand{\Tab}[1]{Table~\ref{#1}}

\DeclareRobustCommand{\Fig}[1]{Fig.~\ref{#1}}
\DeclareRobustCommand{\Figs}[2]{Figs.~\ref{#1} and \ref{#2}}
\DeclareRobustCommand{\Figss}[3]{Figs.~\ref{#1}, \ref{#2}, and \ref{#3}}
\DeclareRobustCommand{\Eq}[1]{Eq.~(\ref{#1})}
\DeclareRobustCommand{\Eqs}[2]{Eqs.~(\ref{#1}) and (\ref{#2})}
\DeclareRobustCommand{\Ref}[1]{Ref.~\cite{#1}}
\DeclareRobustCommand{\Refs}[1]{Refs.~\cite{#1}}

\newcommand{\Nsub}[2]{\tau_{#1}^{(#2)}}
\newcommand{\Nsubnobeta}[1]{\tau_{#1}}

\DeclareFontFamily{U}{wncy}{}
    \DeclareFontShape{U}{wncy}{m}{n}{<->wncyi10}{}
    \DeclareSymbolFont{mcy}{U}{wncy}{m}{n}
    \DeclareMathSymbol{\backwardsN}{\mathord}{mcy}{"49}

\newcommand{\pythia}[1]{\textsc{Pythia\xspace #1}}
\newcommand{\madgraph}[1]{\textsc{MadGraph5\xspace #1}}
\newcommand{\fastjet}[1]{\textsc{FastJet\xspace #1}}
\newcommand{\herwig}[1]{\textsc{Herwig\xspace #1}}

\newcommand{\vincia}[1]{\textsc{Vincia\xspace #1}}

\preprint{MIT-CTP 4825}

\title{New Angles on Energy Correlation Functions}

\author[1,2,3]{Ian Moult,}
\author[3]{Lina Necib,}
\author[3]{and Jesse Thaler}

\affiliation[1]{Berkeley Center for Theoretical Physics, University of California, Berkeley, CA 94720, USA}
\affiliation[2]{Theoretical Physics Group, Lawrence Berkeley National Laboratory, Berkeley, CA 94720, USA}
\affiliation[3]{Center for Theoretical Physics, Massachusetts Institute of Technology, Cambridge, MA 02139, USA}

\emailAdd{ianmoult@lbl.gov}
\emailAdd{lnecib@mit.edu}
\emailAdd{jthaler@mit.edu}

\abstract{Jet substructure observables, designed to identify specific features within jets, play an essential role at the Large Hadron Collider (LHC), both for searching for signals beyond the Standard Model and for testing QCD in extreme phase space regions.
In this paper, we systematically study the structure of infrared and collinear safe substructure observables, defining a generalization of the energy correlation functions to probe $n$-particle correlations within a jet.
These generalized correlators provide a flexible basis for constructing new substructure observables optimized for specific purposes.
Focusing on three major targets of the jet substructure community---boosted top tagging, boosted $W/Z/H$ tagging, and quark/gluon discrimination---we use power-counting techniques to identify three new series of powerful discriminants: $M_i$, $N_i$, and $U_i$. 
The $M_i$ series is designed for use on groomed jets, providing a novel example of observables with improved discrimination power after the removal of soft radiation.
The $N_i$ series behave parametrically like the $N$-subjettiness ratio observables, but are defined without respect to subjet axes, exhibiting improved behavior in the unresolved limit.
Finally, the $U_i$ series improves quark/gluon discrimination by using higher-point correlators to simultaneously probe multiple emissions within a jet.
Taken together, these observables broaden the scope for jet substructure studies at the LHC.
}

\begin{document} 
\maketitle

\section{Introduction}\label{sec:intro}

With the Large Hadron Collider (LHC) rapidly acquiring data at a center-of-mass energy of $13$ TeV, jet substructure observables are playing a central role in a large number of analyses, from Standard Model measurements \cite{Chatrchyan:2012sn,CMS:2013cda,Aad:2015cua,Aad:2015lxa,ATLAS-CONF-2015-035,Aad:2015rpa,Aad:2015hna,ATLAS-CONF-2016-002,ATLAS-CONF-2016-039,ATLAS-CONF-2016-034,CMS-PAS-TOP-16-013,CMS-PAS-HIG-16-004} to searches for new physics \cite{CMS:2011bqa,Fleischmann:2013woa,Pilot:2013bla,TheATLAScollaboration:2013qia,Chatrchyan:2012ku,CMS-PAS-B2G-14-001,CMS-PAS-B2G-14-002,Khachatryan:2015axa,Khachatryan:2015bma,Aad:2015owa,Aaboud:2016okv,Aaboud:2016trl,Aaboud:2016qgg,ATLAS-CONF-2016-055,ATLAS-CONF-2015-071,ATLAS-CONF-2015-068,CMS-PAS-EXO-16-037,CMS-PAS-EXO-16-040,Khachatryan:2016mdm,CMS-PAS-HIG-16-016,CMS-PAS-B2G-15-003,CMS-PAS-EXO-16-017}.\footnote{This is by no means a complete list.  Other studies from the LHC using jet substructure can be found at \url{https://twiki.cern.ch/twiki/bin/view/AtlasPublic} and \url{http://cms-results.web.cern.ch/cms-results/public-results/publications/}.}  As the field of jet substructure matures \cite{Abdesselam:2010pt,Altheimer:2012mn,Altheimer:2013yza,Adams:2015hiv}, observables are being designed for increasingly specific purposes, using a broader set of criteria to evaluate their performance beyond simply raw discrimination power. Continued progress relies on achieving a deeper understanding of the QCD dynamics of jets, allowing for more subtle features within a jet to be exploited. This understanding has progressed rapidly in recent years, due both to advances in explicit calculations of jet substructure observables \cite{Feige:2012vc,Field:2012rw,Dasgupta:2013ihk,Dasgupta:2013via,Larkoski:2014pca,Dasgupta:2015yua,Seymour:1997kj,Li:2011hy,Larkoski:2012eh,Jankowiak:2012na,Chien:2014nsa,Chien:2014zna,Isaacson:2015fra,Krohn:2012fg,Waalewijn:2012sv,Larkoski:2014tva,Procura:2014cba,Bertolini:2015pka,Bhattacherjee:2015psa,Larkoski:2015kga,Dasgupta:2015lxh,Frye:2016okc,Frye:2016aiz,Kang:2016ehg,Hornig:2016ahz} as well as to the development of techniques for understanding the dominant properties of substructure observables using analytic \cite{Walsh:2011fz,Larkoski:2014gra,Larkoski:2014zma} and machine learning \cite{Cogan:2014oua,deOliveira:2015xxd,Almeida:2015jua,Baldi:2016fql,Guest:2016iqz,Conway:2016caq,Barnard:2016qma} approaches.

A particularly powerful method for constructing jet substructure observables is power counting, introduced in \Ref{Larkoski:2014gra}.  Given a basis of infrared and collinear (IRC) safe observables, power counting can identify which combinations are optimally sensitive to specific parametric features within a jet.\footnote{In this paper, we use ``basis'' to refer to any set of observables, even if they do not span the full space of IRC safe observables.} Furthermore, power counting elucidates the underlying physics probed by the observable.  This approach was successfully applied to the energy correlation functions \cite{Larkoski:2013eya}, leading to a powerful 2-prong discriminant called $D_2$ \cite{Larkoski:2014gra}.  Vital to the power counting approach, though, is a sufficiently flexible basis of IRC safe observables to allow the construction of discriminants with specific properties.
 
In this paper, we exploit the known properties of IRC safe observables to systematically identify a useful basis for jet substructure, which we call the generalized energy correlation functions.  These observables---denoted by $\ecfvar{v}{n}{\beta}$ and defined in \Eq{eq:ecf_gen}---are an extension of the original energy correlation functions with a more flexible angular weighting.\footnote{The $\ecfvarnobeta{v}{n}$ notation is inspired by the hypergeometric functions, which are similarly flexible.}  Specially, these new observables correlate $v$ pairwise angles among $n$ particles, whereas the original correlators were restricted to $v$ equaling $n$ choose 2.  Using these generalized correlators, we apply power counting to identify new jet substructure observables for each of the major jet substructure applications at the LHC:  3-prong boosted top tagging, 2-prong boosted $W/Z/H$ tagging, and 1-prong quark/gluon discrimination.  In each case, our new observables exhibit improved performance over traditional observables when tested with parton shower generators.

The flexibility of our basis, combined with insights from power counting, allows us to tailor our observables for specific purposes, beyond those that have been previously considered.  As an interesting example, we are able to specifically design observables for use on groomed jets \cite{Butterworth:2008iy,Ellis:2009su,Ellis:2009me,Krohn:2009th,Dasgupta:2013via,Dasgupta:2013ihk}.  While grooming procedures are heavily used at the LHC to remove jet contamination from initial state radiation, underlying event, and pileup, most LHC analyses apply observables that were designed for use on ungroomed jets.  Here, by understanding the impact of grooming on soft radiation, we introduce a 2-prong discriminant, $\Mobsnobeta{2}$, which exhibits almost no discrimination power on ungroomed jets, but outperforms traditional observables when measured on groomed jets. This observable therefore acts both as a probe of the grooming procedure and as a powerful discriminant.  We also show how the use of groomed observables leads to remarkably stable distributions as a function of the jet mass and $p_T$, even for distributions that are unstable before grooming, such as $D_2$.  This has recently been emphasized as a desirable feature for substructure observables, particularly to facilitate sideband calibration and produce smooth mass distributions for backgrounds \cite{Dolen:2016kst}; observables modified to achieve stability have been used by both ATLAS and CMS \cite{ATL-PHYS-PUB-2015-033,CMS-PAS-EXO-16-030}.

The generalized energy correlation functions allow us to introduce a wide variety of new substructure observables, though we focus on three series with particularly nice properties. 
The first is the $\Mobsnobeta{i}$ series, defined via the ratio
\begin{align}
\Mobs{i}{\beta}=\frac{\ecfvar{1}{i+1}{\beta}}{\ecfvar{1}{i}{\beta}}\,.
\end{align}
These observables identify jets with $i$ hard prongs, but, as mentioned above, are only effective for discrimination on suitably groomed jets. The second is the $\Nobsnobeta{i}$ series, defined via the ratio
\begin{equation}
N_i^{(\beta)} = \frac{\ecfvar{2}{i+1}{\beta}}{(\ecfvar{1}{i}{\beta})^2}\,,
\end{equation}
which are designed to mimic the behavior of the $N$-subjettiness ratio $\Nsubnobeta{i,i-1}$ \cite{Thaler:2010tr,Thaler:2011gf}.  The $\Nobsnobeta{i}$ observables are defined without respect to  subjet axes, and therefore exhibit improved behavior compared to $N$-subjettiness, particularly in the transition to the unresolved region, where the definition of subjet axes becomes ambiguous. The third is the $\Uobsnobeta{i}$ series, defined as
\begin{align}
\Uobs{i}{\beta}=\ecfvar{1}{i+1}{\beta}\,,
\end{align}
which probe multiple emissions within 1-prong jets and can be used to improve quark/gluon discrimination.  In all cases, the parameter $\beta$ controls the overall angular scaling of these observables, and the ${}^{(\beta)}$ superscript will often be dropped when clear from context.

To guide the reader, we summarize the particular applications studied in this paper, so that the (un)interested reader can skip to the relevant section. These observables will be made available in the \texttt{EnergyCorrelator} \fastjet{contrib} \cite{Cacciari:2011ma,fjcontrib} starting in version 1.2.0.
\begin{itemize}
\item \textbf{Boosted Top Tagging} (\Sec{sec:tops}):
	\begin{itemize}
	\item $\Nobsnobeta{3}$: An axes-free observable which reduces to the $N$-subjettiness ratio $\Nsubnobeta{3,2}$ in the resolved limit, but exhibits improved performance in the unresolved limit on groomed jets.
	\end{itemize}
\item \textbf{Boosted $W/Z/H$ tagging}  (\Sec{sec:2prong}):
	\begin{itemize}
	\item $\Mobsnobeta{2}$: A 2-prong discriminant specifically designed for use on groomed jets.
	\item $\Nobsnobeta{2}$: An axes-free observable which reduces to the $N$-subjetttiness ratio $\Nsubnobeta{2,1}$ in the resolved limit, but exhibits improved performance on  both groomed and ungroomed jets.
	\item $\Dobs{2}{\alpha,\beta}$: A generalization of the standard $D_2$ observable \cite{Larkoski:2014gra} specifically designed for groomed jets, which exhibits improved performance when $\alpha=1$, $\beta=2$.
	\end{itemize}
\item \textbf{Quark/Gluon Discrimination}  (\Sec{sec:qvsg}):
	\begin{itemize}
	\item $\Uobsnobeta{i}$: A new series of observables for quark/gluon discrimination which probes the structure of multiple soft gluon emissions from the hard jet core, leading to improved performance over the standard $\Cobsnobeta{1}$ observable \cite{Larkoski:2013eya}. 
	\end{itemize}
\end{itemize}
The specific form of these observables, and the origin of their discrimination power, will be analyzed using power counting.  We verify all power-counting predictions using parton shower generators and compare the performance of our newly introduced observables to traditional observables for each of the above applications.

The remainder of this paper is organized as follows. In \Sec{sec:review}, we review standard substructure and grooming techniques as well as the power counting approach for understanding soft and collinear scaling.  In \Sec{sec:defn}, we discuss the general structure of IRC safe observables and introduce the generalized energy correlation functions, $\ecfvarnobeta{v}{n}$, as well as the $M_i$, $N_i$, and $U_i$ series.  The three key case studies bulleted above appear in \Secss{sec:tops}{sec:2prong}{sec:qvsg}.  We conclude in \Sec{sec:conc} and discuss possible future directions for improving our understanding of jet substructure at the LHC.

\section{Review of Substructure Approaches}\label{sec:review}

In this section, we review a number of standard jet substructure techniques that will be used throughout this paper. We begin in \Sec{sec:ecfs} by defining the energy correlation functions \cite{Larkoski:2013eya} and $N$-subjettiness ratios \cite{Thaler:2010tr,Thaler:2011gf}, both of which are widely used in jet substructure.  In \Sec{sec:soft_drop}, we review the soft drop/modified mass drop \cite{Larkoski:2014wba,Dasgupta:2013ihk,Dasgupta:2013via} algorithm, which we use as our default grooming procedure.  Finally in \Sec{sec:power_counting}, using the 2-point energy correlation function as an example, we review the power-counting approach for analyzing jet substructure observables, which features heavily in later discussions.  Readers familiar with these topics can safely skip to \Sec{sec:defn}, though we recommend reviewing the logic of \Sec{sec:power_counting}.
 
\subsection{Energy Correlation Functions and $N$-subjettiness}\label{sec:ecfs}


The energy correlation functions \cite{Larkoski:2013eya} are a convenient basis of observables for probing multi-prong substructure within a jet.  In this paper, we use the 2-, 3-, and 4-point energy correlation functions, defined as\footnote{We use the normalized dimensionless definition denoted with a lower case $e$ \cite{Larkoski:2014gra}. This is related to the original dimensionful definition in \Ref{Larkoski:2013eya} by $\ecf{n}{\beta} = \text{ECF}(n,\beta)/\left( \text{ECF}(1,\beta) \right)^n.$}
\begin{align}
\label{eq:ECFs}
\ecf{2}{\beta} &= \sum_{1\leq i<j\leq n_J} z_i  z_j \, \theta_{ij}^\beta \, ,\nonumber \\
\ecf{3}{\beta} &= \sum_{1\leq i<j<k\leq n_J} z_i z_{j}z_{k} \, \theta_{ij}^\beta \theta_{ik}^\beta \theta_{jk}^\beta \, , \nonumber \\
\ecf{4}{\beta} &=\sum_{1\leq i<j<k<\ell\leq n_J} z_{i}z_{j}z_{k}z_{\ell} \, \theta_{ij}^\beta \theta_{ik}^\beta \theta_{jk}^\beta  \theta_{i\ell}^\beta \theta_{j\ell}^\beta \theta_{k\ell}^\beta \,,
\end{align}
where $n_J$ is the number of particles in the jet.  The generalization to higher-point correlators is straightforward, though we will not use them here.  For simplicity, we often drop the explicit angular exponent $\beta$, writing the observable as $\ecfnobeta{n}$.  This simplified notation will also be used for other observables introduced in the text.

It is convenient to work with dimensionless observables, written in terms of a generic energy fraction variable, $z$, and a generic angular variable, $\theta$. The precise definitions of the energy fraction and angle can be chosen depending on context and do not affect our power-counting arguments. For the case of $pp$ collisions at the LHC, which is the focus of our later studies, we work with longitudinally boost-invariant variables,
\begin{align}\label{eq:ptratio}  
z_i\equiv\frac{p_{Ti}}{\sum_{j \in \text{jet}} p_{Tj}}\,, \qquad  \theta_{ij}^2\equiv R_{ij}^2 = (\phi_i-\phi_j)^2+(y_i-y_j)^2\,,
\end{align}
where $p_{Ti}$, $\phi_i$, and $y_i$ are the transverse momentum, azimuthal angle, and rapidity of particle $i$, respectively. 
Two other measures intended for $e^+e^-$ collisions are available in the \texttt{EnergyCorrelator} \fastjet{contrib} \cite{Cacciari:2011ma,fjcontrib}.  The first is a definition based strictly on energies and opening angles,
\begin{align}\label{eq:Eratio}
z_i\equiv\frac{E_{i}}{E_{J}}\,,  \qquad \theta_{ij}^2\equiv \Theta_{ij}^2 \,,
\end{align}
where $E_J$ is the total jet energy, and $\Theta_{ij}$ is the Euclidean angle between the 3-momenta $\vec p_i$ and $\vec p_j$.  There is an alternative definition in terms of energies and Mandelstam invariants,
\begin{align}\label{eq:mandelstamratio}
z_i\equiv\frac{E_{i}}{E_{J}}\,,  \qquad \theta_{ij}^2\equiv\frac{2p_i \cdot p_j}{E_i E_j}  \,,
\end{align}
which reduces to \Eq{eq:Eratio} in the collinear limit but is easier for analytic calculations.

From \Eq{eq:ECFs}, we see that the $n$-point energy correlation functions vanish in the soft and collinear limits, and therefore are natural resolution variables for $(n-1)$-prong substructure.  A number of powerful 2-prong discriminants have been formed from the energy correlation functions  \cite{Larkoski:2013eya,Larkoski:2014gra}, namely
\begin{align}
\Cobs{2}{\beta}=\frac{\ecf{3}{\beta}}{(\ecf{2}{\beta})^{2}}\,,\qquad \Dobs{2}{\beta}=\frac{\ecf{3}{\beta}}{(\ecf{2}{\beta})^{3}}\,, \qquad \Dobs{2}{\alpha,\beta}=\frac{\ecf{3}{\alpha}}{(\ecf{2}{\beta})^{3\alpha/\beta}}\,.
\end{align}
Beyond their discrimination power, these observables have nice analytic properties.  First, since they can be written as a sum over particles in the jet without reference to external axes, they are automatically ``recoil-free'' \cite{Catani:1992jc,Dokshitzer:1998kz,Banfi:2004yd,Larkoski:2013eya,Larkoski:2014uqa}. Second, since they have well-defined behavior in various soft and collinear limits, they are amenable to resummed calculations;  in \Ref{Larkoski:2015kga}, $\Dobsnobeta{2}$ was calculated to next-to-leading-logarithmic (NLL) accuracy in $e^+e^-$ for both signal (boosted $Z$) and background (QCD) jets. 

\begin{figure}
\begin{center}
\subfloat[]{\label{fig:D2_ps_review}
\includegraphics[width=6.65cm]{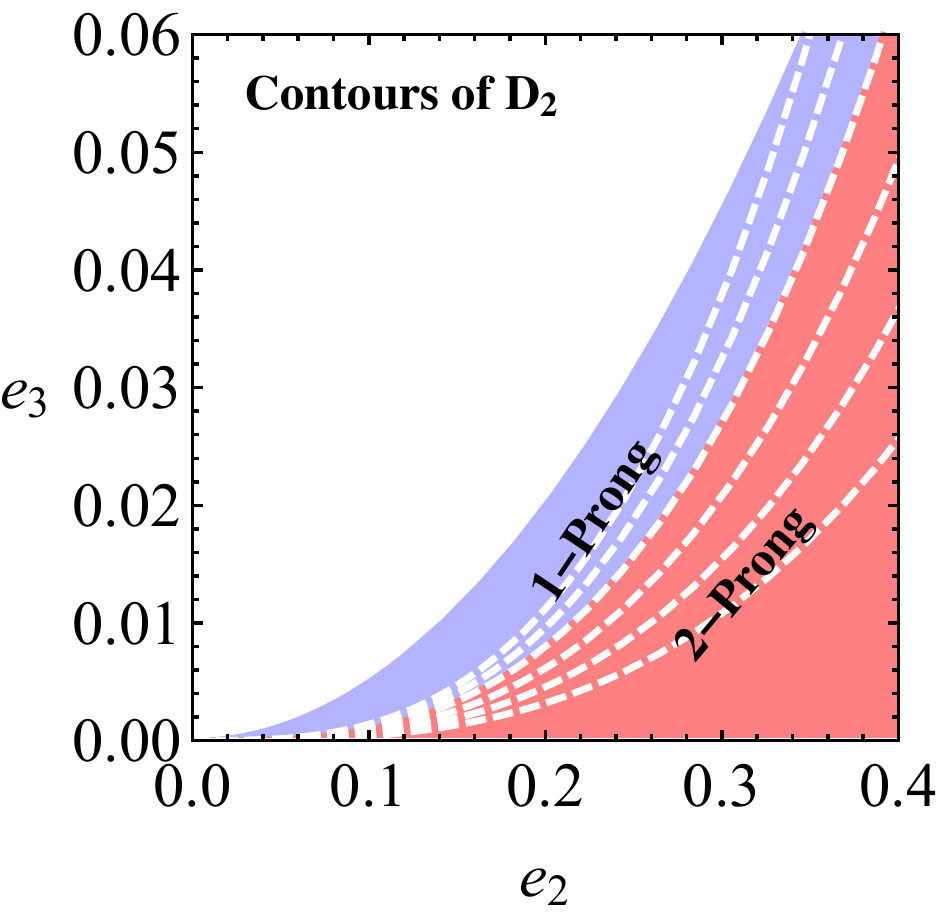}
}\qquad
\subfloat[]{\label{fig:Nsub_ps_review}
\includegraphics[width=6.5cm]{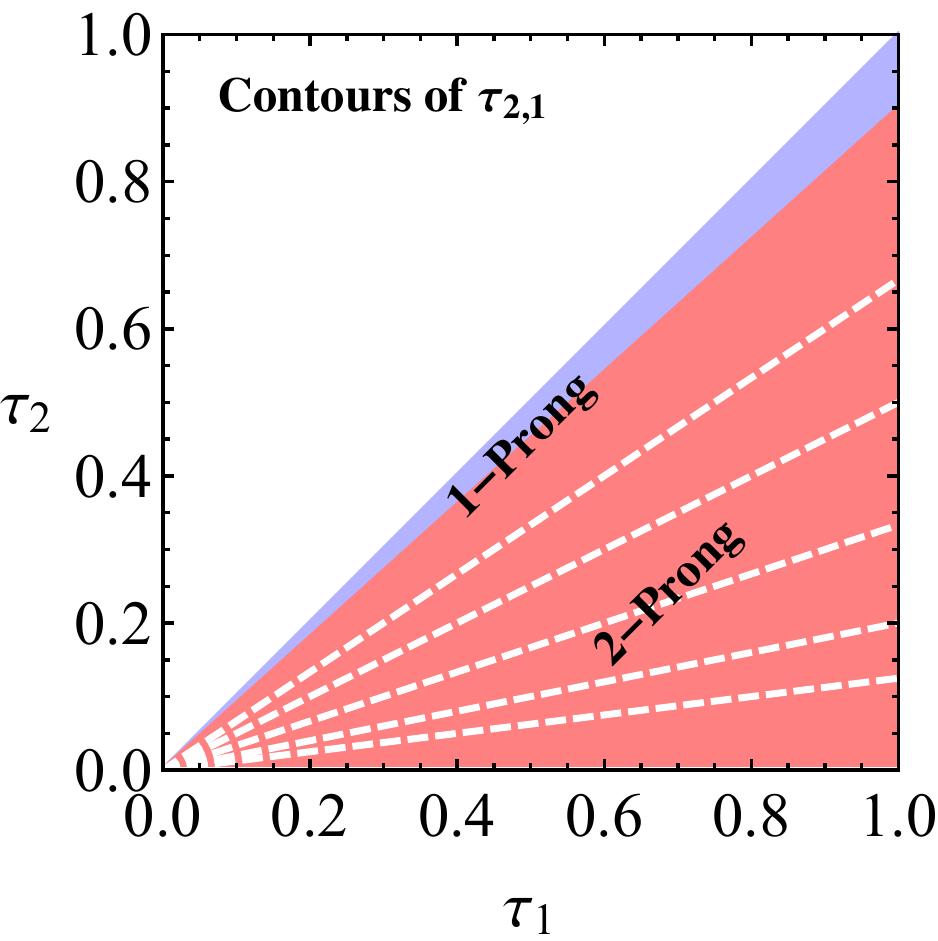}
}
\end{center}
\caption{Schematic depiction of the phase space for (a) the energy correlation functions $\ecfnobeta{2}, \ecfnobeta{3}$ and (b) the $N$-subjettiness observables $\Nsubnobeta{1}, \Nsubnobeta{2}$. In both cases, contours of the relevant ratio observable, $\Dobsnobeta{2}$ or $\Nsubnobeta{2,1}$, are shown as white dashed curves.  These ratios are chosen such that the contours cleanly separate the 1- and 2-prong regions of phase space. 
}
\label{fig:ps_review}
\end{figure}

The basic structure of the $\ecfnobeta{2}$, $\ecfnobeta{3}$ phase space is shown in \Fig{fig:D2_ps_review} and discussed in more detail in \Refs{Larkoski:2014gra,Larkoski:2015kga}. Signal jets which have resolved 2-prong structure live in the region of phase space satisfying $\ecfnobeta{3}\ll (\ecfnobeta{2})^3$, whereas QCD background jets with 1-prong structure live in the phase space region defined by $(\ecfnobeta{2})^3 \ll\ecfnobeta{3}\ll (\ecfnobeta{2})^2$.  The observable $\Dobsnobeta{2}$ is designed to define contours which cleanly separate the 1-prong and 2-prong regions of phase space, and therefore identifies the extent to which a jet is 1- or 2-prong-like.

Observables for boosted top tagging have also been proposed using the energy correlation functions, namely the $C_3$ observable \cite{Larkoski:2013eya}, 
\begin{align}\label{eq:C3_def}
\Cobs{3}{\beta}=\frac{  \ecf{4}{\beta}   \ecf{2}{\beta}   }{ ( \ecf{3}{\beta}  )^2  }\,,
\end{align}
and the $D_3$ observable \cite{Larkoski:2014zma}, 
\begin{align}\label{eq:D3_full_def}
D_3^{(\alpha,\beta,\gamma)}=  \frac{   \ecf{4}{\gamma}   \left ({\ecf{2}{\alpha}}\right)^{\frac{3\gamma}{\alpha}}  }{     \left( \ecf{3}{\beta}\right )^{\frac{3\gamma}{\beta}}     } +x \frac{   \ecf{4}{\gamma}   \left (\ecf{2}{\alpha}\right)^{\frac{2\gamma}{\beta}-1}  }{      \left (\ecf{3}{\beta}\right )^{\frac{2\gamma}{\beta}}    }   +y \frac{   \ecf{4}{\gamma}   \left (\ecf{2}{\alpha}\right)^{\frac{2\beta}{\alpha}-\frac{\gamma}{\alpha}}  }{      \left (\ecf{3}{\beta}\right )^{2}    }\,.
\end{align}
Here, $x$ and $y$ are constants given in \Ref{Larkoski:2014zma} that depend on the jet mass and $p_T$.  The $\Cobsnobeta{3}$ observable does not exhibit particularly good discrimination power, and while $\Dobsnobeta{3}$, which was constructed using the power counting approach, performs well, it has a complicated functional form.  For the boosted top study in \Sec{sec:tops}, we compare to a simplified version of the $\Dobsnobeta{3}$ observable
\begin{align}\label{eq:D3_simp_def}
D_3^{(\beta)}=  \frac{   \ecf{4}{\beta}   \left ({\ecf{2}{\beta}}\right)^{3}  }{     \left( \ecf{3}{\beta}\right )^{3}     } \,,
\end{align}
obtained by setting $x=y=0$, which behaves well on groomed jets. Unlike its more complicated cousin, this simplified $D_3$ has only a single angular exponent.

We also find it interesting to compare our new observables to $N$-subjettiness. The (normalized) $N$-subjettiness observable $\Nsubnobeta{N}$  \cite{Thaler:2010tr,Thaler:2011gf} is defined as\footnote{This observable is based on the global event shape $N$-jettiness \cite{Stewart:2010tn}, which has recently been used to define the \textsc{XCone} jet algorithm \cite{Stewart:2015waa,Thaler:2015xaa}.}
\begin{equation}\label{eq:nsubdef}
\Nsub{N}{\beta} = \sum_{1\leq i \leq n_J} z_{i}\min\left\{
\theta_{i1}^\beta,\dotsc,\theta_{iN}^\beta
\right\} \ .
\end{equation}
Here, the angle $\theta_{iK}$ is measured between particle $i$ and subjet axis $K$ in the jet.  As for the case of the energy correlation functions, a number of different possible measures can be used to define $\theta_{iK}$.  For our LHC studies, we take $\theta_{iK}=R_{iK}$, analogously to \Eq{eq:ptratio}.   

Unlike the energy correlation functions of \Eq{eq:ECFs}, which correlate groups of $n$ particles within the jet, $N$-subjettiness divides a jet into $N$ sectors and correlates the particles in each sector with their corresponding axis.  Thus, implicit in the definition of $N$-subjettiness in \Eq{eq:nsubdef} is the definition of appropriate $N$-subjettiness axes. Different definitions of the axes can lead to different behaviors of the observable, particularly away from the resolved limit \cite{Larkoski:2015uaa}.  A natural definition is to choose the axes that minimize the value of $\Nsubnobeta{N}$ itself \cite{Thaler:2011gf}, as is done for the classic $e^+e^-$ event shape thrust \cite{Farhi:1977sg}.  Exact minimization is computationally challenging, though, so a number of definitions which approximate the minimum are used instead, which are provided in the \texttt{Nsubjettiness} \fastjet{contrib} \cite{Cacciari:2011ma,fjcontrib}.

The relevant $N$-subjettiness ratio observables are
\begin{align}
\Nsub{2,1}{\beta}= \frac{\Nsub{2}{\beta}}{\Nsub{1}{\beta}}, \qquad \Nsub{3,2}{\beta}= \frac{\Nsub{3}{\beta}}{\Nsub{2}{\beta}} \ .
\end{align}
Here, $\Nsubnobeta{2,1}$ is designed to be small when a jet has well-resolved 2-prong substructure, making it useful for boosted $W$/$Z$/$H$ tagging.  Similarly, $\Nsubnobeta{3,2}$ is designed to be small in the 3-prong limit, useful for boosted tops.  The observable $\Nsubnobeta{2,1}$ was calculated in $e^+e^-$ collisions for signal (boosted $Z$) jets at N$^3$LL accuracy \cite{Feige:2012vc}.

The phase space for $\Nsubnobeta{1}$, $\Nsubnobeta{2}$ is shown schematically in \Fig{fig:Nsub_ps_review}, along with contours of constant $\Nsubnobeta{2,1}$. Background QCD jets are defined by the linear scaling $\Nsubnobeta{2}\sim \Nsubnobeta{1}$, whereas signal jets are defined by $\Nsubnobeta{2}\ll \Nsubnobeta{1}$. This phase space structure is different from that of the $e_2$ and $e_3$ observables shown in \Fig{fig:D2_ps_review}, where the phase space for background QCD jets is defined by two boundaries with distinct scalings.  It is this fact which ultimately leads to many of the differences seen between $\Dobsnobeta{2}$ and $\Nsubnobeta{2,1}$, including the fact that the $\Nsubnobeta{2,1}$ distribution is more stable as a function of jet mass and $p_T$.  The phase space for $\Nsubnobeta{3,2}$ is similar to $\Nsubnobeta{2,1}$, to be contrasted with the complicated phase space for $\Dobsnobeta{3}$ \cite{Larkoski:2014zma}.  Using the generalized energy correlation functions, we can define new axes-free observables that mirror the phase space structures of $\Nsubnobeta{2,1}$ and $\Nsubnobeta{3,2}$, thereby exhibiting similar scaling and stability behaviors, particularly for groomed jets.  This will be discussed for $\Nsubnobeta{3,2}$ in \Sec{sec:tops} and for $\Nsubnobeta{2,1}$ in \Sec{sec:2prong}.

\subsection{Soft Drop Grooming}\label{sec:soft_drop}

Two powerful tools which have emerged from the study of jet substructure are groomers \cite{Butterworth:2008iy,Ellis:2009su,Ellis:2009me,Krohn:2009th,Dasgupta:2013via,Dasgupta:2013ihk} and pileup mitigation techniques \cite{Cacciari:2007fd,Alon:2011xb,Soyez:2012hv,Tseng:2013dva,Krohn:2013lba,Cacciari:2014gra,Bertolini:2014bba}, both of which remove soft radiation from a jet.  Groomers have proven to be useful both for removing jet contamination as well as for identifying hard multi-prong substructure within a jet.  In this paper, we use the soft drop \cite{Larkoski:2014wba} groomer with $\beta=0$, which coincides with the modified mass drop procedure \cite{Dasgupta:2013ihk,Dasgupta:2013via} with $\mu = 1$. The soft drop groomer exhibits several theoretical advantages over other groomers; in particular, it removes non-global logarithms \cite{Dasgupta:2001sh} to all orders, and it mitigates the process dependence of jet spectra.  The soft-dropped groomed jet mass has recently been calculated to NNLL accuracy \cite{Frye:2016okc,Frye:2016aiz}.

\begin{figure}
\begin{center}
\subfloat[]{\label{fig:D2_ps_review_groom}
\includegraphics[width=6.65cm]{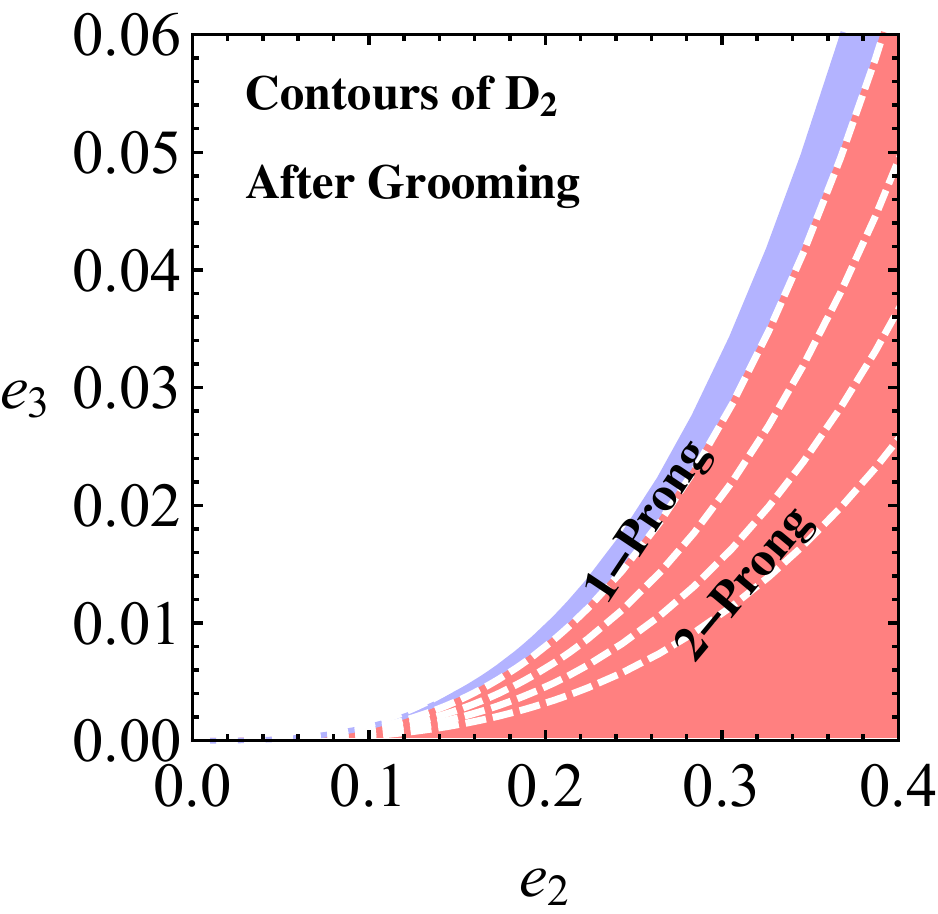}
}\qquad
\subfloat[]{\label{fig:Nsub_ps_review_groom}
\includegraphics[width=6.5cm]{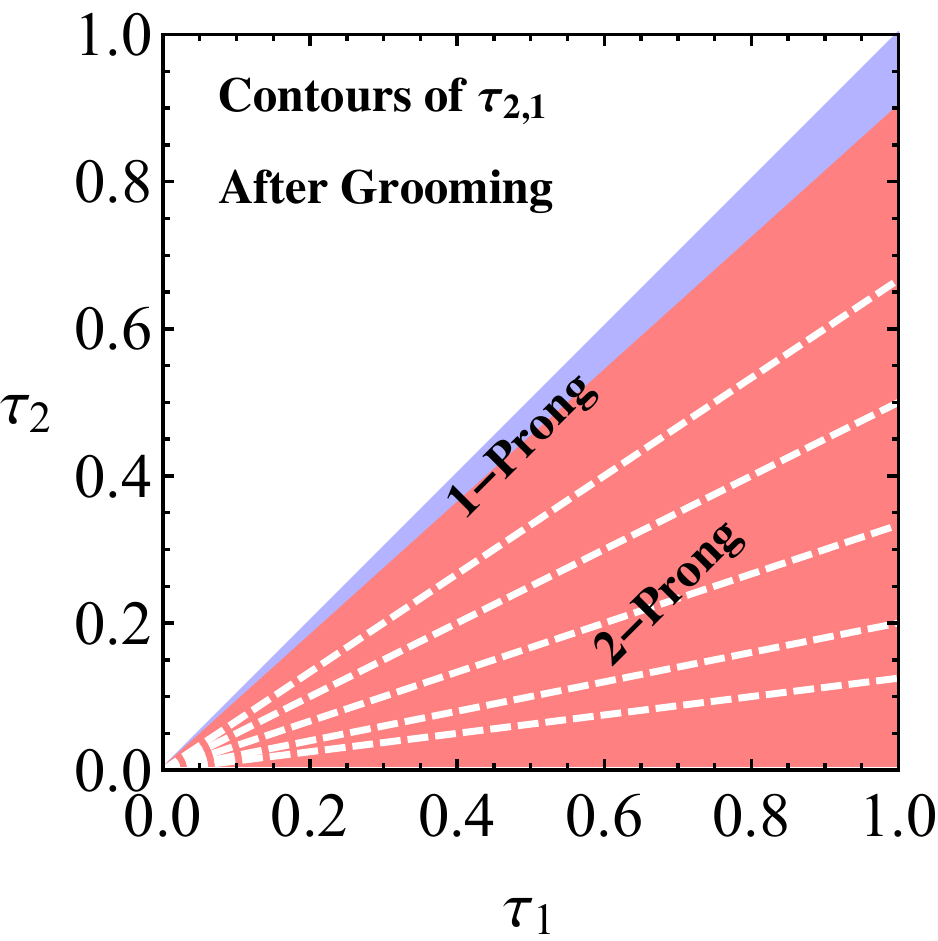}
}
\end{center}
\caption{Same as \Fig{fig:ps_review}, but after applying jet grooming.  The upper-boundary of phase space for $\Dobsnobeta{2}$ is modified by removing soft radiation, while the parametric behavior of $\Nsubnobeta{2,1}$ is unchanged.  This modified phase space for 2-prong discriminants will be discussed in more detail in \Sec{sec:2prong}.}
\label{fig:ps_review_groom}
\end{figure}

Starting from a jet identified with an IRC safe jet algorithm (such as anti-$k_t$ \cite{Cacciari:2008gp}), the soft drop algorithm is defined using Cambridge/Aachen (C/A) reclustering \cite{Dokshitzer:1997in,Wobisch:1998wt,Wobisch:2000dk}.  Specializing to the case of $\beta=0$, the algorithm proceeds as follows:
\begin{enumerate}

\item Recluster the jet using the C/A clustering algorithm, producing an angular-ordered branching history for the jet.

\item Step through the branching history of the reclustered jet.  At each step, check the soft drop condition
\begin{align}\label{eq:sd_cut}
\frac{\min\left[ p_{Ti}, p_{Tj}  \right]}{p_{Ti}+p_{Tj}}> \zcut \,.
\end{align}
Here, $\zcut$ is a parameter defining the scale below which soft radiation is removed.  If the soft drop condition is not satisfied, then the softer of the two branches is removed from the jet.  This process is then iterated on the harder branch.

\item The soft drop procedure terminates once the soft drop condition is satisfied.

\end{enumerate}
Given a jet that has been groomed with the soft drop procedure, we can then measure any IRC safe observable on this jet and it will remain IRC safe.  As we will see, because soft drop removes soft radiation from a jet, power-counting arguments for groomed jets can be dramatically different than those for ungroomed jets.  This is previewed in \Fig{fig:ps_review_groom}, where the phase space for $\Dobsnobeta{2}$ is substantially modified by the removal of soft radiation.  

More general groomers are expected to give rise to similar power-counting modifications.  For example, the soft drop condition in \Eq{eq:sd_cut} can be generalized to include an angular weighting exponent $\beta$, which controls the aggressiveness of the groomer, and we expect deviations away from our default of $\beta = 0$ to yield similar behavior, so long as the groomer continues to remove parametrically soft particles.  We also expect that other groomers such as trimming \cite{Krohn:2009th}, which is used heavily by the ATLAS experiment, will behave similarly for the same value of $\zcut$.  We leave a detailed study of other groomers to future work. 

\subsection{Power Counting the Soft/Collinear Behavior}\label{sec:power_counting}

An efficient approach for studying jet substructure is power counting \cite{Larkoski:2014gra}, which allows one to determine the parametric scaling of observables. This parametric behavior is determined by the soft and collinear limits of QCD and is robust to hadronization or modeling in parton shower generators.  Here, we briefly review the salient features of power counting, using the 2-point energy correlator as an example.  We refer readers interested in a more detailed discussion to the original paper.

High-energy QCD jets are dominated by soft and collinear radiation, a language which will be used frequently throughout this paper.  Since QCD is approximately conformal, there is no intrinsic energy or angular scale associated with this radiation.\footnote{We ignore the scale $\Lambda_{\text{QCD}}$ for this discussion, focusing on regions of phase space dominated by perturbative dynamics. While $\Lambda_{\text{QCD}}$ plays an important role in certain phase space regions, for IRC safe observables it contributes only a power-suppressed contribution away from singular limits.}  By applying a measurement to a jet, though, one introduces a scale, which then determines the scaling of soft and collinear radiation.  The simple observation that all scales are set by the measurement itself allows for a powerful understanding of the jet's energy and angular structure.  Arguments along these lines are ubiquitous in the effective field theory (EFT) community.  For example, in Soft Collinear Effective Theory (SCET) \cite{Bauer:2000ew,Bauer:2000yr,Bauer:2001ct,Bauer:2001yt}, they are used to identify the appropriate EFT modes required to describe a particular set of measurements. 

\begin{figure}
\begin{center}
\subfloat[]{\label{fig:unresolved}
\includegraphics[width=6cm]{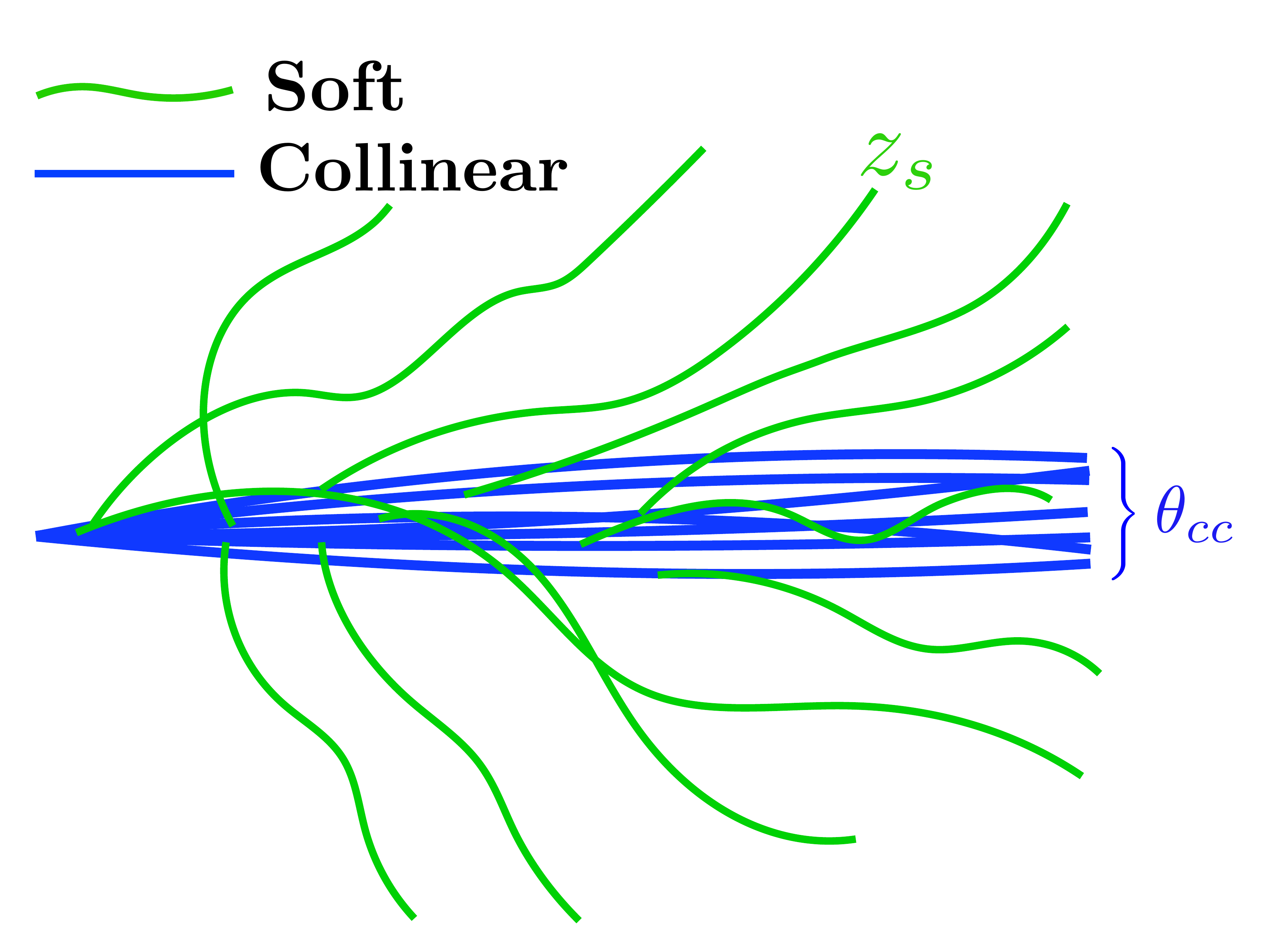}    
}\qquad
\subfloat[]{\label{fig:NINJA}
\includegraphics[width=6cm]{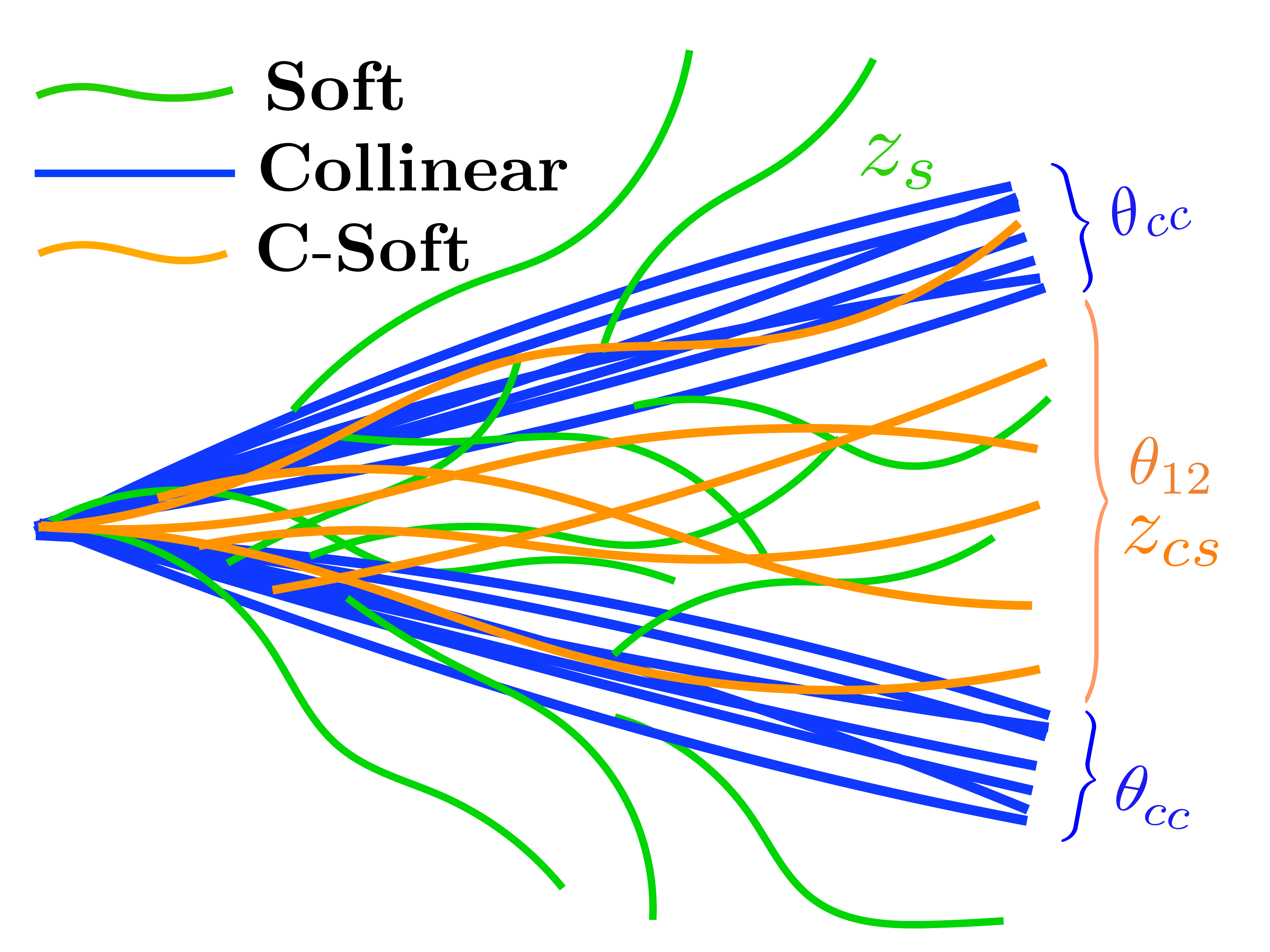}
}
\end{center}
\caption{(a) Schematic of a 1-prong jet, showing the dominant soft (green) and collinear (blue) radiation, as well as the characteristic scales $z_s$ and $\theta_{cc}$.  (b) Schematic of a 2-prong jet, showing the dominant soft (green), collinear (blue), and collinear-soft (orange) radiation, as well as the characteristic scales, $z_s$, $\theta_{cc}$, $z_{cs}$, and $\theta_{12}$. 
}
\label{fig:pics_jets}
\end{figure}

In the context of power counting, soft and collinear emissions are defined by their parametric scalings.  A soft emission, denoted by $s$, is defined by
\begin{equation}
z_s \ll1 \ , \qquad \theta_{sx}\sim 1 \,.
\end{equation}
Here, $z_s$ is the momentum fraction, as defined in \Eq{eq:ptratio}, and $\theta_{sx}$ is the angle to any other particle $x$ in the jet, including other soft particles.  The scaling $\theta_{sx}\sim 1$ means that $\theta_{sx}$ is not assigned any parametric scaling associated with the measurement.  A collinear emission, denoted by $c$, is defined by
\begin{equation}
z_c\sim1 \ , \qquad \theta_{cc}\ll 1\,, \qquad \theta_{cs}\sim 1\, .
\end{equation}
Here, $\theta_{cc}$ is the angle between two collinear particles, while $\theta_{cs}$  is the angle between a collinear particle and a soft particle. In an EFT context, overlaps between soft and collinear regions are systematically removed using the zero-bin procedure \cite{Manohar:2006nz}, but this is not relevant for the arguments here.  The soft and collinear modes are illustrated in \Fig{fig:unresolved} and their scalings are summaried in \Tab{tab:unresolved}.

\begin{table}[t]
\begin{center}
\subfloat[]{
\label{tab:unresolved}
\begin{tabular}{ccc}
\hline
\hline
Mode & Energy  & Angle \\ 
\hline
soft& $z_s$  & $1$\\
collinear & 1 & $\theta_{cc}$ \\
\hline
\hline
\end{tabular}
}
$\qquad$
\subfloat[]{
\label{tab:NINJA}
\begin{tabular}{ccc}
\hline
\hline
Mode & Energy  & Angle \\ 
\hline
soft& $z_s$  & $1$\\
collinear & 1 & $\theta_{cc}$ \\
c-soft & $z_{cs}$ &$\theta_{12}$ \\
\hline
\hline
\end{tabular}
}
\end{center}
\caption{Summary of the modes in \Fig{fig:pics_jets}:  (a) 1-prong jets versus (b) 2-prong jets. 
}
\label{tab:pc_jets}
\end{table}

We now use the simple example of $\ecfnobeta{2}$ to demonstrate how an applied measurement sets the scaling of soft and collinear radiation.\footnote{In this analysis, we do not consider the scale set by the jet radius, $R$. For $R\ll1$, the jet radius must also be considered in the power counting and the scale $R$ appears in perturbative calculations.  For recent work on the resummation of logarithms associated with this scale, see \Refs{Dasgupta:2014yra,Chien:2015cka,Kolodrubetz:2016dzb,Kang:2016mcy}.} The analysis of more general observables proceeds analogously.  Repeating \Eq{eq:ECFs} for convenience, the 2-point energy correlation function is
\begin{equation}\label{eq:2pt_ex}
\ecf{2}{\beta} = \sum_{1\leq i<j\leq n_J} z_i  z_j \, \theta_{ij}^\beta\,.
\end{equation}
If we only consider regions of phase space where $\ecfnobeta{2}\ll1$, such that we have a well-defined collimated jet, all particles in the jet either have small $z_i$ or small $\theta_{ij}$.  In this phase space region, the observable is indeed dominated by soft and collinear emissions. 

To determine the scaling of $z_s$ and $\theta_{cc}$ in terms of the observable, we can consider the different possible contributions to $\ecfnobeta{2}$: soft-soft correlations, soft-collinear correlations, and collinear-collinear correlations.  Parametrically, $\ecfnobeta{2}$ can therefore be written as
\begin{equation}
\ecf{2}{\beta} \sim \sum_{s} z_{s}z_{s} \, \theta_{ss}^\beta + \sum_{s,c} z_{s}z_{c} \, \theta_{cs}^\beta + \sum_{c} z_{c}z_{c} \, \theta_{cc}^\beta  \,.
\end{equation}
Expanding this result to leading order in $z_s$ and $\theta_{cc}$, we find
\begin{equation}
\ecf{2}{\beta} \sim \sum_{s}  z_s +\sum_{c}  \theta_{cc}^\beta  \,.
\end{equation}
For simplicity, we drop the summation symbol, writing
\begin{equation}\label{eq:fact_e2}
\ecf{2}{\beta} \sim  z_s + \theta_{cc}^\beta  \,.
\end{equation}

Since we have only measured a single observable, $\ecfnobeta{2}$, it sets the only scale in the jet, and there is no measurement to further distinguish the scalings of soft and collinear particles. We therefore find the scaling of $z_s$ and $\theta_{cc}$ in terms of the observable, 
\begin{equation}
z_s \sim \ecf{2}{\beta}\,, \qquad   \theta_{cc} \sim \left(\ecf{2}{\beta}\right)^{1/\beta}   \,.
\end{equation}
More generally, after identifying all parametrically different modes that can contribute to a set of measurements, the scaling of those modes is determined by the measured observables.

In this paper, we are interested not only in jets with soft and collinear radiation, but also in jets which have well-resolved substructure.  In addition to the strictly soft and collinear modes which are found in \Fig{fig:unresolved}, a jet with well-resolved substructure also includes radiation emitted from the dipoles within the jet, shown in orange for the particular case of a 2-prong jet in \Fig{fig:NINJA}.  This radiation is referred to as ``collinear-soft" (or just ``c-soft'') as it has a characteristic angle $\theta_{12}$ defined by the opening angle of the subjets, as well as a momentum fraction $z_{cs}\ll1$, both of which are set by the measurement. The appropriate EFT description for multi-prong substructure is referred to as SCET$_+$ \cite{Bauer:2011uc,Procura:2014cba,Larkoski:2015zka,Pietrulewicz:2016nwo}, and the scaling of the collinear-soft mode is summarized in \Tab{tab:NINJA}.     Using the mode structure of multi-prong jets, it is straightforward to apply power-counting arguments to a wide variety of $n$-prong jet substructure observables, as demonstrated in \Secs{sec:tops}{sec:2prong}.

We also apply power-counting arguments to groomed jets after soft drop has been applied.   The effect of the grooming algorithm is not just to remove jet contamination, but also to modify the power counting in interesting, and potentially useful, ways.   As discussed in \Sec{sec:soft_drop}, soft drop with $\beta=0$ is defined with a single parameter $\zcut$, which determines the scale below which soft radiation is removed.  To perform a proper power-counting analysis, one should also incorporate the scale $\zcut$ and consider different cases depending on the relative scaling of $\zcut$ and $z_s$.  For simplicity, we ignore this complication through most of this paper and assume that the soft drop procedure simply removes the soft modes.  That said, the residual soft scaling will matter for the quark/gluon study in \Sec{sec:qvsg}.  For a more detailed discussion, and a proper treatment of the scale $\zcut$ involving collinear-soft modes, see \Refs{Frye:2016okc,Frye:2016aiz}.
 
\section{Enlarging the Basis of Jet Substructure Observables}\label{sec:defn}

An important goal of jet substructure is to design observables that efficiently identify particular features within a jet.  A popular, and theoretically well-motivated, approach is to construct observables from combinations, often ratios, of IRC safe jet shapes.\footnote{These ratios are not themselves IRC safe, but are instead Sudakov safe \cite{Larkoski:2013paa,Larkoski:2015lea}. For a discussion of Sudakov safety for the case of $\Dobsnobeta{2}$, see \Ref{Larkoski:2015kga}.  For this reason, the ratio observables we construct in this paper cannot be written in the form of \Eq{eq:gen_obs}, even though their $\ecfvar{v}{n}{\beta}$ ingredients can.}  Such observables are widely employed at the LHC, and have proven to be both experimentally useful and theoretically tractable. Indeed, the observables reviewed in \Sec{sec:ecfs}---$\Nsubnobeta{2,1}$, $\Nsubnobeta{3,2}$, $\Cobsnobeta{2}$, and $\Dobsnobeta{2}$---are all of this form.

Essential to this approach is a flexible basis of IRC safe observables from which to build discriminants.  While the original energy correlators are indeed a useful basis, they are still somewhat restrictive.  For example, the phase space structure of $e_2$ and $e_3$ in \Fig{fig:D2_ps_review} is completely fixed, as are all of the parametric properties inherited from this structure, such that $D_2$ is the only combination that parametrically distinguishes 1- and 2-prong substructure.  

In this section, we enlarge the basis of jet substructure observables by defining generalizations of the energy correlation functions, allowing for a more general angular dependence than considered in \Eq{eq:ECFs}.  These new observables are flexible building blocks, which we use in the rest of this paper to identify promising tagging observables using power-counting techniques.\footnote{An alternative approach to identifying specific features within jets is machine learning, which has seen significant recent interest \cite{Cogan:2014oua,deOliveira:2015xxd,Almeida:2015jua,Baldi:2016fql,Guest:2016iqz,Conway:2016caq,Barnard:2016qma}.  The contrast between these strategies has been dubbed ``deep thinking'' versus ``deep learning".  In the deep thinking approach pursued here, the goal is to identify the physics principles that lead to discrimination power, focusing on observables with desirable properties for first-principles calculations.  In the deep learning approach, the goal is to use reliable training samples to optimize the discrimination power and, in many cases, visualize the underlying physics.  Ultimately, one would want to merge these two approaches, which could help avoid theoretical blindspots in the cataloging of observables and mitigate modeling uncertainties inherent in training samples. Detailed studies in data, ideally with high purity samples, will also be needed for a complete understanding.}

\subsection{General Structure of Infrared/Collinear Safe Observables}\label{sec:gen_IRC}

\begin{figure}
\begin{center}
\includegraphics[width=6.5cm]{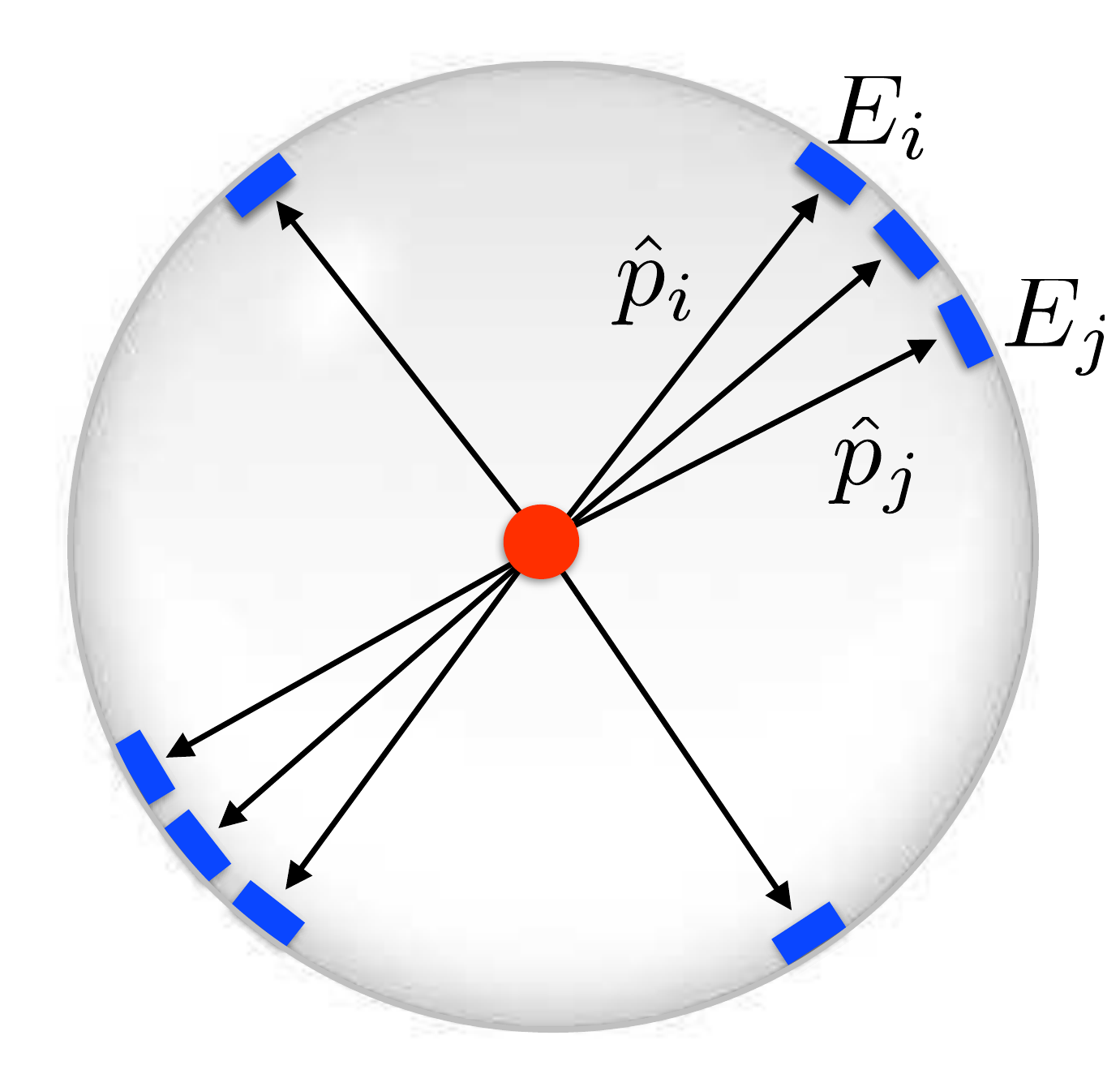}
\end{center}
\caption{Schematic depiction of a hard scattering event.   A general IRC safe observable can be constructed by summing over all energy deposits, $E_i$, in an event, with a symmetric angular weighting function depending on the dimensionless unit vectors $\hat p_i$.
}
\label{fig:sphere_observable}
\end{figure}

In order to engineer the phase space structure of observables to have specific properties, we first need to systematically understand the structure of IRC safe observables that probe $n$-particle correlations. The general structure of an IRC safe observable is shown schematically in \Fig{fig:sphere_observable}, where any IRC safe observable can be constructed from the energy deposits and angular information on the sphere.  In the $pp$ case, of course, one typically uses the longitudinally boost-invariant quantities $p_T$ and $R_{ij}$, but the following argument is insensitive to that coordinate change.

As shown in \Ref{Tkachov:1995kk,Sveshnikov:1995vi,Cherzor:1997ak,Tkachov:1999py}, any IRC safe observable can be constructed from the following (complete) basis of observables\footnote{With a completely generic angular weighting function, $f_N$, this basis is of course overcomplete.} 
\begin{align}
\label{eq:gen_obs}
F_N(\{p_i\})=\sum_{i_1} \sum_{i_2} \ldots \sum_{i_N} E_{i_1} E_{i_2} \ldots E_{i_N} \, f_N( \hat p_{i_1}, \hat p_{i_2}, \ldots,  \hat p_{i_N} )\,,
\end{align}
where $E_i$ is the energy of particle $i$, $\hat p_i$ is a dimensionless unit vector describing its direction, and $f_N$ is a symmetric function of its arguments.  For IRC safety, we must further demand that the function $f_N$ vanishes when any two particles become collinear.   Note that \Eq{eq:gen_obs} is a linear function of the momenta of the particles and a symmetric function of the angles.  This basis of observables are referred to in the literature as $C$-correlators  \cite{Tkachov:1995kk,Sveshnikov:1995vi,Cherzor:1997ak,Tkachov:1999py}.  

Since the above discussion is completely general, it is not immediately obvious that it is useful for jet substructure studies.  Still, \Eq{eq:gen_obs} has the interesting feature that, while the dependence on the energies is fixed by IRC safety, the angular function $f_N$ is much less restricted and can be chosen for specific purposes.  The original energy correlators in \Eq{eq:ECFs} are a specific case of \Eq{eq:gen_obs}, where, up to an overall normalization, the angular weighting function is 
\begin{align}
\label{eq:eNangles}
\ecf{N}{\beta}: \qquad f_N( \hat p_{i_1}, \hat p_{i_2}, \ldots,  \hat p_{i_N} )=\prod_{s < t \in \{i_1, i_2 , \dots, i_n \}} \theta_{st}^\beta\,.
\end{align}
The key observation is that by considering alternative angular weighting functions for $n$-point correlators beyond \Eq{eq:eNangles}, we can define a more flexible basis of observables for  jet substructure studies.

\subsection{New Angles on Energy Correlation Functions: $\ecfvar{v}{n}{\beta}$}
\label{sec:gen_ecf}

There are many known decompositions of the angular function $f_N$---including Fox-Wolfram moments \cite{Fox:1978vu,Fox:1978vw} and orthogonal polynomials on the sphere \cite{GurAri:2011vx}---but these are not necessarily optimal for jet substructure.  The reason is that jets with well-resolved subjets exhibit a hierarchy of distinct angular scales, so we need to design $f_N$ to identify hierarchical---instead of averaged---features within a jet.

As seen in \Eq{eq:eNangles}, the original energy correlation functions do capture multiple angular scales, but they do so all at once; it would be preferable if $f_N$ could identify one angular scale at a time in order to isolate different physics effects.  Furthermore, to make power-counting arguments more transparent, we want $f_N$ to exhibit homogeneous angular scaling, such that each term in \Eq{eq:gen_obs} has a well-defined scaling behavior without having to perform a non-trivial expansion in the soft and collinear limits.

With these criteria in mind, we can now translate the general language of IRC safe observables into a useful basis for jet substructure studies.  The angular function $f_N$ has to be symmetric in its arguments, and the simplest symmetric function that preserves homogeneous scaling is the $\min$ function.\footnote{The appearance of $\min$ can also be viewed as the lowest-order Taylor expansion of a more generic observable, which should be a good approximation in the case of small radius jets.  This can be seen explicitly in \App{app:alt}, where different functional forms are compared that give the same quantitative behavior as the $\min$ version here.  Another motivation for the $\min$ definition is that it naively behaves more similarly to thrust \cite{Farhi:1977sg} or $N$-jettiness \cite{Stewart:2010tn}, though we emphasize that $\ecfvarnobeta{v}{n}$ does not rely on external axes.}  This leads us to the generalized energy correlation functions, which depend on $n$ factors of the particle energies and $v$ factors of their pairwise angles,
\begin{equation}\label{eq:ecf_gen}
\ecfvar{v}{n}{\beta} = \sum_{1 \leq i_1 < i_2 < \dots < i_n \leq n_J} z_{i_1} z_{i_2} \dots z_{i_n} \prod_{m = 1}^{v} \min^{(m)}_{s < t \in \{i_1, i_2 , \dots, i_n \}} \left\{ \theta_{st}^{\beta} \right\},
\end{equation}
where $\min^{(m)}$ denotes the $m$-th smallest element in the list.  For a jet consisting of fewer than $n$ particles, $\ecfvarnobeta{v}{n}$ is defined to be zero.  More explicitly, the three arguments of the generalized energy correlation functions are as follows.
\begin{itemize}
\item The subscript $n$, appearing to the right of the observable, denotes the number of particles to be correlated.  This plays the same role as the $n$ subscript for the standard $e_n$ energy correlators in \Eq{eq:ECFs}. 
\item The subscript $v$, appearing to the left of the observable, denotes the number of pairwise angles entering the product.  By definition, we take $v \leq \binom{n}{2}$, and the minimum then isolates the product of the $v$ smallest pairwise angles.
\item The angular exponent $\beta>0$ can be used to adjust the weighting of the pairwise angles, as in \Eq{eq:ECFs}.
\end{itemize}
For the special case of $v = \binom{n}{2}$, the generalized energy correlators reduce to the standard ones in \Eq{eq:ECFs}, with $\ecfvarnobeta{1}{2}\equiv\ecfnobeta{2}$, $\ecfvarnobeta{3}{3}\equiv\ecfnobeta{3}$, $\ecfvarnobeta{6}{4}\equiv\ecfnobeta{4}$, and so on for the higher-point correlators. 

Compared to the original energy correlators, the generalization in \Eq{eq:ecf_gen} allows more flexibility in the angular scaling; this simplifies the construction of useful ratios and extends the possible applications of energy correlators.  In the case of boosted top tagging, for example, the standard $\ecfnobeta{4}=\ecfvarnobeta{6}{4}$ observable involves six different pairwise angles.  A decaying boosted top quark, however, does not have six characteristic angular scales, so most of these angles are redundant and only serve to complicate the structure of the observable.  This is reflected in the definition of $\Dobsnobeta{3}$ in \Eq{eq:D3_full_def}, which involves three distinct terms \cite{Larkoski:2014zma}.

To make more explicit the definition in \Eq{eq:ecf_gen}, we summarize the particular correlators used in our case studies below.  For boosted 2-prong tagging in \Sec{sec:2prong}, we use the 2-point energy correlation function
\begin{align}\label{eq:explicit_twopointvar}
\ecfvar{1}{2}{\beta}&\equiv\ecf{2}{\beta}=\sum_{1\leq i<j\leq n_J} z_{i}z_{j} \, \theta_{ij}^\beta\ ,
\end{align}
whose definition is unique, since it only involves only a single pairwise angle.  We also need the 3-point correlators, which have three variants probing different angular structures:\footnote{For $\ecfvarnobeta{2}{3}$, note that $\min\{{a,b,c}\} \times \min^{(2)} \{a,b,c\} = \min\{ab,ac,bc\}$.}
\begin{align}\label{eq:explicit_ecfvar}
\ecfvar{1}{3}{\beta}&=\sum_{1\leq i<j<k\leq n_J} z_{i}z_{j}z_{k} \min \left\{ \theta_{ij}^\beta\,,  \theta_{ik}^\beta\,, \theta_{jk}^\beta  \right\} \ , \nonumber \\
\ecfvar{2}{3}{\beta}&=\sum_{1\leq i<j<k\leq n_J} z_{i}z_{j}z_{k} \min \left\{\theta_{ij}^\beta \theta_{ik}^\beta\,, \theta_{ij}^\beta  \theta_{jk}^\beta\,,     \theta_{ik}^\beta \theta_{jk}^\beta    \right\}  \ , \nonumber \\
\ecf{3}{\beta}\equiv\ecfvar{3}{3}{\beta}&=\sum_{1\leq i<j<k\leq n_J} z_{i}z_{j}z_{k} \, \theta_{ij}^\beta \theta_{ik}^\beta \theta_{jk}^\beta \,.
\end{align}
Interestingly, we are able to construct powerful observables from each of these three 3-point correlators, resulting in different tagging properties.

For boosted top tagging in \Sec{sec:tops}, we also need the 4-point correlators.  There are six possible variants, but we only study three of them in the body of the text:
\begin{align}
\ecfvar{1}{4}{\beta} &= \sum_{1\leq i<j<k<\ell\leq n_J} z_{i}z_{j}z_{k}z_{\ell}   \min \left\{ \theta_{ij}^\beta, \theta_{ik}^\beta, \theta_{jk}^\beta, \theta_{i\ell}^\beta, \theta_{j\ell}^\beta, \theta_{k\ell}^\beta \right\} \ , \nonumber\\
\ecfvar{2}{4}{\beta} &= \sum_{1\leq i<j<k<\ell\leq n_J} z_{i}z_{j}z_{k}z_{\ell}  \min \left\{ \theta_{ij}^\beta, \theta_{ik}^\beta, \theta_{jk}^\beta, \theta_{i\ell}^\beta, \theta_{j\ell}^\beta, \theta_{k\ell}^\beta \right\} \nonumber\\
& \hspace{4.5cm} \times \min^{(2)} \left\{ \theta_{ij}^\beta, \theta_{ik}^\beta, \theta_{jk}^\beta, \theta_{i\ell}^\beta, \theta_{j\ell}^\beta, \theta_{k\ell}^\beta \right\}  \ , \nonumber\\
&\hspace{0.2cm}\vdots \nonumber \\
 \ecf{4}{\beta}\equiv\ecfvar{6}{4}{\beta} &= \sum_{1\leq i<j<k<\ell\leq n_J} z_{i}z_{j}z_{k}z_{\ell} \theta_{ij}^\beta \theta_{ik}^\beta \theta_{jk}^\beta  \theta_{i\ell}^\beta \theta_{j\ell}^\beta \theta_{k\ell}^\beta \,,
 \label{eq:4pointcases}
\end{align}
where $\min^{(2)}$ is again the second smallest element in the list.  Here, we see the simplicity in the angular structure of $\ecfvarnobeta{1}{4}$ and $\ecfvarnobeta{2}{4}$, as compared to $\ecfvarnobeta{6}{4}$ which involves all six angles. The vertical dots denote other 4-point correlation functions; we have not found them to be particularly useful, but they might have applications in (and beyond) jet substructure. 

When constructing jet substructure observables, it is often desirable to work with ratios that are approximately boost invariant.  Since the different generalized correlators probe a different number of energy fractions and pairwise angles, each scales differently under Lorentz boosts.  Under a boost $\gamma$ along the jet axis and assuming a narrow jet, the energies and angles scale as
\begin{align}
z_i \to z_i \,, \qquad \theta_{ij} \to \gamma^{-1} \theta_{ij}\,.
\end{align}
This implies that the transformation of $\ecfvarnobeta{v}{n}$ under boosts along the jet axis is determined solely by the $v$ index, 
\begin{align}
\label{eq:boost}
\ecfvar{v}{n}{\beta}\to \gamma^{-v\beta} \ecfvar{v}{n}{\beta}\,.
\end{align}
Therefore, another way of interpreting the different $\ecfvarnobeta{v}{n}$ is as ways of probing $n$ particle correlations with different properties under Lorentz boosts.   The $v$ index therefore broadens the set of boost-invariant combinations that can be formed.

Finally, we remark that the definition in \Eq{eq:ecf_gen} is certainly not unique, and we explore a few alternative definitions in \App{app:alt} that reduce to the $\min$ function in collinear limits.  To further generalize \Eq{eq:ecf_gen} while maintaining homogeneous scaling, one could use different angular exponents depending on the ordering of the angles.  For the cases that we consider, though, we find that $\ecfvarnobeta{v}{n}$ is sufficiently general to provide excellent performance while keeping the form of the observable (relatively) simple.  That said, we expect alternative $f_N$ functions to also be useful, and their performance could be studied using the same power-counting techniques pursued here.

\subsection{New Substructure Discriminants}\label{sec:mandn_series}

Our case studies are based primarily on three series of observables formed from the generalized correlators. We summarize their definitions here, and study  their discrimination power in the forthcoming sections using both power-counting arguments and parton shower generators. 

\subsubsection{The $M_i$ Series}\label{sec:m_series}

The $M_i$ series of observables is defined as
\begin{align}
\Mobs{i}{\beta}=\frac{\ecfvar{1}{i+1}{\beta}}{\ecfvar{1}{i}{\beta}}\,.
\end{align}
This observable is dimensionless, being formed as a ratio of dimensionless observables.  As can be seen from \Eq{eq:boost}, it is also invariant to boosts along the jet axis, since one angular factor appears in both the numerator and denominator.

These observables are constructed to identify $i$ hard prongs, but due to their limited angular structure, they are only effective when acting on suitably groomed jets.  The main example of the $M_i$ series that we will consider explicitly in this paper is
\begin{align}
\Mobs{2}{\beta}=\frac{\ecfvar{1}{3}{\beta}}{\ecfvar{1}{2}{\beta}}\,,
\end{align}
which provides an example of a 2-prong substructure observable that only performs well after grooming. In \App{app:add_plots_M3}, we briefly discuss the behavior of $\Mobsnobeta{3}$ for boosted top tagging, where we argue that a more aggressive grooming strategy would be needed to make $\Mobsnobeta{3}$ perform well. 

\subsubsection{The $N_i$ Series}\label{sec:n_series}

We also define the $N_i$ series of observables  as
\begin{equation}
N_i^{(\beta)} = \frac{\ecfvar{2}{i+1}{\beta}}{(\ecfvar{1}{i}{\beta})^2}.
\end{equation}
As with the $M_i$ series, the $N_i$ series is dimensionless, and from \Eq{eq:boost}, it is boost invariant, as two angular factors appear in both the numerator and denominator.  Indeed, the fact that the 2-point correlation function appears squared in the denominator is fixed by boost invariance.

Two particular examples we find useful for this paper are
\begin{align}
\Nobsnobeta{2}=\frac{ \ecfvar{2}{3}{\beta}  }{   ( \ecfvar{1}{2}{\beta}   )^2}\,,
\end{align}
which is a powerful boosted $W/Z/H$ tagger, and
\begin{align}
\Nobsnobeta{3}=\frac{ \ecfvar{2}{4}{\beta}  }{   ( \ecfvar{1}{3}{\beta}   )^2}\,,
\end{align}
which is a powerful boosted top tagger on groomed jets.  More generally, $N_i$ should be effective as an $i$-prong tagger, as discussed in \App{app:Nsub_Ni}, at least for groomed jets.

The $N_i$ observables take their name from the fact that in the limit of a resolved jet, they behave parametrically like the $N$-subjettiness ratio observables, as discussed in \Secs{sec:tops}{sec:2prong}. Despite their similarity to $N$-subjettiness, the $N_i$ observables achieve their discrimination power in a substantially different manner, which has both theoretical and experimental advantages.

\subsubsection{The $\Uobsnobeta{i}$ Series}\label{sec:qvsg_series}

Finally, we consider the $\Uobsnobeta{i}$ series of observables defined as
\begin{align}
\Uobs{i}{\beta}=\ecfvar{1}{i+1}{\beta}\,,
\end{align}
which are designed for quark/gluon discrimination. Note that unlike $\Mobsnobeta{i}$ and $\Nobsnobeta{i}$, the $\Uobsnobeta{i}$ observables are not boost invariant. For the case $i=1$, $U_1$ coincides with the usual quark/gluon discriminants formed from the energy correlation functions \cite{Larkoski:2013eya}, namely
\begin{align}
\Uobs{1}{\beta}=\Cobs{1}{\beta}=\ecfvar{1}{2}{\beta}=\ecf{2}{\beta}\,,
\end{align}
which probe single soft particle correlations within the jet. For $i>1$, the $\Uobsnobeta{i}$ observables probe multi-particle correlations within the jet in a specific way that is useful for quark/gluon discrimination.

\section{Simplifying Observables for Boosted Top Tagging}\label{sec:tops}

Boosted top tagging has achieved significant attention at the LHC, with a large number of proposed observables to distinguish 3-prong hadronic top jets from the QCD background \cite{Kaplan:2008ie,Thaler:2008ju,Almeida:2008yp,Almeida:2008tp,Plehn:2009rk,Plehn:2010st,Almeida:2010pa,Thaler:2010tr,Thaler:2011gf,Jankowiak:2011qa,Soper:2012pb,Larkoski:2013eya,Anders:2013oga,Freytsis:2014hpa,Larkoski:2015yqa,Kasieczka:2015jma,Lapsien:2016zor}.  In addition to $b$-tagging one of the subjets \cite{ATLAS-CONF-2012-100,CMS:2013vea,ATL-PHYS-PUB-2014-014,ATL-PHYS-PUB-2015-035,CMS-PAS-BTV-15-001,CMS-PAS-BTV-15-002,ATLAS-CONF-2016-001,ATLAS-CONF-2016-002} and requiring the (groomed) jet mass to be close to $m_t \simeq 172~\GeV$, two of the most effective tagging observables are shower deconstruction \cite{Soper:2011cr,Soper:2012pb,Soper:2014rya} and the $N$-subjettiness ratio $\Nsubnobeta{3,2}$ \cite{Thaler:2010tr,Thaler:2011gf}. Shower deconstruction works by testing the compatibility of a QCD shower model with the observed shower pattern; it is an extremely powerful discriminant, particularly at lower efficiencies. Jet shapes like $N$-subjettiness are also powerful discriminants, particularly at higher efficiencies.  For a detailed discussion and experimental study, see, for example, \Refs{Aad:2016pux,CMS-PAS-JME-15-002}.  

In this section, we use the generalized energy correlation functions to construct $N_3$, a simple but powerful boosted top tagger designed for use on groomed jets.  Unlike $\Nsubnobeta{3,2}$, $N_3$ is defined without reference to external axes, allowing it to achieve better background rejection at high signal efficiencies.  Interestingly, in the limit of well-resolved subjets and acting on groomed jets, $N_3$ has identical power counting to $N$-subjettiness.  The behavior on ungroomed jets is discussed in \App{app:add_plots_N3}.

\subsection{Constructing the $N_3$ Observable}
\label{sec:tops_makeobs}

\begin{figure*}[t]
\centering
\subfloat[]{\label{fig:triple_NINJA}  
\includegraphics[width=6.5cm]{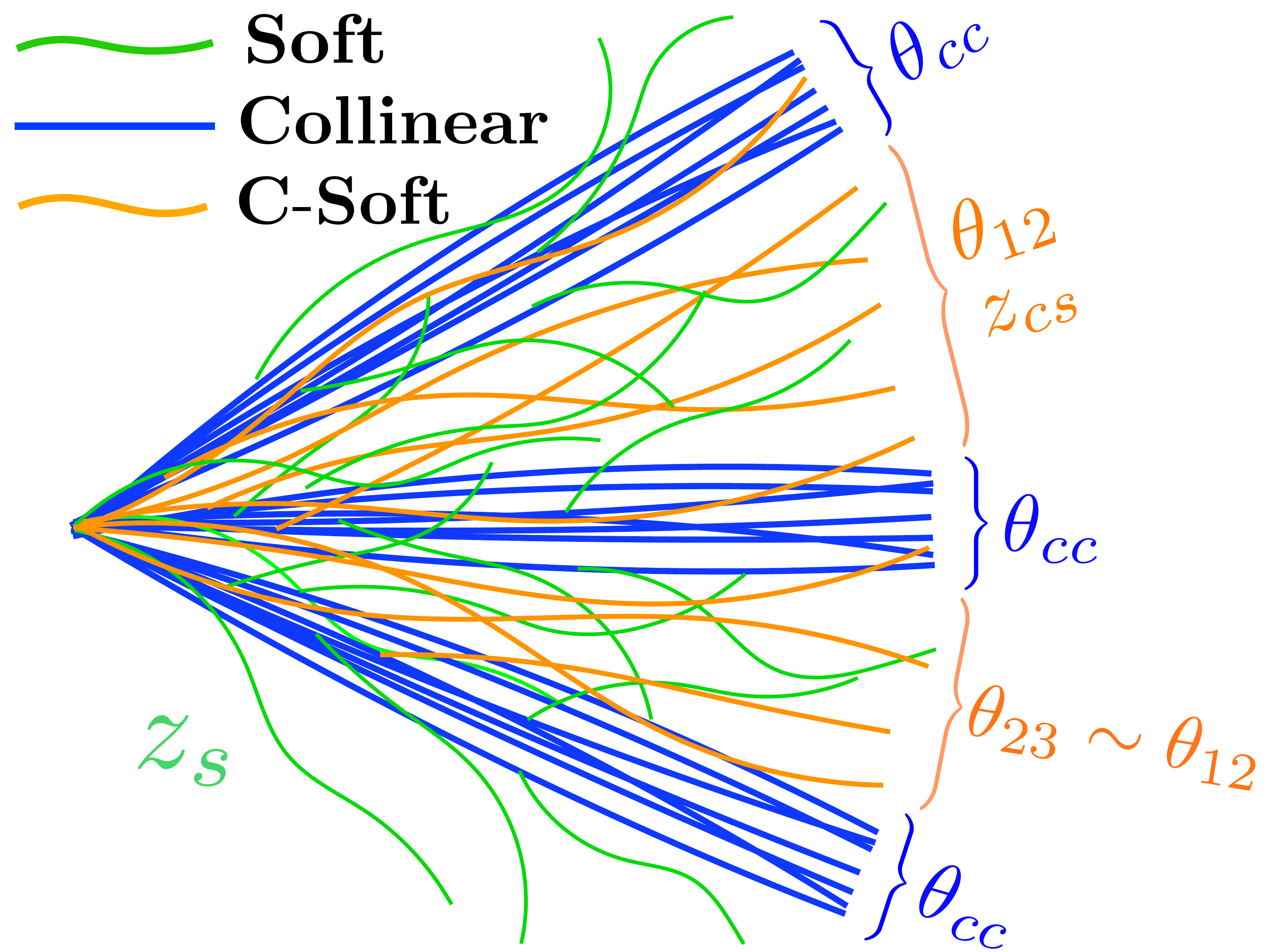}  
} \ \ 
\subfloat[]{\label{fig:p_NINJA}
\includegraphics[width=6.5cm]{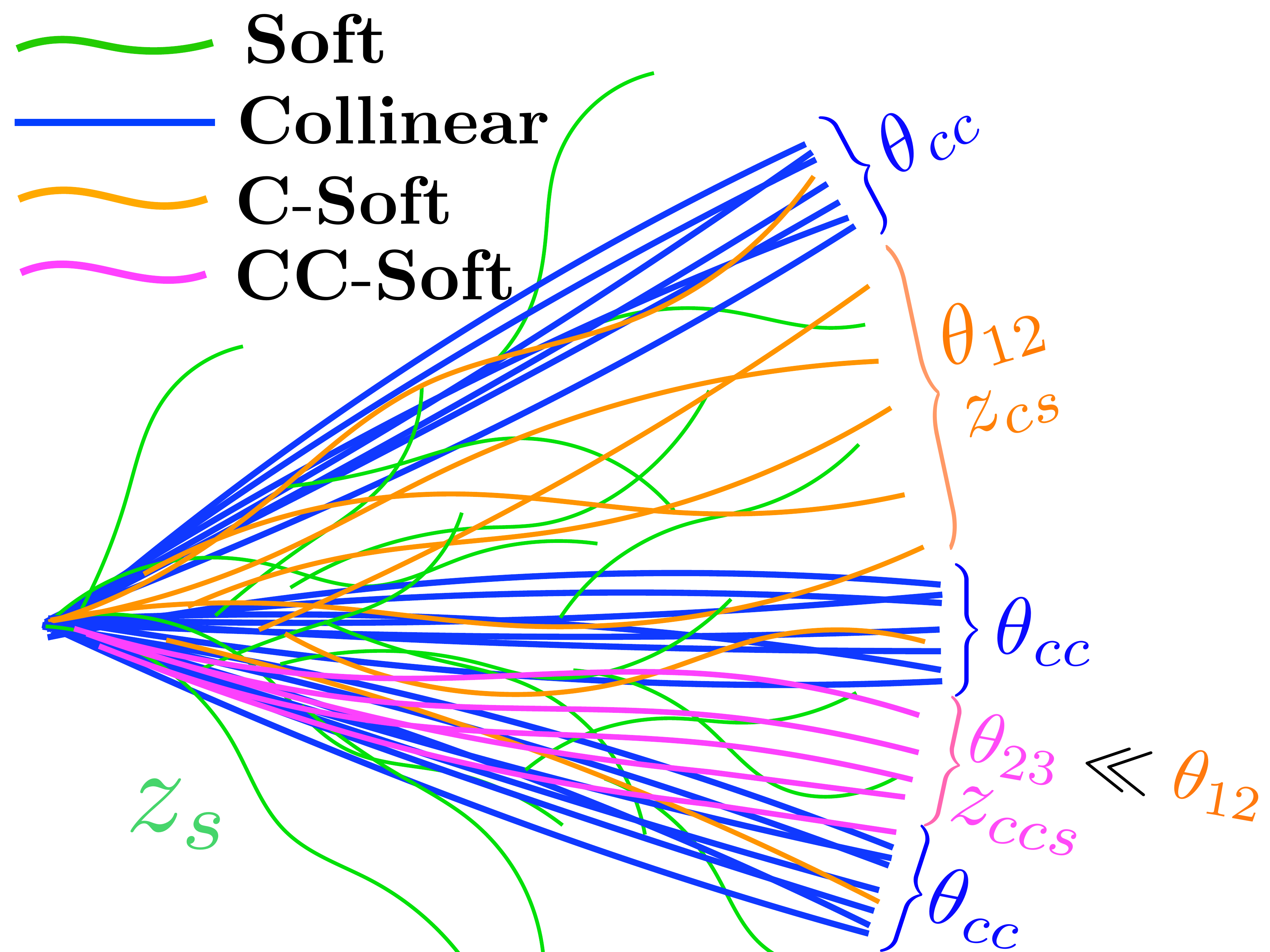} 
}
\caption{Configurations used in the power-counting analysis for $\Nobsnobeta{3}$, showing the modes and scales entering the description of the jets.  In (a), the three subjets carry equal energies, and there is no hierarchy between the angles.  In (b), each of the subjets carries equal energies, but there is a hierarchy in the opening angles of the jets, requiring an extra collinear-collinear-soft mode, shown in magenta, in the power-counting analysis. 
}
\label{fig:triple_NINJAS}
\end{figure*}

\begin{table}[t]
\begin{center}
\begin{tabular}{ccc}
\hline
\hline
Mode & Energy  & Angle \\ 
\hline
soft& $z_s$  & $1$\\
collinear & 1 & $\theta_{cc}$ \\
c-soft & $z_{cs}$ &$\theta_{12}$ \\
cc-soft& $z_{ccs}$ & $\theta_{23}$ \\
\hline
\hline
\end{tabular}
\end{center}
\caption{A summary of the modes in \Fig{fig:p_NINJA} which enter the power-counting analysis for boosted top quarks. 
}
\label{tab:pc_top}
\end{table}

To detect boosted top jets with hard 3-prong substructure, we can use combinations of 2-point, 3-point, and 4-point correlators. Due to the large number of possible combinations, the power counting approach becomes essential to systematically study the behavior of these observables.

In order to use power counting to probe the boundary between the 3-prong (signal) and 2-prong (background) regions of phase space, one must analyze signal configurations which approach this boundary.  For this reason, we consider not only the case of three subjets with equal energies and opening angles, as shown in \Fig{fig:triple_NINJA}, but also the strongly-ordered limit, shown in \Fig{fig:p_NINJA}, where two of the three prongs become collinear.  When the opening angles are hierarchical, the emission modes for each of the dipoles are distinct and must be treated separately, as discussed in \Ref{Larkoski:2014zma}. For lack of a better name, we call these additional modes collinear-collinear-soft modes (shown in magenta in \Fig{fig:p_NINJA}) to distinguish them from collinear-soft modes (shown in orange). A summary of these different modes, and the scaling of their angles and energies, are given in \Tab{tab:pc_top}. These modes satisfy the relations
\begin{align}
\label{eq:3prongscaling}
z_{cs}\ll z_{ccs} \ll 1\,, \qquad \theta_{cc} \ll \theta_{23} \ll \theta_{12} \ll 1\,.
\end{align}
Note the reversal of the energy and angle hierarchies:  collinear-collinear-soft modes have smaller angles but higher energies than collinear-soft modes.  With this slight modification, the power-counting analysis proceeds identically to the simpler case shown in \Fig{fig:NINJA}.

Many experimental analyses use jet shapes as measured on groomed jets, even if the original jet shapes were proposed without grooming.  Grooming has the advantage of making jet properties resistant to pileup contamination and it also leads to observables that are more stable as the jet mass and $p_T$ are varied.  More generally, grooming techniques minimize sensitivity to low momentum particles and the corresponding experimental uncertainties associated with their reconstruction. It is also possible to use a combination of groomed and ungroomed (or lightly groomed) substructure discriminants \cite{gregory_talk,gregory_paper}.  Here, we design our observable specifically for use on groomed jets, since it will help us identify discriminants that are both stable and high-performing.  From the perspective of power counting, grooming simplifies the scaling properties of observables, since we can ignore regions of phase space with soft wide-angle subjets.  In the past, such regions caused complications in designing top tagging observables based on energy correlators \cite{Larkoski:2014zma}, as seen in the definition of $D_3$ in \Eq{eq:D3_full_def}.  After jet grooming, we can drop soft radiation (shown in green in \Fig{fig:triple_NINJAS}) for the purposes of power counting.

From \Eq{eq:4pointcases}, we have six 4-point correlators we could use to form ratio observables with the 2- and 3-point correlators.  To reduce the number of possibilities, we restrict our attention to boost-invariant combinations, but this still leaves many ratios to test.  In \App{app:N3_identify}, we outline a systematic strategy to isolate the most promising 3-prong discriminants using power counting.   Here, we focus on the best performing observable,
\begin{align}
\Nobs{3}{\beta}=\frac{ \ecfvar{2}{4}{\beta}  }{ (\ecfvar{1}{3}{\beta})^2   }\,,
\end{align}
which was presented in \Sec{sec:n_series} as a member of the $N_i$ series.

To understand why $\Nobsnobeta{3}$ is a powerful discriminant on groomed jets, we need to contrast the phase space for 3-prong signal jets versus 2-prong background jets.  For the 3-prong top signal, it is sufficient to study the strongly-ordered limit in \Fig{fig:p_NINJA}, since the balanced case of \Fig{fig:triple_NINJA} can be obtained by setting $z_{ccs} =  z_{cs}$ and $\theta_{23} =  \theta_{12}$.  Using the methods of \Sec{sec:power_counting} on the modes from \Tab{tab:pc_top}, we find the following parametric scaling: 
\begin{align}
\text{3-prong signal (groomed):}\qquad \ecfvar{1}{3}{\beta}&\sim  \theta_{23}^\beta\,, \nn \\
\ecfvar{2}{4}{\beta}& \sim z_{cs} \theta_{12}^\beta \theta_{23}^\beta + z_{ccs} \theta_{23}^{2\beta}   + \theta_{23}^\beta \theta_{cc}^\beta \,.
\end{align}
The dominant background to boosted top quarks are gluon and quark jets, particularly bottom quarks when subjet $b$-tagging is used \cite{ATLAS-CONF-2012-100,CMS:2013vea,ATL-PHYS-PUB-2014-014,ATL-PHYS-PUB-2015-035,CMS-PAS-BTV-15-001,CMS-PAS-BTV-15-002,ATLAS-CONF-2016-001,ATLAS-CONF-2016-002}.  While we ordinarily think of these as being 1-prong backgrounds (see \Fig{fig:unresolved}), they are mainly relevant when they feature 2-prong substructure from a hard parton splitting.  Therefore, the phase space configuration we have to consider for the background is that of \Fig{fig:NINJA}.  Using the modes from \Tab{tab:NINJA}, we find 
\begin{align}
\text{2-prong background (groomed):}\qquad\ecfvar{1}{3}{\beta}& \sim z_{cs} \theta_{12}^\beta  + \theta_{cc}^\beta\,, \nn \\
\ecfvar{2}{4}{\beta}& \sim  z_{cs}^2 \theta_{12}^{2\beta} + z_{cs} \theta_{12}^\beta \theta_{cc}^\beta  +  \theta_{cc}^{2\beta}\,.
\end{align}
From these power-counting relations, we now want to derive the scaling of $\ecfvarnobeta{2}{4}$ versus $\ecfvarnobeta{1}{3}$.

\begin{figure}
\begin{center}
\subfloat[]{\label{fig:N3_ps}
\includegraphics[width=6.95cm]{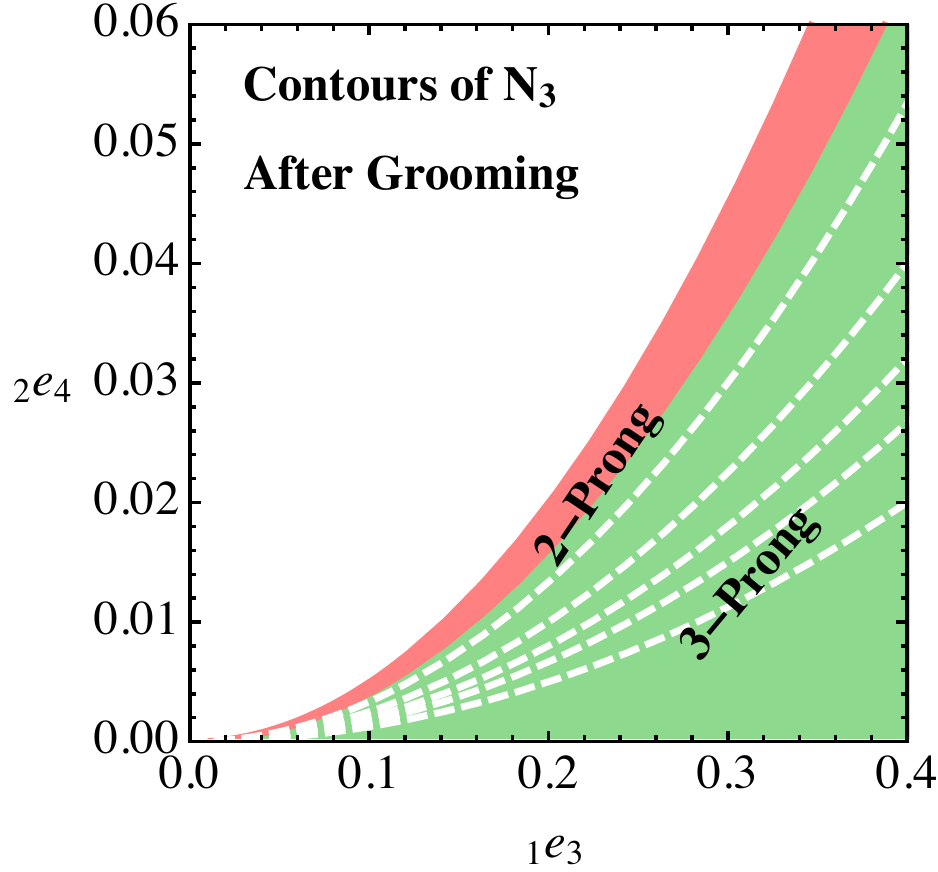}
}\qquad
\subfloat[]{\label{fig:tau32_ps}
\includegraphics[width=6.5cm]{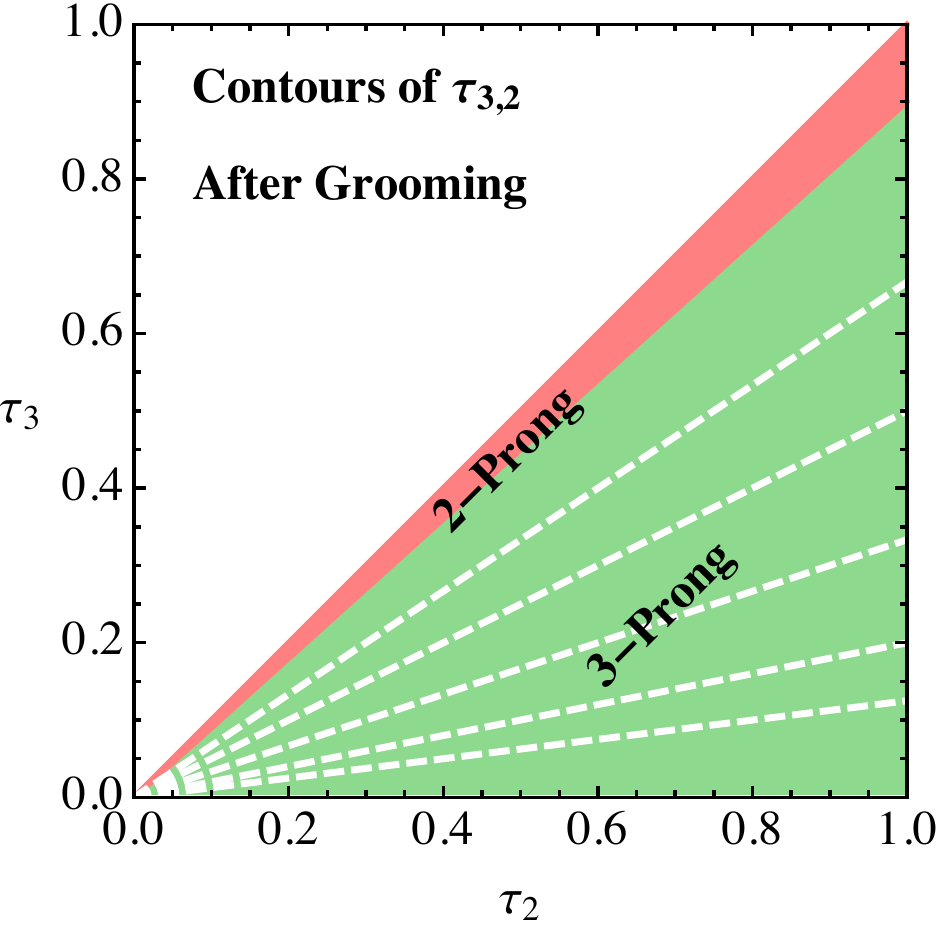}
}
\end{center}
\caption{Comparison of the phase space for (a) $\Nobsnobeta{3}$ and (b) $\Nsubnobeta{3,2}$ after grooming has been applied.  The phase space structure in the two cases is similar, with the background restricted to a single scaling at the upper boundary. 
}
\label{fig:ps_tops}
\end{figure}

For signal jets, $\ecfvarnobeta{2}{4}$ is always smaller than $\ecfvarnobeta{1}{3}$, since they share a factor of $\theta_{23}^\beta$, but each term in $\ecfvarnobeta{2}{4}$ is also multiplied by parametrically small quantity.  In particular, $\theta_{cc} \ll \theta_{23}$ by the assumption of \Eq{eq:3prongscaling}, so we have the parametric relation
\begin{align}
\text{3-prong signal (groomed):}\qquad  \ecfvar{2}{4}{\beta} \ll (\ecfvar{1}{3}{\beta})^2\,.
\end{align}
A much more detailed derivation of this scaling, and an illustration of how it can be identified systematically, is presented in \App{app:N3_pc}.
For background jets, each term in $\ecfvarnobeta{2}{4}$ is the product of two terms in $\ecfvarnobeta{1}{3}$, so we have the relation
\begin{align}
\text{2-prong background (groomed):}\qquad \ecfvar{2}{4}{\beta} \sim (\ecfvar{1}{3}{\beta})^2\,.
\end{align}
This shows that the particular combination chosen to define $\Nobsnobeta{3}$ is indeed appropriate, since we can isolate the top signal region by making a cut of $\Nobsnobeta{3} \ll 1$.  These phase space relations are shown in \Fig{fig:N3_ps}.

\begin{figure*}[t]
\centering
\subfloat[]{\label{fig:NINJA_ECF}
\includegraphics[width=6.5cm]{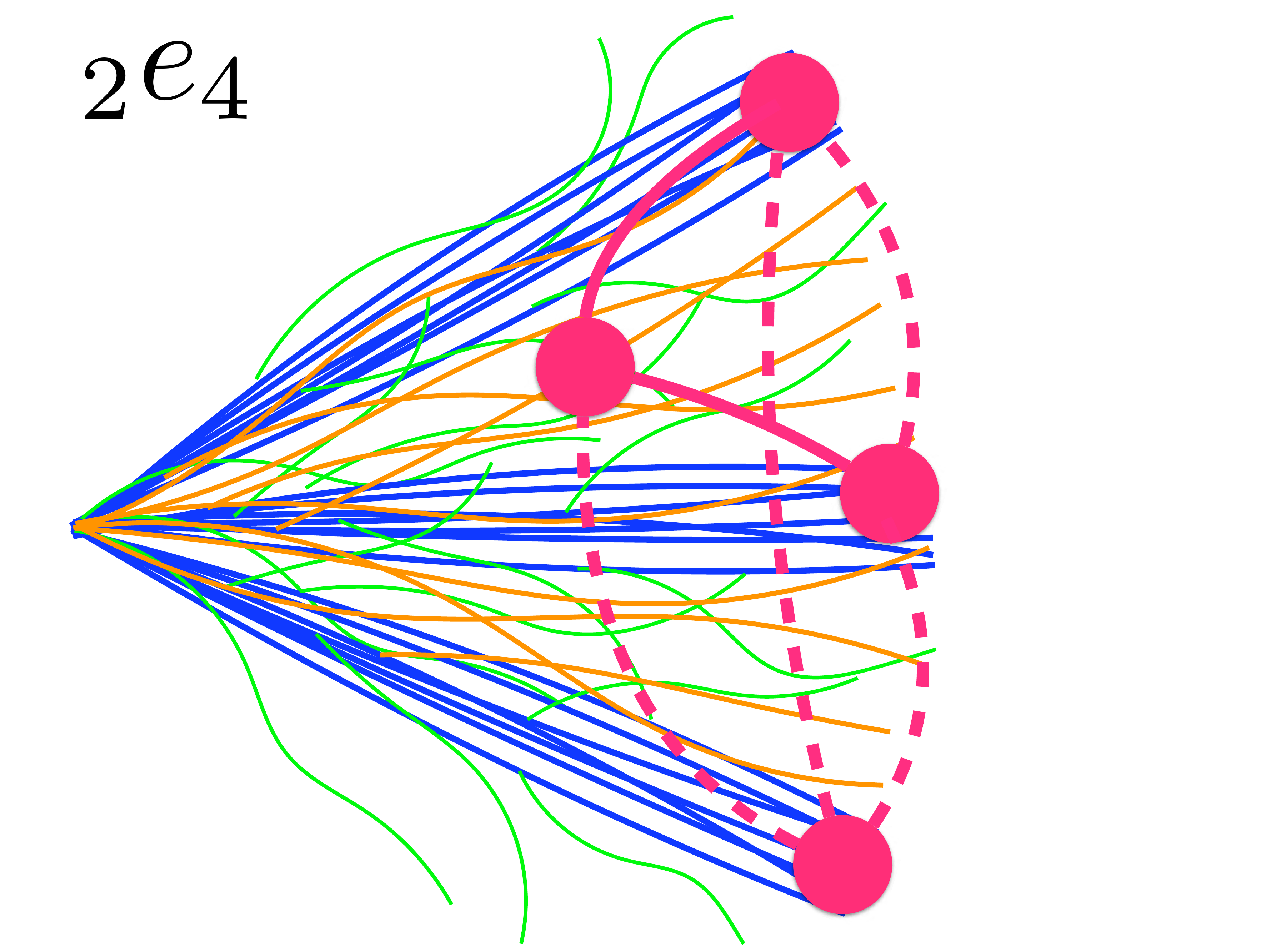}    
} \ \ 
\subfloat[]{\label{fig:NINJA_Nsub}
\includegraphics[width=6.5cm]{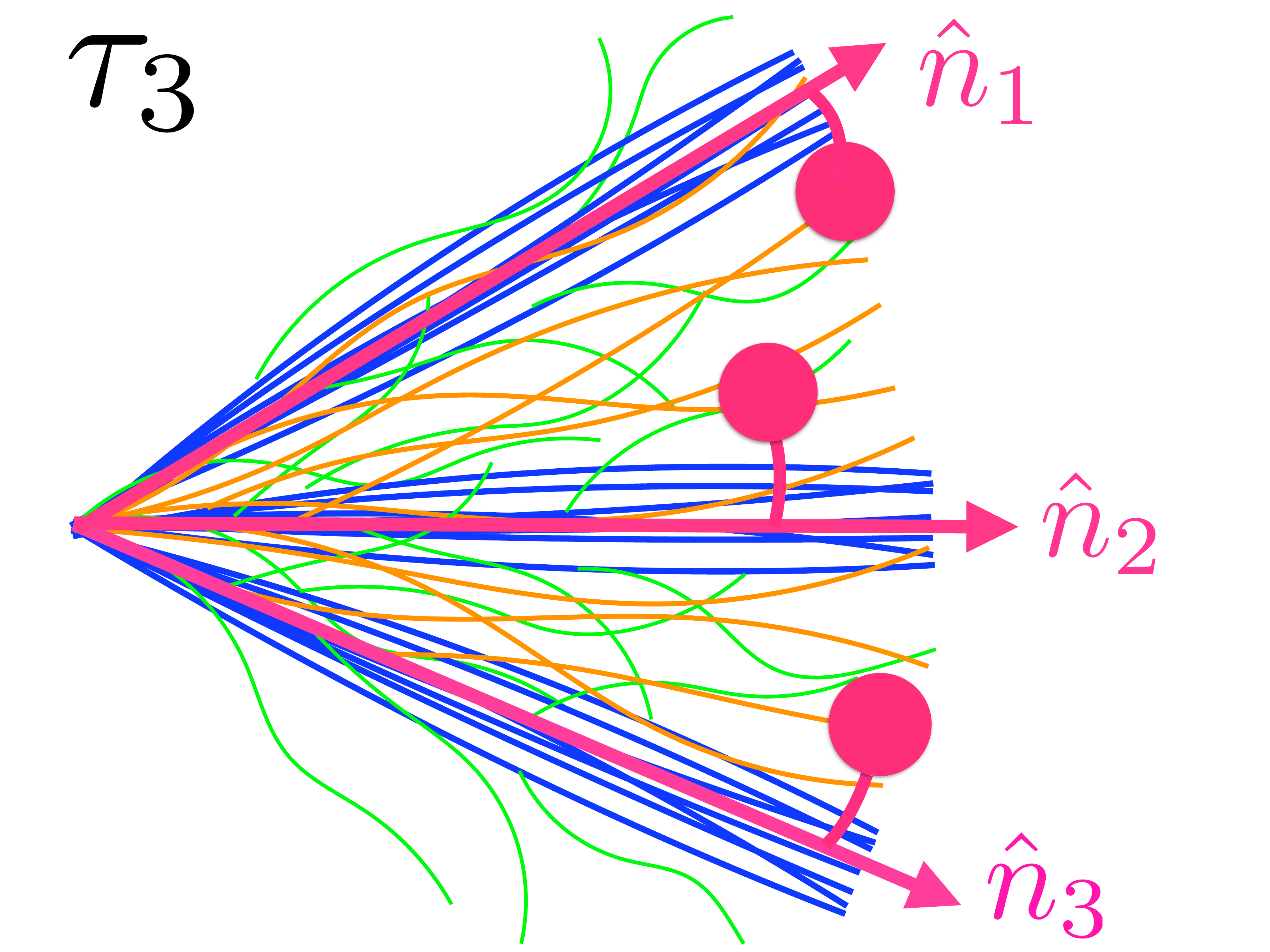} 
}
\caption{Comparison of the functional structure of (a) $\ecfvarnobeta{2}{4}$ and (b) $\Nsubnobeta{3}$. The $\ecfvarnobeta{2}{4}$ observable correlates quadruplets of particles (and two of their six pairwise angles), while the $\Nsubnobeta{3}$ observable correlates particles with axes.  
}
\label{fig:NINJA_obs}
\end{figure*}

To further improve our understanding, it is instructive to compare this with the $N$-subjettiness ratio $\Nsubnobeta{3,2}$, whose phase space is shown in \Fig{fig:tau32_ps}. For strongly-ordered 3-prong substructure, we find 
\begin{align}
\text{3-prong signal (groomed):}\qquad  \Nsub{2}{\beta}&\sim \theta_{23}^\beta \sim \ecfvar{1}{3}{\beta}\,, \nn \\
\Nsub{3}{\beta}&\sim  z_{cs} \theta_{12}^\beta  + z_{ccs} \theta_{23}^\beta  +  \theta_{cc}^\beta \sim  \frac{\ecfvar{2}{4}{\beta}}{ \theta_{23}^\beta}\,.
\end{align}
For 2-prong background jets, we find 
\begin{equation}
\text{2-prong background (groomed):}\qquad \Nsub{2}{\beta} \sim \Nsub{3}{\beta} \sim z_{cs} \theta_{cs}^\beta  + \theta_{cc}^\beta \sim \ecfvar{1}{3}{\beta}\,.
\end{equation}
Remarkably, in both cases, this leads to the relations
\begin{align}
\ecfvar{1}{3}{\beta}&\sim \Nsub{2}{\beta}\,, \nn \\
\ecfvar{2}{4}{\beta}&\sim  \Nsub{2}{\beta}  \Nsub{3}{\beta} \label{eq:2e4factorize}\,.
\end{align}
Therefore, on groomed jets, the $\Nobsnobeta{3}$ and $\Nsubnobeta{3,2}$ observables are parametrically identical:  
\begin{align}\label{eq:N3_related_Nsub}
\Nobs{3}{\beta}=\frac{\ecfvar{2}{4}{\beta}}{(\ecfvar{1}{3}{\beta})^2}\sim  \frac{ \Nsub{2}{\beta}  \Nsub{3}{\beta}  }{ ( \Nsub{2}{\beta})^2  } =  \frac{  \Nsub{3}{\beta}  }{  \Nsub{2}{\beta}  }\,.
\end{align}
This result is quite surprising.  By summing over groups of four particles and taking double products of their pairwise angles, we have achieved an observable that behaves parametrically like an $N$-subjettiness ratio. 

The observables $N_3$ and $\Nsubnobeta{3,2}$ achieve their discrimination power in substantially different ways, as shown schematically in \Fig{fig:NINJA_obs}.  Each term in $\ecfvarnobeta{2}{4}$ is sensitive to multiple energies and angles and contains cross terms like $ \theta_{12}^\beta \theta_{cc}^\beta$.  By contrast, $N$-subjettiness does not contain such cross terms; after determining the axes, each term in the $N$-subjettiness sum is independent of the presence of other subjets.  Despite these differences,  \Eq{eq:2e4factorize} shows that the 4-point correlation function factorizes into a product of lower-point $N$-subjettiness observables, yielding the same parametric behavior in the resolved limit.

While there are no parametric difference between $N_3$ and $\Nsubnobeta{3,2}$, our parton shower study will show that $\Nobsnobeta{3}$ exhibits improved discrimination power on groomed jets, particularly at high efficiencies.  Part of the reason this occurs is because $N_3$ is defined without respect to subjet axes.  This not only offers the practical advantage of not needing to specify an axes-finding algorithm, but it also has an effect on the behavior of $N_3$ away from the power-counting regime.  Recall that $N$-jettiness was originally designed to isolate regions of phase space where there are $N$ well-resolved jets \cite{Stewart:2010tn}. In this limit, the axes are well defined and independent of the particular axes definition up to power corrections.  When used in jet substructure, however, $N$-subjettiness is used both in the limit of well-resolved subjets as well as in the limit of unresolved subjets. Indeed, in many substructure analyses, relatively loose requirement are placed on $N$-subjettiness, such that the $\Nsubnobeta{3,2}$ cut is placed precisely in the unresolved region. Here, $N$-subjettiness can exhibit pathological behavior related to the axes choice \cite{Larkoski:2015uaa}.  By contrast, the $N_3$ observable, being composed simply as sums over the jet constituents, is well behaved throughout the entire jet spectrum, and this will be reflected in its improved performance.

\subsection{Performance in Parton Showers}\label{sec:N3_MC}

\begin{figure}
\begin{center}
\subfloat[]{\label{fig:N3_uu}
\includegraphics[width=6.5cm]{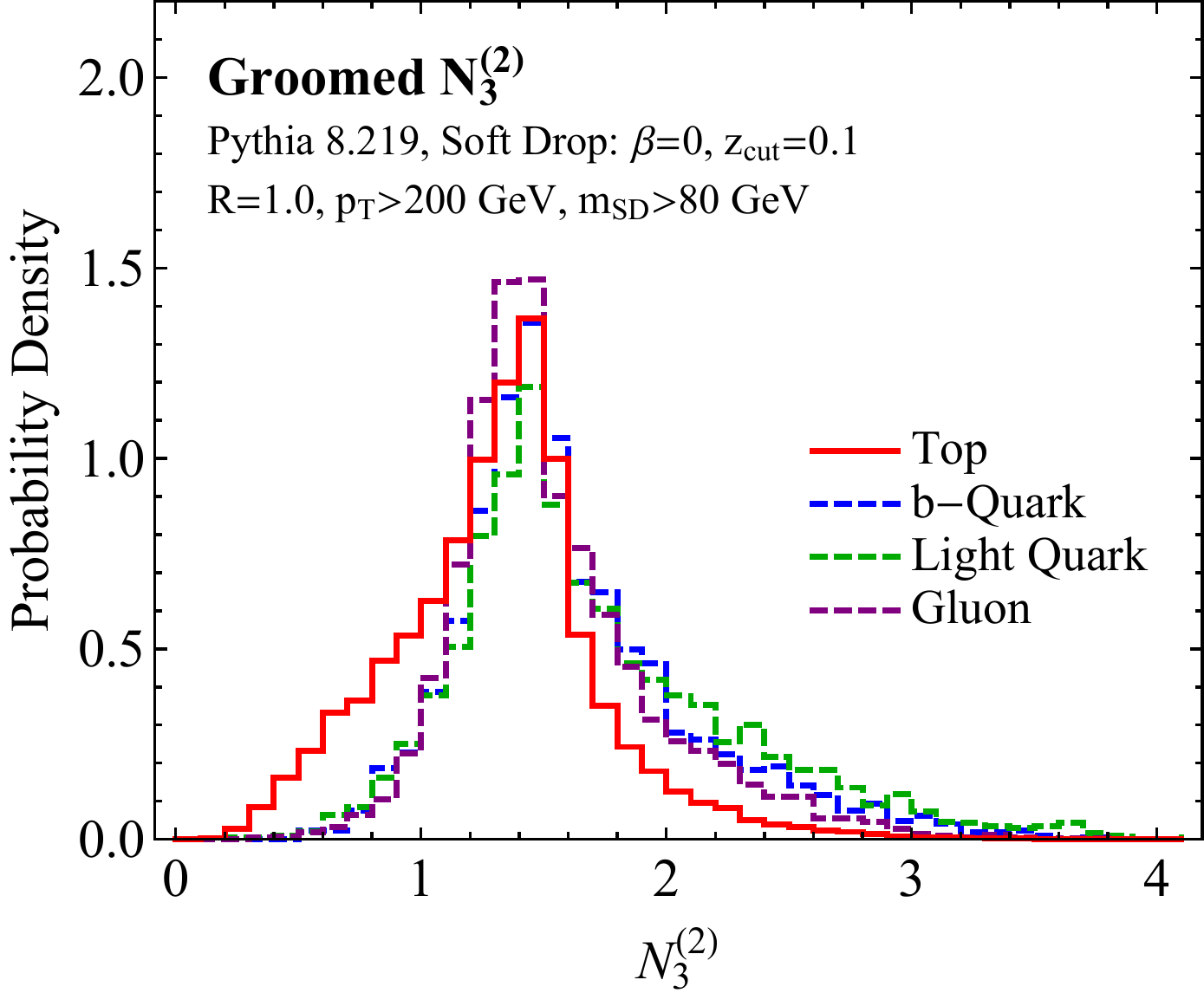}
}\qquad
\subfloat[]{\label{fig:N3_uu_boosted}
\includegraphics[width=6.5cm]{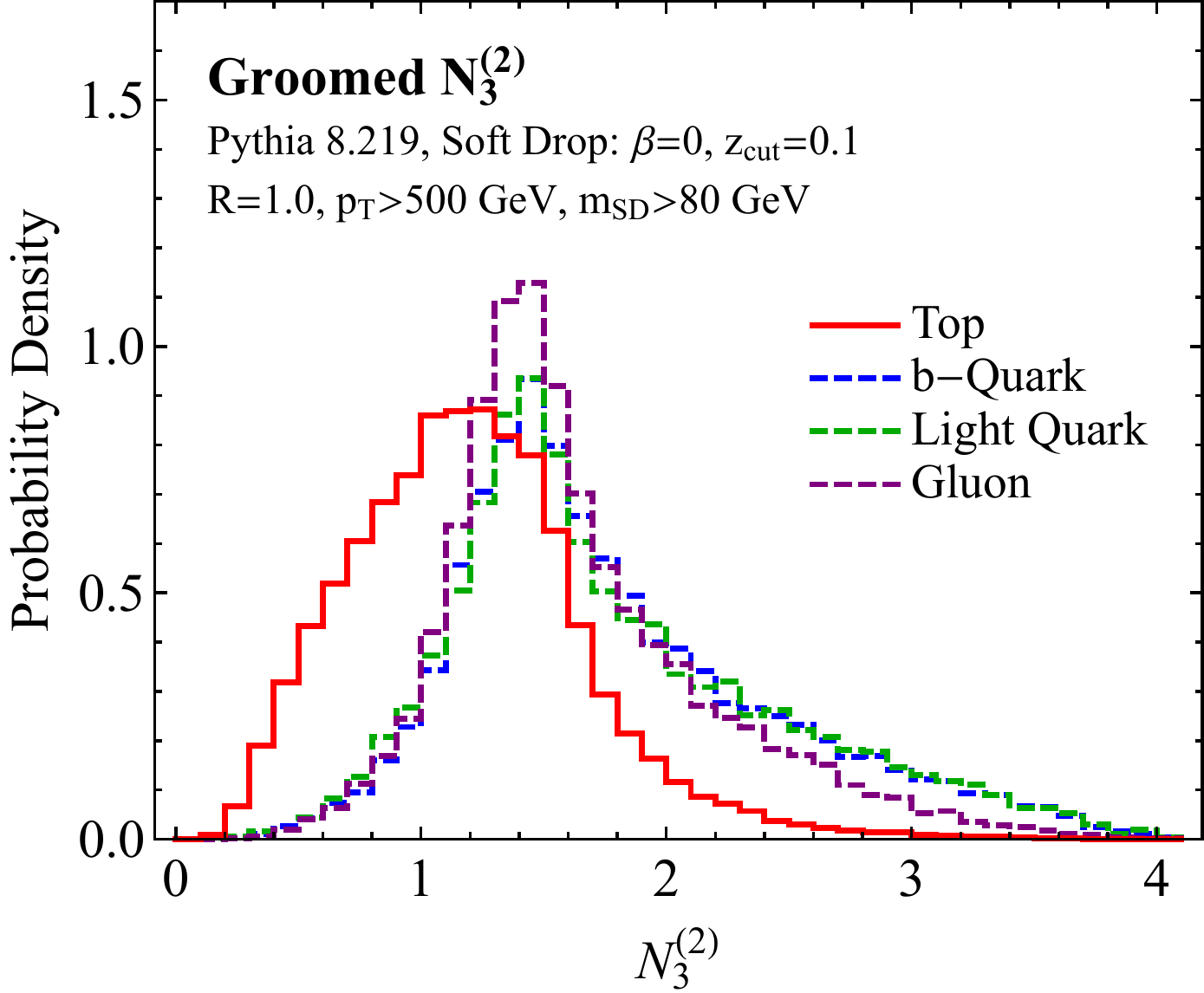}
}
\end{center}
\caption{Distributions of $N_3$ on groomed jets for (a) $p_{TJ}=200$ GeV and (b) $p_{TJ}=500$ GeV, comparing signal top jets to background QCD jets initiated from $b$-quarks, light quarks, and gluons. 
}
\label{fig:3prong_N3}
\end{figure}

\begin{figure}
\begin{center}
\subfloat[]{\label{fig:N3_b_linear}
\includegraphics[width=6.5cm]{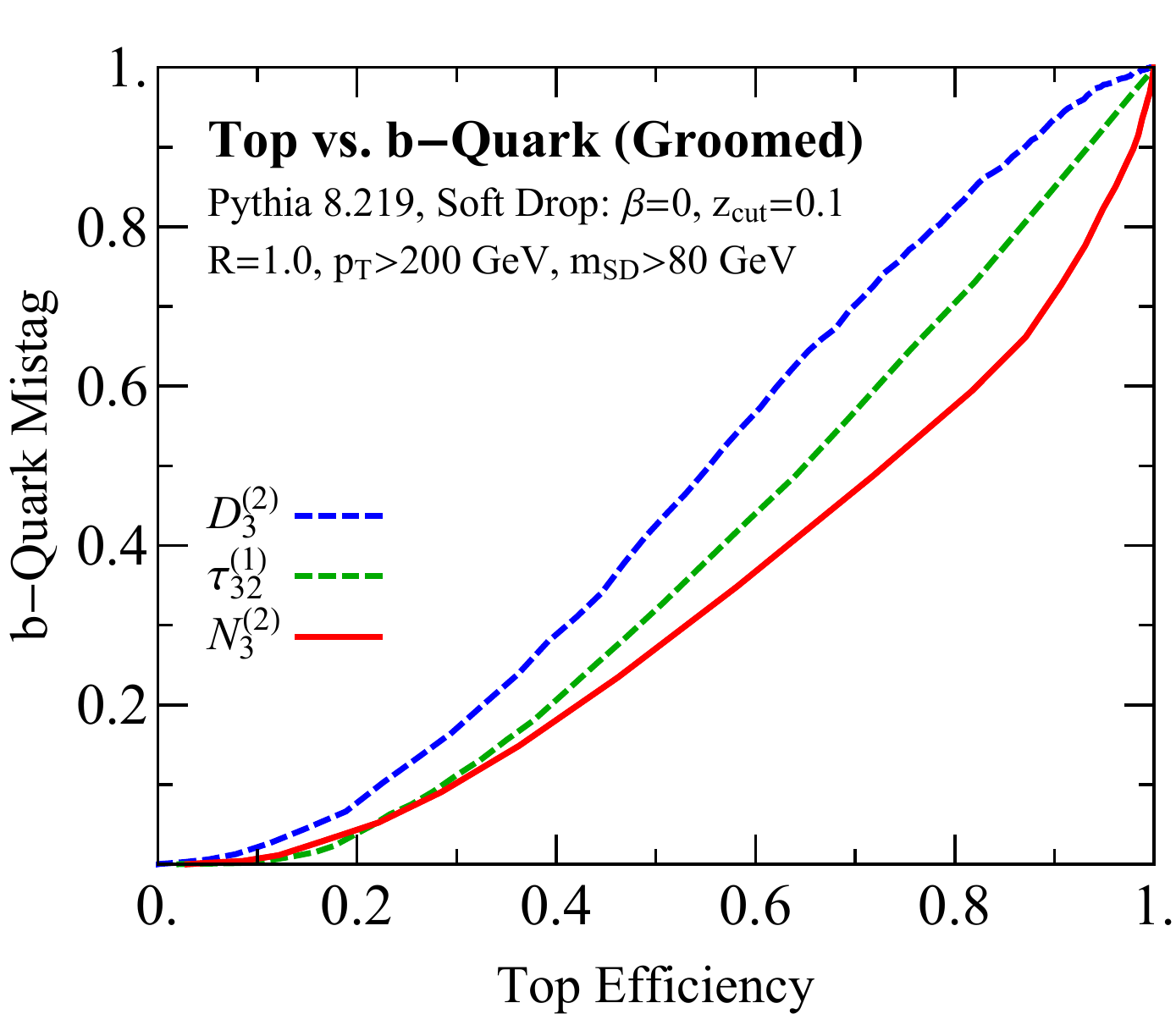}
}\qquad
\subfloat[]{\label{fig:N3_b_linear_200}
\includegraphics[width=6.5cm]{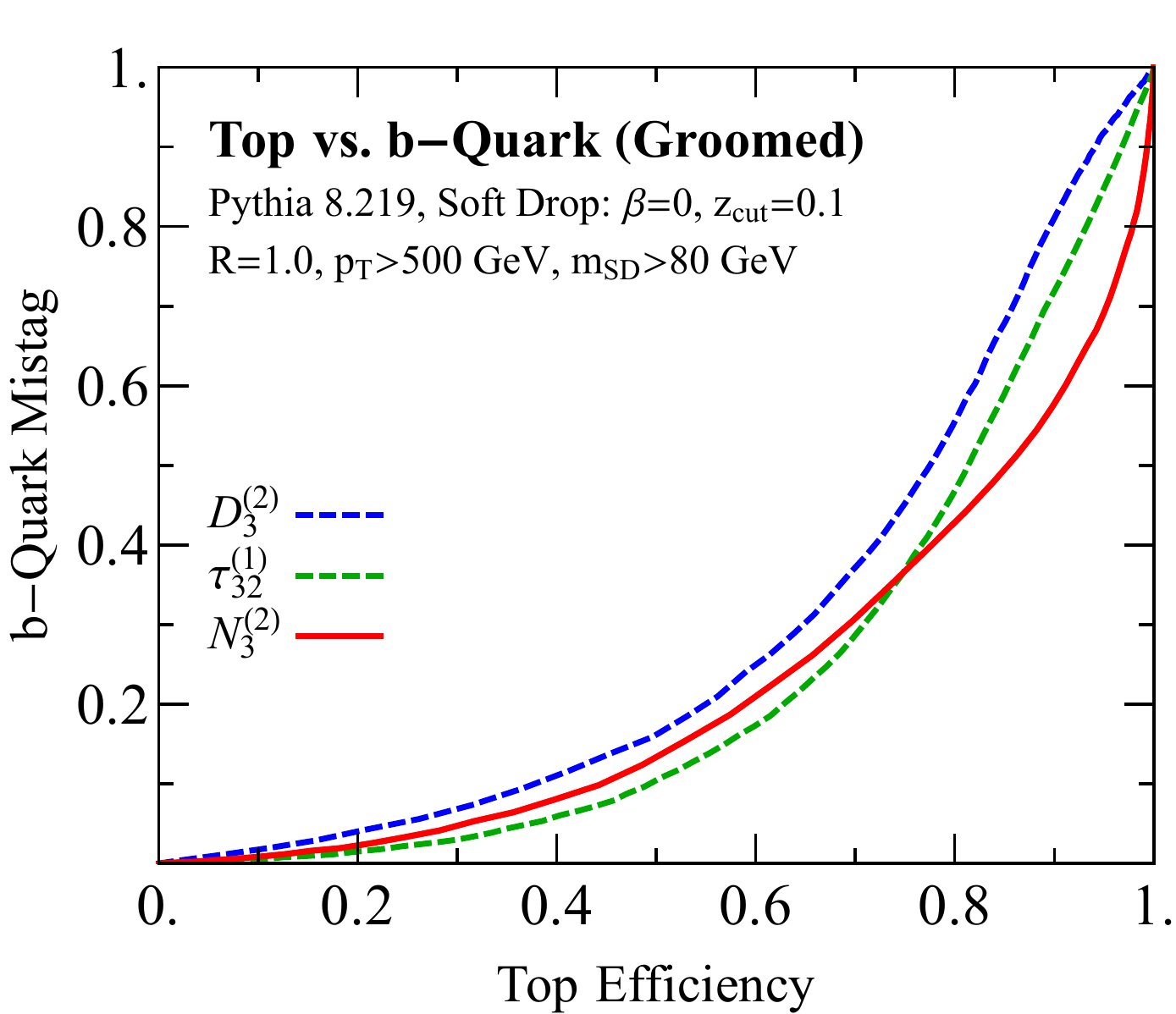}
}\nonumber\\
\subfloat[]{\label{fig:N3_quark_linear}
\includegraphics[width=6.5cm]{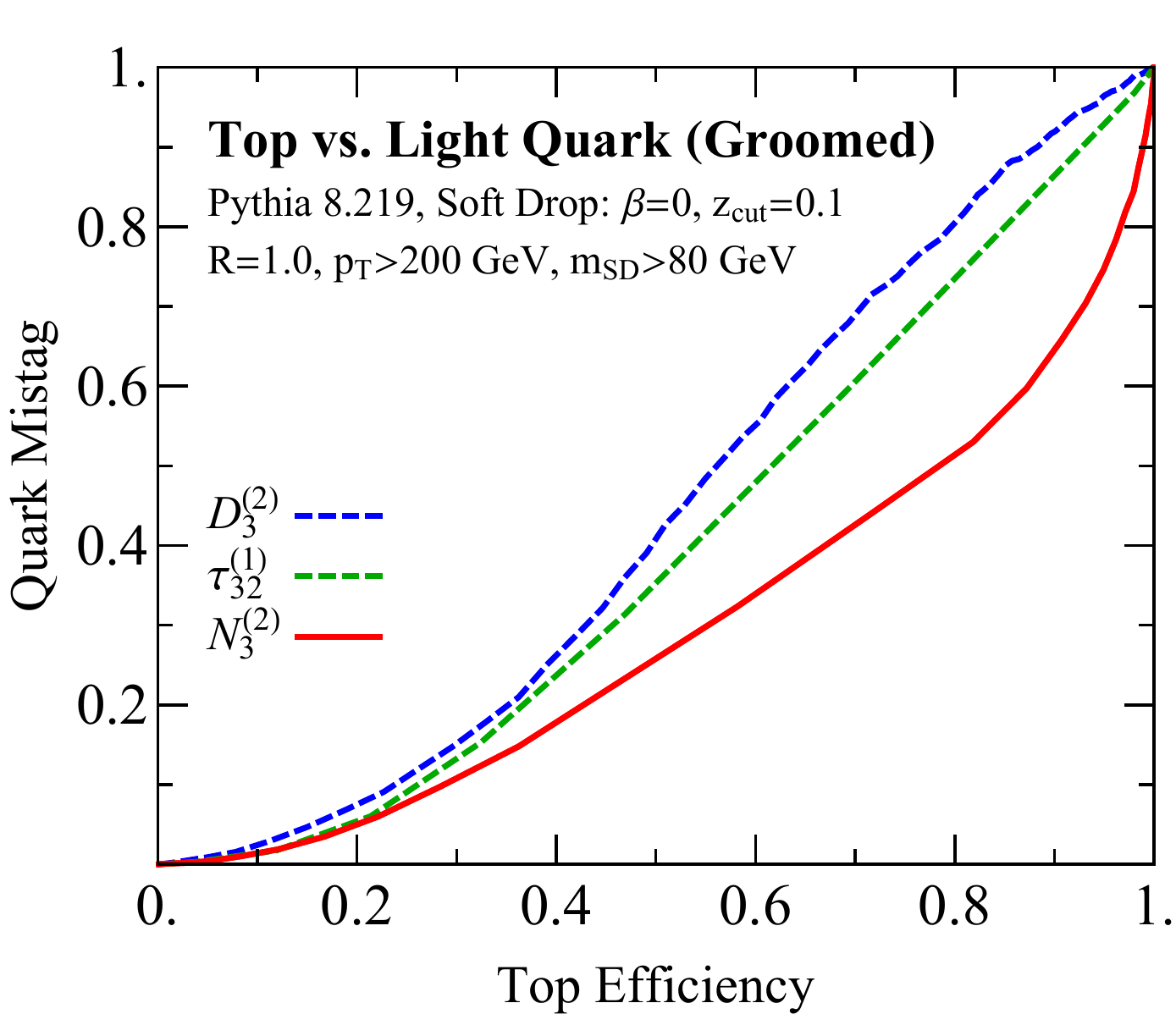}
}\qquad
\subfloat[]{\label{fig:N3_quark_linear_200}
\includegraphics[width=6.5cm]{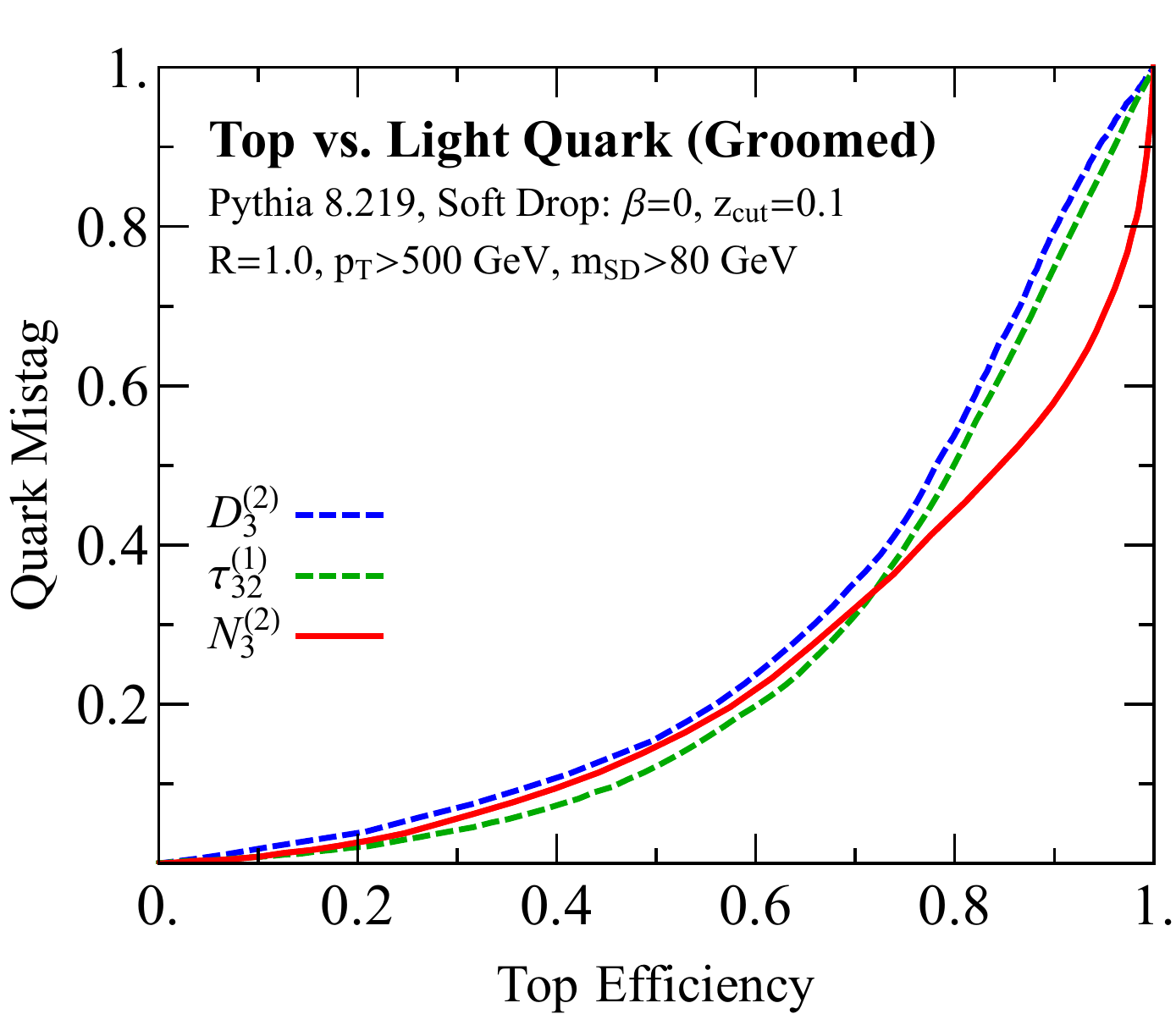}
}\nonumber\\
\subfloat[]{\label{fig:N3_gluon_linear}
\includegraphics[width=6.5cm]{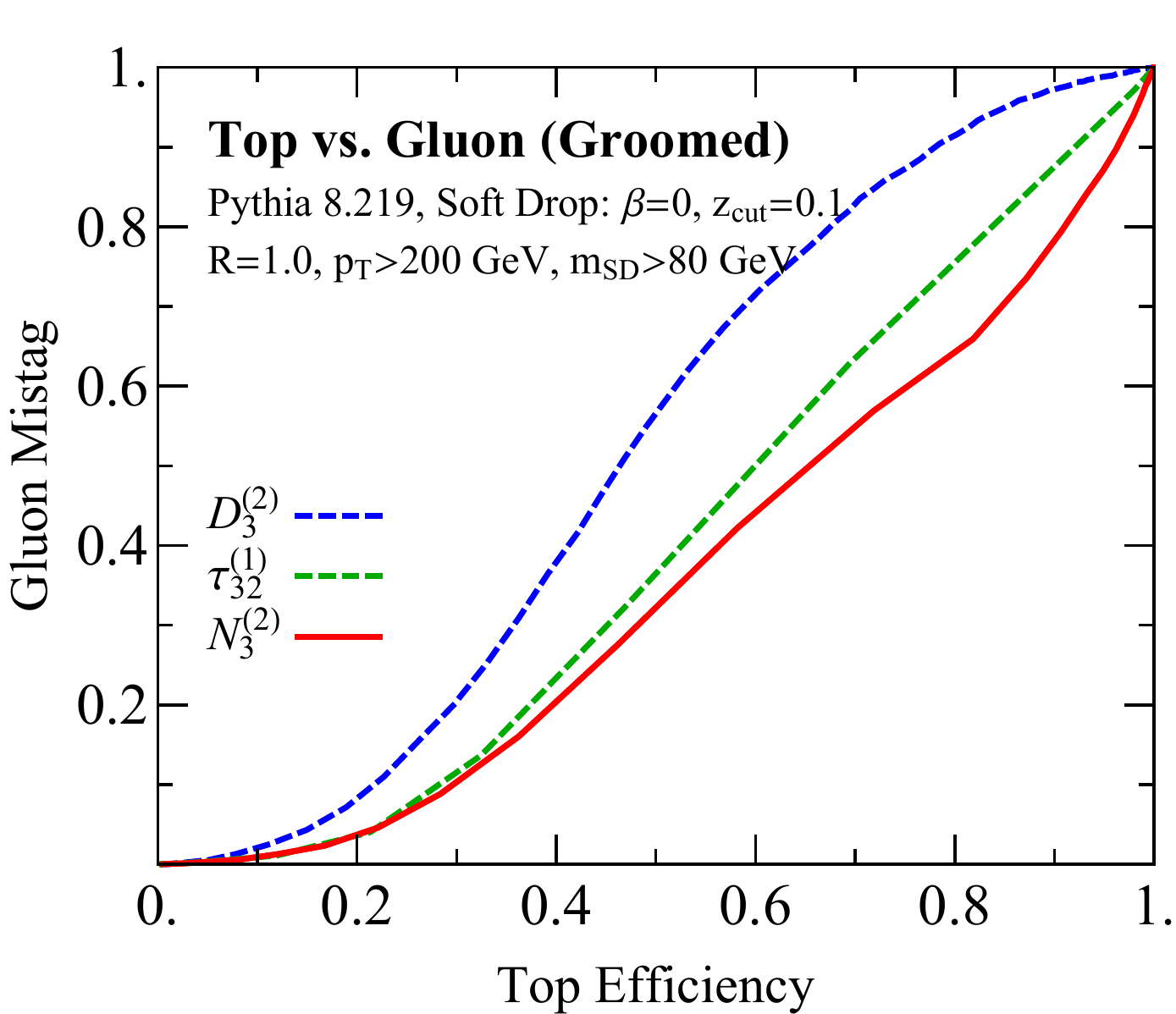}
}\qquad
\subfloat[]{\label{fig:N3_gluon_linear_200}
\includegraphics[width=6.5cm]{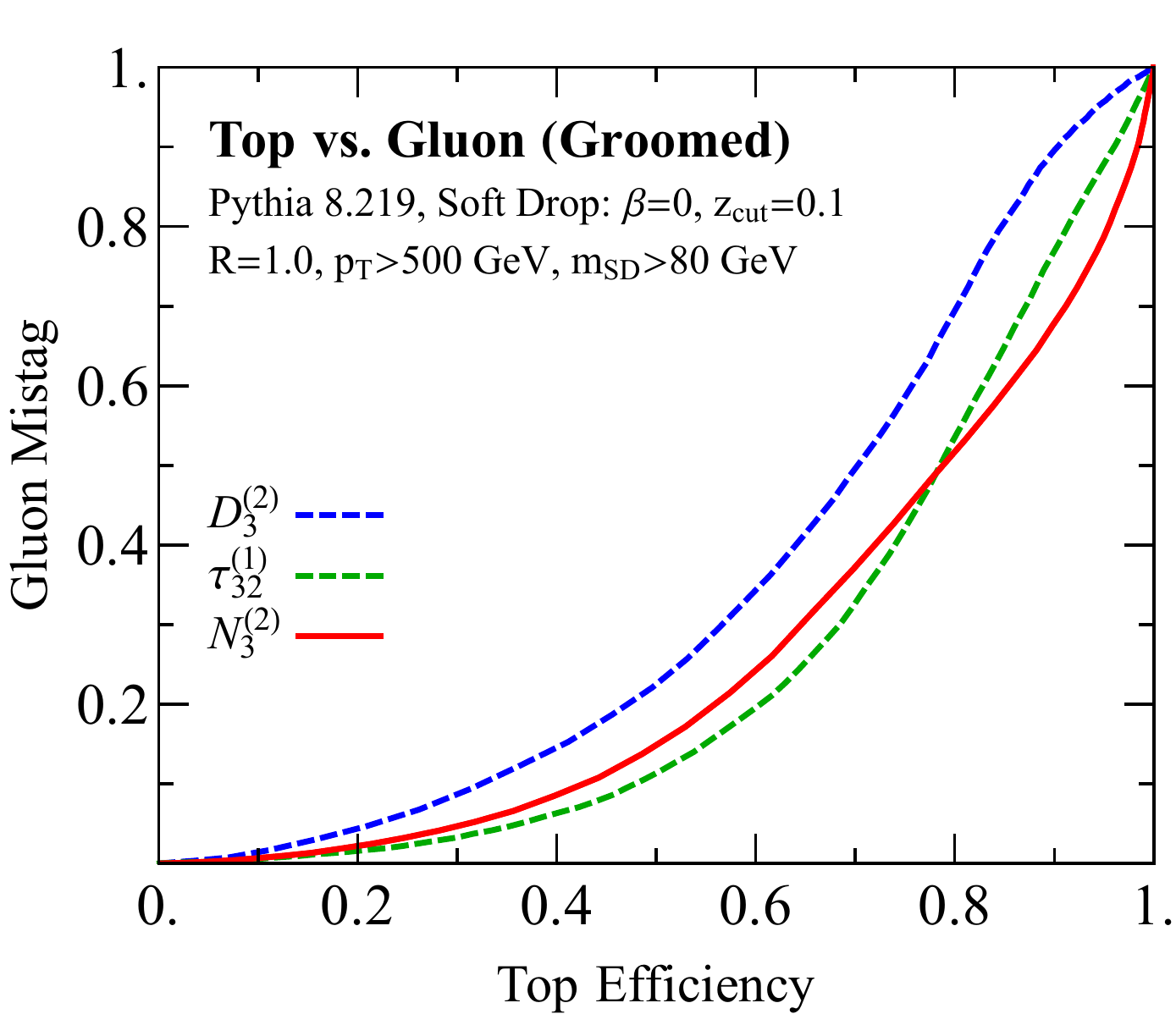}
}
\end{center}
\caption{ROC curves comparing the groomed observables $\Nobsnobeta{3}$, $\Nsubnobeta{3,2}$, and $\Dobsnobeta{3}$ for (left column) $p_{TJ}=200$ GeV and (right column) $p_{TJ}=500$ GeV.  The discrimination is shown for boosted top quarks against (top row) $b$-jets, (middle row) light quark jets, and (bottom row) gluon jets.  In all cases, soft drop grooming has been applied, and the selection efficiency is after a groomed jet mass cut of $m_{\rm SD} > 80~\GeV$. The $\Nobsnobeta{3}$ observable offers improved discrimination power, particularly at high signal efficiencies. 
}
\label{fig:N3_ROC_3prong_linear}
\end{figure}

Having understood the power counting of $\Nobsnobeta{3}$ on groomed jets, we now study its behavior in parton shower generators, comparing $\Nobsnobeta{3}$ with both $\Nsubnobeta{3,2}$ and the simplified version of $\Dobsnobeta{3}$ defined in \Eq{eq:D3_simp_def}.  The comparison to $\Nsubnobeta{3,2}$ is particularly interesting, since the parametrics in \Eq{eq:N3_related_Nsub} suggest it should perform similarly to $\Nobsnobeta{3}$ in the resolved limit.

For our parton shower study, we generate background QCD jets from $pp\to jj$ events, where we consider separately the cases of $j=g$ (gluon) and $j=u$ (representative of light quarks).  We also consider the case of $b$-quark backgrounds, which are interesting to treat separately due to recent advances in $b$-tagged substructure \cite{ATLAS-CONF-2012-100,CMS:2013vea,ATL-PHYS-PUB-2014-014,ATL-PHYS-PUB-2015-035,CMS-PAS-BTV-15-001,CMS-PAS-BTV-15-002,ATLAS-CONF-2016-001,ATLAS-CONF-2016-002}; heavy quarks were generated from the process $pp\to b \bar b$.  The boosted top signal is generated from $pp\to t \bar t$ events, with both tops decaying hadronically.  

Events were generated with \madgraph{2.3.3} \cite{Alwall:2014hca} at the $13$ TeV LHC and showered with \pythia{8.219} \cite{Sjostrand:2006za,Sjostrand:2007gs} with underlying event and hadronization implemented with the default settings. Anti-$k_T$ \cite{Cacciari:2008gp} jets with radius $R=1.0$ were clustered in \fastjet{3.2.0} \cite{Cacciari:2011ma} using the Winner Take All (WTA) recombination scheme \cite{Larkoski:2014uqa,Larkoski:2014bia}.\footnote{WTA axes align with a hard prong within the jet.  They are nice theoretically, as they avoid recoil due to soft emissions \cite{Catani:1992jc,Dokshitzer:1998kz,Banfi:2004yd,Larkoski:2013eya,Larkoski:2014uqa}. For low $p_T$ tops, however, the use of WTA axes can potentially lead to lopsided axes.  We explicitly checked that are our results are unmodified if standard $E$-scheme recombination is used instead.} The energy correlation functions and $N$-subjettiness ratio observables were calculated using the \texttt{EnergyCorrelator} and \texttt{Nsubjettiness} \fastjet{contrib}s \cite{Cacciari:2011ma,fjcontrib}. For $N$-subjettiness, we use one-pass WTA minimization with $\beta=1$. As a concrete example of a groomer, we use $\beta=0$ soft drop  \cite{Larkoski:2014wba} (a.k.a.~modified mass drop with $\mu =1$ \cite{Dasgupta:2013ihk,Dasgupta:2013via}) with $\zcut=0.1$, though our general observations should be independent of the particular choice of groomer.

As discussed in \Sec{sec:tops_makeobs}, we focus on the behavior of the observables on groomed jets, where $\Nobsnobeta{3}$ was designed to perform well and where $\Nobsnobeta{3}$ behaves parametrically like $\Nsubnobeta{3,2}$. In \App{app:add_plots_N3}, we study boosted top tagging without grooming, where $\Nobsnobeta{3}$ is still a reasonably powerful discriminant on ungroomed jets, but not as strong as  $\Nsubnobeta{3,2}$.  We also discuss the behavior of $\Mobsnobeta{3}$ in \App{app:add_plots_M3}, using power-counting arguments to show why it is a poor discriminant with standard groomers, but might perform better with a more aggressive grooming strategy. 

In \Fig{fig:3prong_N3}, we show distributions for groomed $\Nobsnobeta{3}$, comparing the top jet signal to the backgrounds of $b$-quark, light quark, and gluon jets.  A groomed mass cut of $m_{\rm SD}>80$ GeV is applied, following a recent ATLAS study \cite{ATL-PHYS-PUB-2015-053}.  
Here, we use $\beta=2$ as the angular exponent for $\Nobsnobeta{3}$; power counting does not, in this case, predict a preferred value of $\beta$, so it could be optimized for experimental performance.  The behavior of these distributions is quite interesting, particularly for $p_{TJ}>500$ GeV in \Fig{fig:N3_uu_boosted}, where the top quarks are truly boosted.  The signal distribution drops off sharply above $\Nobsnobeta{3} \simeq 1.5$, while the background distribution extends to larger values for all three samples. This behavior leads to excellent performance at high signal efficiencies, and is quite different than for $\Nsubnobeta{3,2}$ (see \Fig{fig:app_tau32} in \App{app:add_plots_N3}).  Note that these distributions are calculated after the soft drop mass cut, so the region where $\Nobsnobeta{3}$ exhibits improved performance is the one directly relevant for LHC searches.

\begin{figure}
\begin{center}
\subfloat[]{\label{fig:N3_stabilityA}
\includegraphics[width=6.5cm]{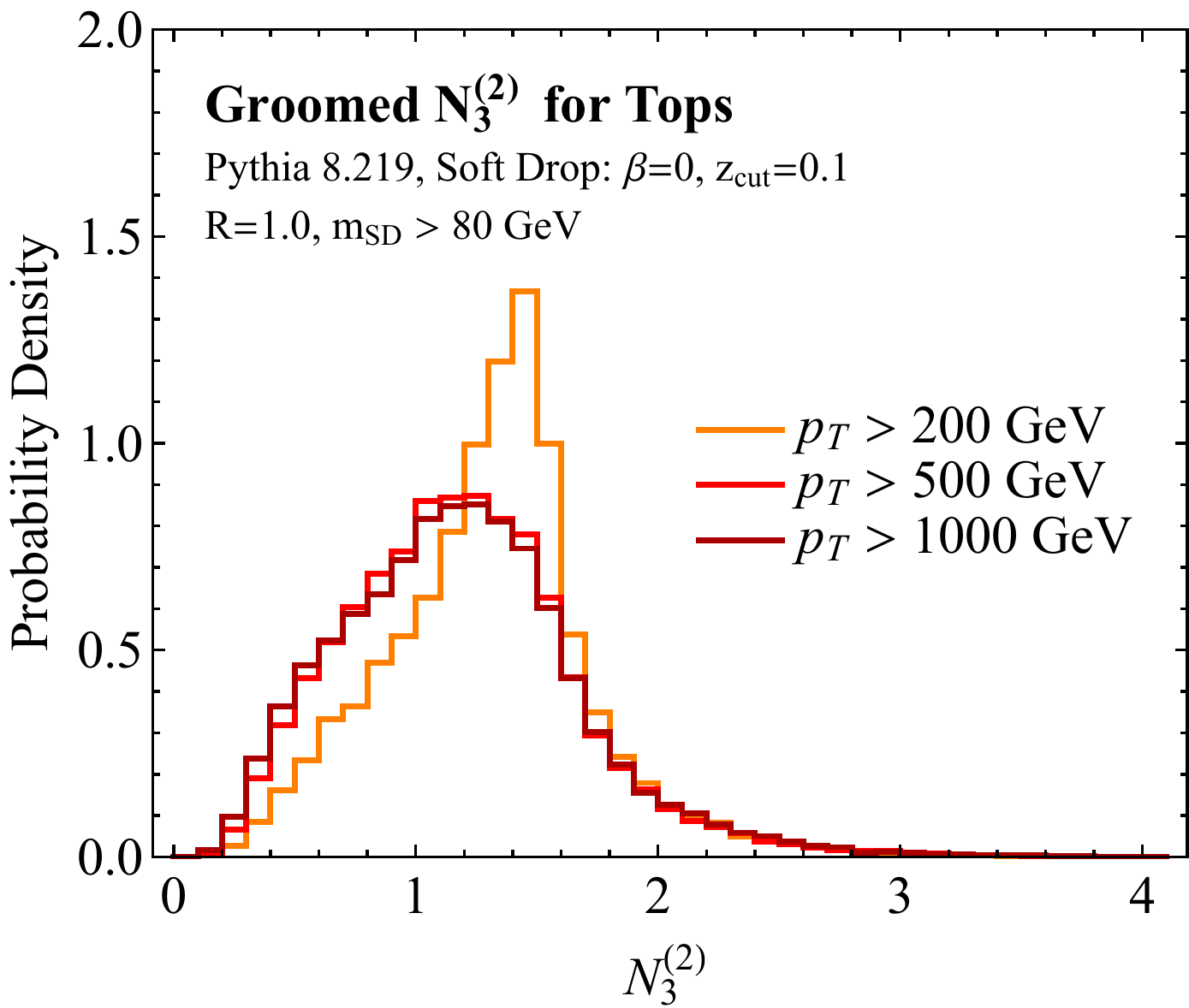}
}\qquad
\subfloat[]{\label{fig:N3_stabilityB}
\includegraphics[width=6.5cm]{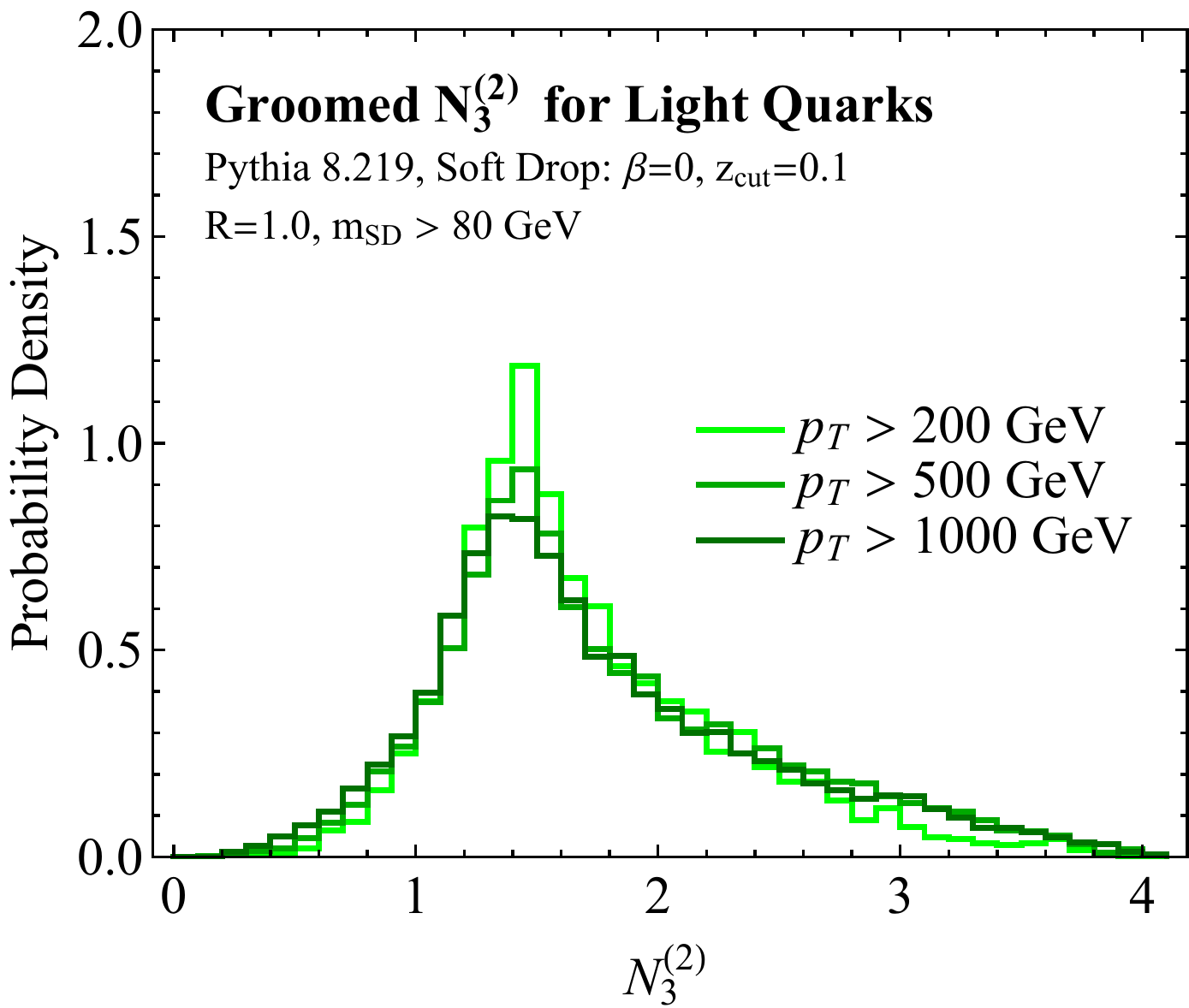}
}
\end{center}
\caption{Stability of the groomed $\Nobsnobeta{3}$ observable as a function of $p_{TJ}$ for (a) signal and (b) background distributions.  Here, light quark jets are used as representative of the background; gluons and $b$-jets behave similarly. The shift in the signal distribution in the lowest $p_{TJ}$ bin is due to a high fraction of top quarks whose decay products are not fully captured by the $R=1.0$ jet radius.
}
\label{fig:N3_stability}
\end{figure}

In \Fig{fig:N3_ROC_3prong_linear}, we show signal efficiency versus background rejection (ROC) curves for boosted top discrimination against $b$-quark, light quark, and gluon jets.  In these and all subsequent ROC curves, the efficiency and mistag rates are given \emph{after} applying a baseline mass requirement, in order to show just the gain in performance from adding a substructure cut.  The baseline efficiencies for the $m_{\rm SD}>80$ GeV mass selection are
\begin{align}
\text{$p_{TJ}=200$ GeV}: & \quad \mathcal{E}_t =61\%, \quad \mathcal{E}_b = 2.4\%, \quad \mathcal{E}_q = 2.8\%,  \quad \mathcal{E}_g = 6.6\%, \\
\text{$p_{TJ}=500$ GeV}: & \quad \mathcal{E}_t =87\%, \quad \mathcal{E}_b = 10\%,\phantom{.} \quad \mathcal{E}_q = 10\%,  \quad \mathcal{E}_g = 19\%,
\end{align}
and the final efficiencies and mistag rates are obtained by multiplying these baseline values by those shown in \Fig{fig:N3_ROC_3prong_linear}.  Comparing $p_{TJ}=200$ GeV and $p_{TJ}=500$ GeV, we conclude that the behavior of $N_3$ is reasonably robust as a function of $p_{TJ}$ (see \Fig{fig:N3_stability} for higher $p_{TJ}$ values). The simplified version of $D_3$ with this choice of angular exponent gives rather poor discrimination power, especially for gluon jets; the apparent negative discrimination power for certain ROC curves in \Fig{fig:N3_ROC_3prong_linear} is due to the use of a (non-optimal) one-sided cut.  It is also satisfying to see the behavior predicted from the power-counting analysis.  At lower top efficiencies, where there are well-resolved jets, $N_3$ and $\Nsubnobeta{3,2}$ exhibit similar discrimination power, but at higher efficiencies, where there are not well-resolved jets, the structure of $N_3$ leads to considerably improved performance.  It would be interesting to see whether these parton shower predictions remain true in LHC data. 

Finally, another important feature of the soft-dropped $N_3$ observable is its stability as the mass and $p_{T}$ of the jet are varied. This has recently been emphasized in \Ref{Dolen:2016kst} as a highly desirable feature of jet substructure observables, as it removes mass sculpting.  In \Fig{fig:N3_stability}, we show the signal and background distributions for three different values of the jet $p_{TJ}$, namely $p_{TJ}= \{200, 500, 1000\}$ GeV following  \Ref{ATL-PHYS-PUB-2015-053}.  Remarkable stability of the $N_3$ distribution is seen, with the main distortion appearing for the top sample in the lowest $p_{TJ}$ bin, where the $R = 1.0$ jet radius is not always large enough to capture all of the top decay products.  Between $p_{TJ}=500$ GeV and $1000$ GeV, there are almost no changes to either the signal or background distributions.  Though not shown here, $\Nsubnobeta{3,2}$ exhibits comparable stability to $N_3$, as expected since they share the same power counting. We conclude that soft-dropped $N_3$ is a powerful boosted top tagger that exhibits many experimentally desirable features.

\section{New Observables for 2-prong Substructure}
\label{sec:2prong}

Jet substructure techniques have played an increasingly important role in recent LHC searches, especially for new resonances with decays involving boosted $W/Z/H$ bosons \cite{Aad:2015owa,Aaboud:2016okv,Aaboud:2016trl,Aaboud:2016qgg,ATLAS-CONF-2016-055,ATLAS-CONF-2015-071,ATLAS-CONF-2015-068,ATLAS-CONF-2016-082,ATLAS-CONF-2016-083,Khachatryan:2015bma,CMS:2016pfl,CMS:2016mvc,CMS:2016wev}.  In order to understand any possible hint of new physics in diboson analyses, it is essential to have exceptional control over the behavior of jet substructure discriminants, to allay concerns about possible analysis artifacts \cite{Goncalves:2015yua,Martin:2016jdw}.  In our view, echoing the perspective of \Ref{Dolen:2016kst}, properties like stability with jet $p_T$ and resilience to mass sculpting are just as important as (and perhaps more so than) absolute tagging performance.

In this section, we use the generalized correlators to construct 2-prong taggers that are robust and perform well.  This is an application where the original energy correlators have already proven useful through the $\Cobsnobeta{2}$ and $\Dobsnobeta{2}$ ratios \cite{Larkoski:2013eya,Larkoski:2014gra}.  Here, we propose three new ratios: 
\begin{align}
\Mobs{2}{\beta}=\frac{\ecfvar{1}{3}{\beta}}{\ecfvar{1}{2}{\beta}}\,, \qquad \Nobs{2}{\beta}=\frac{ \ecfvar{2}{3}{\beta}  }{   ( \ecfvar{1}{2}{\beta}   )^2}\,, 
\qquad \Dobs{2}{\alpha,\beta}=\frac{\ecfvar{3}{3}{\alpha}}{(\ecfvar{1}{2}{\beta})^{3\alpha/\beta}}\,,
\end{align}
corresponding to the three variants of the 3-point correlator in \Eq{eq:explicit_ecfvar}.   Each of these observables is sensitive to different angular structures within the jet and therefore achieves its discrimination power in a different manner.  This fact is highlighted in their different behavior under grooming,  where $\Mobsnobeta{2}$ and $\Dobs{2}{1,2}$ were constructed to only perform well on groomed jets.  Therefore, these observables are probes not only of 2-prong jet substructure but also of any grooming procedure applied to the jet.\footnote{See also \Ref{Larkoski:2015npa} for an example of an observable designed specifically to probe the grooming procedure by measuring non-global correlations, and \Ref{Dasgupta:2015yua} for an example of improving discrimination power by understanding the behavior of the grooming procedure.} 

\subsection{Power-Counting Analysis and Observable Phase Space}\label{sec:twoprong_PC}

The power counting for 2-prong discriminants follows straightforwardly from \Sec{sec:power_counting}, using the modes summarized in \Fig{fig:pics_jets} and \Tab{tab:pc_jets}.  Since the phase space is much simpler than in the 3-prong case, we can study the behavior of $\Mobsnobeta{2}$, $\Nobsnobeta{2}$, and $\Dobs{2}{\alpha,\beta}$ both before and after jet grooming.  

To begin, we consider the 1-prong background in \Tab{tab:pc} and power count the contributions to $\ecfnobeta{2}$ and $\ecfvarnobeta{v}{3}$ from every possible triplet of soft and collinear modes.  We do the same for the 2-prong signal in \Tab{tab:pc_ninja}, where we also have to consider collinear-soft modes, though we do not show the power-suppressed triplets for brevity.  These tables show that while the 3-point correlators have similar behavior for soft particles, they have different behavior for correlations among collinear particles (cf.\ the first row of \Tab{tab:pc} and the second and third row of \Tab{tab:pc_ninja}).  This is expected given the different number of pairwise angles in the definition of each  $\ecfvarnobeta{v}{3}$.  We discuss the consequences of this power counting for each of the proposed ratios in the following subsections.

\begin{table}[t]
\begin{center}
\begin{minipage}[t]{.4\linewidth}
\vspace{0pt}
\includegraphics[width=5cm]{figures/unresolved_jet.pdf} 
\end{minipage}%
\begin{minipage}[t]{.5\linewidth}
\vspace{15pt}
\begin{tabular}[b]{ccccc}
\hline
\hline
Modes&$\ecf{2}{\beta}$ &$\ecfvar{1}{3}{\beta}$&$\ecfvar{2}{3}{\beta}$&$\ecfvar{3}{3}{\beta}$ \\ 
\hline
$CCC$ & $\theta_{cc}^{\beta}$ &$\theta_{cc}^{\beta}$ & $\theta_{cc}^{2\beta}$& $\theta_{cc}^{3\beta}$ \\
$CCS$ & $z_s + \theta_{cc}^{\beta}$ & $z_s \theta_{cc}^{\beta}$ & $z_s \theta_{cc}^{\beta}$ &$z_s \theta_{cc}^{\beta}$ \\
$CSS$ & $z_s$ & $z_s^2$ & $z_s^2$ & $z_s^2$ \\
$SSS$ & $z_s^2$  & $z_s^3$ & $z_s^3$ & $z_s^3$\\
\hline
\hline
\end{tabular}
\end{minipage}
\end{center}
\caption{Parametric contributions to $\ecfnobeta{2}$ and the 3-point correlators, $\ecfvarnobeta{v}{3}$, in the case of a jet with 1-prong substructure.  The different contributions arise from correlations among soft ($S$) and collinear ($C$) radiation. 
}
\label{tab:pc}
\end{table}

\begin{table}[t]
\begin{center}
\begin{minipage}[t]{.4\linewidth}
\vspace{0pt}
\includegraphics[width=5cm]{figures/NINJA.pdf} 
\end{minipage}%
\begin{minipage}[t]{.5\linewidth}
\vspace{20pt}
\begin{tabular}[b]{ccccc}
\hline
\hline
Modes &$\ecf{2}{\beta}$&$\ecfvar{1}{3}{\beta}$&$\ecfvar{2}{3}{\beta}$&$\ecfvar{3}{3}{\beta}$ \\ 
\hline
$C_1C_2\,S$ &$\theta_{12}^{\beta}$ &$z_s \theta_{12}^{\beta}$ &$z_s \theta_{12}^{\beta}$ &$z_s \theta_{12}^{\beta}$  \\
$C_1C_2\, C$ & $\theta_{12}^{\beta}$ & $  \theta_{cc}^{\beta}$ & $\theta_{12}^{\beta} \theta_{cc}^{\beta}$ & $\theta_{12}^{2\beta} \theta_{cc}^{\beta}$  \\
$C_1C_2\,C_s$ & $\theta_{12}^{\beta}$ & $z_{cs}\theta_{12}^{\beta}$ & $z_{cs}\theta_{12}^{2\beta}$ &$z_{cs} \theta_{12}^{3\beta}$  \\
\hline
\hline
\end{tabular}
\end{minipage}
\end{center}
\caption{Same as \Fig{tab:pc}, but for a jet with a resolved 2-prong substructure.  The different contributions arise from correlations among soft ($S$), collinear ($C_i$), and collinear-soft ($C_s$) radiation. Power-suppressed contributions are not shown.
}
\label{tab:pc_ninja}
\end{table}

\subsubsection{$\Mobsnobeta{2}$}

The observable $\Mobsnobeta{2}$ is based on $\ecfvarnobeta{1}{3}$:
\begin{align}
\Mobs{2}{\beta}=\frac{\ecfvar{1}{3}{\beta}}{\ecfvar{1}{2}{\beta}}\,.
\end{align}
We first consider its behavior on 1- and 2-prong jets without grooming.  For 1-prong background jets from \Tab{tab:pc}, we have 
\begin{align}
\text{1-prong background (ungroomed):} \quad \ecf{2}{\beta} & \sim  z_s + \theta_{cc}^\beta  \,, \nn \\
\ecfvar{1}{3}{\beta} & \sim  z_s^2 +  \theta_{cc}^{\beta} \,.
\label{eq:1prongM2scaling}
\end{align}
This exhibits a non-trivial phase space with boundaries $\ecfvarnobeta{1}{3} \sim (\ecfnobeta{2})^2$ when the jet is dominated by soft radiation, and $\ecfvarnobeta{1}{3} \sim \ecfnobeta{2}$ when the jet is dominated by collinear radiation. For 2-prong signal jets from \Tab{tab:pc_ninja}, we have 
\begin{align}
\text{2-prong signal (ungroomed):} \quad \ecf{2}{\beta} & \sim  \theta_{12}^\beta \,, \nn \\
\ecfvar{1}{3}{\beta} &\sim z_{cs} \theta_{12}^{\beta}  +  \theta_{cc}^{\beta}  \,.
\end{align}
From the fact that $z_{cs}\ll 1$, and $\theta_{cc} \ll \theta_{12}$, one therefore finds the inequality $\ecfvarnobeta{1}{3} \ll \ecfnobeta{2}$. 

\begin{figure}
\begin{center}
\subfloat[]{\label{fig:1D2_ps}
\includegraphics[width=6.5cm]{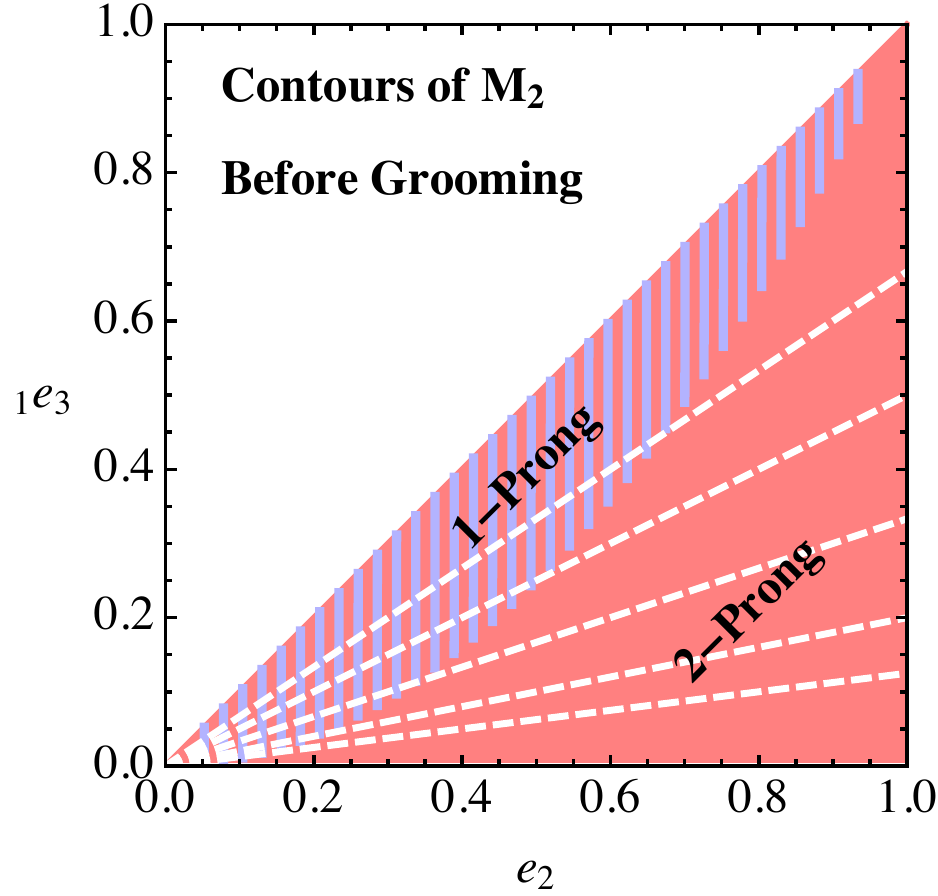}
}\qquad
\subfloat[]{\label{fig:1D2_ps_groomed}
\includegraphics[width=6.5cm]{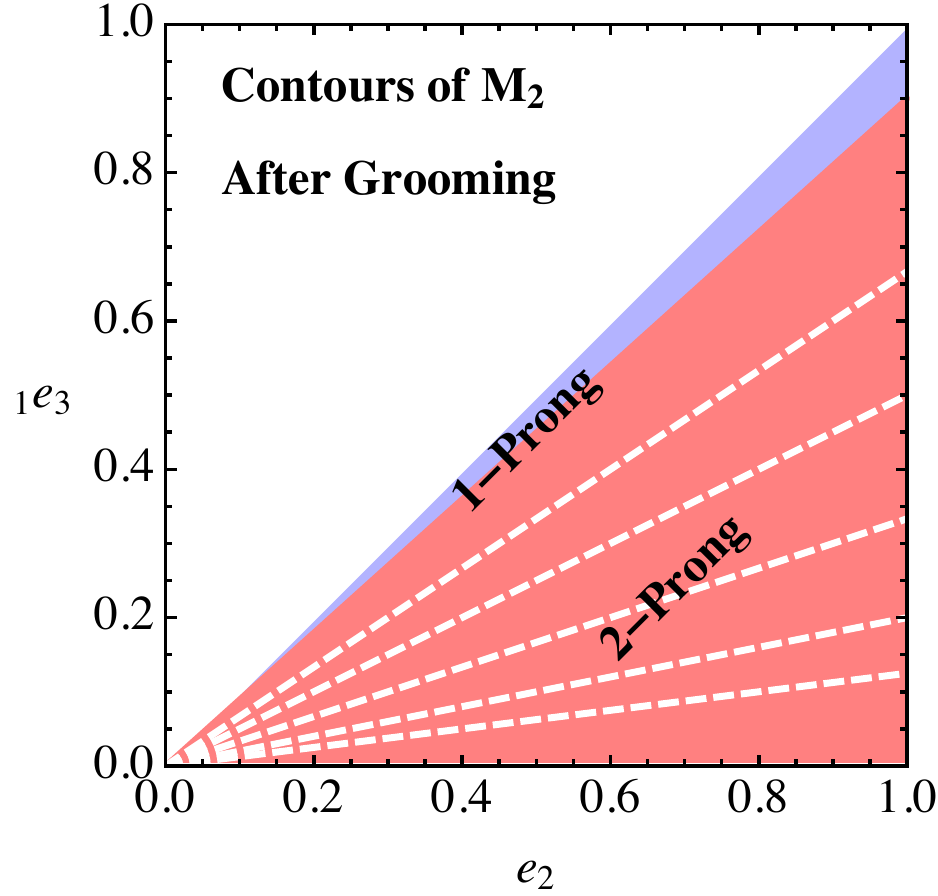}
}
\end{center}
\caption{Parametric phase space for the $\Mobsnobeta{2}$ observable (a) before grooming and (b) after grooming. The grooming procedures removes wide-angle soft radiation, pushing 1-prong jets to the upper boundary of the phase space. 
}
\label{fig:1D2_ps_global}
\end{figure}

The phase space for $\Mobsnobeta{2}$ is shown in \Fig{fig:1D2_ps}.  This power-counting analysis demonstrates that before any grooming has been applied, there is considerable overlap between the parametric phase space regions occupied by 1- and 2-prong jets.  Therefore, $\Mobsnobeta{2}$ has limited discrimination power on ungroomed jets.  The power-counting analysis also makes clear why $\Mobsnobeta{2}$ performs so poorly:  1-prong jets are dominated by soft radiation with scaling $\ecfvarnobeta{1}{3} \sim (\ecfnobeta{2})^2$, which overlaps with the 2-prong signal region with $\ecfvarnobeta{1}{3} \ll \ecfnobeta{2}$. The fact that this overlap is caused only by soft radiation also suggests that it can be eliminated by applying a jet grooming procedure to remove soft radiation. 

In \Fig{fig:1D2_ps_groomed}, we show the phase space for $\Mobsnobeta{2}$ after grooming.  Soft drop removes the $z_s$ contributions from \Eq{eq:1prongM2scaling}, which pushes 1-prong background jets to the upper boundary of the phase space with $\ecfvarnobeta{1}{3} \sim \ecfnobeta{2}$.  By contrast, the parametric scaling of the signal jets is unaffected by the soft drop procedure.\footnote{As stated at the end of \Sec{sec:power_counting}, for simplicity we do not power count the grooming parameter $z_{\rm cut}$.  It is well understood how to properly incorporate $z_{\rm cut}$ into the power-counting analysis (see e.g.~\cite{Frye:2016aiz,Frye:2016okc}), but this has a negligible impact for understanding the qualitative behavior of 2-prong discriminants.}  This yields a triangular phase space that resembles the case of $\Nsubnobeta{2,1}$ in \Fig{fig:Nsub_ps_review}, where 1-prong background jets live on the upper boundary and 2-prong signal jets live in the bulk.  Perhaps counterintuitively, the soft drop procedure pushes the background to larger values of $\Mobsnobeta{2}$, achieving better discrimination power.

The $\Mobsnobeta{2}$ observable therefore provides an interesting example of a discriminant that only performs well after grooming.  It emphasizes the parametric effect that grooming procedures can have on radiation within a jet, beyond simply removing jet contamination.  For this reason, we expect precision calculations of the $\Mobsnobeta{2}$ distribution to provide useful insights into the behavior of such grooming procedures.

\subsubsection{$\Nobsnobeta{2}$}

\begin{figure}
\begin{center}
\subfloat[]{\label{fig:2D2_ps}
\includegraphics[width=6.5cm]{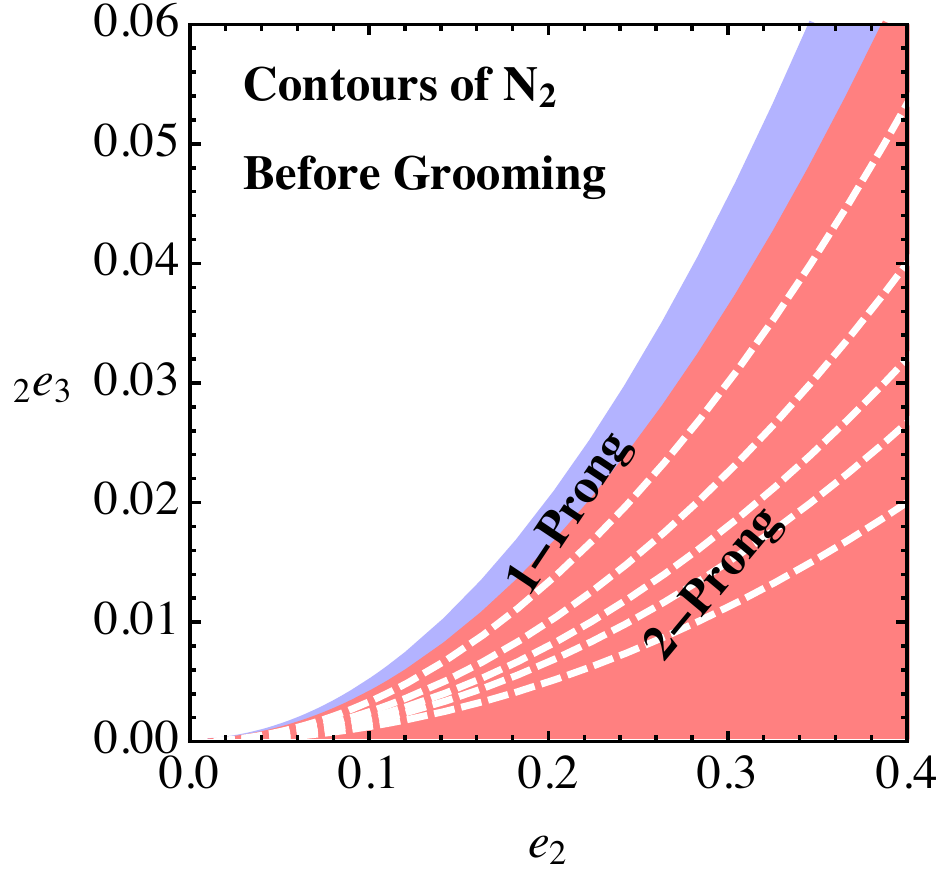}
}\qquad
\subfloat[]{\label{fig:2D2_ps_groomed}
\includegraphics[width=6.5cm]{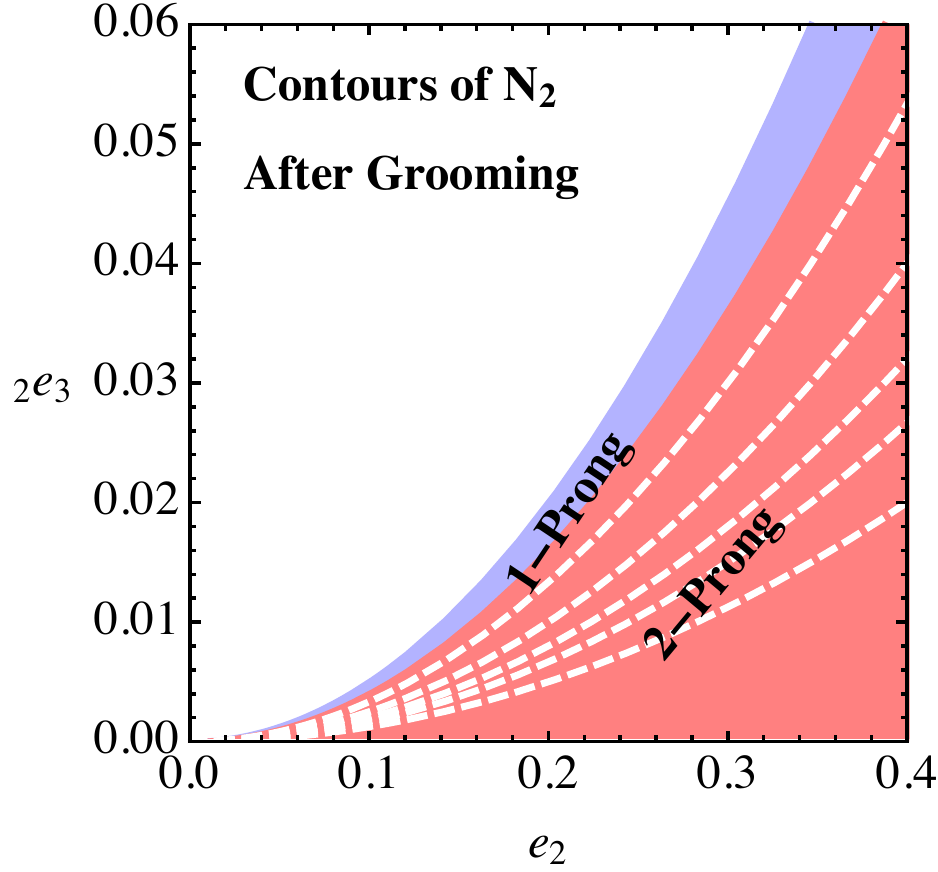}
}
\end{center}
\caption{Same as \Fig{fig:1D2_ps_global} but for $\Nobsnobeta{2}$. The grooming procedure does not modify the scaling of the phase space, so (a) and (b) are identical as far as power counting is concerned. Therefore, the $\Nobsnobeta{2}$ observable exhibits good discrimination power both before and after grooming is applied. 
}
\label{fig:2D2_ps_global}
\end{figure}

The observable $\Nobsnobeta{2}$ is based on $\ecfvarnobeta{2}{3}$,
\begin{align}
\Nobs{2}{\beta}=\frac{\ecfvar{2}{3}{\beta}}{ (\ecfvar{1}{2}{\beta})^2 }\,.
\end{align}
The power-counting argument for $\Nobsnobeta{2}$ closely parallels $\Mobsnobeta{2}$.  We will see that the phase space for $\Nobsnobeta{2}$ is parametrically unmodified by the grooming procedure, making it perform well on both groomed and ungroomed jets.

We again begin by analyzing the parametric behavior of the observable on ungroomed jets.  Using  \Tab{tab:pc} for 1-prong background jets, we find
\begin{align}\label{eq:N2_singleprong}
\text{1-prong background (ungroomed):} \quad \ecf{2}{\beta} & \sim z_s +  \theta_{cc}^\beta  \,, \nn \\
\ecfvar{2}{3}{\beta} & \sim z_s^2 + z_s \theta_{cc}^\beta   + \theta_{cc}^{2\beta} \,.
\end{align}
In contrast to $\Mobsnobeta{2}$, the 1-prong background jets exhibit only a single scaling, $\ecfvarnobeta{2}{3} \sim (\ecfnobeta{2})^2$, for jets dominated by either soft or collinear radiation.  Using  \Tab{tab:pc_ninja} for 2-prong signal jets, we find 
\begin{align}\label{eq:N2_twoprong}
\text{2-prong signal (ungroomed):} \quad \ecf{2}{\beta} & \sim  \theta_{12}^\beta \,, \nn \\
\ecfvar{2}{3}{\beta} & \sim z_s \theta_{12}^{\beta} + z_{cs} \theta_{12}^{2\beta} + \theta_{cc}^{\beta} \theta_{12}^{\beta}  \,.
\end{align}
Signal jets satisfy the inequality $\ecfvarnobeta{2}{3} \ll (\ecfnobeta{2})^2$, explaining the definition of the $N_2$ observable.  The phase space before grooming is summarized  in \Fig{fig:2D2_ps}, where there is clear separation between 1-prong background jets, which live on the upper boundary of the phase space, and 2-prong signal jets, which live in the bulk of the phase space, again resembling the case of $\Nsubnobeta{2,1}$ in \Fig{fig:Nsub_ps_review}.

Because the 1-prong background jets have a single scaling, removing $z_s$ from \Eq{eq:N2_singleprong} has no effect on the parametric phase space.  Similarly, removing $z_s$ from \Eq{eq:N2_twoprong} does not change the parametrics of the 2-prong signal.  Therefore, $\Nobsnobeta{2}$ behaves more similarly to other 2-prong discriminants in the literature, since its discrimination power does not come entirely from the grooming procedure.   The power-counting analysis also suggests that $\Nobsnobeta{2}$ should be a powerful 2-prong discriminant both before and after grooming is applied; this will be verified in the parton shower studies below.

It is also interesting to contrast the $\Nobsnobeta{2}$ phase space in \Fig{fig:2D2_ps} with that of $\Dobsnobeta{2}$ in \Fig{fig:D2_ps_review}.  For $\Dobsnobeta{2}$, the background jets are bounded by two different scaling behaviors:
\begin{equation}
(e_2^{(\beta)})^3 < e_3^{(\beta)} < (e_2^{(\beta)})^2.
\end{equation}
For $\Nobsnobeta{2}$, by contrast, the background jets exhibit a single scaling and therefore live entirely on the boundary of phase space:
\begin{equation}
_2e_3^{(\beta)} \sim (e_2^{(\beta)})^2.
\end{equation}
Since this boundary is purely geometric, the $\Nobsnobeta{2}$ distributions are remarkably insensitive to the mass or $p_T$ of the jet, even before grooming is applied. 

Just as $N_3$ is related to $\Nsubnobeta{3,2}$ (see \Sec{sec:tops_makeobs}), $N_2$ behaves parametrically like $\Nsubnobeta{2,1}$ in the resolved limit.  The power-counting analysis proceeds identically as for $N_3$ and will not be repeated here; see \App{app:Nsub_Ni} for the general argument relating $N_i$ to $\Nsubnobeta{i,i-1}$.  We want to emphasize again that, in analogy to \Fig{fig:NINJA_obs}, $N_2$ exhibits $\Nsubnobeta{2,1}$-like behavior without reference to any axes within the jet.  It therefore does not exhibit the axes pathologies that arise for $N$-subjettiness in the limit of unresolved substructure, and $N_2$ can therefore be expected to have improved performance compared to $\Nsubnobeta{2,1}$, particularly at high efficiencies. 

\subsubsection{$\Dobs{2}{1,2}$}

Our final example of a 2-prong discriminant is based on $\ecfvarnobeta{3}{3} = \ecfnobeta{3}$, where we reconsider the $\Dobsnobeta{2}$ observable with two distinct angular exponents,
\begin{equation}\label{eq:D2_def_genangle}
\Dobs{2}{\alpha, \beta} \equiv \frac{\ecf{3}{\alpha}}{\left(\ecf{2}{\beta}\right)^{3\alpha/\beta}} \,.
\end{equation}
The case of $\alpha=\beta$ was first defined in \Ref{Larkoski:2014gra} and analytically calculated in \Ref{Larkoski:2015kga}.  While the phase space for $\Dobs{2}{\alpha, \beta}$ was discussed in detail in \Ref{Larkoski:2015kga}, we focus on the impact that $\alpha\not=\beta$ has on groomed jet discrimination.

\begin{figure}
\begin{center}
\subfloat[]{\label{fig:3D2_ps}
\includegraphics[width=6.5cm]{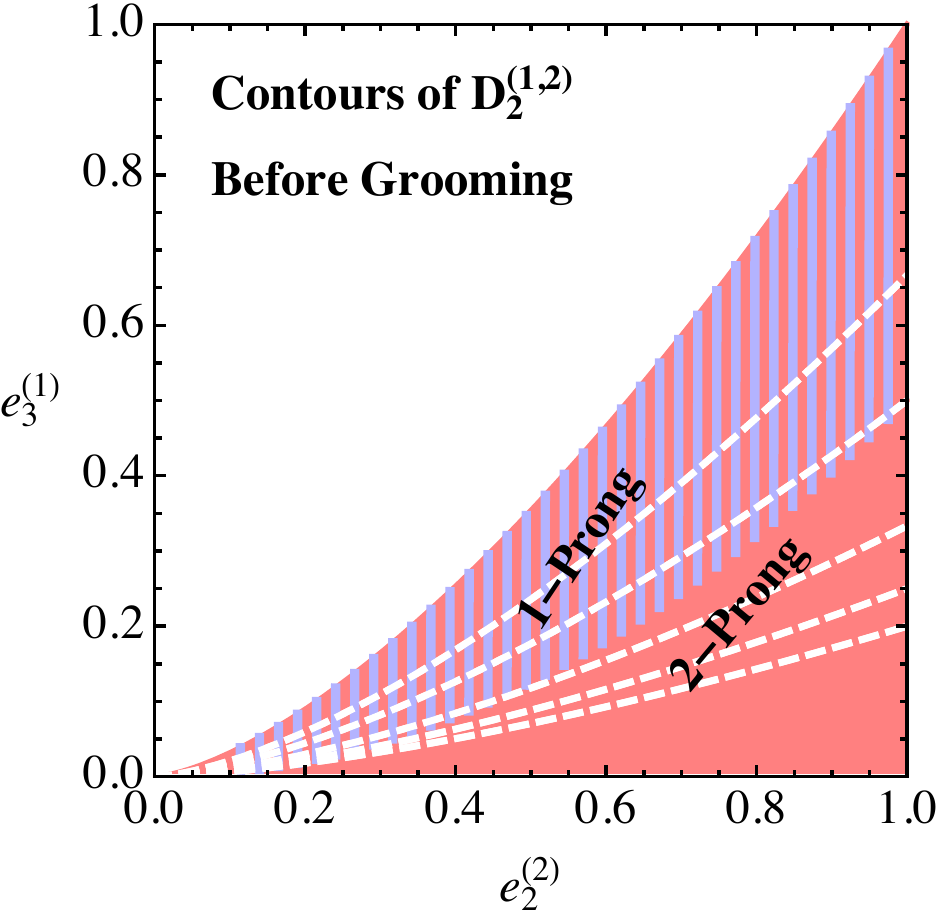}
}\qquad
\subfloat[]{\label{fig:3D2_ps_groomed}
\includegraphics[width=6.5cm]{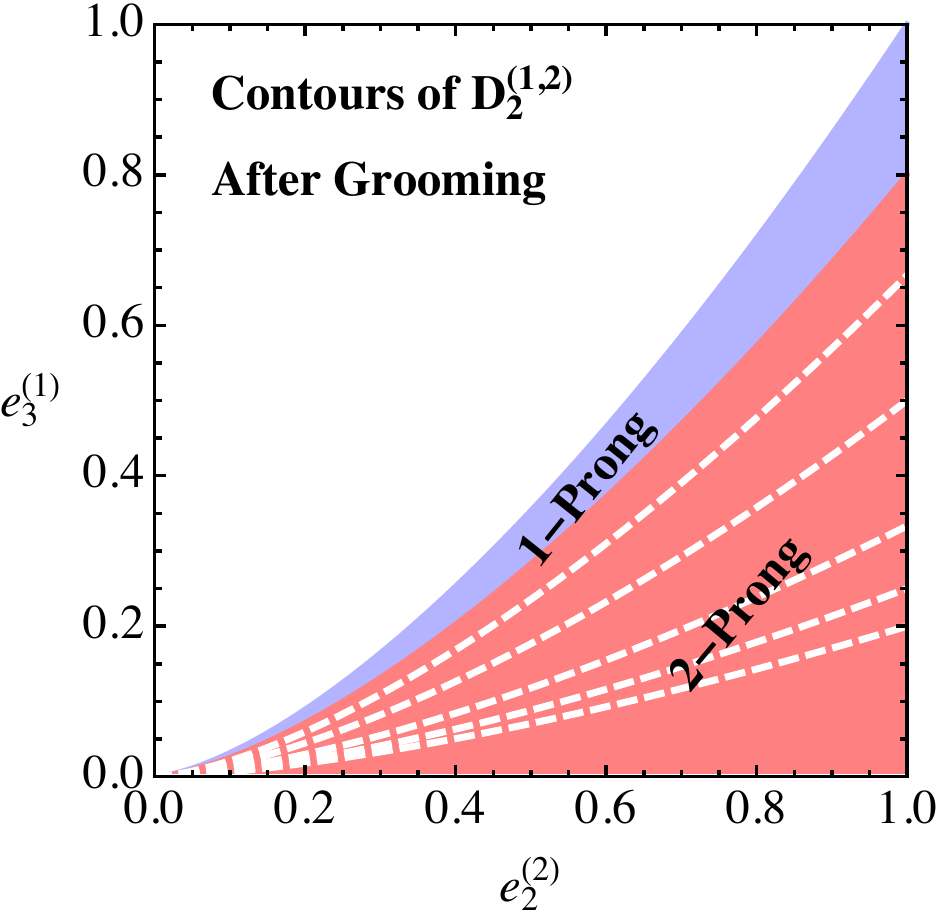}
}
\end{center}
\caption{Same as \Fig{fig:1D2_ps_global} but for $\Dobs{2}{1,2}$.  Only after grooming is applied can the 1- and 2-prong regions of phase space be separated. 
}
\label{fig:3D2_ps_global}
\end{figure}

With distinct angular exponents $\alpha$ and $\beta$, 1-prong background jets exhibit the scaling 
\begin{align}
\text{1-prong background (ungroomed):} \quad \ecf{2}{\beta} & \sim z_s +  \theta_{cc}^\beta  \,, \nn \\
\ecf{3}{\alpha} & \sim z_s^2 + z_s \theta_{cc}^\alpha  + \theta_{cc}^{3\alpha} \,.
\end{align}
The background therefore occupies a non-trivial phase space with boundaries $\ecf{3}{\alpha} \sim (\ecf{2}{\beta})^2$, when the jet is dominated by soft radiation, and $\ecf{3}{\alpha} \sim (\ecf{2}{\beta})^{3\alpha/\beta}$, when the jet is dominated by collinear radiation. The 2-prong signal has the parametric scaling 
\begin{align}
\text{2-prong signal (ungroomed):} \quad \ecf{2}{\beta} & \sim  \theta_{12}^\beta \,, \nn \\
\ecf{3}{\alpha} & \sim z_s \theta_{12}^{\alpha} +z_{cs} \theta_{12}^{3\alpha} + \theta_{12}^{2\alpha} \theta_{cc}^{\alpha}\,,
\end{align}
from which one can derive the relation $\ecf{3}{\alpha} \ll (\ecf{2}{\beta} )^{3\alpha/\beta}$.  This demonstrates that the definition of $\Dobs{2}{\alpha, \beta}$ in \Eq{eq:D2_def_genangle} is indeed appropriate for 2-prong substructure, confirming the expectation from boost invariance (see~\Eq{eq:boost}).

As discussed in \Ref{Larkoski:2015kga} for ungroomed jets, the observable $\Dobs{2}{\alpha, \beta} $ only provides good discrimination between 1-prong and 2-prong jets for 
\begin{align}\label{eq:constraint_alpha}
3\alpha >2\beta\,.
\end{align} 
When this relation is violated, the phase space regions for signal and background jets overlap.  This is shown in \Fig{fig:3D2_ps}, where contours of $\Dobs{2}{\alpha, \beta}$ cannot separate the 1- and 2-prong regions when \Eq{eq:constraint_alpha} is violated.  After a grooming procedure is applied, though, the overlapping phase space region is removed, as shown schematically in \Fig{fig:3D2_ps_groomed}.  Now the constraint in \Eq{eq:constraint_alpha} no longer applies, and the angular exponents can be chosen with a particular focus on discrimination power on groomed jets.

The choice of $(\alpha,\beta)$ exponents could be tuned to optimize performance, but we advocate that $\alpha=1$, $\beta=2$ is a natural choice for groomed 2-prong discrimination.  This choice explicitly violates \Eq{eq:constraint_alpha}, so $\Dobs{2}{1, 2}$ can only have good performance after grooming.  The choice of $\beta=2$ is motivated by the relation
\begin{align}\label{eq:e2_mass_relation}
\ecf{2}{2} \simeq \frac{p_{TJ}^2}{m_J^2}\,,
\end{align}
such that a cut on the jet mass, or $p_T$, is effectively a cut on $\ecf{2}{2}$.  Without grooming, one would typically take $\alpha = 2$, but with grooming, one can lower the angular exponent $\alpha$ to $1$ to more directly probe collinear emissions.  Importantly, by considering the observable with separate $\alpha$ and $\beta$ exponents, we are able to satisfy both the requirement that it behaves sensibly under a mass cut, as well as improve its sensitivity to collinear emissions.  This is not possible with the $\alpha=\beta$ version of the $\Dobsnobeta{2}$ observable, and indeed, $\Dobs{2}{1, 2}$ leads to improved discrimination power on groomed jets.  From an analytic perspective, the choice of $\alpha=1$, $\beta=2$ simplifies calculations, hopefully facilitating precision calculations of $\Dobs{2}{1, 2}$ on groomed jets at the LHC.

\subsection{Performance in Parton Showers}\label{sec:twoprong_MC}

We now perform a parton shower study to verify the predictions of the above power-counting analysis.   It is useful to briefly summarize our robust predictions regarding the behavior of $\Mobsnobeta{2}$, $\Nobsnobeta{2}$, and $\Dobs{2}{1,2}$ as boosted 2-prong taggers:
\begin{itemize}
\item The $\Mobsnobeta{2}$ observable should provide little discrimination power before grooming, but will act as a powerful discriminant after the removal of wide-angle soft radiation.
\item The $\Nobsnobeta{2}$ observable will act as a powerful discriminant both before and after grooming, matching the behavior of $\Nsubnobeta{2,1}$ in the resolved limit. 
\item The $\Dobs{2}{1,2}$ observable will behave similarly to $\Mobsnobeta{2}$, providing good discrimination power only after grooming has been applied.
\end{itemize}
These predictions rely only on parametric scalings and are therefore independent of the implementation details of the perturbative parton shower or the hadronization model. For conciseness, we only  show results generated with \pythia{8.219}, though we used \vincia{2.0.01} \cite{Giele:2007di,Giele:2011cb,GehrmannDeRidder:2011dm,Ritzmann:2012ca,Hartgring:2013jma,Larkoski:2013yi,Fischer:2016vfv} to check that the same results could be obtained with an alternative perturbative shower.  We have not yet studied hadronization uncertainties, but we expect them to be small, particularly for groomed jets.

\begin{figure}
\begin{center}
\subfloat[]{\label{fig:M2_gluonplot}
\includegraphics[width=6.5cm]{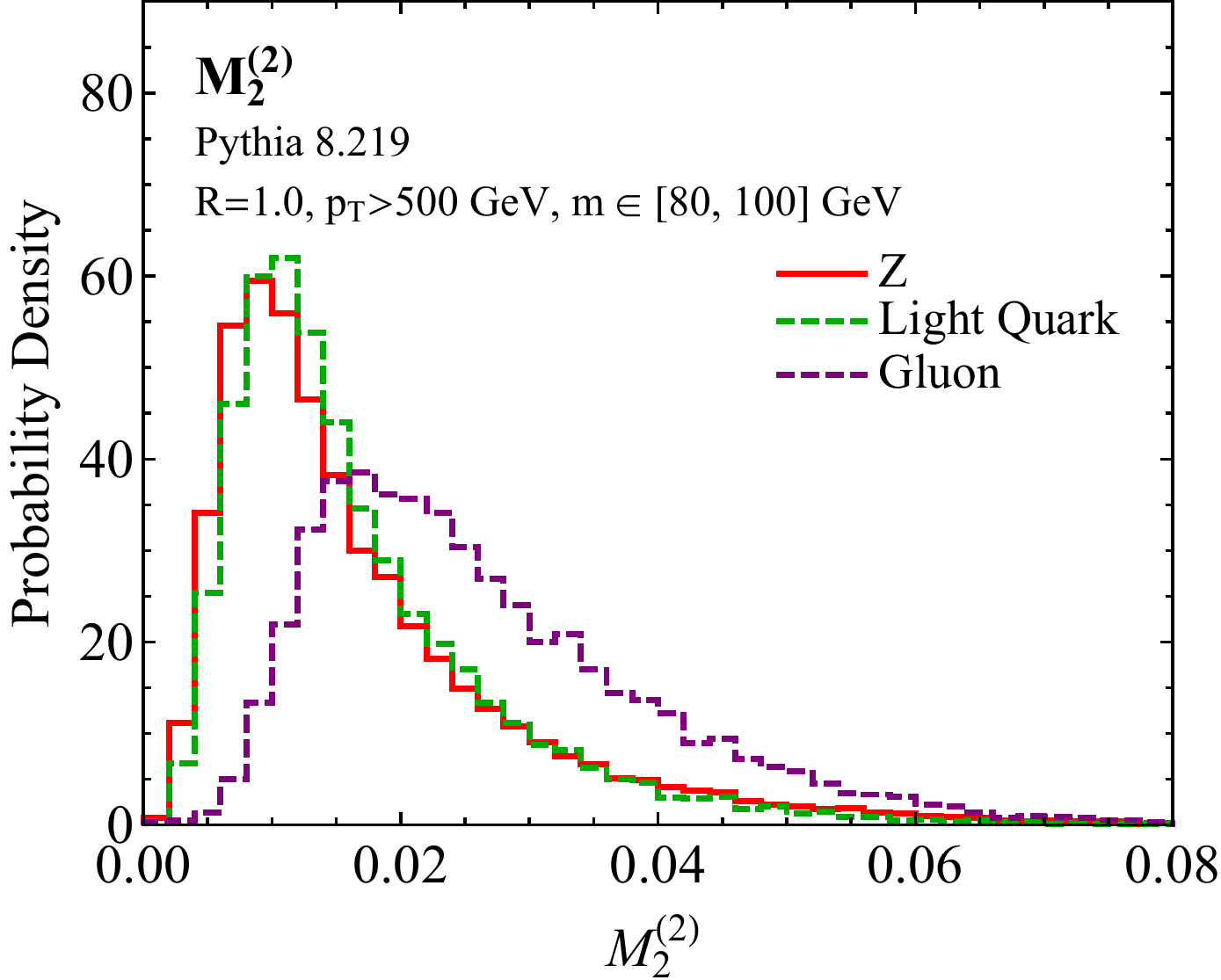}
}\qquad
\subfloat[]{\label{fig:M2_gluonplot_groomed}
\includegraphics[width=6.5cm]{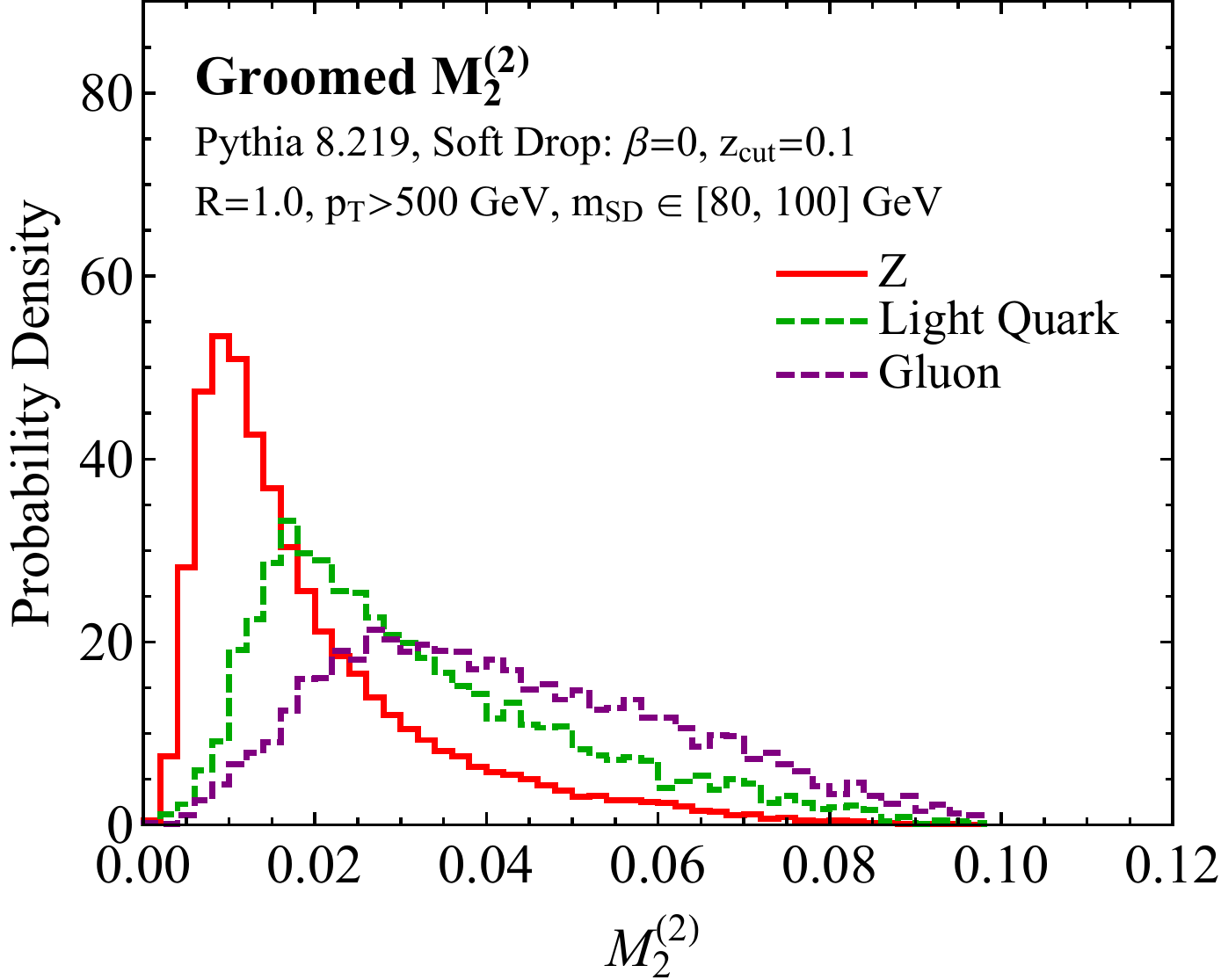}
}\nonumber \\
\subfloat[]{\label{fig:N2_gluonplot}
\includegraphics[width=6.5cm]{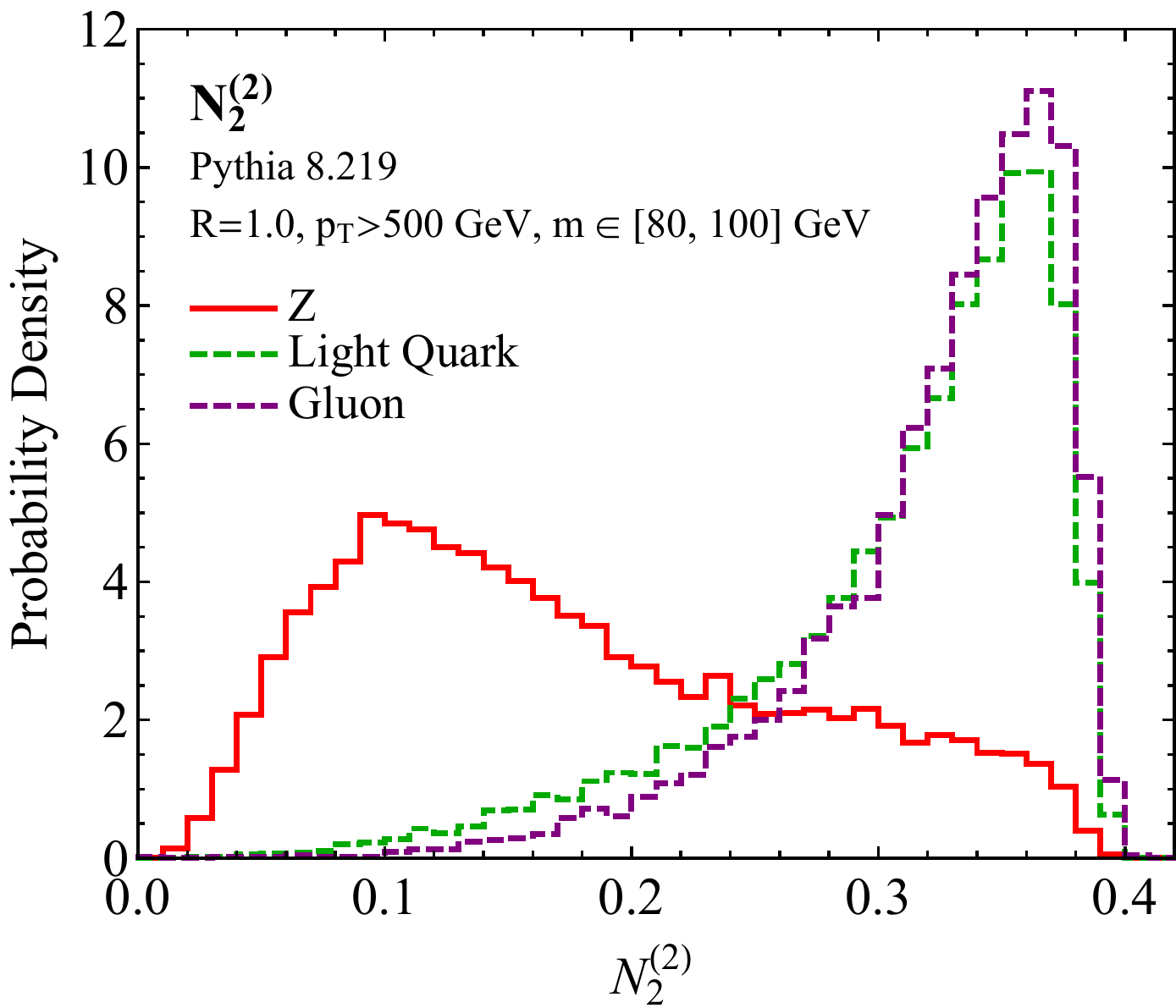}
}\qquad
\subfloat[]{\label{fig:N2_gluonplot_groomed}
\includegraphics[width=6.5cm]{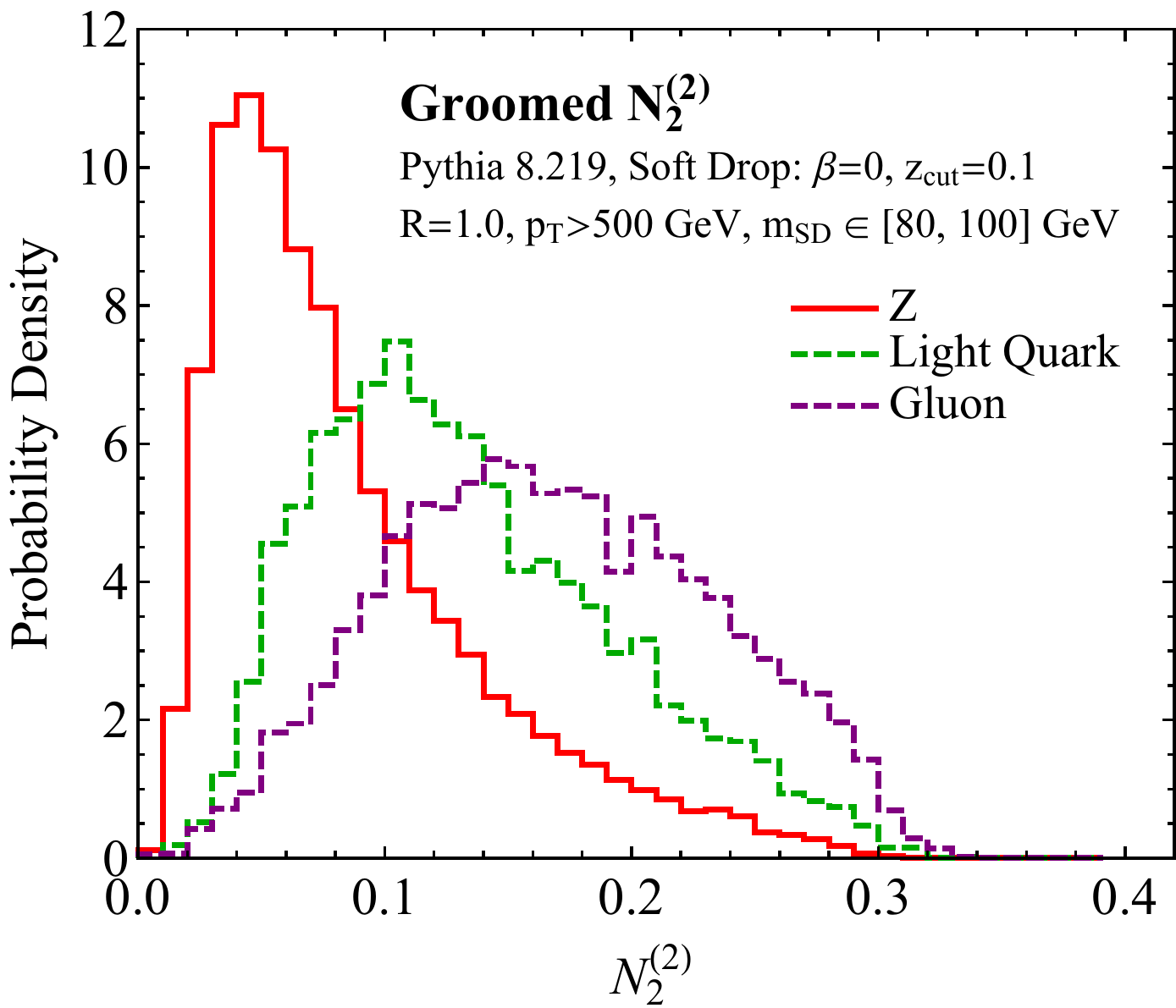}
}\nonumber \\
\subfloat[]{\label{fig:D2_gluonplot}
\includegraphics[width=6.5cm]{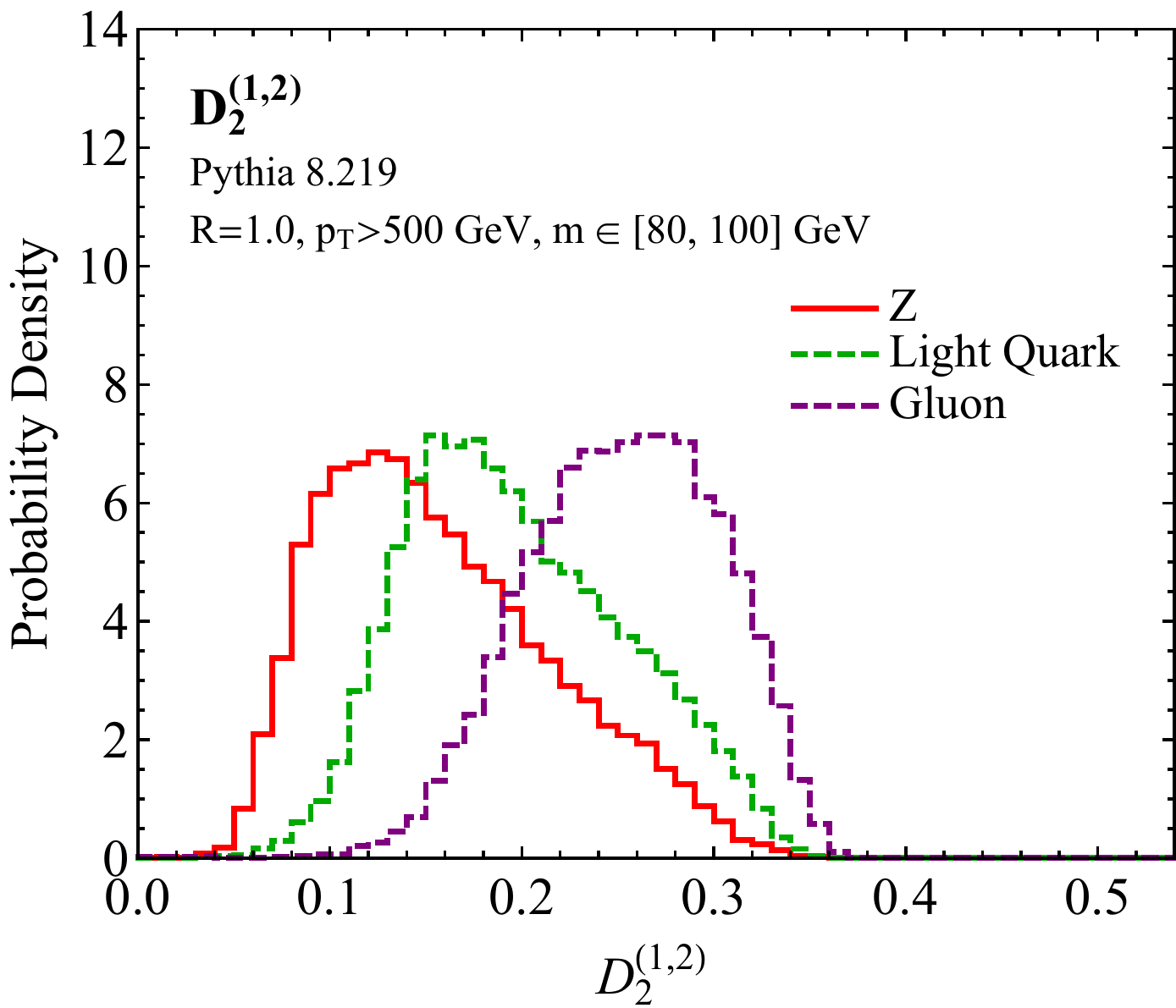}
}\qquad
\subfloat[]{\label{fig:D2_gluonplot_groomed}
\includegraphics[width=6.5cm]{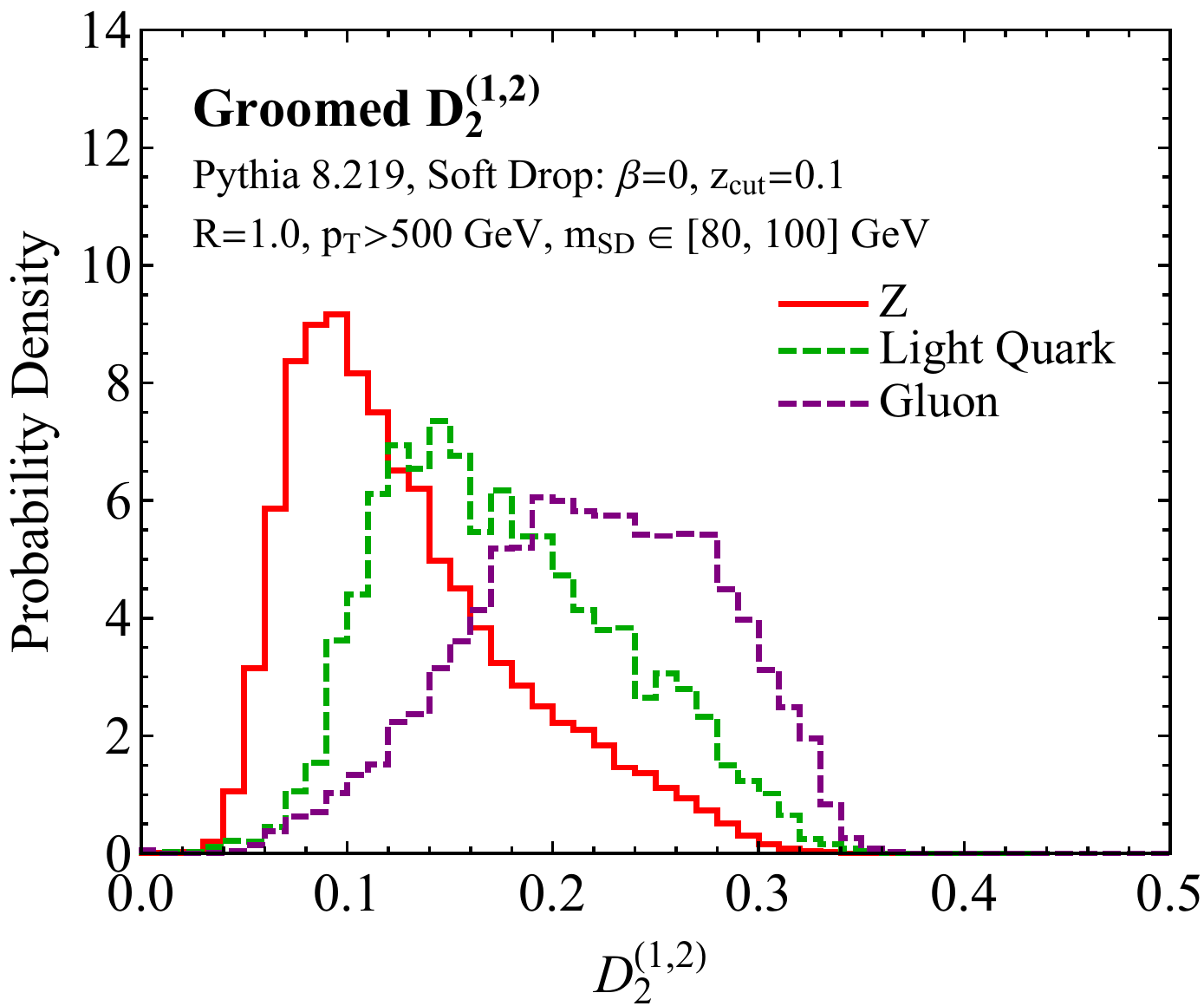}
}
\end{center}
\caption{Distributions for (top row) $\Mobsnobeta{2}$, (middle row) $\Nobsnobeta{2}$, and (bottom row) $\Dobs{2}{1,2}$ measured on boosted $Z$ and quark/gluon jets. The results are shown (left column) before grooming and (right column) after grooming.
}
\label{fig:2prong_obs_gluon}
\end{figure}

To verify these power-counting predictions, we use the same analysis and generation strategy as \Sec{sec:N3_MC}, again using a jet radius of $R = 1.0$. We generate background QCD jets from $pp\to Zj$ events, where we consider separately the cases of $j=g$ (gluon) and $j=u,d,s$ (light quark), letting the $Z$ decay leptonically to avoid additional hadronic activity. The 2-prong signal of boosted $Z$ bosons are generated from $pp\to ZZ$ events, with one $Z$ decaying leptonically, and the other to light quarks, $q=u,d,s$. We do not address in this paper the issue of sample dependence and the impact of color connections to the rest of the event.  While it would be interesting to compare the discrimination power of $\Nobsnobeta{2}$ against the more-prevalent $pp\to jj$ background, we expect the conclusions from $pp\to Zj$ to be robust, especially after grooming has been applied. 

For concreteness, we always set the angular exponent in the energy correlator to $\beta=2$, such that a mass cut directly corresponds to a cut on the denominator of the observable, see \Eq{eq:e2_mass_relation}.  While this is a nice theoretical feature, it is by no means necessary, and the value of $\beta$ could be optimized for experimental performance.  To focus on the phase space where tagging performance actually matters, we place a cut of $m \in [80,100]~$GeV for all of the ungroomed distributions and a cut of $m_{\rm SD} \in [80,100]~$GeV for all of the groomed distributions.  We only present distributions with a cut of $p_T > 500$ GeV, though other $p_T$ ranges exhibit similar behaviors.

In \Fig{fig:2prong_obs_gluon}, we show normalized distributions of $\Mobsnobeta{2}$, $\Nobsnobeta{2}$, and $\Dobs{2}{1,2}$ before and after soft drop grooming.  Despite all being derived from 3-point correlators, they exhibit rather different behaviors.  As expected, $\Mobsnobeta{2}$ is a poor discriminant before grooming is applied; amusingly, the distributions of the $Z$ boson signal and quark jet background are essentially identical.  As predicted by the power-counting analysis, the soft drop grooming procedure pushes the background $\Mobsnobeta{2}$ distributions to larger values while leaving the signal distribution largely unmodified.  We are not aware of another substructure discriminant with such a dramatic shift in behavior after jet grooming.

Turning to $\Nobsnobeta{2}$, it exhibits good discrimination power both before and after grooming is applied, even though the shapes of the distributions are substantially modified by grooming.  Before grooming, the $\Nobsnobeta{2}$ distribution exhibits a sharp edge at its upper boundary.  This arises because 1-prong background jets have a single parametric scaling and are therefore compressed along the upper boundary of the phase space (see \Fig{fig:2D2_ps_global}).  After grooming, $\Nobsnobeta{2}$ remains a powerful discriminant, as the phase space is parametrically unchanged by the grooming procedure.  As expected, the peak values of the distributions decrease as soft radiation is groomed away, but the range spanned by the distribution remains approximately constant.  This highlights the fact that parametric arguments give robust predictions about the boundaries of phase space but not the specific shapes of the distributions.  In \App{app:Nsub_Ni}, we also verify that $\Nobsnobeta{2}$ and $\Nsubnobeta{2,1}$ exhibit the same parametric behaviors in the resolved limit.

Finally, the $\Dobs{2}{1,2}$ observable, while only a fair discriminant before grooming, exhibits good discrimination power after soft drop is applied. Therefore, we have seen that all of the power-counting predictions are observed in the parton shower generators, suggesting that parametric scalings dominate the behavior of these observables, at least for the purposes of 2-prong substructure tagging. From \Fig{fig:2prong_obs_gluon}, we see that some of the observables behave quite differently for the quark and gluon samples.  We revisit the possibilities of using $\ecfvarnobeta{v}{3}$ for quark/gluon discrimination in \Sec{sec:qvsg}, where we introduce the $U_2$ observable, which is based on $\ecfvarnobeta{1}{3}$, similar to $\Mobsnobeta{2}$.

\begin{figure}
\begin{center}
\subfloat[]{\label{fig:nD2_ps_quark_pythia}
\includegraphics[width=6.5cm]{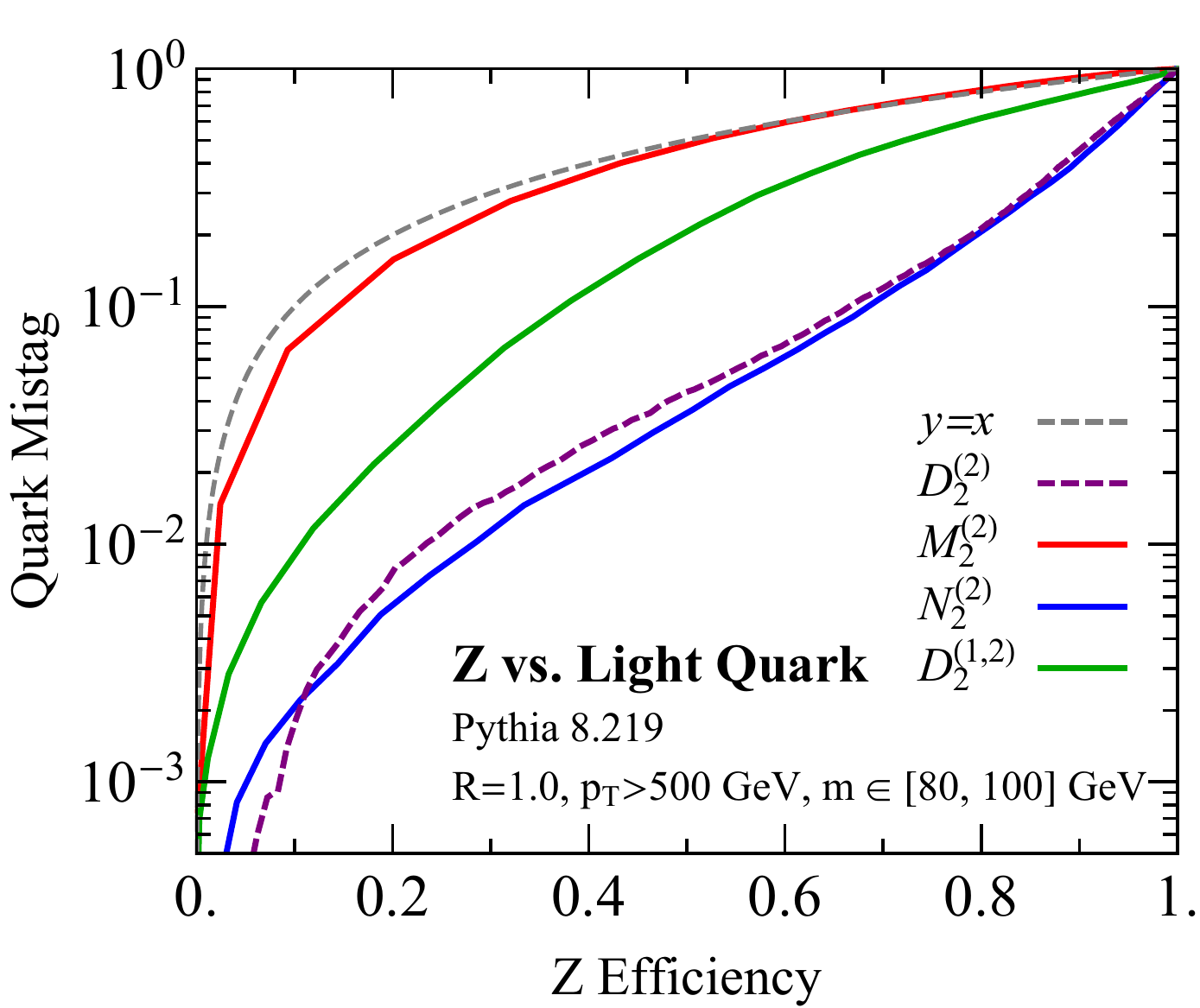}
}\qquad
\subfloat[]{\label{fig:nD2_ps_quark_pythia_SD}
\includegraphics[width=6.5cm]{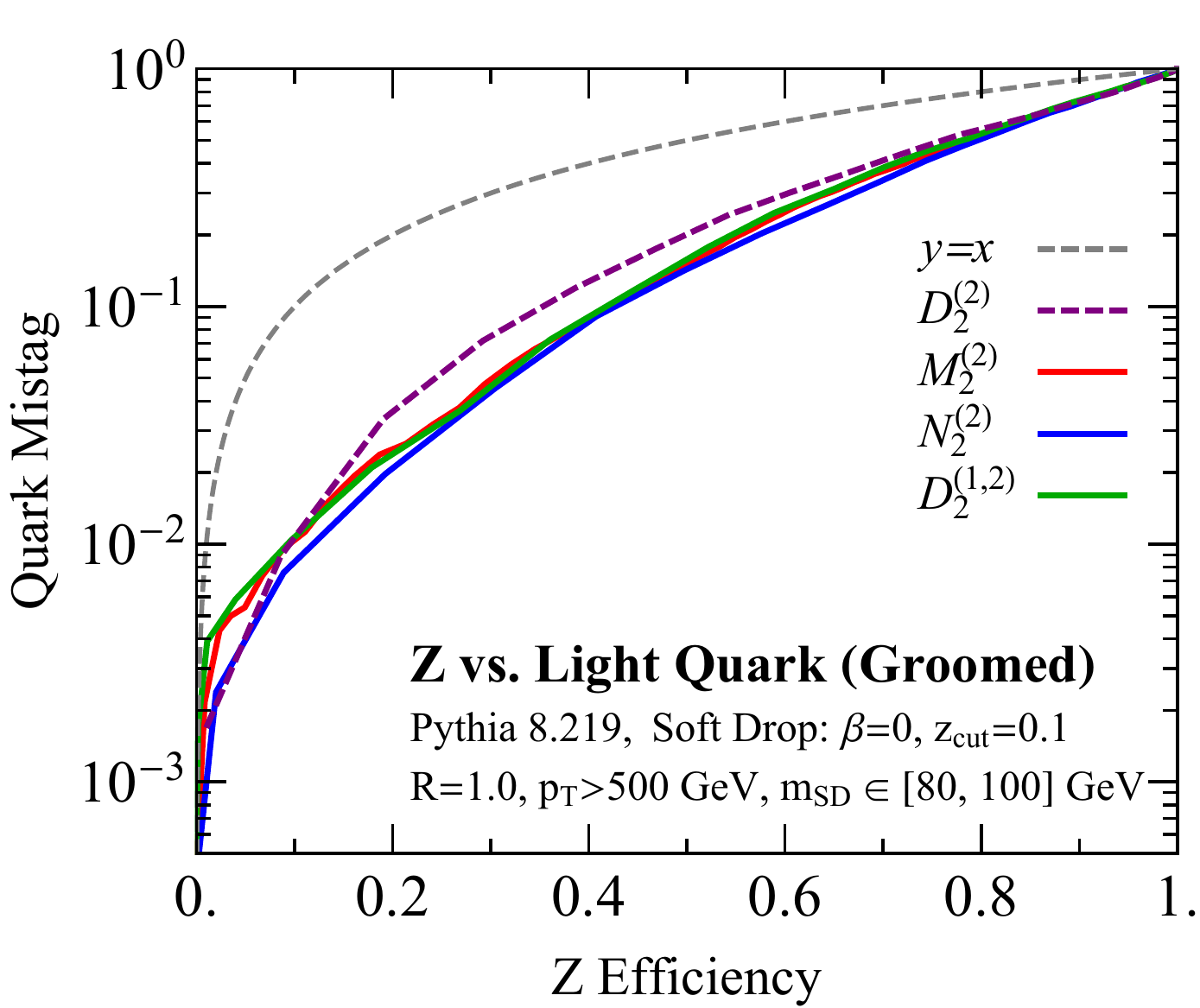}
}\qquad
\subfloat[]{\label{fig:nD2_ps_gluon_pythia}
\includegraphics[width=6.5cm]{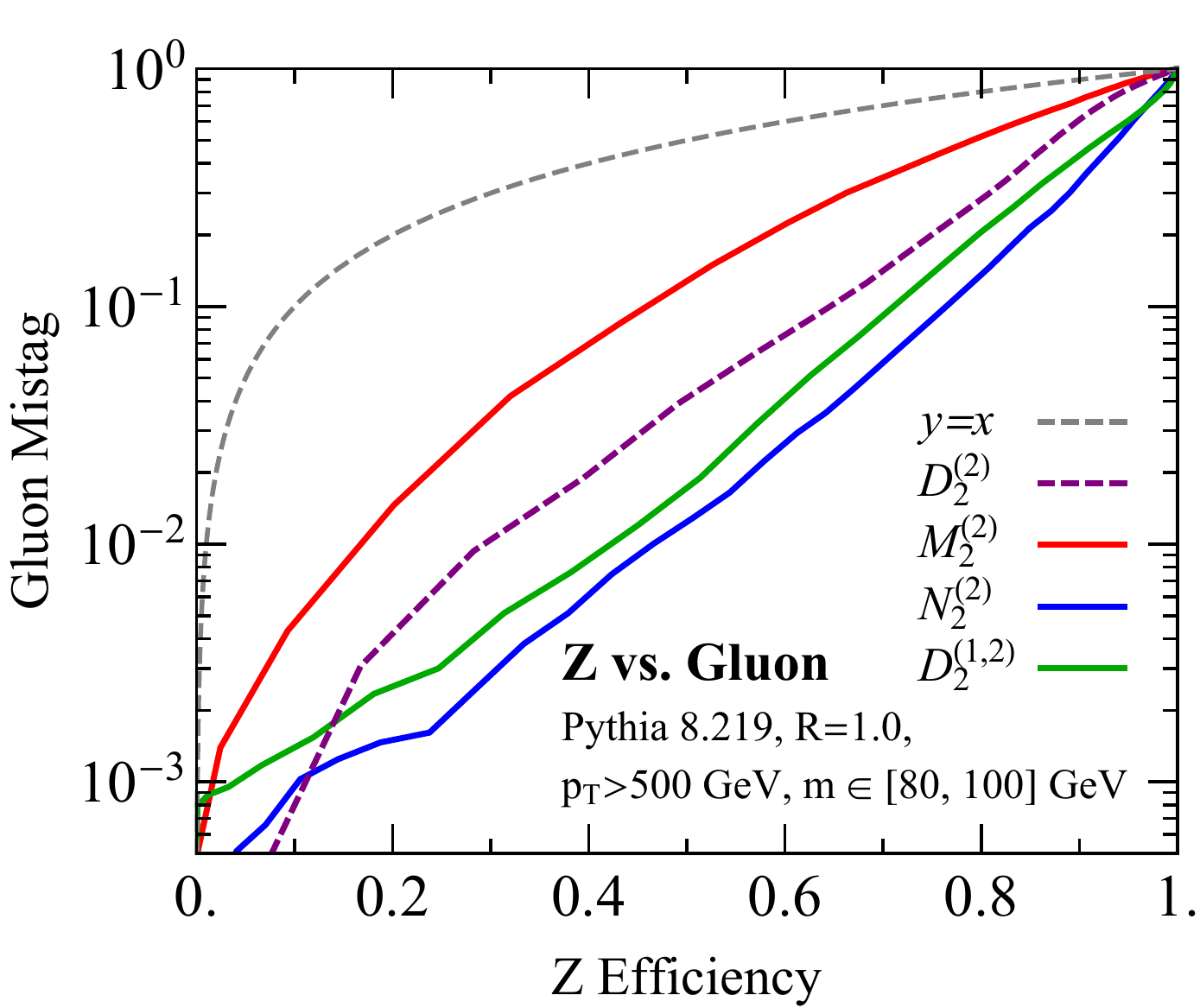}
}\qquad
\subfloat[]{\label{fig:nD2_ps_gluon_pythia_SD}
\includegraphics[width=6.5cm]{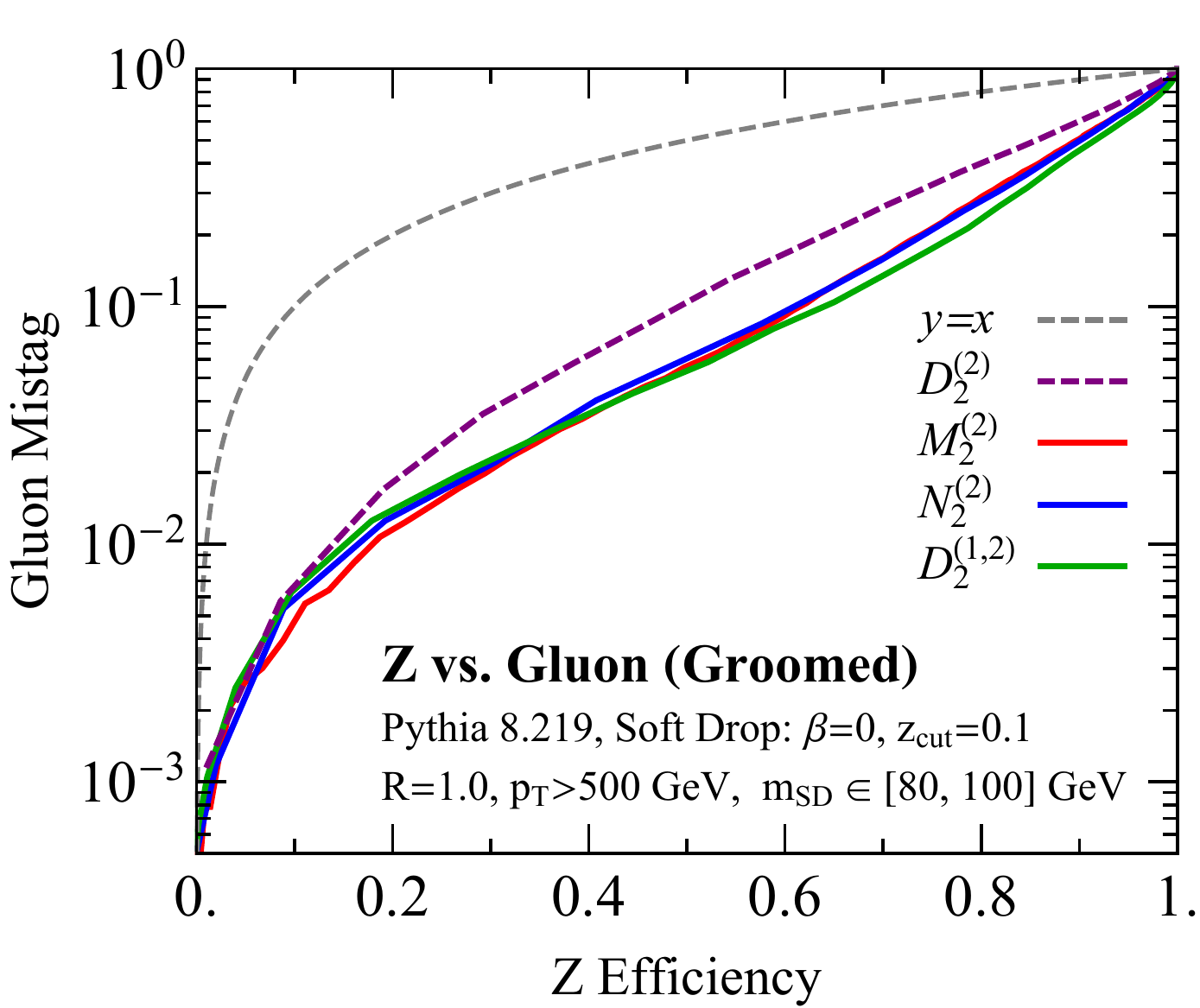}
}
\end{center}
\caption{ROC curves for boosted $Z$ boson (left column) before grooming and (right column) after grooming.  The discrimination power is shown against (top row) quark jets and (bottom row) gluon jets.  As a point of reference, we show the ROC curve for $D_2^{(2)}$, which is currently used by the ATLAS experiment, in dashed purple.  As predicted by power counting, the application of grooming greatly modifies the relative performance of the different observables. Note that an ungroomed mass cut is applied in the left column, while a groomed mass cut is applied  in the right column.  Efficiencies from these mass cuts are given in \Eq{eq:2prong_eff}. See \Fig{fig:app_compare_Nsub_ROC} in \App{app:Nsub_Ni} for a comparison to $\Nsubnobeta{2,1}$, and see \Fig{fig:hybrid_obs} in \App{app:hybrid} for a hybrid strategy using a groomed mass cut but ungroomed discriminants.
}
\label{fig:nD2_ROC_2prong_quark}
\end{figure}

To study the discrimination power more quantitatively, we show ROC curves before and after grooming in \Fig{fig:nD2_ROC_2prong_quark}, considering the quark and gluon backgrounds separately.  The baseline efficiencies for the ungroomed and groomed mass selections are
\begin{align}
\text{$m \in [80,100]~$GeV}: & \quad \mathcal{E}_{Z} = 27\%, \quad \mathcal{E}_q = 17\%,\phantom{.}  \quad \mathcal{E}_g = 15\%, \nonumber \\
\text{$m_{\rm SD} \in [80,100]~$GeV}: & \quad \mathcal{E}_{Z} = 37\%, \quad \mathcal{E}_q = 2.6\%,  \quad \mathcal{E}_g = 4.3\%, \label{eq:2prong_eff}
\end{align}
where we again normalize the ROC curves to show only the gains from the new 2-prong discriminants.\footnote{Note the improved signal significance in the groomed case, which offsets the apparent decrease in discrimination performance when comparing the ungroomed and groomed ROC curves.}  We use $\Dobsnobeta{2}$ (with $\beta = 2$) as a standard reference, since it is currently used by the ATLAS experiment for its excellent tagging performance \cite{ATLAS-CONF-2015-035,Aad:2015rpa,ATLAS-CONF-2015-068,ATLAS-CONF-2015-071,ATLAS-CONF-2015-073,Aaboud:2016trl,ATLAS-CONF-2016-016,ATLAS-CONF-2016-039,ATLAS-CONF-2016-055,ATLAS-CONF-2016-082,ATLAS-CONF-2016-083}.\footnote{Note that ATLAS uses $\Dobsnobeta{2}$ after jet trimming \cite{Krohn:2009th}, which has a similar parametric behavior to $\Dobsnobeta{2}$ after soft drop in the region we are considering.}

Of the three new observables, only $\Nobsnobeta{2}$ is designed to act as a discriminant on ungroomed jets.  In both \Figs{fig:nD2_ps_quark_pythia}{fig:nD2_ps_gluon_pythia}, we see that $\Nobsnobeta{2}$ outperforms the standard $\Dobsnobeta{2}$ observable in discriminating against both quark and gluon jets.  From power-counting arguments, we cannot predict the relative performance between the quark and gluon samples, but the fact that $\Nobsnobeta{2}$ sees significant performance gains on the gluon sample is very encouraging.  As discussed in \Sec{sec:twoprong_PC}, the discrimination power of $\Nobsnobeta{2}$ is closely related to $\Nsubnobeta{2,1}$ in the resolved limit, but with an improved behavior in the transition to the unresolved region.  We discuss this relation in more detail in \App{app:Nsub_Ni}, showing that $\Nobsnobeta{2}$ has slightly improved performance compared to $\Nsubnobeta{2,1}$ on ungroomed jets, but considerably improved performance after grooming.

After jet grooming, shown in \Figs{fig:nD2_ps_quark_pythia_SD}{fig:nD2_ps_gluon_pythia_SD}, all three new observables offer improved discrimination power over $\Dobsnobeta{2}$.  Comparing the results before and after grooming, we see  dramatic gains in performance for $\Mobsnobeta{2}$ and $\Dobs{2}{1,2}$, as expected from power counting.  It is rather curious that after grooming, all three observable offer comparable discrimination power, even though they are based on $\ecfvarnobeta{v}{3}$ correlators with different characteristic behaviors.  It would be interesting to study the correlations between these observables to see if they are probing complementary physics effects.  Such correlations go beyond the power-counting analysis of this paper, so we leave a study to future work.

Thus far, we have only considered observables measured entirely on either groomed or ungroomed jets.  Experimentally, though, it may be desirable to measure ungroomed observables after the application of a groomed mass cut (see e.g. \cite{CMS-PAS-JME-14-002}); we refer to this as a ``hybrid" strategy.   In \App{app:hybrid}, we present ROC curves for $\Mobsnobeta{2}$, $\Nobsnobeta{2}$, $\Dobsnobeta{2}$, and $\Nsubnobeta{2,1}$ using this hybrid strategy and analyze their behavior using power counting.  We leave a more detailed study of the optimal use of mixed groomed/ungroomed observables to future work.

\subsection{Stability in Parton Showers}\label{sec:twoprong_MC_stable}

In addition to their absolute performance, our new 2-prong discriminants exhibit stable behavior, especially after grooming.  As recently emphasized in \Ref{Dolen:2016kst}, stability of background distributions as a function of mass and $p_T$ cuts is an important consideration when designing jet substructure observables.  Excessive dependence on jet mass and $p_T$ can lead to mass sculpting, which can increase systematic uncertainties in sideband fits, counteracting gains from improved tagging performance.

To illustrate how the phase space structure controls the stability of the observable, it is interesting to study the stability of $D_2$, $M_2$, and $N_2$ before and after grooming.  These three observables represent the three scaling behaviors we have encountered in this paper.  Prior to grooming, we have:
\begin{itemize}
\item $D_2$ in \Fig{fig:D2_ps_review}:  The background occupies a non-trivial phase space region that does not overlap with the signal.
\item $M_2$ in \Fig{fig:1D2_ps_global}:  The background occupies a non-trivial phase space region overlapping with the signal.
\item $N_2$ in \Fig{fig:2D2_ps_global}:  The background is confined to a single scaling on the boundary of phase space.
\end{itemize}
The $\Dobs{2}{1,2}$ observable has a similar phase space structure to $M_2$, and will therefore behave similarly, so we do not show it explicitly in this section.  Note that $\Nsubnobeta{2,1}$ has the same phase space structure as $N_2$, so it exhibits related stability properties.

\begin{figure}
\begin{center}
\subfloat[]{\label{fig:D2qmass}
\includegraphics[width=6.85cm]{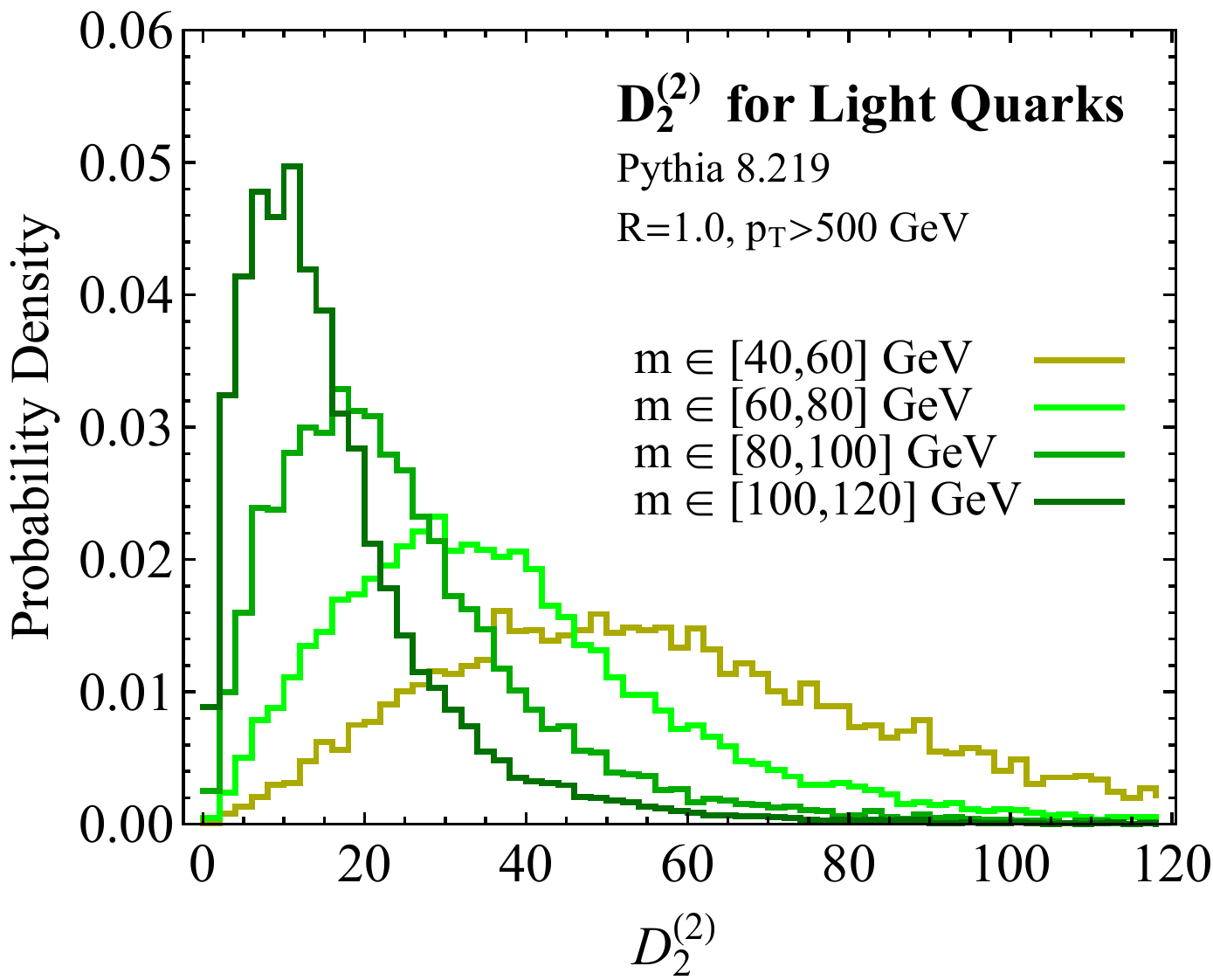}
}\qquad
\subfloat[]{\label{fig:D2qmass_groomed}
\includegraphics[width=6.5cm]{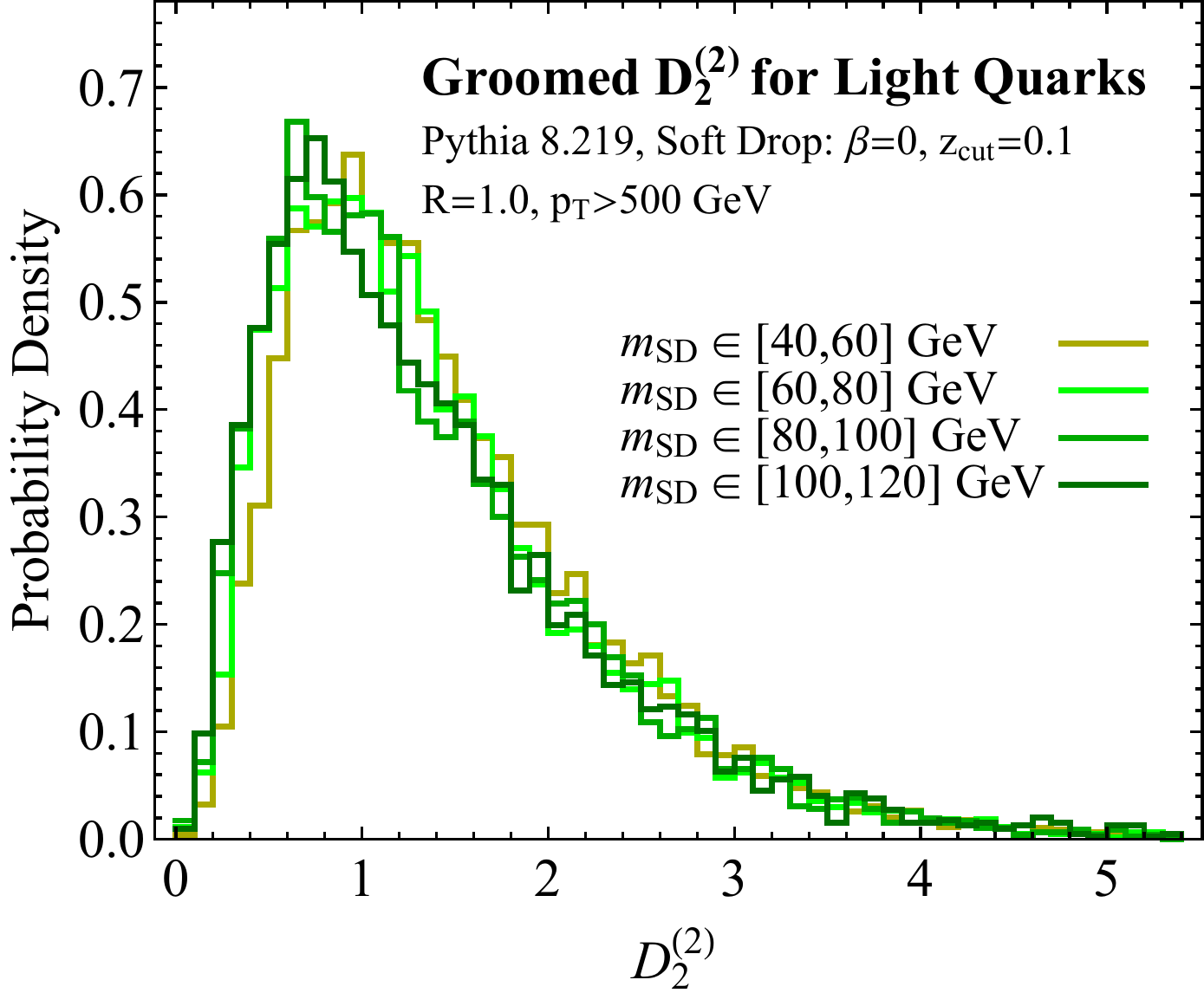}
}\nonumber \\
\subfloat[]{\label{fig:M2qmass}
\includegraphics[width=6.6cm]{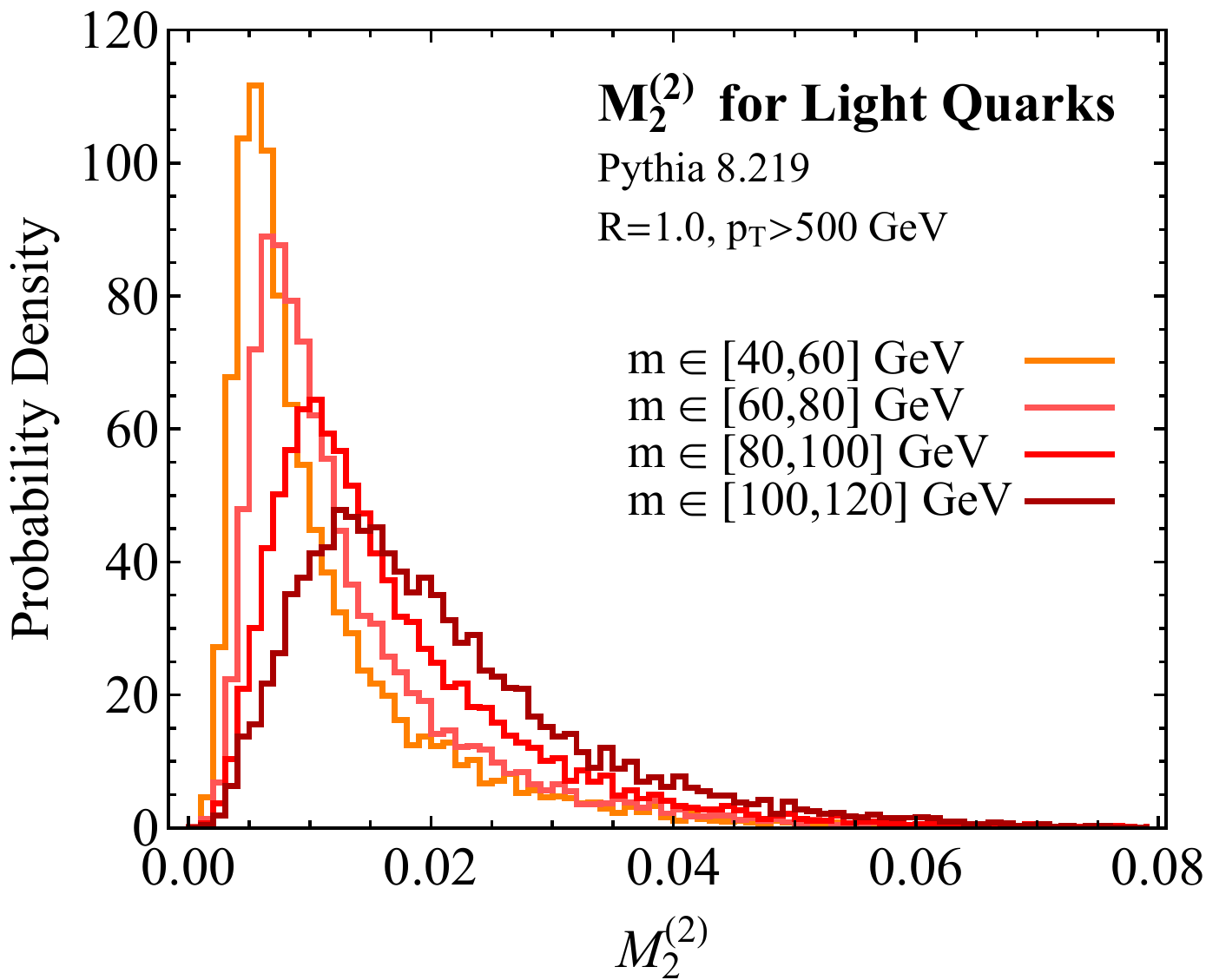}
}\qquad
\subfloat[]{\label{fig:M2qmass_groomed}
\includegraphics[width=6.5cm]{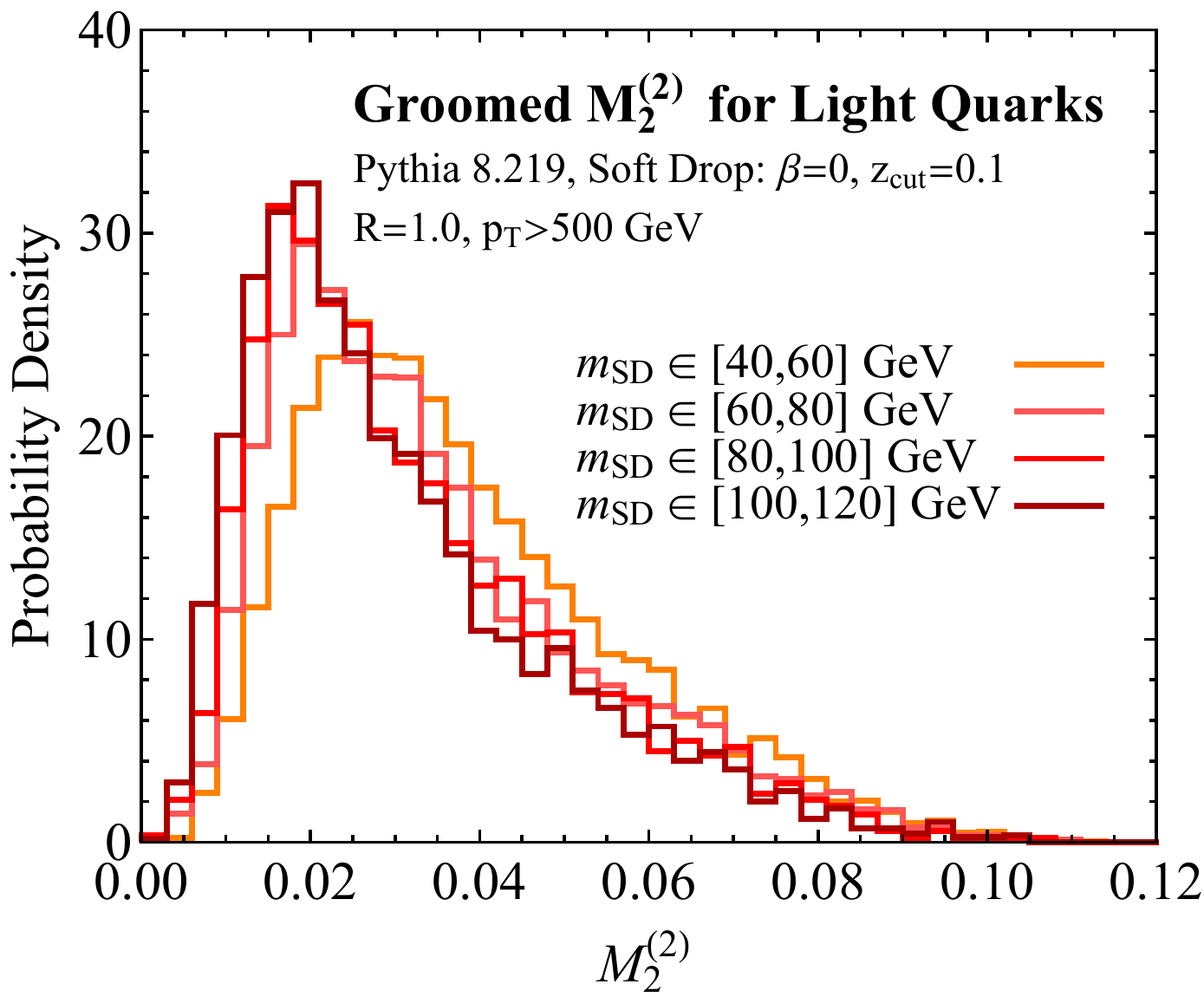}
}\nonumber \\
\subfloat[]{\label{fig:N2qmass}
\includegraphics[width=6.5cm]{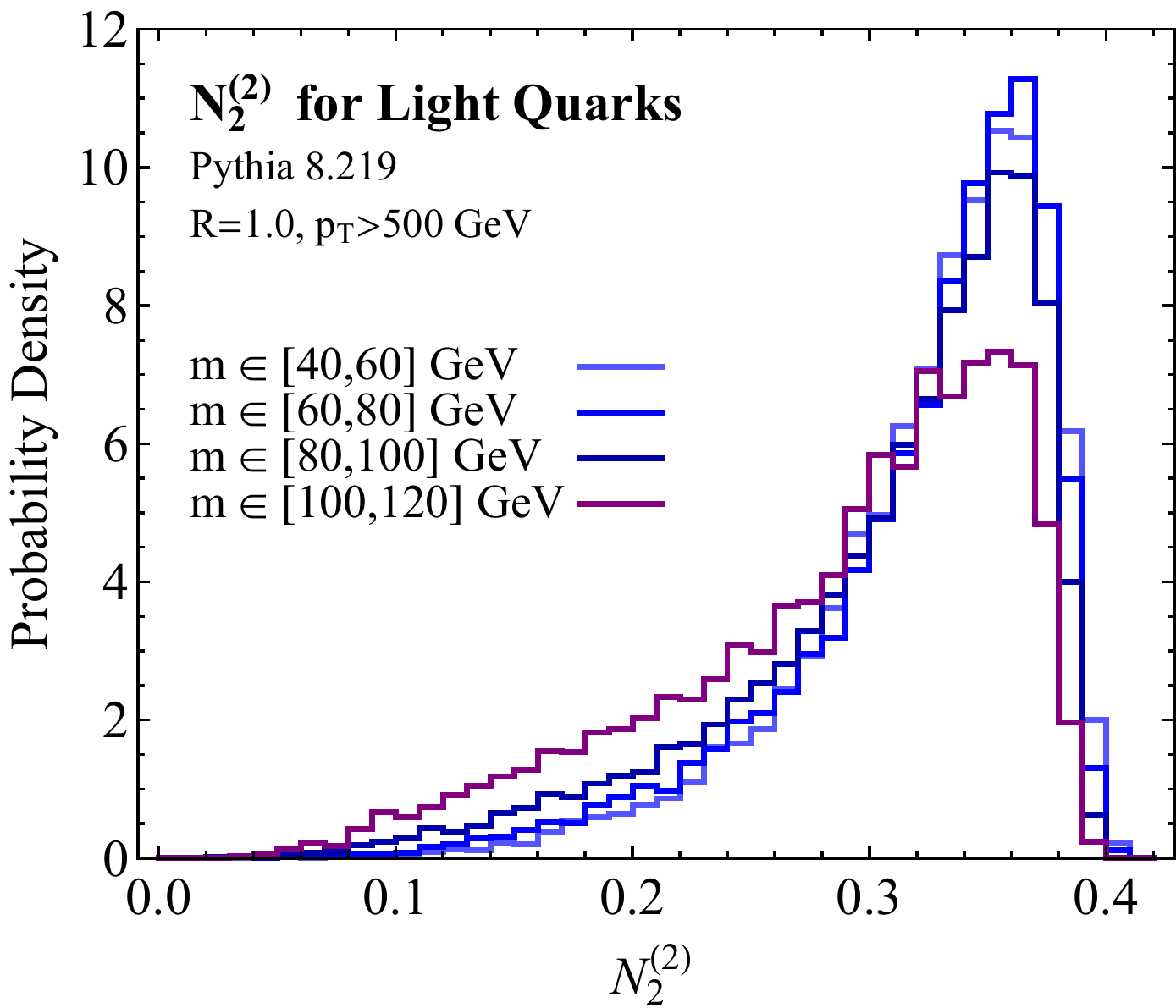}
}\qquad
\subfloat[]{\label{fig:N2qmass_groomed}
\includegraphics[width=6.5cm]{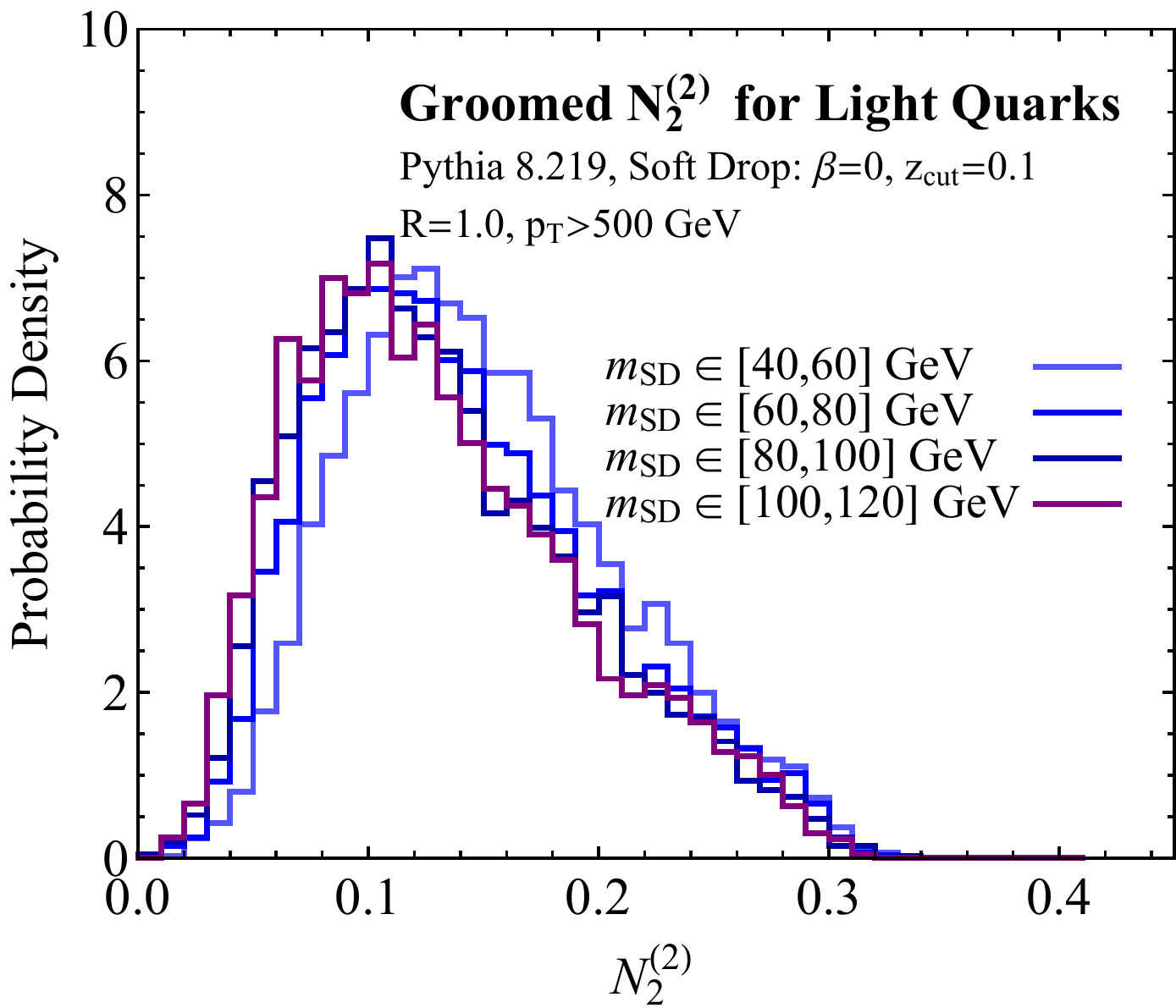}
}
\end{center}
\caption{Stability of the light-quark background distributions for (top row) $D_2$, (middle row) $M_2$, and (bottom row) $N_2$ as a function of the jet mass cut, comparing (left column) ungroomed jets to (right column) groomed jets.  The different structure of the phase space for $N_2$ leads to improved stability before grooming is applied.  After grooming, all observables exhibit excellent stability. Similar results are found for gluon jets as well.
}
\label{fig:2prong_mass}
\end{figure}

In \Fig{fig:2prong_mass}, we use parton showers to test the stability of $D_2$, $M_2$, and $N_2$ on the light quark background as the jet mass cut is varied.\footnote{We could alternatively vary the cut on the jet $p_{T}$. From the power-counting analysis, all stability properties are determined by functions of the ratio $m_J/p_{TJ}$, and therefore it is straightforward to understand the $p_{TJ}$ dependence from the $m_J$ dependence.}  Prior to grooming, only the $N_2$ observable exhibits any degree of stability on the background.  After grooming, all three observables have a nicely stable peak position and shape, and the residual variation could be compensated using the decorrelation technique of \Ref{Dolen:2016kst}. We can now use a power-counting analysis to demonstrate how these behaviors are dictated by the form of the phase space. Although we focus on light quark jets in \Fig{fig:2prong_mass}, similar stability properties are observed for gluon jets.  This is also emphasized by the power-counting argument, which is insensitive to the quark or gluon nature of the jet.

We begin by considering the observables before grooming. For $D_2$ in \Fig{fig:D2_ps_review}, the background region is defined by two different scalings, one of which defines the upper boundary of the phase space and one of which defines the scaling of the boundary between the signal and background, and therefore the scaling of the desired cut value for discrimination. The upper boundary of the phase space is defined by the scaling $\ecfnobeta{3} \sim  (\ecfnobeta{2})^2$, leading to the maximum value 
\begin{align}
D^{ \text{max}}_2\sim \frac{\ecfnobeta{3}}{(\ecfnobeta{2})^3} \sim \frac{ (\ecfnobeta{2})^2}{(\ecfnobeta{2})^3}  \sim \frac{1}{\ecfnobeta{2}}\,.
\end{align}
Simplifying to the case of $\beta=2$, and using \Eq{eq:e2_mass_relation}, we have
\begin{align}\label{eq:D2max}
D^{(2), \text{max}}_2\sim \frac{p_{TJ}^2}{m_J^2}\,,
\end{align}
which depends sensitively on $m_J$ and $p_{TJ}$.  This behavior can be clearly seen in \Fig{fig:D2qmass}, where the $D_2$ distribution shifts dramatically with the jet mass cut, an undesirable feature for the purposes of sideband calibration.

For $M_2$ with a phase space given in \Fig{fig:1D2_ps_global}, we see quite different behavior.  In this case, the upper boundary of the phase space is defined by  $\ecfvarnobeta{1}{3} \sim \ecfnobeta{2}$, and therefore $M_2$ has a maximum value
\begin{align}\label{eq:M2max}
M^{ \text{max}}_2 \sim \frac{\ecfvarnobeta{1}{3}}{\ecfnobeta{2}} \sim \frac{\ecfnobeta{2}}{\ecfnobeta{2}} \sim \text{const}\,,
\end{align}
which is largely independent of the jet mass, $p_T$, and the angular exponent $\beta$.  Stability of the maximal value (endpoint), though, is not sufficient to guarantee stability of the distribution. Indeed, the scaling of the lower boundary of the phase space for the background is $\ecfvarnobeta{1}{3} \sim (\ecfnobeta{2})^2$, so we expect a sharp drop in the background, and therefore a peak in the distribution, around
\begin{align}\label{eq:M2peak}
M^{ \text{peak}}_2 \sim \frac{\ecfvarnobeta{1}{3}}{\ecfnobeta{2}} \sim \frac{(\ecfnobeta{2})^2}{\ecfnobeta{2}} \sim \ecfnobeta{2}\,.
\end{align}
Simplifying again to the case of $\beta=2$, and using \Eq{eq:e2_mass_relation}, we have
\begin{align}
M^{(2), \text{peak}}_2\sim \frac{m_{J}^2}{p_{TJ}^2}\,,
\end{align}
which depends sensitively on $m_J$ and $p_{TJ}$, but in exactly the opposite way as $D_2$. This behavior is observed in \Fig{fig:M2qmass}.

Finally, for $N_2$ shown in \Fig{fig:2D2_ps_global}, the background region is defined by a single scaling, namely $\ecfvarnobeta{2}{3} \sim (\ecfnobeta{2})^2$, which defines the upper boundary.  Since there is a single scaling, we expect the peak for the background distribution to be defined by the same scaling.  This means that $N_2$ has a maximum value and a peak location that both scale like
\begin{align}\label{eq:N2max}
N^{ \text{max,peak}}_2 \sim \frac{\ecfvarnobeta{2}{3}}{(\ecfnobeta{2})^2} \sim \frac{(\ecfnobeta{2})^2}{(\ecfnobeta{2})^2} \sim \text{const}\,,
\end{align}
which is largely independent of the jet mass, $p_T$, and the angular exponent $\beta$.  This is well verified in the parton shower analysis, as shown in \Fig{fig:N2qmass}. Thus, we see that by carefully engineering the phase space of an observable, one can achieve properties, such as stability, that are important experimentally.  In this specific case, the stability of the full $N_2$ distribution gives further evidence that $N_2$ is a promising 2-prong tagger, even without grooming.

After grooming away soft radiation, we see from \Figss{fig:D2qmass_groomed}{fig:M2qmass_groomed}{fig:N2qmass_groomed} that all the distributions are stable, and from our power counting analysis, it is easy to understand why this is true.  For $D_2$, grooming has a dramatic impact (note the change in the $x$-axis range), since it removes the region of phase space that leads to the undesired scaling behavior in \Eq{eq:D2max} (see also \Fig{fig:D2_ps_review_groom}). In this way, the endpoint for groomed $D_2$ (as well as the whole distribution) becomes remarkably robust to the jet mass cut.  For the $M_2$ observable, the grooming removes the background in the bulk of phase space and pushes it to the upper boundary, as shown in \Fig{fig:1D2_ps_global}, stabilizing the peak of the $M_2$ distribution but leaving the endpoint largely unchanged.  After jet grooming, the parametric phase space for $N_2$ is unmodified, so the endpoint and peak scaling in \Eq{eq:N2max} should not change.  Comparing \Figs{fig:N2qmass}{fig:N2qmass_groomed}, we see that the specific value of the $N_2$ endpoint and peak is modified, but the stability with varying mass cut is robust.

Therefore, in all cases after grooming, we have
\begin{align}
\text{groomed}: \quad D^{ \text{max,peak} }_2 \sim \text{const}\,, \quad  M^{ \text{max, peak}}_2\sim  \text{const} \,, \quad   N^{ \text{max,peak}}_2 \sim  \text{const}\,.
\end{align}
This demonstrates three distinct ways of generating a stable distribution:  engineering the background phase space to directly have the desired boundary (e.g.~$N_2$), or grooming soft radiation to the stabilize the boundary (e.g.~$D_2$) or the peak (e.g.~$M_2$) of the background distribution. It is important to emphasize that the power-counting analysis can only identify the power-law scaling of the distribution in $m_J$ or $p_{TJ}$.  Removing this power-law scaling does not, however, guarantee complete numerical stability of the distribution.  For this, techniques such as designing decorrelated taggers (DDT) \cite{Dolen:2016kst} can be used.  We expect that methods like DDT will be most powerful when applied to variables that are already naturally stable, but we leave a study to future work.

\section{Improving Quark/Gluon Discrimination}\label{sec:qvsg}

A major challenge in the field of jet substructure is reliable quark/gluon discrimination. Despite its many potential applications, there has been significant difficulty both in understanding the behavior of quark/gluon discriminants in parton showers, as well as in developing analytically-tractable observables which surpass the Casimir scaling limit (see \Eq{eq:casimir_scaling} below).  For detailed discussions of these issues, we refer the reader to \Refs{Gallicchio:2011xc,Gallicchio:2011xq,Gallicchio:2012ez,Larkoski:2013eya,Larkoski:2014pca,Badger:2016bpw}, as well as to studies in data \cite{CMS:2013wea,CMS:2013kfa,Aad:2014gea,ATLAS-CONF-2016-034}.

Quark/gluon discrimination has mostly been studied using IRC safe observables, such as the angularities \cite{Berger:2003iw,Almeida:2008yp} or 2-point energy correlation functions $C_1 = \ecfnobeta{2}$ \cite{Larkoski:2013eya}, which are set by a single emission at LL accuracy.\footnote{Important exceptions are (IRC unsafe) multiplicity-based observables, which have a long history in QCD \cite{Brodsky:1976mg,Konishi:1978yx,Mueller:1983cq,Malaza:1984vv,Gaffney:1984yd,Malaza:1985jd,Catani:1991pm,Catani:1992tm,Dremin:1993vq,Dremin:1994bj,Capella:1999ms,Bolzoni:2012ii} (see \cite{Aad:2016oit} for a recent experimental study), and more recently, shower deconstruction \cite{FerreiradeLima:2016gcz}.} At LL order, and ignoring nonperturbative effects, one can show that the discrimination power of such observables is set by the Casimir scaling relation
\begin{align}\label{eq:casimir_scaling}
\text{disc}(x)=x^{C_A/C_F}=x^{9/4}\,,
\end{align}
where $x$ is the fraction of quarks retained by the cut and $\text{disc}(x)$ is the fraction of gluons retained.  In this way, discrimination power is capped by the ratio of the gluon and quark color charges, $C_A/C_F = 9/4$.  Casimir scaling arises because after a single emission, the discrimination power is set only by the color factor associated with the hard jet core, independent of the particular details of the observable.

Beyond LL accuracy, where one is sensitive to physics beyond the leading emission, improved discrimination power is observed.  In \Ref{Larkoski:2013eya}, an analytic calculation of $C_1$ was performed at NLL accuracy, and a noticeable increase in discrimination power beyond the Casimir limit was found for $\beta<1$ (though not confirmed in an ATLAS study \cite{Aad:2014gea}).  For small values of $\beta$, however, one is highly sensitive to nonperturbative effects, which must be modeled or extracted from data.  Particularly for gluon jets, which are not well constrained by LEP event shape data \cite{Abdallah:2003xz,Heister:2003aj,Achard:2004sv,Abbiendi:2004qz}, this leads to significant discrepancies between distributions obtained from different parton shower generators.\footnote{This has been coined the ``\pythia{}-\herwig{} sandwich'', with LHC data as the filling.}  This in turn leads to rather large uncertainties in the predicted quark/gluon efficiencies; see \Refs{Larkoski:2014pca,Badger:2016bpw} for detailed studies.

Given the Casimir scaling limit of single-emission observables, a promising approach for improving quark/gluon discrimination is to design observables that are directly sensitive to multiple emissions within the jet, even at lowest order.  In this section, we  define a series of observables $\Uobsnobeta{i}$ specifically intended for this purpose.  Since these observables exhibit different behavior from standard single-emission observables, they may also prove useful in improving the parton shower description of quark and gluon jets.  We will particularly emphasize the stability of their discrimination power as a function of the angular exponent $\beta$, which could be helpful for disentangling perturbative and nonperturbative effects.

\subsection{Probing Multiple Emissions with $\Uobsnobeta{i}$}\label{sec:qvsg_setup}

A standard observable for quark/gluon discrimination is the 2-point energy correlation function $\ecfnobeta{2}$, whose scaling was derived already in \Eq{eq:fact_e2} for 1-prong jets: 
\begin{align}\label{eq:e2pc_qvsg}
\ecf{2}{\beta}\sim z_s +  \theta_{cc}^\beta\,.
\end{align}
As discussed, $\ecfnobeta{2}$ is set at LL accuracy by a single emission from the hard core.  Note that the scaling is the same for quarks and gluons, since $C_F = 4/3$ versus $C_A = 3$ is not a parametric difference between the samples.

To go beyond this single-emission behavior, we consider the 3-point correlators, $\ecfvarnobeta{v}{3}$, which explicitly probe two emissions from the hard jet core.  Using the modes in \Tab{tab:unresolved}, we derive the following scalings (which were already given in \Sec{sec:twoprong_PC}): 
\begin{align}\label{eq:pc_qvsg}
\ecfvar{1}{3}{\beta} &\sim z_s^2+\theta_{cc}^\beta \,, \nonumber\\
\ecfvar{2}{3}{\beta} &\sim z_s^2+ z_s \theta_{cc}^\beta  +\theta_{cc}^{2\beta}\,, \nonumber\\
 \ecfvar{3}{3}{\beta} &\sim z_s^2+ z_s \theta_{cc}^\beta  +\theta_{cc}^{3\beta}\,.
\end{align}
We can draw a number of interesting conclusions from \Eq{eq:pc_qvsg}.  First, in the majority of phase space there is a direct relationship between the last two 3-point correlators and the 2-point correlator: $\ecfvarnobeta{2}{3}\sim (\ecfnobeta{2})^2$ and $\ecfvarnobeta{3}{3}\sim (\ecfnobeta{2})^2$.\footnote{Because of the $\theta_{cc}^{3\beta}$ term, this parametric relation is strictly speaking not true for $\ecfvarnobeta{3}{3}$, but the difference is power suppressed in much of the phase space.}  We therefore do not expect $\ecfvarnobeta{2}{3}$ or $\ecfvarnobeta{3}{3}$ to yield improved quark/gluon discrimination power compared to $\ecfnobeta{2}$; this illustrates the importance of understanding parametric correlations between different observables.  By contrast, $\ecfvarnobeta{1}{3}$ does not obey such a relation to $\ecfnobeta{2}$, since only for $\ecfvarnobeta{1}{3}$ is the cross term $\theta_{cc}^\beta z_s$ power suppressed.   Since $\ecfvarnobeta{1}{3}$ directly probes the double-soft limit of a jet, without soft/collinear cross talk at leading power, we can expect it to carry more information about the flavor of the jet's initiating parton. This intuition will be verified in our parton shower study.

Another interesting feature of $\ecfvarnobeta{1}{3}$ is the relative scaling between the collinear and soft modes, as can be seen from comparing \Eq{eq:pc_qvsg} to \Eq{eq:e2pc_qvsg}.  To improve quark/gluon efficiency with $\ecfnobeta{2}$, one typically needs to use small values of the angular exponent $\beta$.  Since $\ecfvarnobeta{1}{3}$ already has a suppressed soft scaling, it can achieve good quark/gluon discrimination at comparatively higher values of the angular exponent.  In the parton shower study below, we will find that the performance of $\ecfvarnobeta{1}{3}$ with $\beta=2$ is comparable to $\ecfnobeta{2}$ with $\beta=0.2$. This relative scaling also modifies the structure of nonperturbative corrections, although we will not discuss this aspect further in this paper.\footnote{Our reluctance to weigh in on nonperturbative corrections is because a standard shape function analysis \cite{Korchemsky:1999kt,Korchemsky:2000kp,Bosch:2004th,Hoang:2007vb,Ligeti:2008ac}, which is applicable for $\ecfnobeta{2}$, does not hold for $\ecfvarnobeta{1}{3}$.  In future work, we might hope to extend the shape function logic to non-additive observables like $\ecfvarnobeta{1}{3}$.}  Note that the discrimination power as a function of $\beta$ is not a prediction of power counting and can only be obtained by explicit calculations (or measurements) of the distributions.

Seeing the potential of $\ecfvarnobeta{1}{3}$, it is natural to consider higher-point correlators. For an $n+1$ point correlator, we have
\begin{align}\label{eq:U_pc}
\ecfvar{1}{n+1}{\beta} \sim z_s^n+ \theta_{cc}^\beta  \,,
\end{align}
which probes the $n$-soft limit, again without soft/collinear cross talk at leading power.  We are therefore led to define the $\Uobsnobeta{i}$ series of observables,
\begin{align}
\Uobs{i}{\beta}=\ecfvar{1}{i+1}{\beta}\,,
\end{align}
for quark/gluon discrimination. The reason one might expect $\Uobsnobeta{i}$ to perform better with increasing $i$ is that higher-point correlators can effectively ``count'' more emissions than lower-point correlators.  Since gluon jets generate more emissions than quark jets, on average by a factor of $C_A/C_F$, one expects improved quark/gluon contrast with each additional emission probed; this intuition will be borne out in the parton shower study below.  More generally, we hope that these observables will prove useful for probing the structure of the QCD shower. 

From the power counting in \Eq{eq:U_pc}, we see that the scaling of the soft modes for $\Uobsnobeta{i}$ depends on the index $i$ as $z_s^i$. One might therefore naively think that after grooming is applied, all the $\Uobsnobeta{i}$ observables would be identical.  This is not the case for a fixed value of $\zcut$, however, since the soft scale increases as a function of $i$.  To emphasize this point, the average values of $\Uobsnobeta{i}$ are typically $\langle \Uobsnobeta{2} \rangle =0.05$ and $\langle \Uobsnobeta{3} \rangle =0.01$ (see \Fig{fig:2prong_obs_gluon} from our parton shower study below).  By \Eq{eq:U_pc}, these correspond to $z_s$ values of $z_s\simeq 0.25$ and $z_s\simeq 0.4$, respectively, both of which are well above the $\zcut=0.1$ scale that we use as our grooming benchmark.  Therefore, the emissions that dominate the $\Uobsnobeta{2}$ and $\Uobsnobeta{3}$ distributions are not actually removed by our grooming procedure.  Thus, the behavior of $U_i$ is expected to be more resilient to grooming for larger values of $i$.

\begin{figure}
\begin{center}
\subfloat[]{\label{fig:app_qvsg1a}
\includegraphics[width=6.5cm]{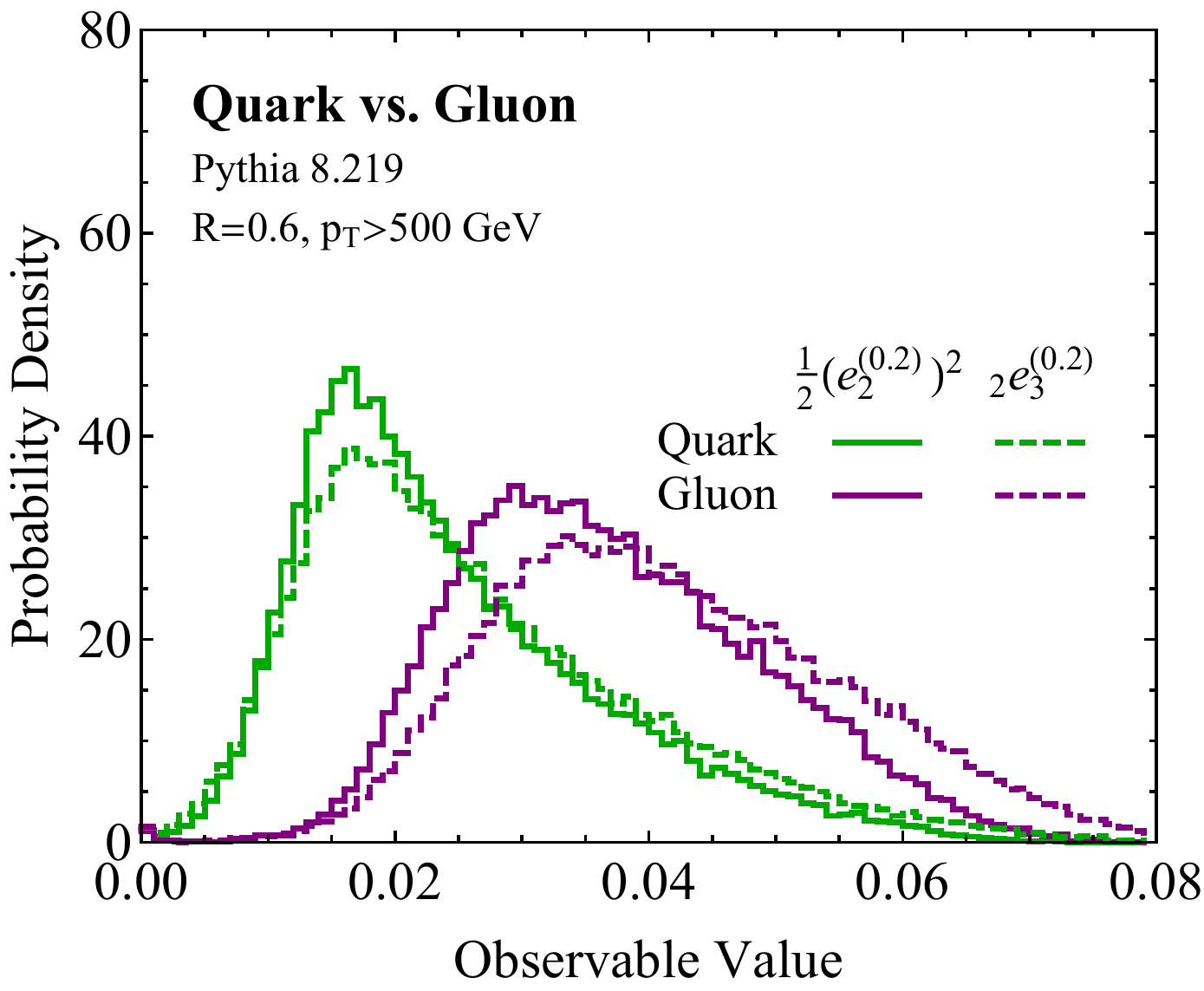}
}\qquad
\subfloat[]{\label{fig:app_qvsg1b}
\includegraphics[width=6.5cm]{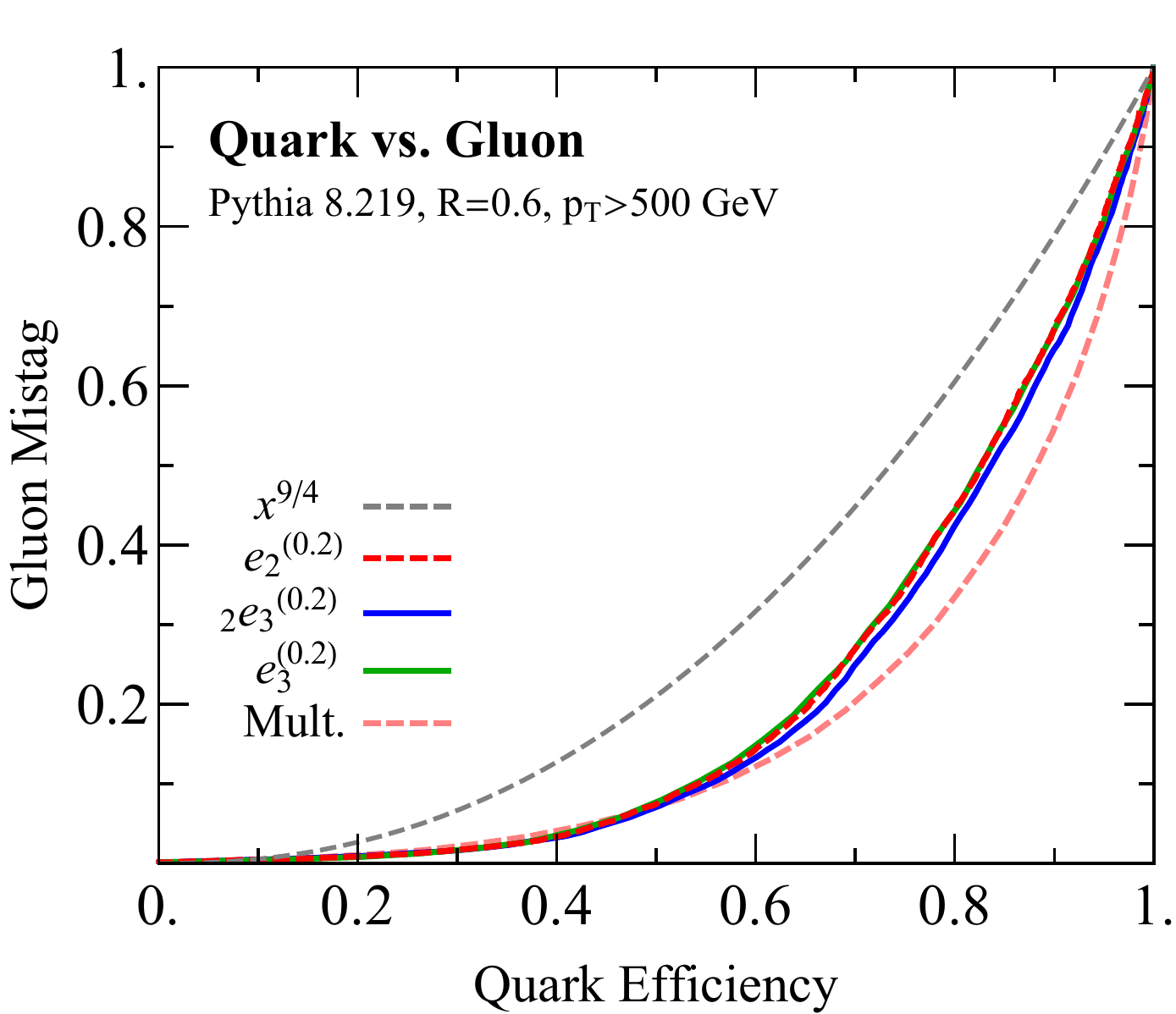}
}
\end{center}
\caption{Comparison of the 3-point correlators, $\ecfvarnobeta{2}{3}$ and $\ecfvarnobeta{3}{3}$, with the 2-point correlator, $\ecfnobeta{2}=\Uobsnobeta{1}$.  The fact that these observables are related by power counting is shown by verifying (a) the scaling relation $\ecfvarnobeta{2}{3} \sim \frac{1}{2}(\ecfnobeta{2})^2$ at the level of distributions and (b) the nearly identical quark/gluon discrimination power using a ROC curve.  The prediction from Casimir scaling and the result for hadron multiplicity are shown for reference. 
}
\label{fig:app_qvsg1}
\end{figure}

\subsection{Performance in Parton Showers}\label{sec:qvsg_mc}

We now use a parton shower study to verify the above power-counting predictions and to assess quantitatively the potential improvements in quark/gluon discrimination power achievable using higher-point correlators.  For reasons of computational time we restrict our study of the $\Uobsnobeta{i}$ series to $i=1,2,3$.\footnote{For a jet with $n_J$ particles, the computational cost of $\Uobsnobeta{i}$ scales like $n_J^{i+1}$. On a typical laptop, the analysis of a single jet takes around $\{0.14~\text{ms}, 0.86~\text{ms}, 11~\text{ms}\}$ for $\{U_1,U_2,U_3\}$.} The quark and gluon jets are generated from the same \textsc{Pythia} $pp \to Z+j$ samples described in \Sec{sec:twoprong_MC}, and the same overall analysis strategy applies, though no cut is placed on jet masses.  Furthermore, we use a smaller jet radius of $R=0.6$.  Given known parton shower uncertainties, it would be interesting to study different shower and hadronization algorithms to understand the degree to which LHC measurements of $\Uobsnobeta{i}$ could provide insight into quark/gluon tagging; we leave such studies to future work.

We begin by verifying the power-counting argument of \Eq{eq:pc_qvsg}, which suggested that $\ecfvarnobeta{2}{3}$ and $\ecfnobeta{3}$ should be highly correlated with $U_1 = C_1 = \ecfnobeta{2}$.  Even though $\ecfvarnobeta{2}{3}$ and $\ecfnobeta{3}$ probe three particle correlations, they have a fixed scaling relation with respect to $\ecfnobeta{2}$, and are therefore not expected to provide new information for quark/gluon tagging.  Taking $\ecfvarnobeta{2}{3}$ as a representative example in \Fig{fig:app_qvsg1a}, we compare the distributions of $\ecfvarnobeta{2}{3}$ and $\frac{1}{2}(\ecfnobeta{2})^2$; they are remarkably similar so we conclude that power counting is indeed capturing the dominant scaling relation.  From the ROC curves in \Fig{fig:app_qvsg1b}, we see that the discrimination power of $\ecfnobeta{2}$, $\ecfvarnobeta{2}{3}$, and $\ecfvarnobeta{3}{3}$ are very similar for the same value of $\beta$, with limited improvement observed by including 3-particle correlations.   This emphasizes that probing multi-particle correlations does not, in and of itself, improve quark/gluon discrimination, since higher-point correlation functions can be correlated with lower-point correlation functions.

\begin{figure}
\begin{center}
\subfloat[]{\label{fig:e3_gluonplot}
\includegraphics[width=6.5cm]{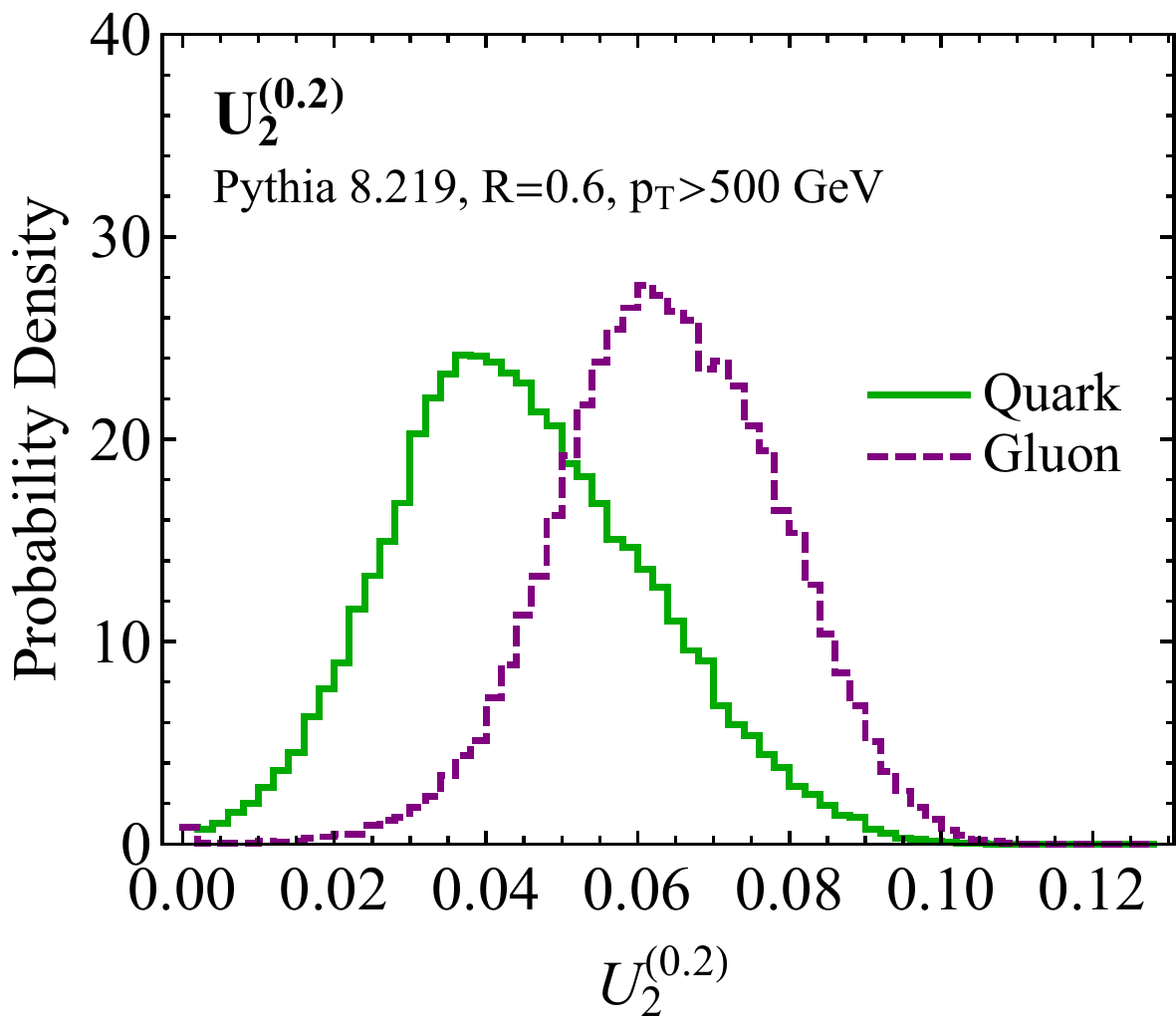}
}\qquad
\subfloat[]{\label{fig:e4_gluonplot_b}
\includegraphics[width=6.95cm]{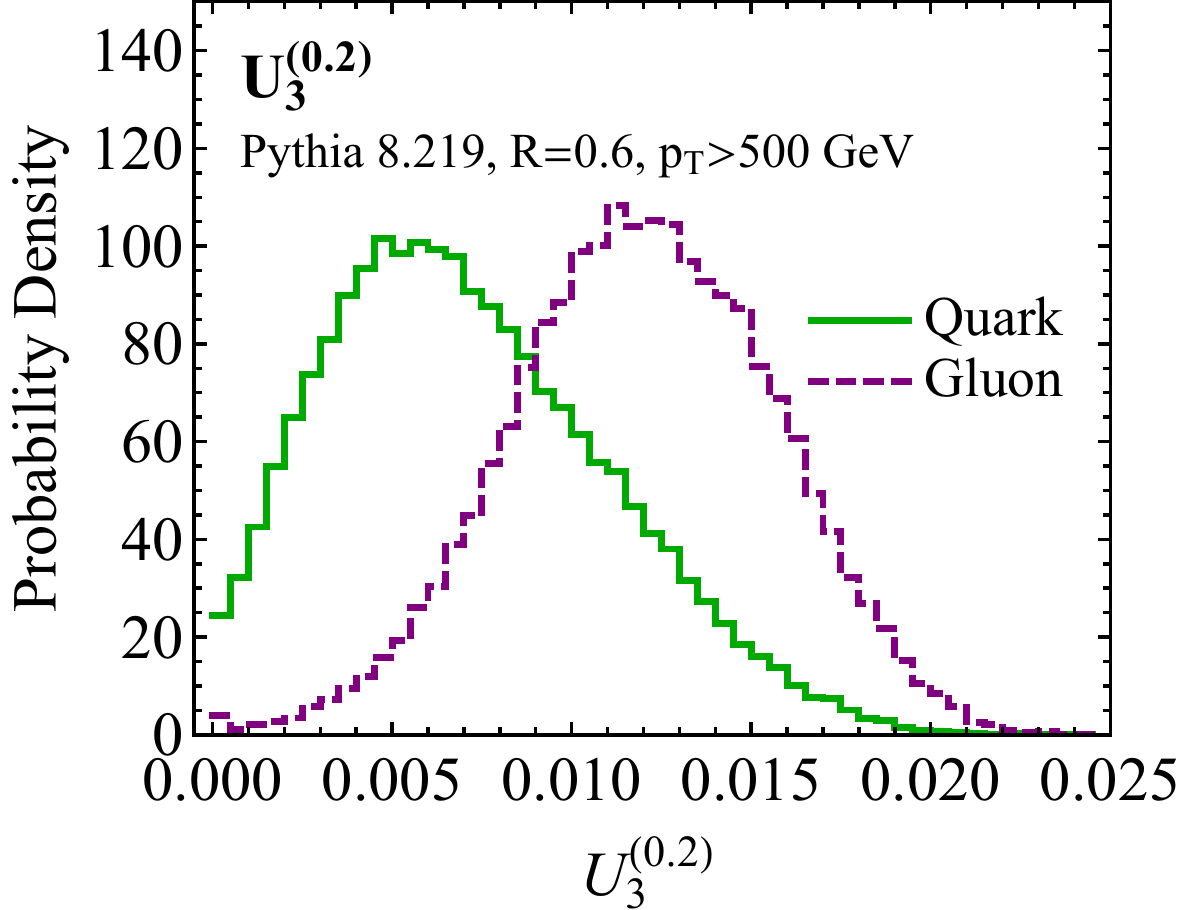}
}
\end{center}
\caption{Distributions of (a) $\Uobsnobeta{2}$ and (b) $\Uobsnobeta{3}$ for $\beta=0.2$, as measured on quark and gluon jets. 
}
\label{fig:e3_e4_quarkgluon}
\end{figure}

\begin{figure}
\begin{center}
\subfloat[]{\label{fig:2prong_obs_gluon_diffbeta_b}
\includegraphics[width=6.5cm]{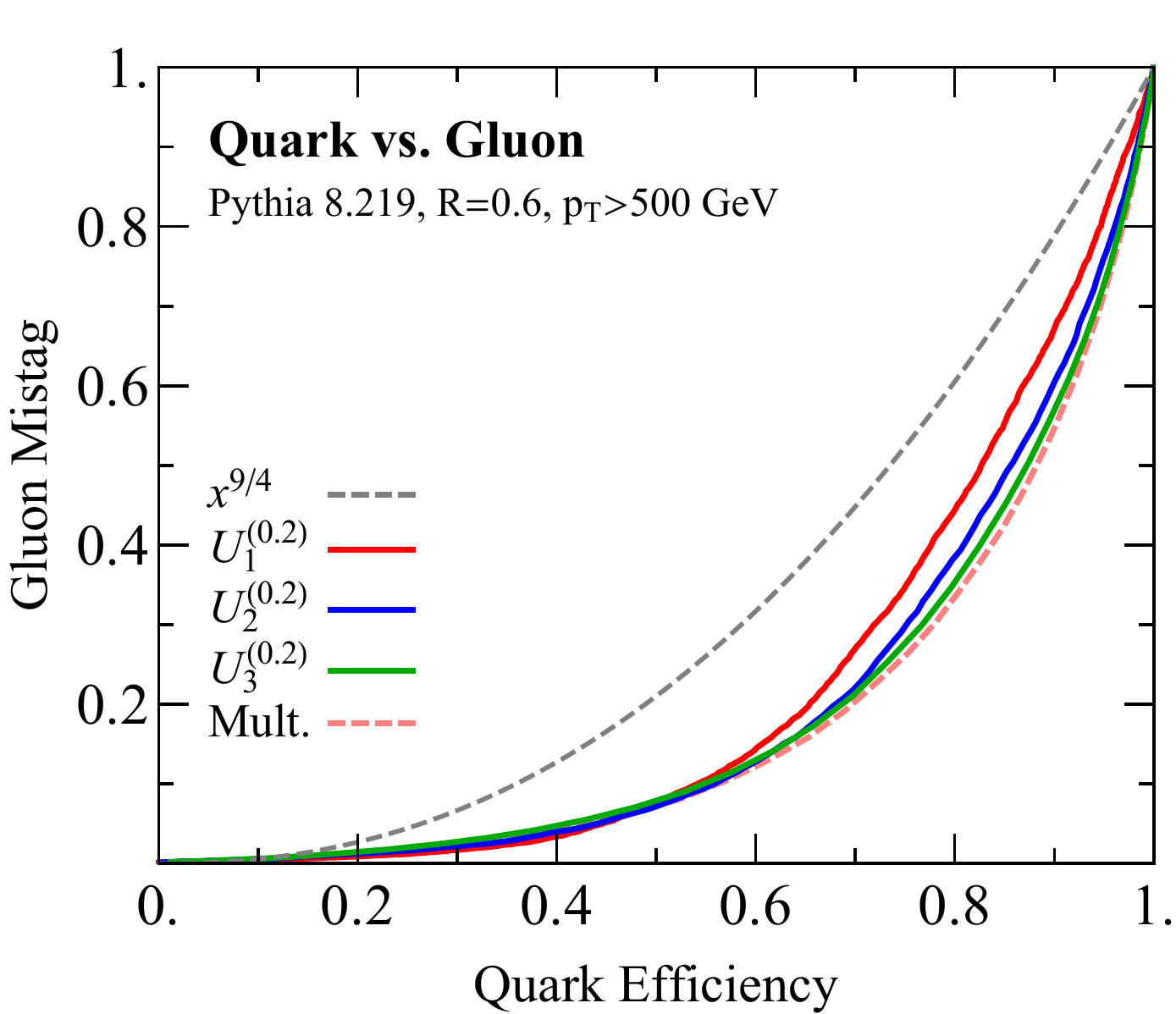}
}\qquad
\subfloat[]{\label{fig:2prong_obs_gluon_diffbeta_a}
\includegraphics[width=6.5cm]{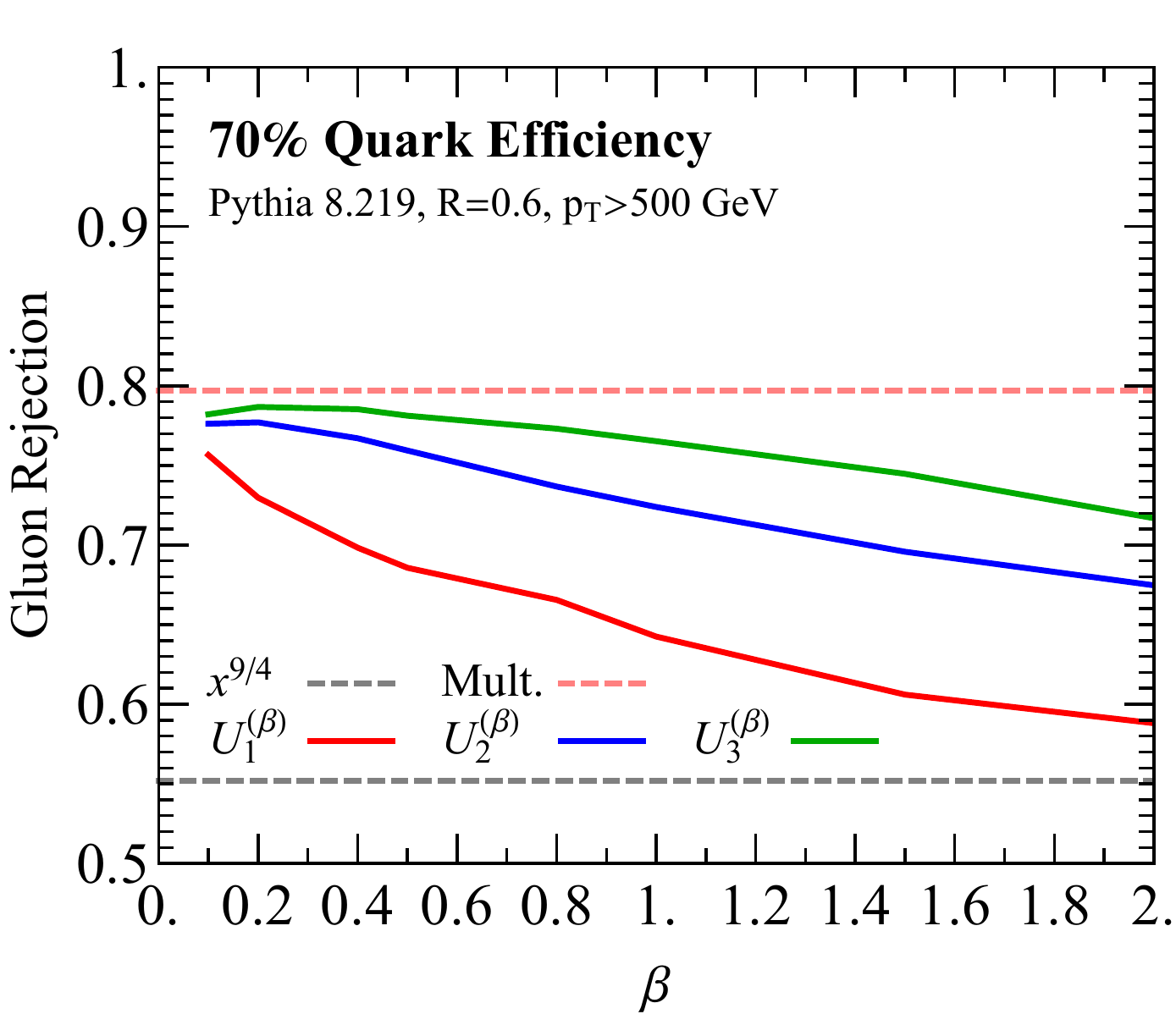}
}
\end{center}
\caption{Comparison of the quark/gluon discrimination power for $\Uobsnobeta{1}$, $\Uobsnobeta{2}$, and $\Uobsnobeta{3}$ to the prediction from Casimir scaling and the result for hadron multiplicity.  (a)  ROC curves demonstrating the improvement in performance as more emissions are probed.  (b) Gluon rejection at $70\%$ quark efficiency as a function of the angular exponent $\beta$.  The performance of the $\Uobsnobeta{i}$ observables appears to asymptote to hadron multiplicity as $i$ is increased.
}
\label{fig:2prong_obs_gluon_diffbeta}
\end{figure}

We now consider the behavior of $\Uobsnobeta{2}$ and $\Uobsnobeta{3}$, which were designed to exploit multi-particle correlations to improve quark/gluon discrimination. In \Fig{fig:e3_e4_quarkgluon}, we show distributions of $\Uobsnobeta{2}$ and $\Uobsnobeta{3}$ with $\beta=0.2$, indicating good separation of the quark and gluon samples.  This is quantified in \Fig{fig:2prong_obs_gluon_diffbeta_b}, which shows ROC curves for $\Uobsnobeta{i}$ comparing $i=1,2,3$.  Recall that $\Uobsnobeta{1}= C_1 = \ecfnobeta{2}$ is a standard quark/gluon discriminant and a useful baseline to assess performance gains (even if \textsc{Pythia} itself skews optimistic about quark/gluon separation power \cite{Larkoski:2013eya,Larkoski:2014pca}).  Going from $\Uobsnobeta{1}$ to $\Uobsnobeta{2}$ to $\Uobsnobeta{3}$, the discrimination power at high efficiencies does increase with more emissions being probed, though the change is relatively small going from $i = 2$ to $i = 3$.  

Beyond absolute performance gains, it is also interesting to study the relative performance of $\Uobsnobeta{i}$ as a function of the angular exponent $\beta$.  In \Fig{fig:2prong_obs_gluon_diffbeta_a}, we show the gluon rejection at $70\%$ quark efficiency as a function of $\beta$.\footnote{We chose 70\% quark efficiency as a benchmark, since it was used in the recent study of \Ref{ATLAS-CONF-2016-034}, though the features emphasized in the text are largely independent of this particular choice.  At very low quark efficiencies, deep in the nonperturbative regime, the different $U_i$ behaviors merge.}  Unlike for $U_1 = \ecfnobeta{2}$, where the discrimination power falls off rapidly with increasing $\beta$, for $\Uobsnobeta{2}$, and even more so for $\Uobsnobeta{3}$, the discrimination power remains well above the Casimir scaling limit, even into the large $\beta$ regime where $\Uobsnobeta{i}$ should be amenable to fixed-order or resummed perturbative calculations.   We find this much flatter behavior of the discrimination power with respect to $\beta$ to be one of the most interesting features of these observables, suggestive that multiple soft emissions are just as important as hard collinear emissions for discriminating quarks from gluons.  Full ROC curves for different values of the angular exponents are provided in \App{app:add_qg}.

It would be interesting to see if there is asymptotic behavior as $i \to \infty$, though this is likely only meaningful in the context of a comparative study of parton shower generators, since it depends sensitively on the assumptions made for correlated soft emissions.  As a first step in this direction, in \Fig{fig:e3_e4_quarkgluon} we compare $U_i$ to hadron multiplicity, which is known to be a powerful quark/gluon discriminant. Remarkably, the performance of the $\Uobsnobeta{i}$ observables appears to asymptote to multiplicity as $i$ is increased, both in the shape of the ROC curves as well as in the behavior as a function of $\beta$. It would be interesting to understand whether this connection can be made formal, and whether the $\Uobsnobeta{i}$ observables can be used to give an IRC safe definition of a multiplicity-like observable.

\begin{figure}
\begin{center}
\subfloat[]{\label{fig:qvsg_groom_dist_a}
\includegraphics[width=6.3cm]{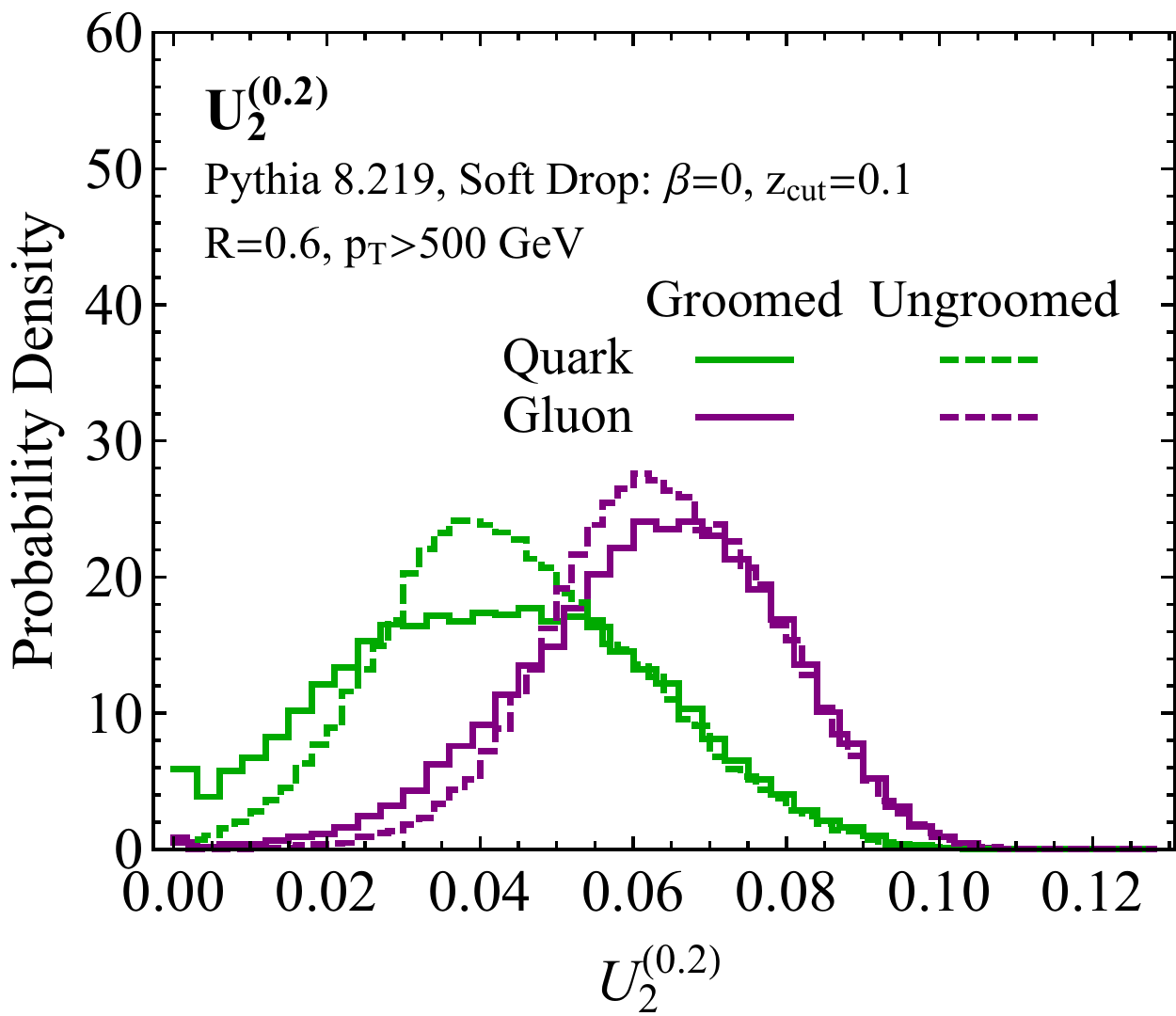}
}\qquad
\subfloat[]{\label{fig:qvsg_groom_dist_b}
\includegraphics[width=6.75cm]{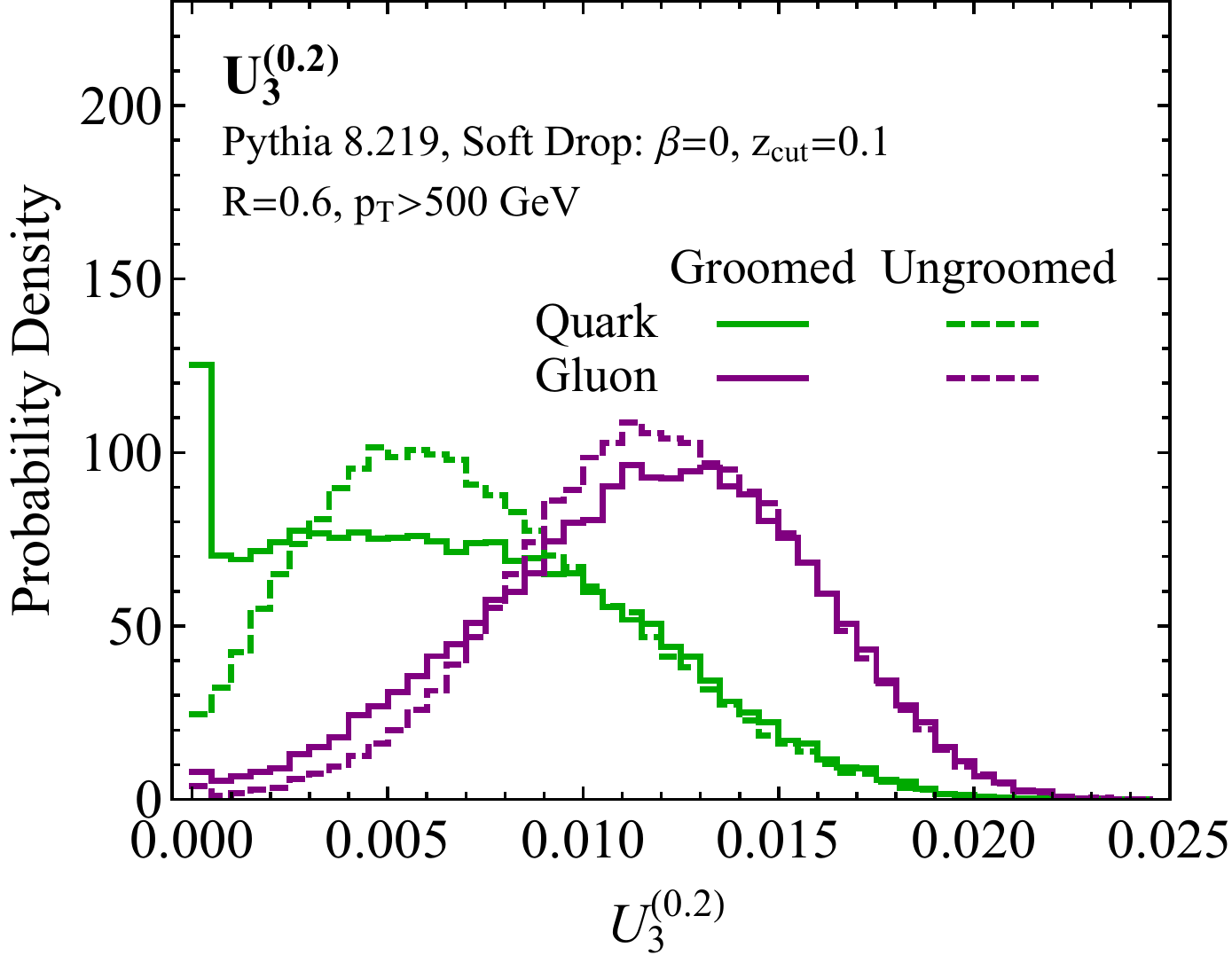}
}
\end{center}
\caption{Distributions of (a) $\Uobsnobeta{2}$ and (b) $\Uobsnobeta{3}$ for $\beta=0.2$, before and after grooming.  At large values of the observable, grooming has no impact on either the quark or gluon distribution, as expected.  The corresponding groomed ROC curves are given in \Fig{fig:2prong_obs_gluon_diffbeta_groom}. In (b), the bin at zero is due to jets that have three or fewer particles after grooming.}
\label{fig:qvsg_groom_dist}
\end{figure}

Finally, we want to test whether this improvement in quark/gluon discrimination power is robust to grooming.  In \Fig{fig:qvsg_groom_dist}, we compare the $\Uobsnobeta{2}$ and $\Uobsnobeta{3}$ distributions before and after grooming has been applied.  At large values of the observables, relatively little difference is observed for our baseline grooming parameters, as expected from the power-counting analysis of \Sec{sec:qvsg_setup}.  At smaller values of the observables, there is a distortion in the distributions due to the fact that grooming substantially decreases the overall particle multiplicity.  In particular, there are expected features at $\Uobsnobeta{2} = 0$ ($\Uobsnobeta{3} = 0$), from when the grooming gives less than three (four) particles in the jet.  In this regime, power-counting arguments are no longer applicable since the distribution is dominated by nonperturbative effects.  That said, as shown in \App{app:add_qg}, the ROC curves after grooming exhibit the same features as in the ungroomed case, with $\Uobsnobeta{2}$ and $\Uobsnobeta{3}$ outperforming $\Uobsnobeta{1}$, indicating that this parametric prediction is still robust.

It would be of great interest to perform explicit calculations of $\Uobsnobeta{2}$ to understand its exact dependence on the color Casimirs, as well as on the angular exponent $\beta$.  A resummed calculation, in particular, would shed light onto the all-orders structure of multiple-emission observables, which have not been widely explored in the literature.\footnote{See \Refs{Bertolini:2015pka,Larkoski:2015kga} for discussions of factorization and resummation of such observables, and \Refs{Field:2012rw,Larkoski:2015uaa} for fixed-order studies.}  It would also be useful to understand whether the measurement of multiple $\Uobsnobeta{i}$ observables with different $\beta$ values could be used to improve quark/gluon discrimination. The multi-differential cross section for $\Uobsnobeta{1}=\ecfnobeta{2}$ with two different angular exponents was calculated in \Refs{Larkoski:2014tva,Procura:2014cba} and the gains in performance for quark/gluon discrimination were studied in \Ref{Larkoski:2014pca} from the perspective of mutual information.  In preliminary investigations, we find that correlations among the $U_i$ are indeed helpful, but we leave a detailed study to future work.

\section{Conclusions}\label{sec:conc}

Continued progress in jet substructure relies on the ability to devise observables that can probe increasingly detailed aspects of jets.
In this paper, we used the known structures imposed by IRC safety to motivate the generalized energy correlation functions, $\ecfvarnobeta{v}{n}$, a flexible basis for constructing new substructure discriminants.  These generalized correlators incorporate an angular weighting function, allowing them to probe different angular structures within a jet.   We presented a number of case studies of relevance to the jet substructure community---boosted top tagging, boosted $W/Z/H$ tagging, and quark/gluon discrimination---demonstrating the power of power-counting techniques to design discriminants for specific purposes.  In each case, our newly-developed observables outperform standard jet shapes in parton shower studies.

The three series of observables introduced in this paper---$\Mobsnobeta{i}$, $\Nobsnobeta{i}$, and $\Uobsnobeta{i}$---exhibit new ways to probe the soft and collinear limits of QCD.  The $\Mobsnobeta{i}$ series is designed for tagging groomed jets, showing that the removal of soft radiation can dramatically change the phase space of $i$-prong discriminants.  The $\Nobsnobeta{i}$ series is designed to mimic $N$-subjettiness in the limit of resolved substructure, showing how to probe radiation patterns around collinear prongs without requiring external axes.  Finally, the $\Uobsnobeta{i}$ series is designed to evade the usual quark/gluon limitations imposed by Casimir scaling, showing the importance of multiple soft emissions for quark/gluon radiation patterns.  Taken together, these observables widen the scope for jet substructure investigations, allowing more handles to optimally use jets at the LHC.

Given their tagging performance, it would be interesting to calculate these observables from first principles.  This would provide insights into the impact of jet grooming on multi-prong observables, the difference between axes-based and axes-free observables, and the structure of multiple emissions within quark/gluon jets.  We are particularly interested in the differences between groomed and ungroomed distributions, since jet grooming not only changes the power counting of observables, but it also changes the logarithmic structure and power corrections in analytic calculations \cite{Dasgupta:2013ihk,Dasgupta:2013via,Larkoski:2014wba,Dasgupta:2015yua,Frye:2016okc,Frye:2016aiz}.  Beyond jet substructure, we suspect that the generalized correlators could eventually be useful as a tool for performing NNLO calculations; powerful slicing schemes have been devised using $N$-jettiness  \cite{Boughezal:2015dva,Gaunt:2015pea} and $\ecfvarnobeta{v}{n}$-based slicing could potentially be valuable in regimes where axes are inappropriate or cumbersome.

One aspect of jet substructure that has not been studied here is the correlations between discriminants.  We did apply power-counting techniques to identify correlations among basis elements to define optimal discriminants, but we did not consider whether power-counting could reveal parametric relationships between different proposed discriminants. Along similar lines, we did not consider in detail the hybrid strategy of using both groomed and ungroomed observables.  In preliminary investigations, we find that, not surprisingly, discriminants with the same power counting are highly correlated.  When discriminants have different power counting, though, there appears to be additional information gained through multi-variate combinations. At the moment, our application of power counting does not tell us what these multi-variate correlations are or whether we can robustly predict high-performing combinations.  We look forward to developing more sophisticated power-counting strategies to exploit these correlations in the future.

Finally, we want to emphasize the importance of first-principles calculations and unfolded experimental measurements of $U_1$, $U_2$, and $U_3$.  While the expected tagging performance of 2- and 3-prong discriminants---like $\Mobsnobeta{2}$, $\Nobsnobeta{2}$, $\Dobs{2}{1,2}$, and $\Nobsnobeta{3}$---can be seen directly from power-counting arguments, this is not the case for quark/gluon discriminants, since $C_F$ and $C_A$ are not parametrically different quantities.  For 1-prong jets, power counting can tell us which soft/collinear features are probed by the $U_i$ series, but it cannot reliably predict their expected parametric behavior or relative performance.  In parton shower studies, we do find that $\Uobsnobeta{2}$ and $\Uobsnobeta{3}$ exhibit improved performance over naive Casimir scaling, even in the larger $\beta$ regime where they are under better perturbative control, suggesting that the $U_i$ series is a sensitive probe of the QCD shower.  Therefore, measurements of the $U_i$ series, along with comparisons to parton shower (and eventually analytic) predictions, are likely to lead to deeper understanding of jets in QCD.

\begin{acknowledgments}
We thank Philip Harris, Andrew Larkoski, Simone Marzani, Ben Nachmann, Sid Narayanan, Duff Neill, Sal Rappoccio, and Nhan Tran for helpful discussions, and we thank Matteo Cacciari, Gavin Salam, and Gregory Soyez for help developing the \texttt{EnergyCorrelator} \fastjet{contrib}. IM is supported by the U.S. Department of Energy (DOE) under cooperative research agreement DE-SC0011090.  The work of LN and JT is supported by the DOE under grant contract numbers DE-SC-00012567 and  DE-SC0015476.  JT is also supported by a Sloan Research Fellowship from the Alfred P. Sloan Foundation.  This work was performed in part at the Aspen Center for Physics, which is supported by National Science Foundation grant PHY-1066293.
\end{acknowledgments}

\appendix

\section{Alternative Angular Weighting Functions}
\label{app:alt}

As discussed in \Sec{sec:gen_IRC}, any symmetric function of the angles, $f_N( \hat p_{i_1}, \hat p_{i_2}, \ldots,  \hat p_{i_N} )$, that vanishes in the collinear limits can in principle be used in \Eq{eq:gen_obs}.  While we argued in \Sec{sec:gen_ecf} that the $\text{min}$ function is particularly effective due to its ability to isolate hierarchical angular structures, other functional forms can certainly be used.  In this appendix, we study two alternate definitions of the angular weighting function, which, from a power counting perspective, are identical to those considered in the text.

For concreteness, we study variants of the $N_2$ observable from \Sec{sec:2prong}, which was based on a 3-point correlator:
\begin{equation}
\ecfvar{2}{3}{\beta}=\sum_{1\leq i<j<k\leq n_J} z_{i}z_{j}z_{k} \min \left\{\theta_{ij}^\beta \theta_{ik}^\beta\,, \theta_{ij}^\beta  \theta_{jk}^\beta\,,     \theta_{ik}^\beta \theta_{jk}^\beta    \right\} \qquad \Rightarrow \qquad N_2^{(\beta)}=\frac{_2e_3^{(\beta)}}{( \ecfvar{1}{2}{\beta})^2}.
\end{equation}  
One variant is to consider an angular weighting function that smoothly approximates the $\text{min}$ function.\footnote{The $r$ notation is motivated by the resistance formula for a set of parallel resistors.}  
\begin{align}
\label{eq:e3j}
_2r_3^{(\beta)} = \sum_{1\leq i<j<k\leq n_J} z_{i} z_{j} z_{k} \left( \frac{1}{\theta_{ij} \theta_{ik}} + \frac{1}{\theta_{ij}  \theta_{jk}} + \frac{1}{\theta_{ik} \theta_{jk} } \right)^{-\beta} \qquad \Rightarrow \qquad R_2^{(\beta)}=\frac{_2r_3^{(\beta)}}{( \ecfvar{1}{2}{\beta})^2}\,.
\end{align}
Another variant is to use the geometric fact that in the collinear limit, the minimum product of pairwise distances is parametrically the same as the area of the triangle spanned by the three points\footnote{To mimic the behavior of $\ecfvarnobeta{1}{3}$, one could consider the triangle area divided by its perimeter, which is parametrically related to the smallest distance in the collinear limit.}
\begin{align}
_2a_3^{(\beta)} = \sum_{1\leq i<j<k\leq n_J} z_i z_j z_k   \left(\sqrt{s (s-\theta_{ij}) (s-\theta_{ik})(s-\theta_{jk})} \right)^\beta  \qquad \Rightarrow \qquad  A_2^{(\beta)}=\frac{_2a_3^{(\beta)}}{( \ecfvar{1}{2}{\beta})^2}\,,
\end{align}
where $s = (\theta_{ij}+\theta_{jk}+\theta_{ik})/2$ comes from Heron's formula.  While $A_2$ is parametrically identical to $N_2$, it has the interesting property that it vanishes when the vectors defining the three particles are coplanar, similar to dipolarity introduced in \Ref{Hook:2011cq}.

\begin{figure}
\begin{center}
\subfloat[]{\label{fig:app_diff_observables_a}
\includegraphics[width=6.5cm]{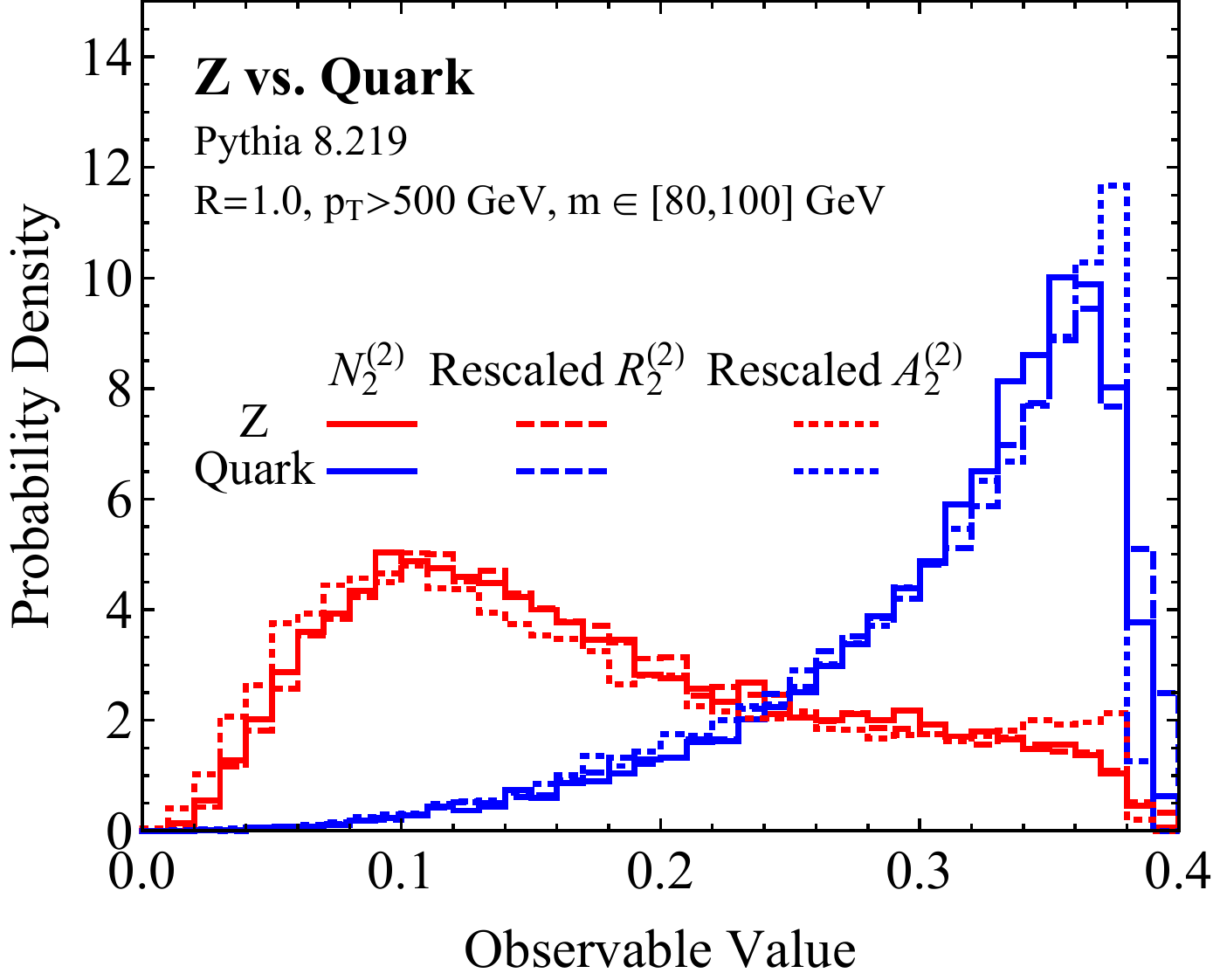}
}\qquad
\subfloat[]{\label{fig:app_diff_observables_b}
\includegraphics[width=6.5cm]{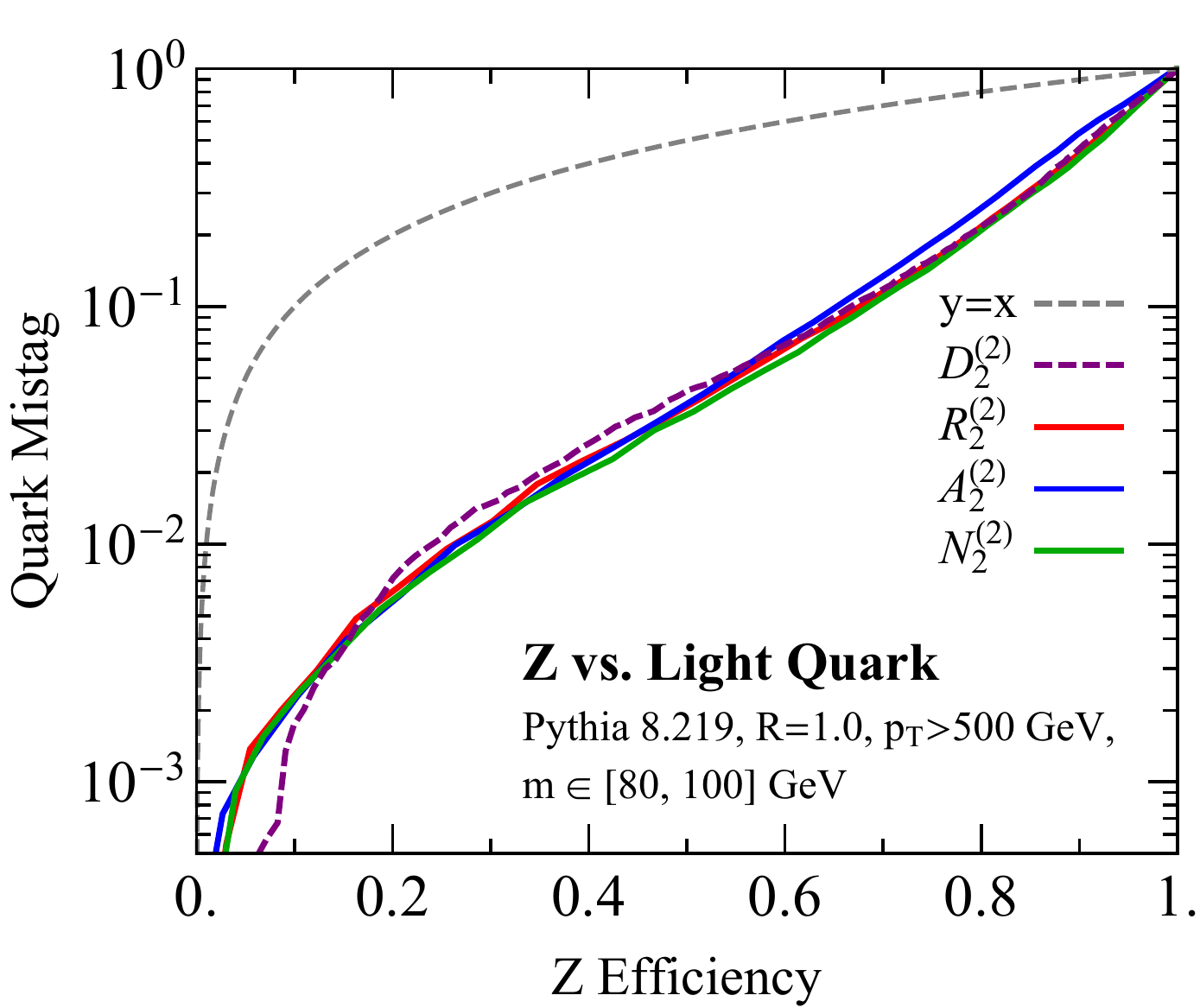}
}
\end{center}
\caption{Two alternative definitions of the $N_2$-style observable---$R_2$ and $A_2$.  (a) Distributions on $Z$ and quark samples. (b) ROC curves for discrimination performance.  Since the alternative definitions have identical power counting to $N_2$, they exhibit parametrically similar behavior. 
}
\label{fig:app_diff_observables}
\end{figure}

Even though the $\Nobsnobeta{2}$, $R_2$, and $A_2$ observables have identical power counting, their distributions could in principle differ by $\mathcal{O}(1)$ numbers, possibly allowing for improved discrimination power.  In \Fig{fig:app_diff_observables_a} we compare the distributions of these three observables in \textsc{Pythia}, showing that they are rather similar. To aid the eye, we have rescaled the $R_2$ and $A_2$ distributions to match the $N_2$ distribution.  Turning to the $Z$ versus quark ROC curve in \Fig{fig:app_diff_observables_b}, the performance is nearly identical.  This further emphasizes that the behavior of the observables is dominated by parametric scalings.  Since we did not find any gains from using these more complicated variants, we restricted the study in the text to the definition given in \Eq{eq:ecf_gen}.

It is still an interesting question whether other choices of angular weighting functions might lead to improved performance in more complicated jet substructure applications.  It seems unlikely, however, since for small radius jets, one can Taylor expand the angular function in the small $\theta$ limit, and observables with the same power counting must have the same lowest-order expansion.  In practice, the use of smoother definitions which approximate the $\text{min}$ function might be useful for performing perturbative calculations.

\section{Aspects of 3-prong Tagging}

\subsection{Challenges for $M_3$}\label{app:add_plots_M3}

In \Sec{sec:m_series}, we defined the general series of $\Mobsnobeta{i}$ observables.  We saw in \Sec{sec:2prong} that the $\Mobsnobeta{2}$ observable was an effective boosted $W/Z/H$ tagger on groomed jets.  One might therefore consider the $\Mobsnobeta{3}$ observable,
\begin{align}
\Mobs{3}{\beta}=\frac{\ecfvar{1}{4}{\beta}}{\ecfvar{1}{3}{\beta}}\,,
\end{align}
as a possible boosted top tagger.  

We can see from a power-counting analysis, however, that even with grooming, $\Mobsnobeta{3}$ will not perform well.  Following the notation of \Sec{sec:tops_makeobs}, a strongly-ordered 3-prong jet has 
\begin{align}
\text{3-prong signal (groomed):} \qquad \ecfvar{1}{3}{\beta}&\sim \theta_{23}^\beta\,, \nn \\
\ecfvar{1}{4}{\beta}&\sim z_{ccs}  \theta_{23}^\beta +  \theta_{cc}^\beta \,,
\end{align}
while a 2-prong background jet has 
\begin{align}\label{eq:M3_pc_bkg}
\text{2-prong background (groomed):} \qquad \ecfvar{1}{3}{\beta}&\sim    z_{cs} \theta_{cs}^\beta +\theta_{cc}^\beta \,, \nn \\
\ecfvar{1}{4}{\beta}&\sim   z_{cs}^2 \theta_{cs}^\beta +  \theta_{cc}^\beta \,.
\end{align}
For signal jets, we have the relation $\ecfvarnobeta{1}{4}\ll \ecfvarnobeta{1}{3}$, so we would like the background to satisfy $\ecfvarnobeta{1}{4}\sim \ecfvarnobeta{1}{3}$.   That desired relation is violated, though, by contributions of the collinear-soft modes to $\ecfvarnobeta{1}{4}$, due to the different $z_{cs}$ scalings in \Eq{eq:M3_pc_bkg}.  We therefore predict from power counting that $\Mobsnobeta{3}$ should be a poor discriminant. 

\begin{figure}
\begin{center}
\includegraphics[width=6.5cm]{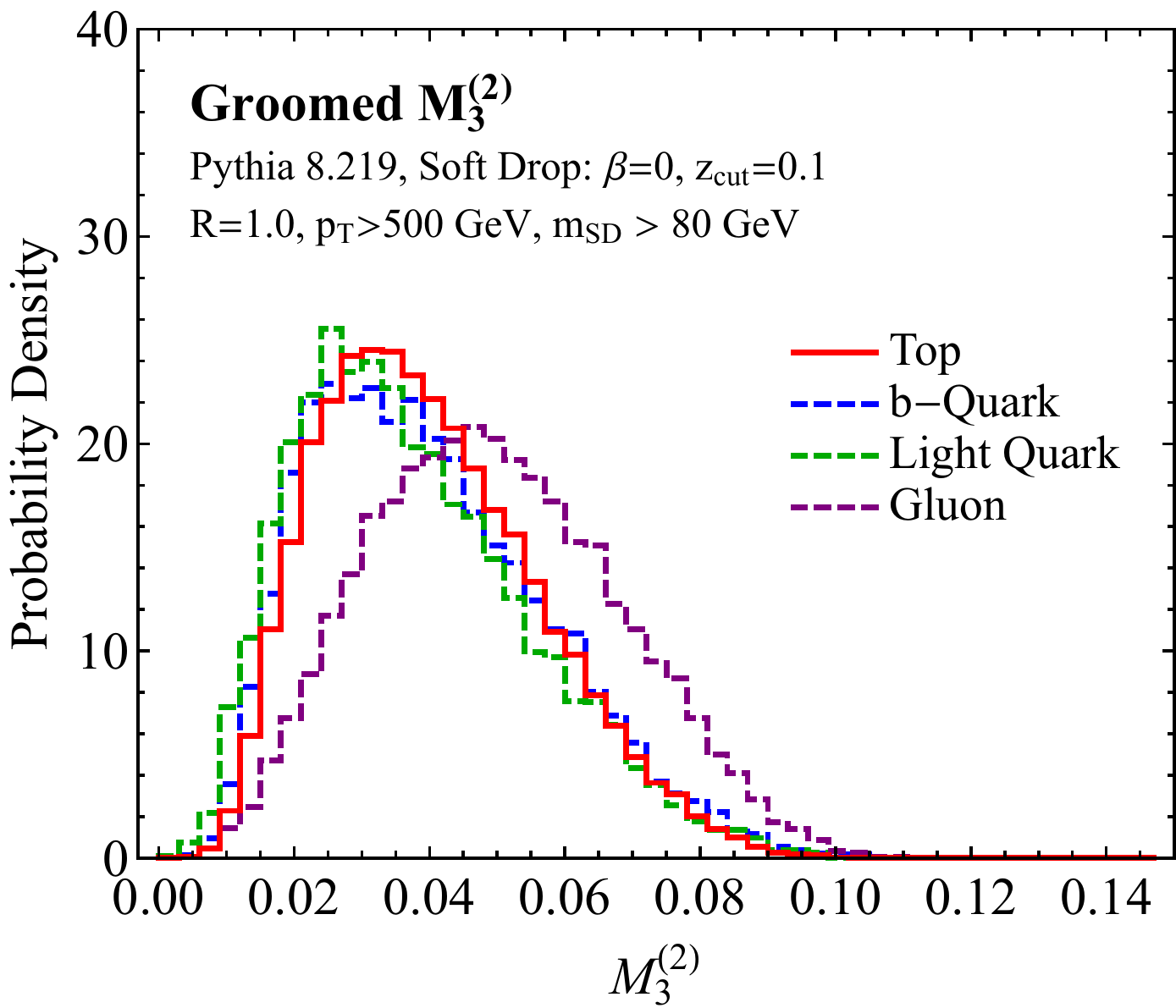}
\end{center}
\caption{Distributions of $\Mobsnobeta{3}$, comparing the signal of boosted top jets to the background of $b$-quarks, light quarks, and gluons.  As expected from power counting, limited discrimination is observed.  
}
\label{fig:app_M3}
\end{figure}

In \Fig{fig:app_M3}, we show the distribution of $\Mobsnobeta{3}$ for boosted top jets compared to those from QCD jet backgrounds, where little discrimination power is observed.   Similar to how ordinary grooming was required for $\Mobsnobeta{2}$ to become an effective discriminant in the 2-prong case, it is likely that another layer of grooming is be needed to remove the undesired collinear-soft contributions to $\Mobsnobeta{3}$ and make it an effective 3-prong tagger.  While we do not pursue $\Mobsnobeta{3}$ further in this paper, it would be interesting to consider alternative grooming methods designed to isolate 3-prong structure and mitigate both soft and collinear-soft radiation.  As a starting point, one could consider doubly-soft-dropped boosted top jets, where after an initial application of soft drop, one reapplies soft drop to the two remaining prongs.

\subsection{$N_3$ Without Grooming}\label{app:add_plots_N3}

\begin{figure}
\begin{center}
\subfloat[]{\label{fig:app_N3_a}
\includegraphics[width=6.5cm]{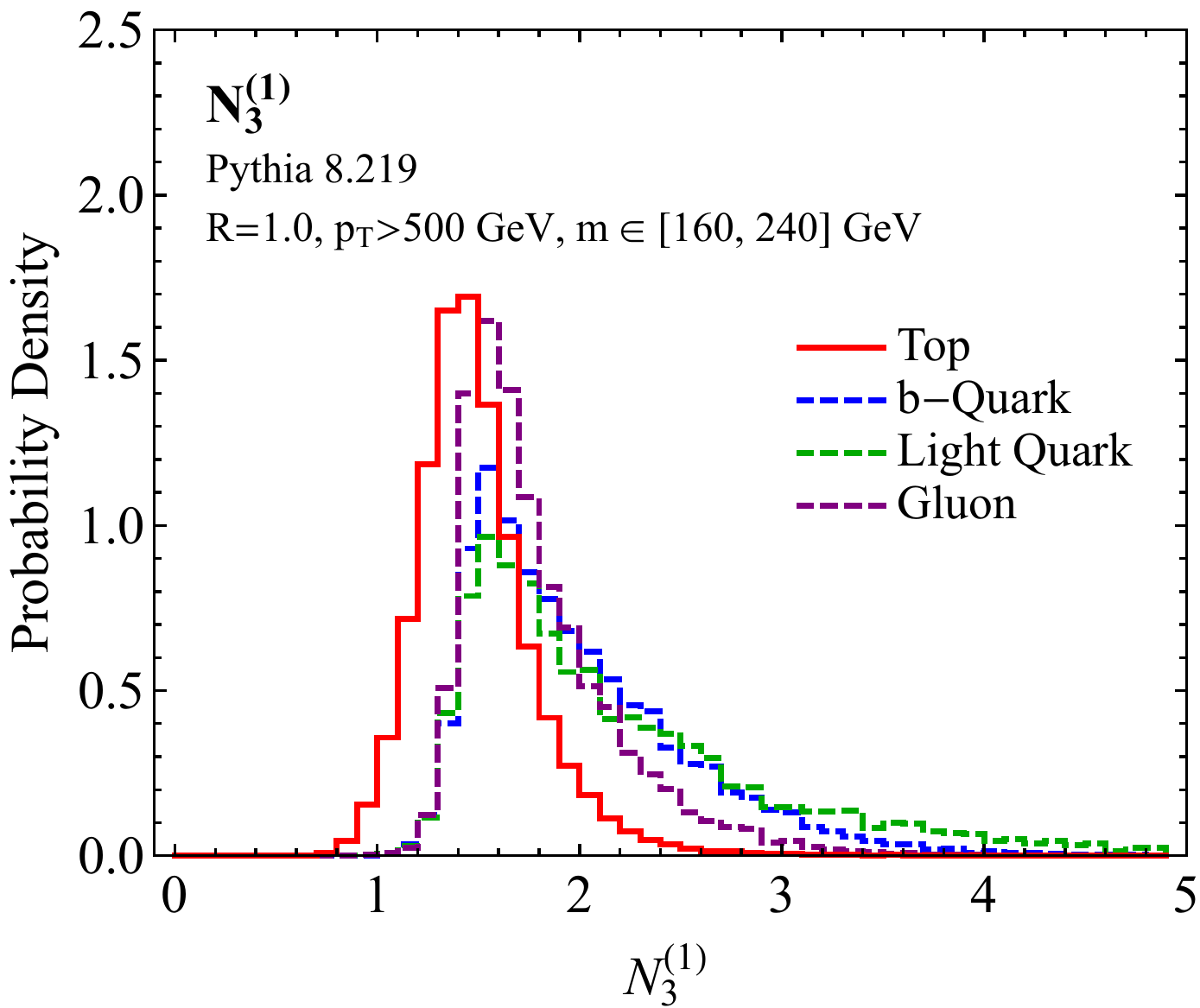}
}
\qquad
\subfloat[]{\label{fig:app_N3_d}
\includegraphics[width=6.5cm]{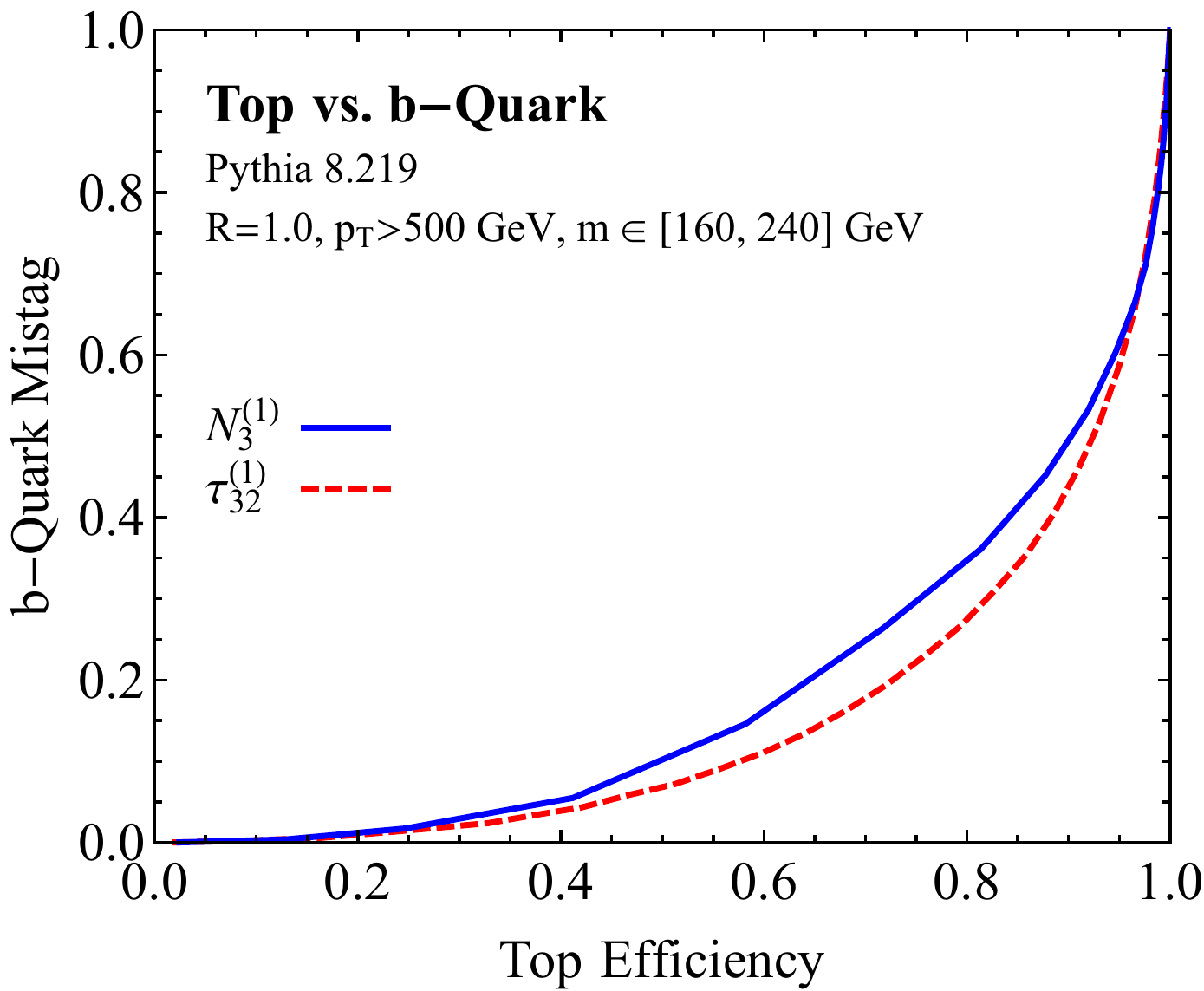}
}
\qquad
\subfloat[]{\label{fig:app_N3_b}
\includegraphics[width=6.5cm]{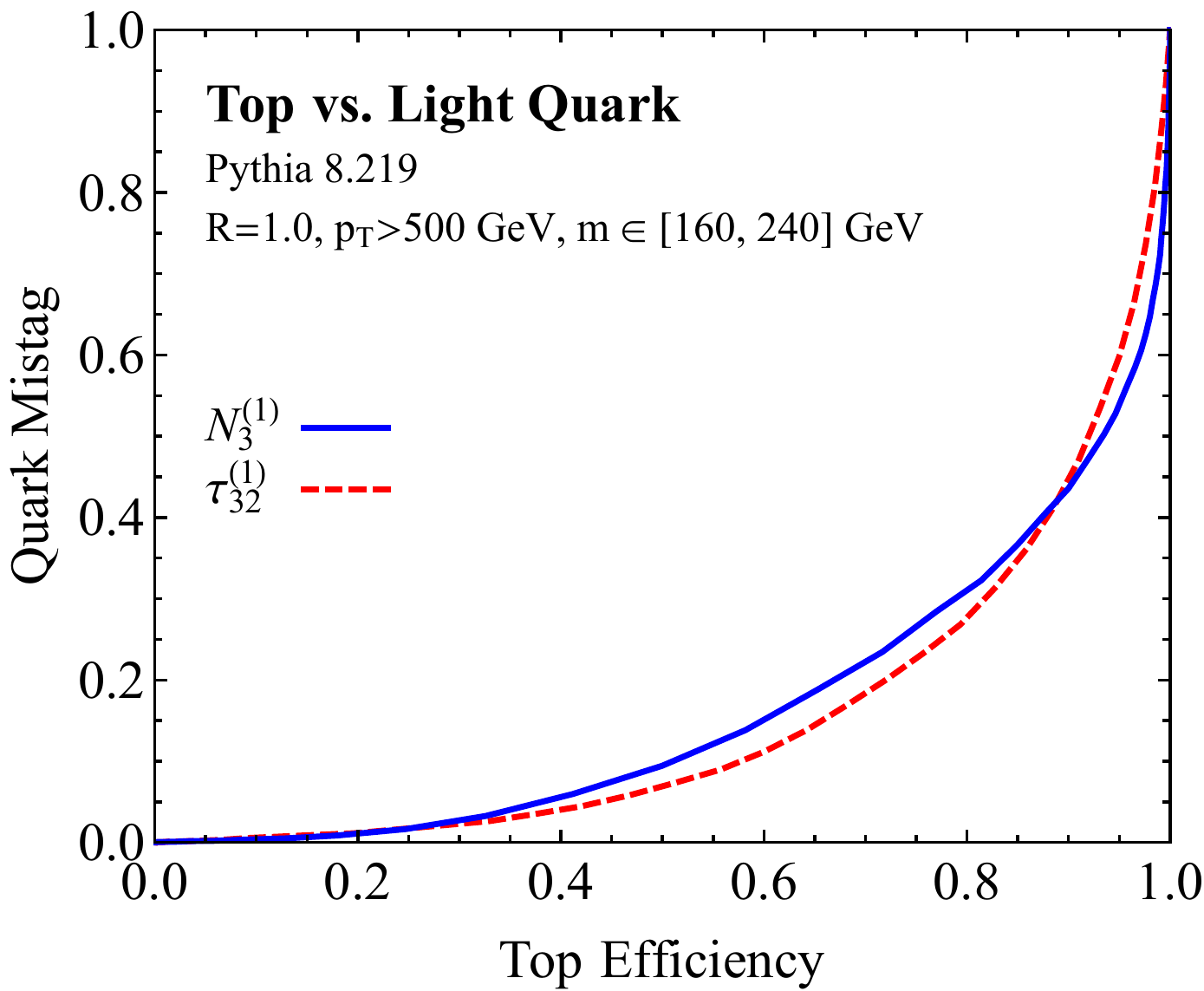}
}
\qquad
\subfloat[]{\label{fig:app_N3_c}
\includegraphics[width=6.5cm]{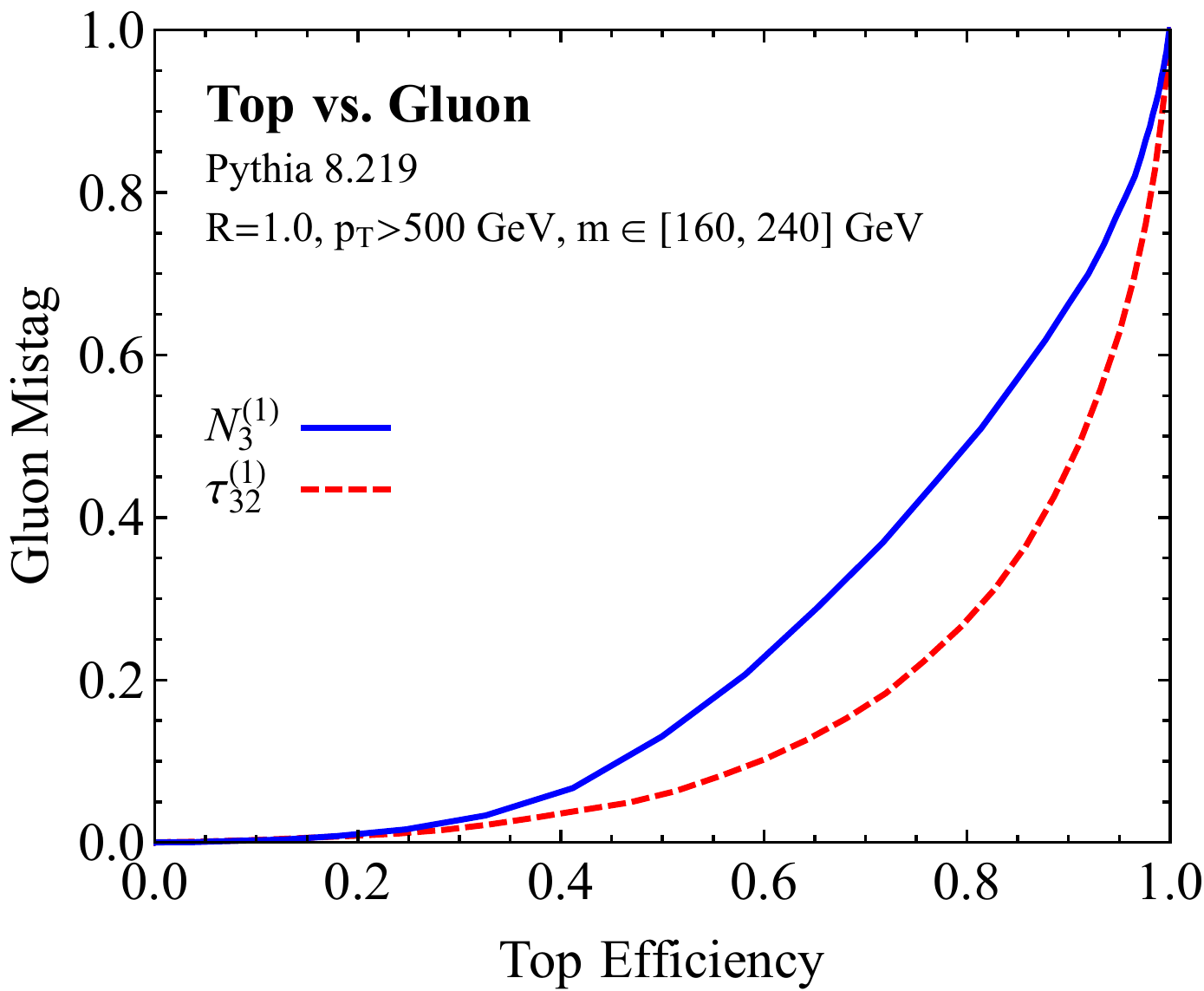}
}
\end{center}
\caption{Behavior of $\Nobsnobeta{3}$ without grooming. (a) Distributions of $\Nobsnobeta{3}$ for boosted top signals compared to $b$-quark, light quark, gluon backgrounds.  (b,c,d) ROC curves comparing the discrimination power of $\Nobsnobeta{3}$ versus $\Nsubnobeta{3,2}$ against the different QCD backgrounds. }
\label{fig:app_N3}
\end{figure}

In \Sec{sec:tops_makeobs}, we argued that on groomed jets with well-resolved substructure, $\Nobsnobeta{3}$ behaves parametrically like $\Nsubnobeta{3,2}$, but exhibits improved discrimination power in the transition to the unresolved region.  On ungroomed jets, however, $\Nobsnobeta{3}$ behaves differently from $\Nsubnobeta{3,2}$, and in particular, it does not provide good discrimination in regions of phase space where there is a soft wide-angle subjet.   This same issue was discussed in detail for the case of $D_3$ in \Ref{Larkoski:2014zma}; the treatment of the soft subjet region of phase space required the addition of two extra terms to $D_3$, leading to the complicated form shown in \Eq{eq:D3_full_def}.  To avoid the soft subjet issue, and to advocate for the stability of groomed observables, we explicitly focused on the case of groomed top jets in \Sec{sec:tops_makeobs}.  

Here, we compare $\Nobsnobeta{3}$ and $\Nsubnobeta{3,2}$ on ungroomed jets. Though $\Nobsnobeta{3}$ was not designed for use on ungroomed jets, it still provides reasonably good discrimination power, though not as good as $\Nsubnobeta{3,2}$.  Distributions of ungroomed $\Nobsnobeta{3}$ are shown in \Fig{fig:app_N3_a}, where we use an alternative mass window cut of $m \in [160, 240]$ GeV.  The discrimination performance for the top signal against the $b$-quark, light quark, and gluon jet backgrounds are shown in \Figss{fig:app_N3_d}{fig:app_N3_b}{fig:app_N3_c}, respectively.   The best performance is seen in rejecting quark jets, although ungroomed $N_3$ has worse performance on gluon jets.  Interestingly, similar quark/gluon differences were seen for $D_3$ in \Ref{Larkoski:2015yqa}, although the nature of this behavior is not understood and is not necessarily connected in any way to the use of energy correlators.

Though $\Nobsnobeta{3}$ was designed for use on groomed jets, we believe that $\Nobsnobeta{3}$ is a sufficiently good discriminant on ungroomed jets to merit further investigations.  At minimum, ungroomed $\Nobsnobeta{3}$ distributions could be measured as a baseline to test the impact of jet grooming.  We offer a bounty to the first group that identifies an axes-free observable with the same power counting as ungroomed $\Nsubnobeta{3,2}$.

\begin{figure}
\begin{center}
\includegraphics[width=6.5cm]{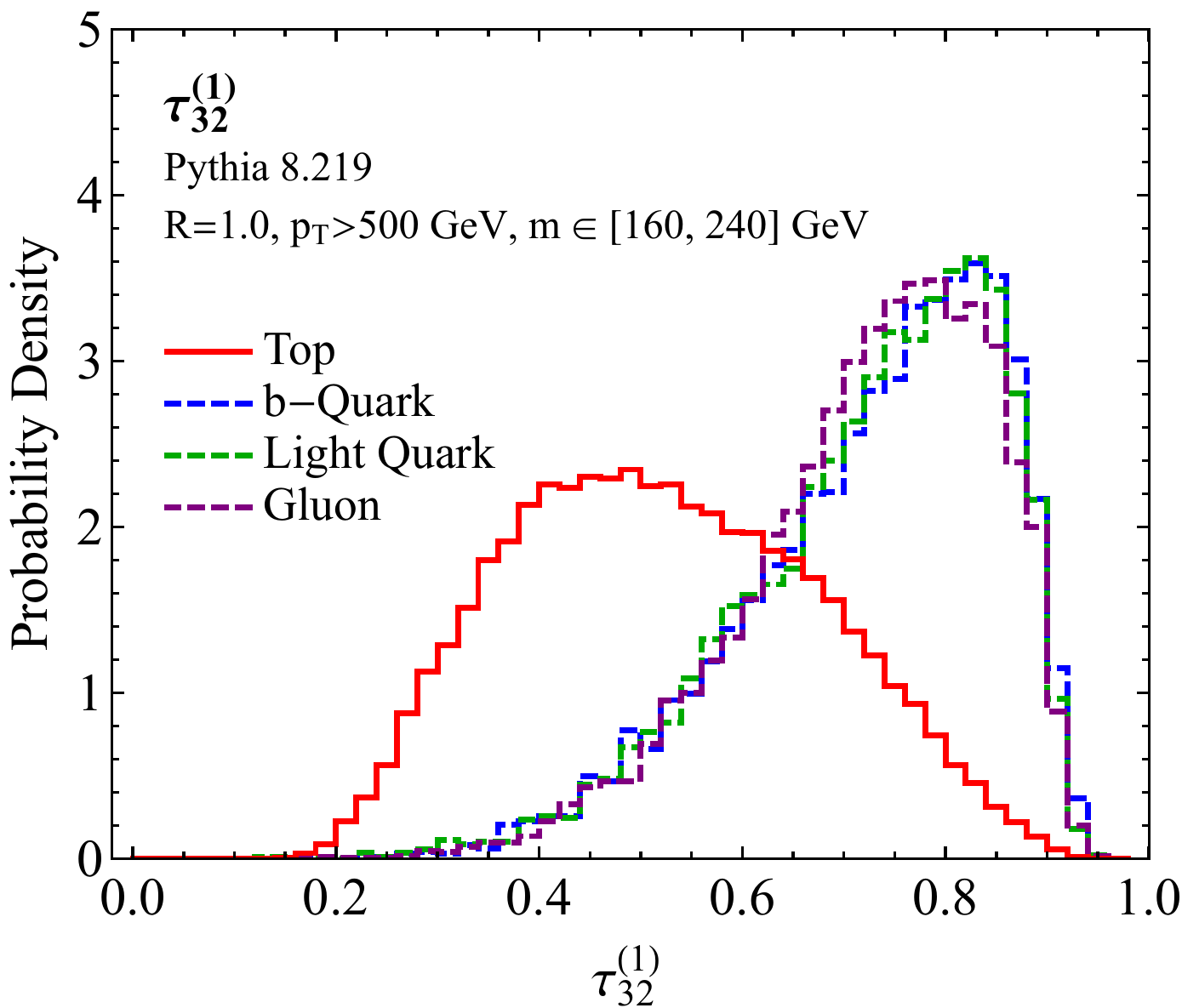}
\qquad
\includegraphics[width=6.5cm]{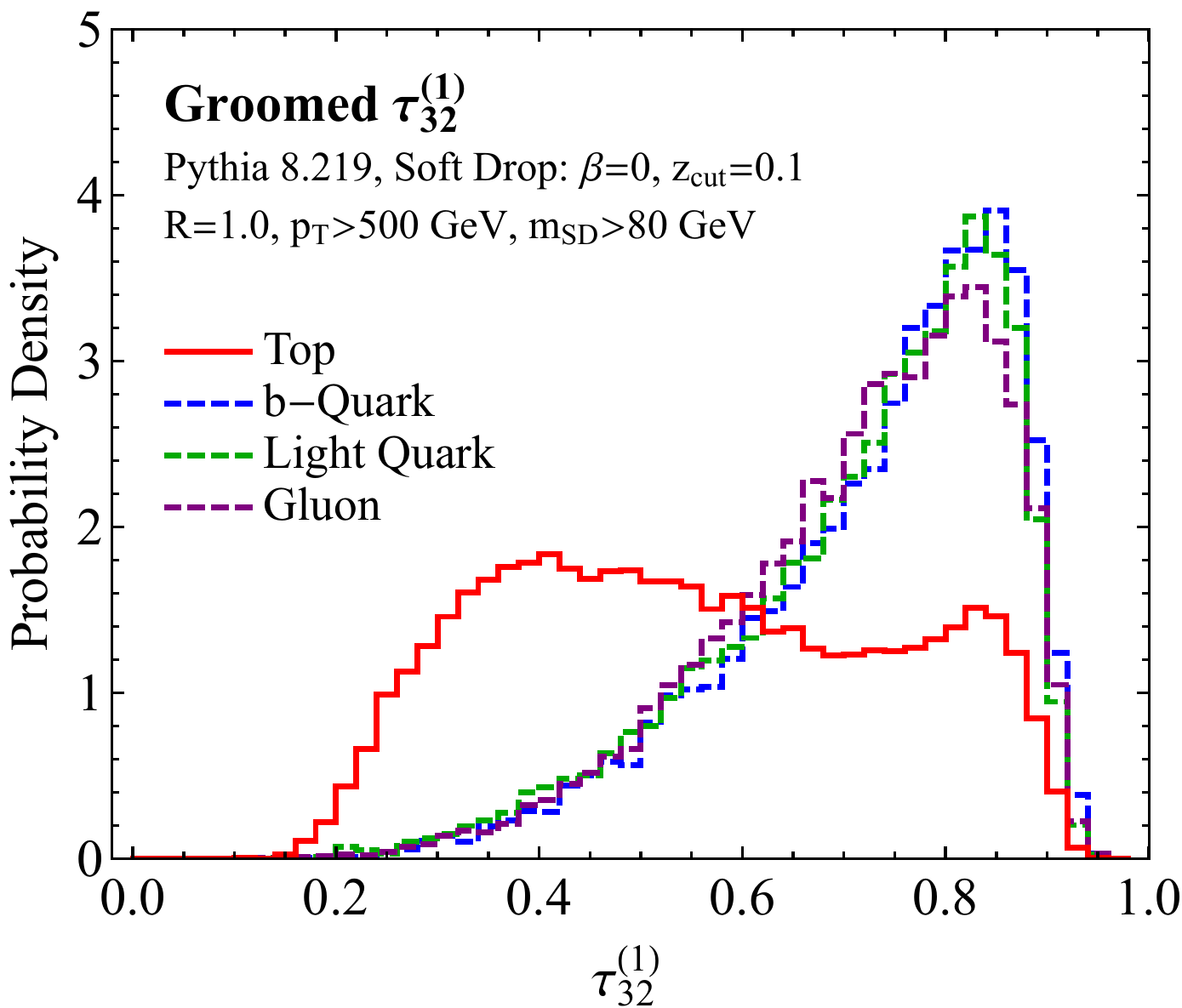}
\end{center}
\caption{Distributions of the $N$-subjettiness ratio $\Nsubnobeta{3,2}$ (a) before grooming and (b) after grooming for both the boosted top signal and the different QCD backgrounds. 
}
\label{fig:app_tau32}
\end{figure}

For completeness, in \Fig{fig:app_tau32}, we show the $N$-subjettiness observable $\Nsubnobeta{3,2}$ as measured on the same samples, both before and after grooming.  As expected, excellent discrimination power is observed is observed before grooming.  After grooming, the discrimination power is worsened primarily due to the behavior in the unresolved region, namely as $\Nsubnobeta{3,2}\to 1$.  It is in this region that $\Nobsnobeta{3}$ exhibits improved performance, as seen already in the behavior of the distributions in \Fig{fig:3prong_N3} and the performance in the ROC curve in \Fig{fig:N3_ROC_3prong_linear}. 

\subsection{Identifying $\Nobsnobeta{3}$}
\label{app:N3_identify}

In \Sec{sec:tops_makeobs}, we considered the observable $\Nobsnobeta{3}$ defined as
\begin{align}
\Nobs{3}{\beta}=\frac{\ecfvar{2}{4}{\beta}}{(\ecfvar{1}{3}{\beta})^2}\,.
\end{align}
There are, however, a large number of other possible observables that could be formed from combinations of the different 2- , 3-, and 4-point correlators.  In this appendix, we describe in more detail the justification for our focus on $\Nobsnobeta{3}$.  It is interesting that this process happens to identify an observable with the same parametric behavior as the $N$-subjettiness ratio $\Nsubnobeta{3,2}$.  As discussed in the text, we focus on the case of groomed jets.  This means that we can ignore soft radiation for our power counting analysis.

For groomed boosted top jets, it is sufficient to consider a 3-prong configuration with hierarchical angles, as illustrated in \Fig{fig:p_NINJA}.  In particular, we do not have to consider the soft subjet phase space region from \Ref{Larkoski:2014zma}, which has hierarchical energies, since those configurations are removed by the grooming procedure. 
For the 3-prong signal, the scaling of the 2-point correlator is
\begin{align}
\ecf{2}{\beta}\sim \theta_{12}^\beta\,,
\end{align}
the scalings of different 3-point correlators are
\begin{align}
\ecfvar{3}{3}{\beta}&\sim \theta_{23}^\beta \theta_{12}^{2\beta}\,, \nn \\
\ecfvar{2}{3}{\beta}&\sim \theta_{23}^\beta \theta_{12}^{\beta}\,, \nn \\
\ecfvar{1}{3}{\beta}&\sim \theta_{23}^\beta \,.
\end{align}
and the scalings of the different 4-point correlators are
\begin{align}\label{eq:app_pc_4point}
\ecfvar{6}{4}{\beta}&\sim z_{cs} \theta_{12}^{5\beta} \theta_{23}^{\beta}  +z_{ccs}  \theta_{12}^{3\beta} \theta_{23}^{3\beta} 
+\theta_{23}^\beta \theta_{cc}^\beta \theta_{12}^{4\beta} \,, \nn \\
\ecfvar{5}{4}{\beta}&\sim  z_{cs} \theta_{12}^{4\beta} \theta_{23}^{\beta} + z_{ccs} \theta_{12}^{2\beta} \theta_{23}^{3\beta} 
+ \theta_{23}^\beta \theta_{cc}^\beta \theta_{12}^{3\beta}\,, \nn \\
\ecfvar{4}{4}{\beta}&\sim  z_{cs} \theta_{12}^{3\beta} \theta_{23}^{\beta} + z_{ccs} \theta_{12}^{1\beta} \theta_{23}^{3\beta} 
+\theta_{23}^\beta \theta_{cc}^\beta \theta_{12}^{2\beta}\,, \nn \\
\ecfvar{3}{4}{\beta}&\sim z_{cs} \theta_{12}^{2\beta} \theta_{23}^{\beta}  +z_{ccs} \hphantom{\theta_{12}^{3\beta}}   \theta_{23}^{3\beta}
+\theta_{23}^\beta \theta_{cc}^\beta \theta_{12}^{\beta}\,, \nn \\
\ecfvar{2}{4}{\beta}&\sim z_{cs} \theta_{12}^{\beta} \theta_{23}^{\beta}  + z_{ccs} \hphantom{\theta_{12}^{3\beta}}   \theta_{23}^{2\beta} 
+\theta_{23}^\beta \theta_{cc}^\beta \hphantom{\theta_{12}^{4\beta}}\,, \nn \\
\ecfvar{1}{4}{\beta}&\sim 0\hphantom{z_c \theta_{12}^{\beta} \theta_{23}^{\beta} }
+z_{ccs} \hphantom{\theta_{12}^{5\beta}}  \theta_{23}^{\beta} 
+\hphantom{\theta_{23}^\beta} \theta_{cc}^\beta \hphantom{\theta_{12}^{4\beta}}\,,
\end{align}
where the alignment and zero in the last line are there just to help guide the eye.  

For 2-prong background jets, all we need is the scaling of the 2-point correlator,
\begin{align}
\ecf{2}{\beta}\sim \theta_{cs}^\beta\,,
\end{align}
and the scalings of the different 3-point correlators, 
\begin{align}
\ecfvar{3}{3}{\beta}&\sim z_{cs} \theta_{cs}^{3\beta} +\theta_{cs}^{2\beta} \theta_{cc}^\alpha\,, \nn \\
\ecfvar{2}{3}{\beta}&\sim z_{cs} \theta_{cs}^{2\beta} +\theta_{cs}^{\beta} \theta_{cc}^\alpha\,, \nn \\
\ecfvar{1}{3}{\beta}&\sim z_{cs}\theta_{cs}^{\beta} + \theta_{cc}^\alpha \,.
\end{align}
While the background scalings of the 4-point correlators would be needed to verify signal/background separation, as was done in \Sec{sec:tops_makeobs}, they are not needed to restrict the combinations under consideration.  Since their form is not particularly illuminating, we do not show them here.

While there are a large number of observables listed above, the analysis can be simplified by noting that for both the signal and background, the information contained in the 3-point correlators $\ecfvarnobeta{2}{3}$ and $\ecfvarnobeta{3}{3}$ is redundant, since it can be expressed in terms of $\ecfnobeta{2}$ and $\ecfvarnobeta{1}{3}$.  Furthermore, any observable derived from power counting will be linear in the 4-point correlator and will have the 3-point correlator appearing in the denominator raised to some power.   Finally, from \Eq{eq:app_pc_4point}, we see that $\theta_{23}$ appears at most raised to the third power; it therefore suffices to consider $\ecfvarnobeta{1}{3}$ raised at most to the third power.  The power of $\ecfnobeta{2}$ is then fixed by Lorentz invariance.

The above logic allows us to write down a parametrically complete set of potential 3-prong observables,
\begin{align}
\mathcal{O}_{v,y} = \frac{\ecfvar{v}{4}{\beta}  \left( \ecf{2}{\beta} \right)^{y-v} }{  \left( \ecfvar{1}{3}{\beta} \right)^y}\,, \qquad v\in \{1,2,3,4,5,6\}\,, \qquad y\in \{1,2,3\}\,, \qquad y\leq v\,.
\end{align}
At this point, one can then either power count each of these options explicitly to test for background isolation, or simply evaluate their performance in a parton shower generator.  To limit the number of options to consider, one can apply the further constraint that $\ecfnobeta{2}$ should not appear explicitly in the observable, to mitigate correlations with the jet mass.  This is equivalent to setting $y=v$, and gives
\begin{align}
T_{v}=\frac{\ecfvar{v}{4}{\beta}   }{  \left( \ecfvar{1}{3}{\beta} \right)^v}\,.
\end{align}
Note that $v=1$ gives $\Mobsnobeta{3}$ and $v=2$ gives $\Nobsnobeta{3}$.  Among all of the $\mathcal{O}_{v,y}$ observables, we found that the best performing one in \textsc{Pythia} was $\Nobsnobeta{3}$, which then became the focus of our boosted top study.

\subsection{Power Counting $\Nobsnobeta{3}$}
\label{app:N3_pc}

While the identification of the parametrically optimal discriminant is usually fairly straightforward given the parametric expressions for the observables, confusions can arise when the scalings have multiple terms.  Here, we present more details for the signal analysis of the $\Nobsnobeta{3}$ observable from \Sec{sec:tops_makeobs}, to illustrate how power counting can be performed systematically. This allows one to avoid potential confusions when there are competing parametric relations.  This same approach can be used in the other examples studied in the paper, though for the 1- and 2-prong case studies, we find that the more heuristic treatment in the text is just as illuminating as the systematic strategy.

We begin by recalling the power counting for $\ecfvarnobeta{1}{3}$ and $\ecfvarnobeta{2}{4}$, considering the signal with (hierarchical) 3-prong substructure: 
\begin{align}
\ecfvar{1}{3}{\beta}&\sim \theta_{23}^\beta\,, \nn \\
\ecfvar{2}{4}{\beta}&\sim z_{cs} \theta_{12}^\beta\theta_{23}^\beta  +z_{ccs} \theta_{23}^{2\beta}  + \theta_{23}^\beta \theta_{cc}^\beta \,.
\label{eq:appB4scaling}
\end{align}
We next need to identify which of the parameters---$\theta_{12}$, $\theta_{23}$, $\theta_{cc}$, $z_{cs}$, and $z_{ccs}$---are set by which measurements.  Since most boosted top analyses apply a mass cut, we assume that $\theta_{12}$ is set by a mass measurement.  This is not crucial, however, and the argument below can be generalized without the fixed-mass assumption.  This leaves us with the task of determining the parametric relationship between $\{\theta_{23},\theta_{cc},z_{cs},z_{ccs}\}$ and $\{\ecfvarnobeta{1}{3},\ecfvarnobeta{2}{4}\}$.  Clearly, the measurement of $\ecfvarnobeta{1}{3}$ sets $\theta_{23}$.  By assumption, there is no hierarchy between the three terms in $\ecfvarnobeta{2}{4}$, yielding the following scaling of the kinematic variables:
\begin{align}
\label{eq:appB4pararelation}
\theta_{23}^\beta \sim \ecfvar{1}{3}{\beta}\,, \qquad \theta_{cc}^\beta \sim \frac{\ecfvar{2}{4}{\beta}}{\ecfvar{1}{3}{\beta}}\,, \qquad z_{cs}\sim \frac{\ecfvar{2}{4}{\beta}}{\theta_{12}^\beta \ecfvar{1}{3}{\beta}}    \,, \qquad z_{ccs}\sim \frac{\ecfvar{2}{4}{\beta}}{(\ecfvar{1}{3}{\beta})^2}\,.
\end{align}

Now, we want to derive an observable which distinguishes 3-prong jets from jets with fewer than 3 prongs.  This can be accomplished by identifying the regions in phase space where the 3-prong EFT description breaks down, and translating that into constraints on the relationship between $\ecfvarnobeta{1}{3}$ and $\ecfvarnobeta{2}{4}$.  As given in \Eq{eq:3prongscaling} and illustrated in \Fig{fig:p_NINJA}, 3-prong phase space is defined by the following four conditions:   
\begin{alignat}{2}
&(a)~~ \theta_{23} \ll \theta_{12}\ll 1\,, \qquad &
(b)~~ \theta_{cc} \ll \theta_{23}\,, \nn \\
&(c)~~ z_{ccs}\ll 1\,, &
(d)~~ z_{cs} \ll z_{ccs}\,, 
\end{alignat}
Plugging \Eq{eq:appB4pararelation} into condition $(a)$, we find 
\begin{align}
(a)\implies \ecfvar{1}{3}{\beta} \ll \ecf{2}{\beta}\ll 1\,.
\end{align}
This just defines the region of validity of our analysis, but is not helpful in determining a relationship between $\ecfvarnobeta{1}{3}$ and $\ecfvarnobeta{2}{4}$. 
Turning to condition $(b)$, we find
\begin{align}
(b) \implies \frac{\ecfvar{2}{4}{\beta}}{\ecfvar{1}{3}{\beta}} \ll \ecfvar{1}{3}{\beta}\implies \frac{\ecfvar{2}{4}{\beta}}{(\ecfvar{1}{3}{\beta})^2} \ll 1\,.
\end{align}
Note that the constraint $\theta_{cc}\ll1$ does not give as strong a bound.
Condition $(c)$ gives the same constraint,
\begin{align}
(c) \implies   \frac{\ecfvar{2}{4}{\beta}}{(\ecfvar{1}{3}{\beta})^2} \ll 1\,.
\end{align}
Finally, we see that condition $(d)$ is already satisfied by condition $(a)$,
\begin{align}
(d) \implies \frac{\ecfvar{2}{4}{\beta}}{\theta_{12}^\beta \ecfvar{1}{3}{\beta}} \ll  \frac{\ecfvar{2}{4}{\beta}}{(\ecfvar{1}{3}{\beta})^2} \implies \ecfvar{1}{3}{\beta} \ll \theta_{12}^\beta.
\end{align}
and provides no extra information.

From this analysis, one finds that the strongest constraint on the breakdown of the 3-prong EFT is 
\begin{align}
\frac{\ecfvar{2}{4}{\beta}}{(\ecfvar{1}{3}{\beta})^2} \ll 1\,,
\end{align}
leading to the definition of the $\Nobsnobeta{3}$ observable,
\begin{align}
\Nobs{3}{\beta}=\frac{\ecfvar{2}{4}{\beta}}{(\ecfvar{1}{3}{\beta})^2}\,.
\end{align}
With practice, one can immediate infer this result from the scaling of the observables in \Eq{eq:appB4scaling}, without having to explicitly consider each EFT constraint, but this example illustrates how the procedure can be performed systematically when confusions arises.

\section{Relationship Between $N_i$ and $N$-subjettiness}\label{app:Nsub_Ni}

\begin{figure*}[t]
\centering
\subfloat[]{\label{fig:app_obs_illus_b}
\includegraphics[width=6.5cm]{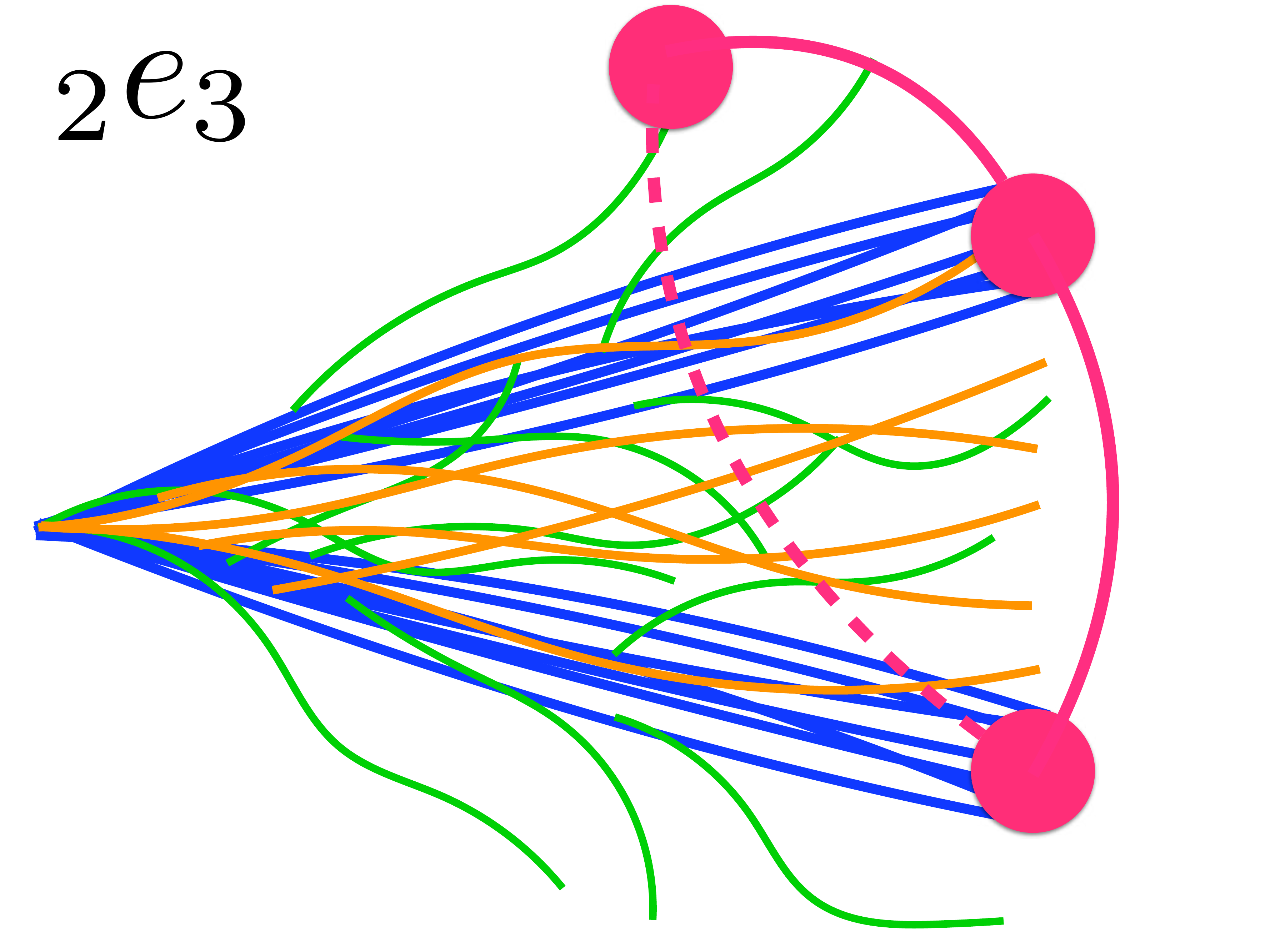}    
} \ \ 
\subfloat[]{\label{fig:app_obs_illus_a}
\includegraphics[width=6.5cm]{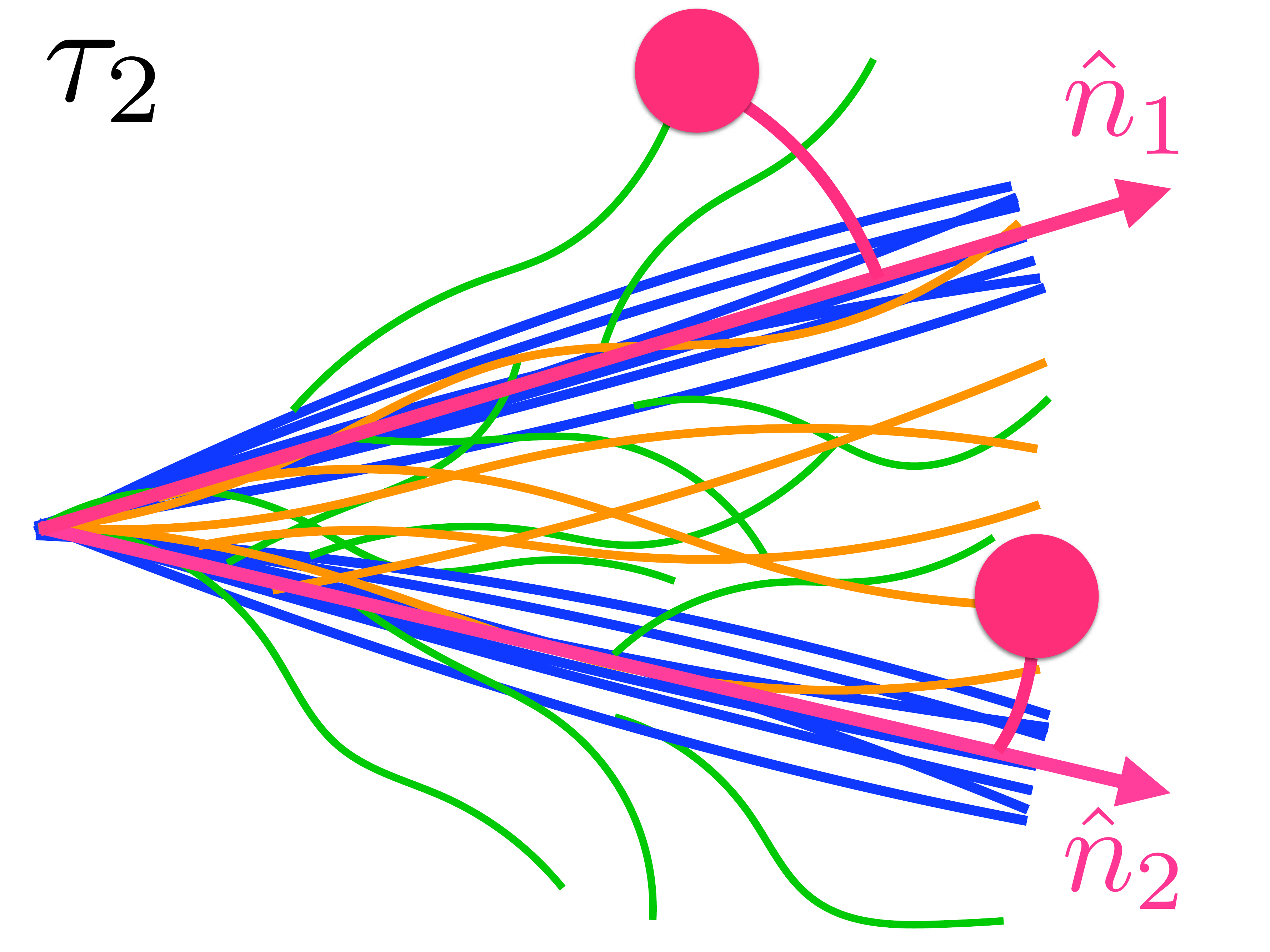} 
}
\caption{Comparison of the functional structure of (a) $\ecfvarnobeta{2}{3}$ and (b) $\Nsubnobeta{2}$. The $\ecfvarnobeta{2}{3}$ observable correlates triplets of particles (and two of their three pairwise angles), while the $\Nsubnobeta{2}$ observable correlates particles with axes.  
}
\label{fig:app_obs_illus}
\end{figure*}

\begin{figure}
\begin{center}
\subfloat[]{\label{fig:app_compare_Nsub_a}
\includegraphics[width=6.5cm]{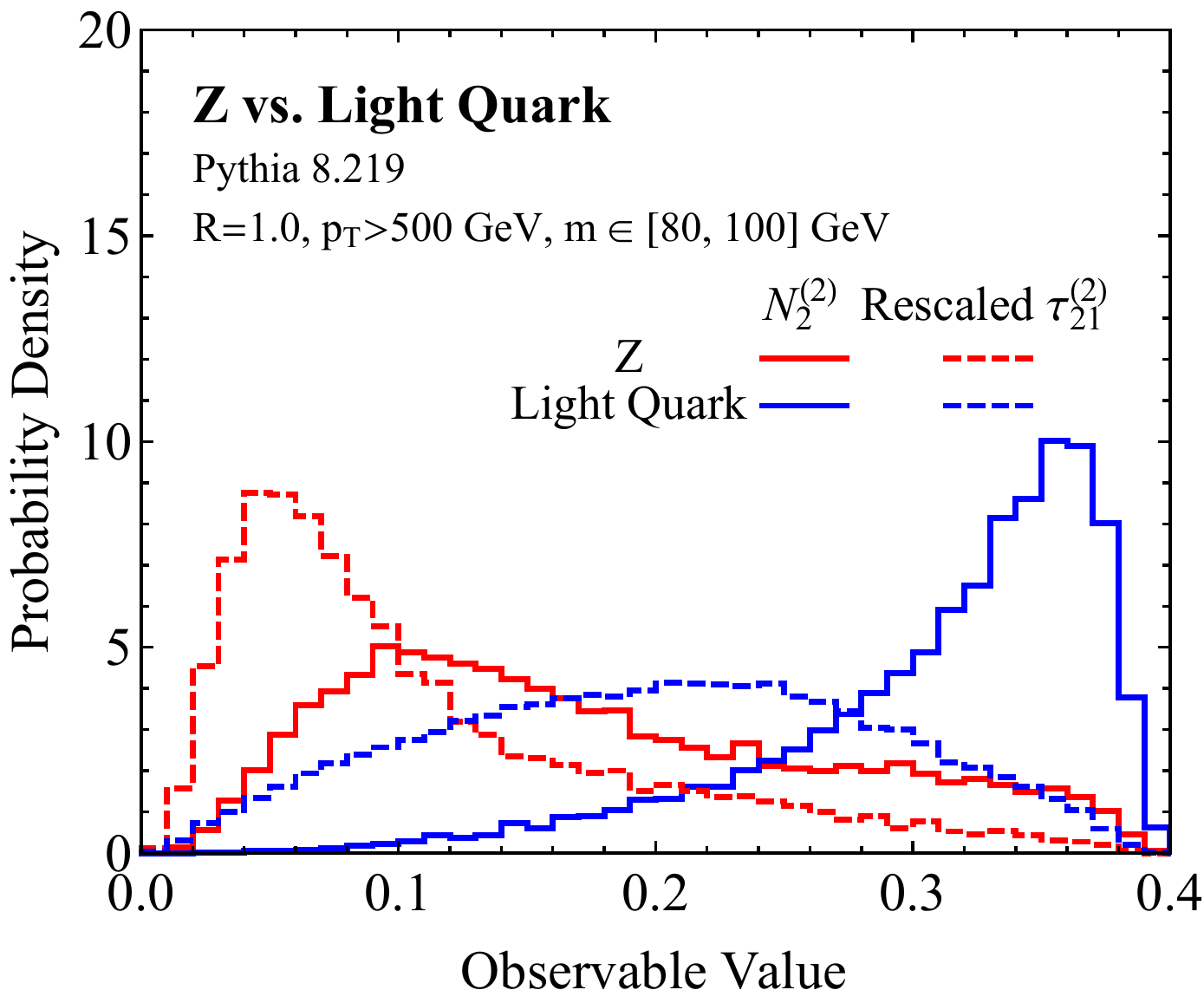}
}\qquad
\subfloat[]{\label{fig:app_compare_Nsub_b}
\includegraphics[width=6.5cm]{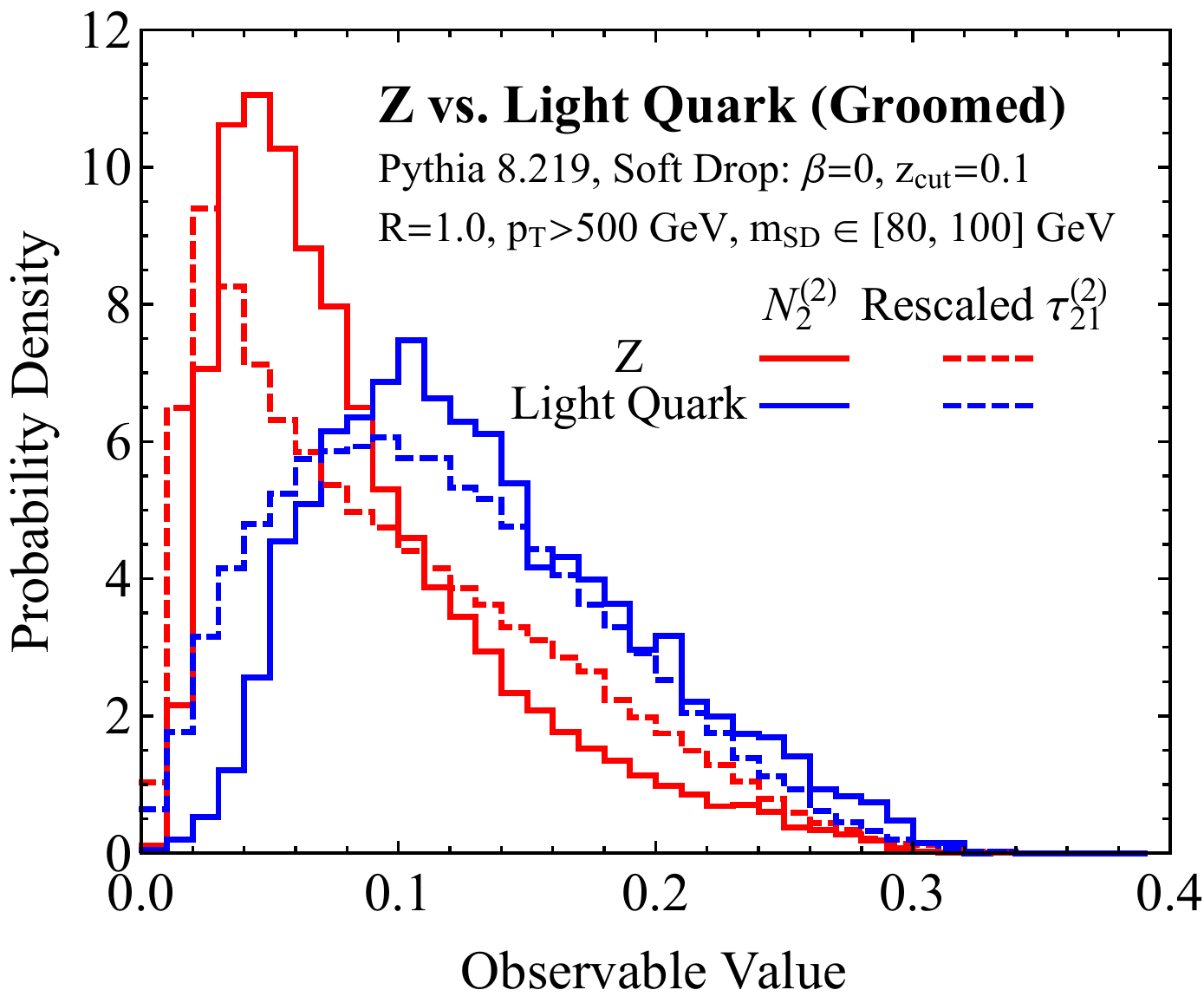}
}
\end{center}
\caption{Distributions of $\Nobsnobeta{2}$ and $\Nsubnobeta{2,1}$ on the $Z$ signal and quark background (a) before grooming and (b) after grooming. To aid visual comparison, $\Nsubnobeta{2,1}$ has been rescaled to match the endpoint of $\Nobsnobeta{2}$.  
}
\label{fig:app_compare_Nsub}
\end{figure}

In \Sec{sec:n_series}, we claimed that the $\Nobsnobeta{i}$ observables and the $N$-subjettiness ratio observables are related for groomed jets.   This was shown explicitly for the case of $i=3$ in \Sec{sec:tops_makeobs}.  In this appendix, we show that this is generically true, suggesting that $\Nobsnobeta{i}$ is indeed an appropriate observable for identifying $i$-prong substructure on groomed jets.

Since we work with groomed jets, we do not have to consider soft subjet configurations (i.e.~$i$-prong jets with hierarchical energies).  Instead, the power counting is determined by the generalization of \Fig{fig:p_NINJA} with hierarchical angles, where a jet has $i$ subjets, two of which become collinear and approach an $(i-1)$-subjet configuration.  We label the two subjets that approach each other by $1$ and $2$, such that $\theta_{12}$ denotes the angle between them.  By assumption, $\theta_{12}$ is smaller than the angles between any other subjets (which we power count as $\theta_{st} \sim 1$), but larger than the typical collinear scale $\theta_{cc}$.

By considering the contributions from collinear modes aligned along subjets $1$ and $2$, we find the parametric relation
\begin{align}
\label{eq:appC_1angle}
\ecfvar{1}{i}{\beta} \sim \theta_{12}^\beta \sim \Nsub{i-1}{\beta}\,,
\end{align}
where all other pairwise combinations of modes are power suppressed.  Here, we are assuming that the $N$-subjettiness axes are defined such that one axis is aligned with subjet 1 or 2, with the remaining $i-2$ axes aligned along the other subjets; this is indeed the configuration that minimizes $\Nsubnobeta{i-1}$ in the small $\theta_{12}$ limit, assuming balanced energies.  Adding an extra axis yields
\begin{equation}
\Nsub{i}{\beta} \sim \theta_{cc}^\beta,
\end{equation}
where now the $i$ axes align with the $i$ subjets.

For the correlator involving two angles, the power-counting analysis yields
\begin{align}
\label{eq:appC_2angle}
\ecfvar{2}{i+1}{\beta} &\sim \theta_{12}^\beta \left( \theta_{cc}^\beta +\ldots \right) \sim \Nsub{i-1}{\beta} \cdot \Nsub{i}{\beta}\,,
\end{align}
where the ellipses denote contributions from collinear-soft modes, which depend on the other angles between the subjets.  To understand the appearance of $\theta_{12}^\beta \theta_{cc}^\beta $, note that the largest contribution to $\ecfvarnobeta{2}{i+1}$ comes from selecting two collinear modes from one subjet and one collinear mode from each of the remaining $i-1$ subjets; for that configuration, the two smallest pairwise angles are indeed $\theta_{cc}$ and $\theta_{12}$.

Generalizing the argument in \App{app:N3_pc}, \Eqs{eq:appC_1angle}{eq:appC_2angle} imply $\ecfvarnobeta{2}{i+1} \ll (\ecfvarnobeta{1}{i})^2$ on $i$-prong signal jets, such that the appropriate $i$-prong discriminant is
\begin{align}
\Nobs{i}{\beta}=\frac{\ecfvar{2}{i+1}{\beta}}{(\ecfvar{1}{i}{\beta})^2}\sim\frac{\Nsub{i}{\beta}}{\Nsub{i-1}{\beta}}\,,
\end{align} 
where the last relation should be understood in the power-counting sense.  Therefore, as advertised, the $N_i$ observable is indeed related to the $N$-subjettiness ratio $\Nsubnobeta{i,i-1}$, and both are expected to be good $i$-prong discriminants.

\begin{figure}
\begin{center}
\subfloat[]{\label{fig:app_compare_ROC_Nsub_a}
\includegraphics[width=6.5cm]{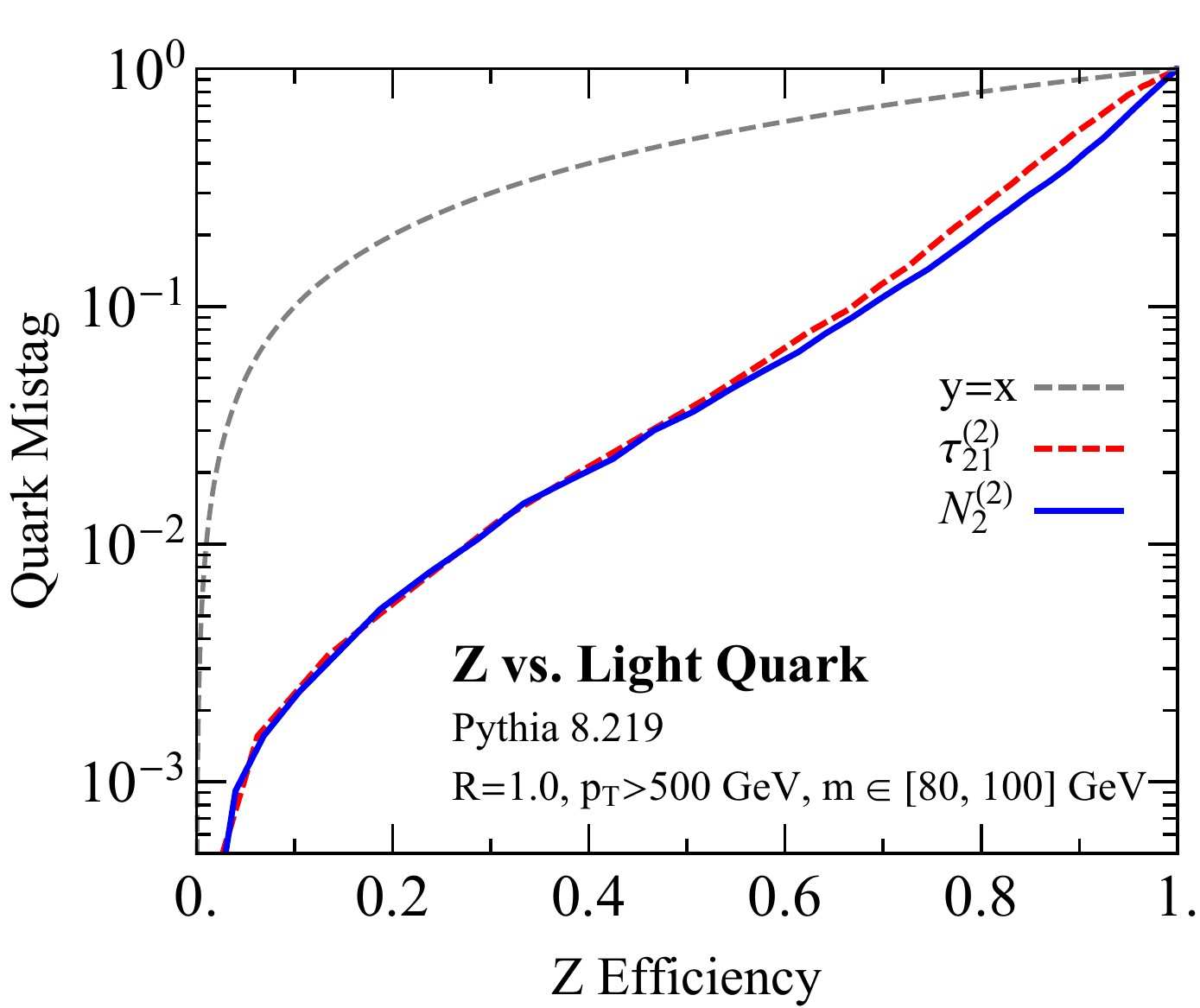}
}\qquad
\subfloat[]{\label{fig:app_compare_ROC_Nsub_b}
\includegraphics[width=6.5cm]{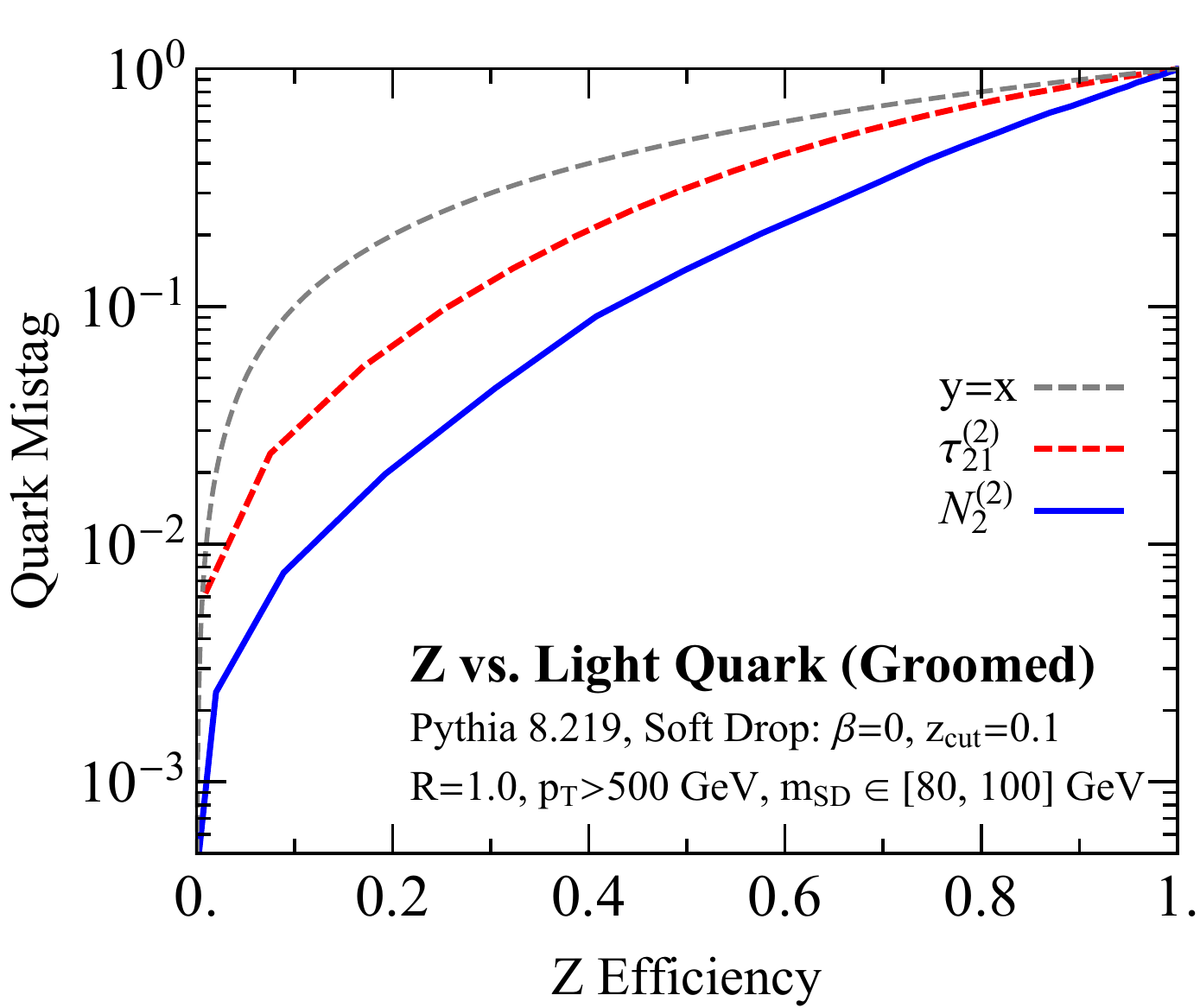}
}\\
\subfloat[]{\label{fig:app_compare_ROC_Nsub_c}
\includegraphics[width=6.5cm]{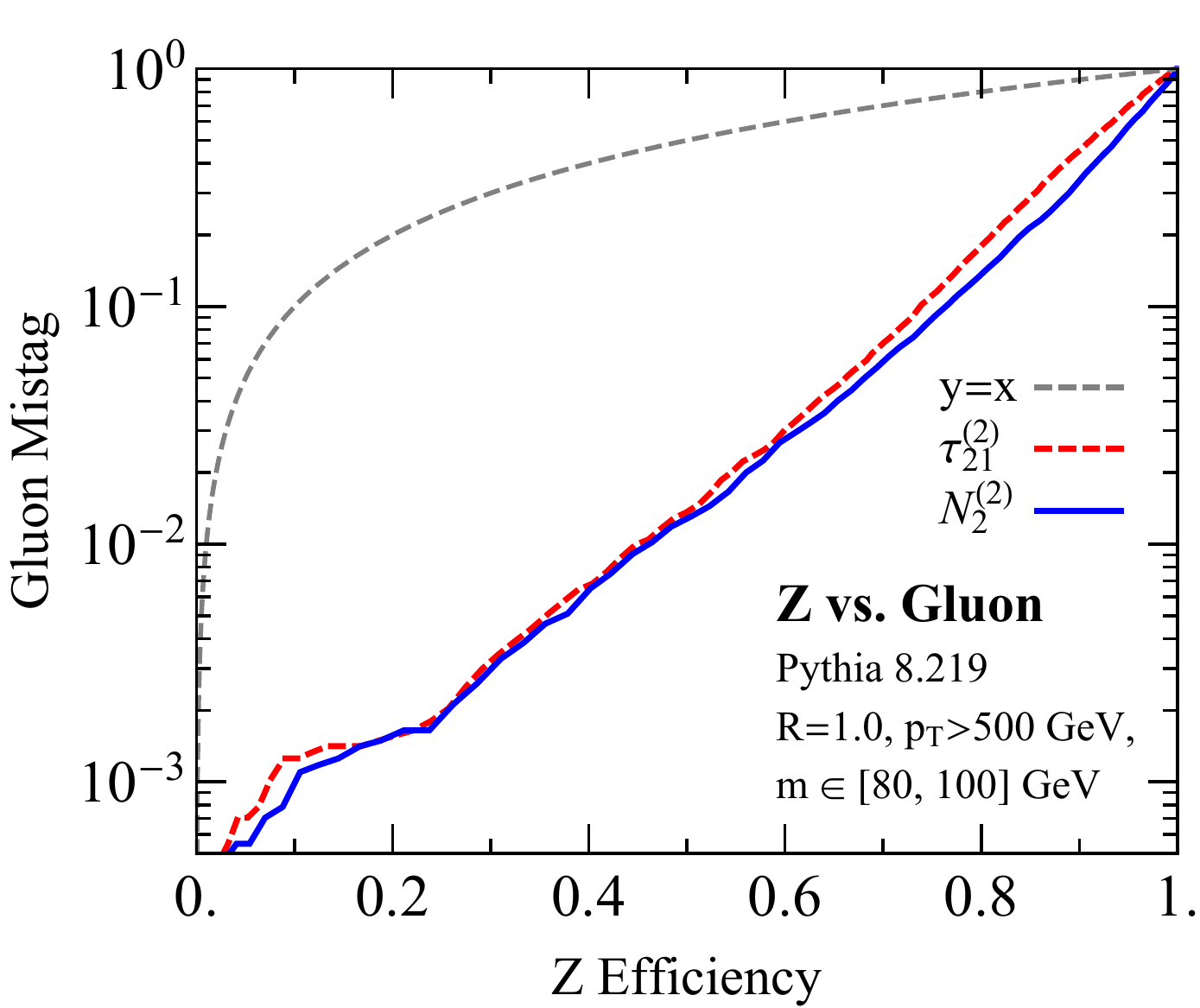}
}\qquad
\subfloat[]{\label{fig:app_compare_ROC_Nsub_d}
\includegraphics[width=6.5cm]{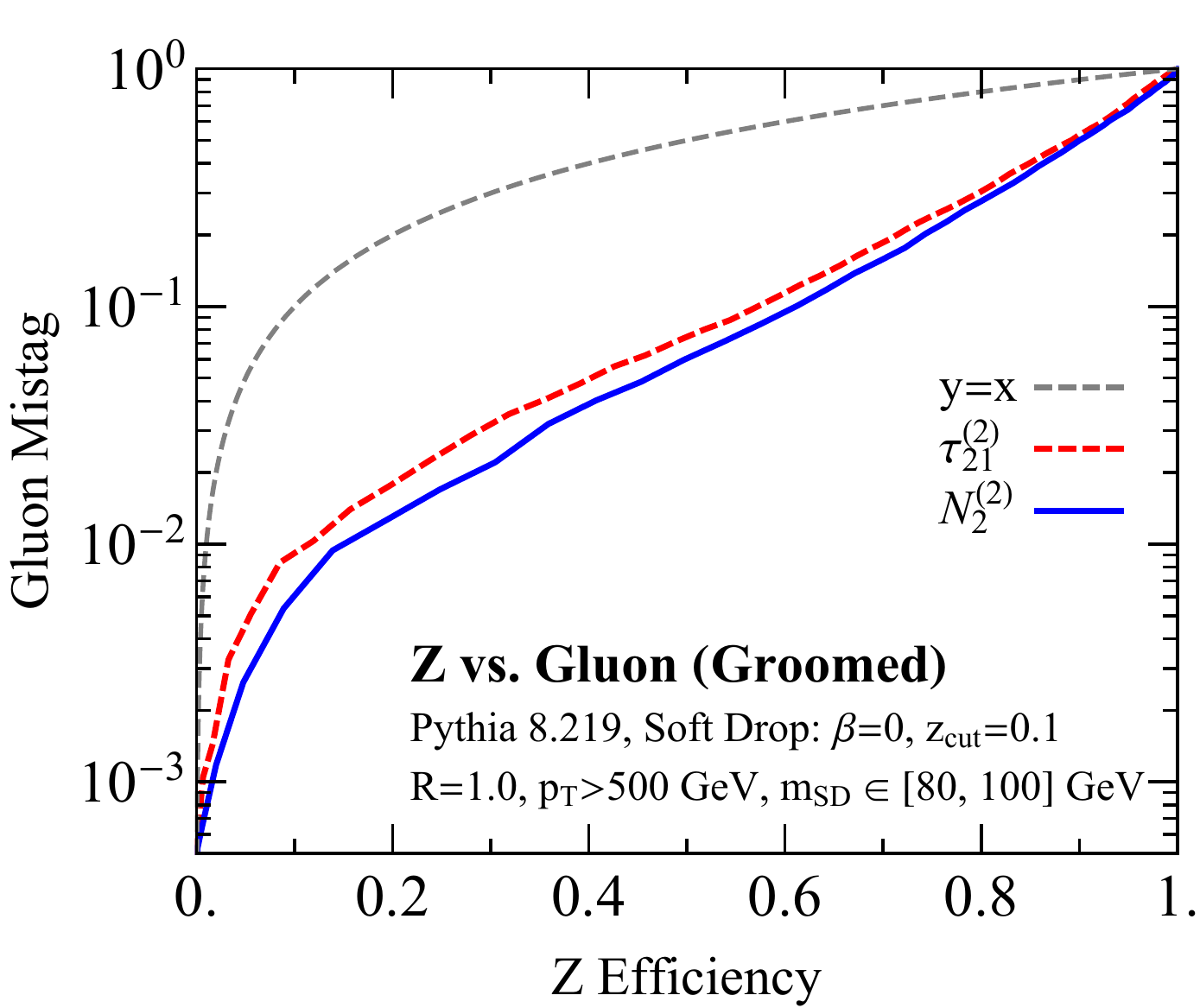}
}
\end{center}
\caption{Same as \Fig{fig:nD2_ROC_2prong_quark} but comparing $\Nobsnobeta{2}$ and $\Nsubnobeta{2,1}$.
}
\label{fig:app_compare_Nsub_ROC}
\end{figure}

As an example to demonstrate this parametric relation, we consider the case $i = 2$, which was alluded to in \Sec{sec:2prong}. The relevant observables are shown schematically in \Fig{fig:app_obs_illus}. In \Fig{fig:app_compare_Nsub}, we show distributions of $\Nsubnobeta{2,1}$ and $\Nobsnobeta{2}$ before and after grooming for $\beta=2$, taking quarks as representative of the background.   To aid in a visual comparison, we have rescaled the $\Nsubnobeta{2,1}$ distributions by a common factor to match the $\Nobsnobeta{2}$ endpoint.  Before grooming, the shapes of the two distributions are quite different, with $\Nobsnobeta{2}$ being much more peaked towards the endpoint for the background.  After soft drop has been applied, the distributions for the two observables are quite similar, as predicted by the power-counting discussion above. 

Still, there is a non-parametric difference between the $\Nsubnobeta{2,1}$ and $\Nobsnobeta{2}$ distributions, which leads to improved tagging performance for $\Nobsnobeta{2}$.  This can be seen by eye in the groomed plot in \Fig{fig:app_compare_Nsub_b}, where the background distribution for $\Nobsnobeta{2}$ is pushed to higher values while the signal distribution is more rapidly falling toward the endpoint.   More quantitatively, we can consider the ROC curves in \Fig{fig:app_compare_Nsub_ROC}.  For the ungroomed case, the discrimination power is similar, with $\Nobsnobeta{2}$ showing slightly improved behavior at higher efficiencies.  For the groomed case, there are significant gains to be had in using $\Nobsnobeta{2}$ instead of $\Nsubnobeta{2,1}$.\footnote{While it is possible that different axes choices for $N$-subjettiness could provide improved performance, it seems to us that any axes definition will be ambiguous in the unresolved region.  This also highlights the nice property that $\Nobsnobeta{2}$ is defined without respect to subjet axes.}

\section{Hybrid Strategies for 2-prong Observables}\label{app:hybrid}

Throughout the text, we focused on discriminants formed from combinations (often ratios) of either groomed or ungroomed observables.  It is also interesting to consider discriminants formed from mixtures of groomed and ungroomed observables \cite{gregory_talk,gregory_paper}, which we will refer to as a hybrid strategy.  While we will not explore this topic in detail, we take as a simple example ungroomed 2-prong observables after the application of a groomed mass cut. 

\begin{figure}
\begin{center}
\subfloat[]{\label{fig:hybrid_obs_a}
\includegraphics[width=6.5cm]{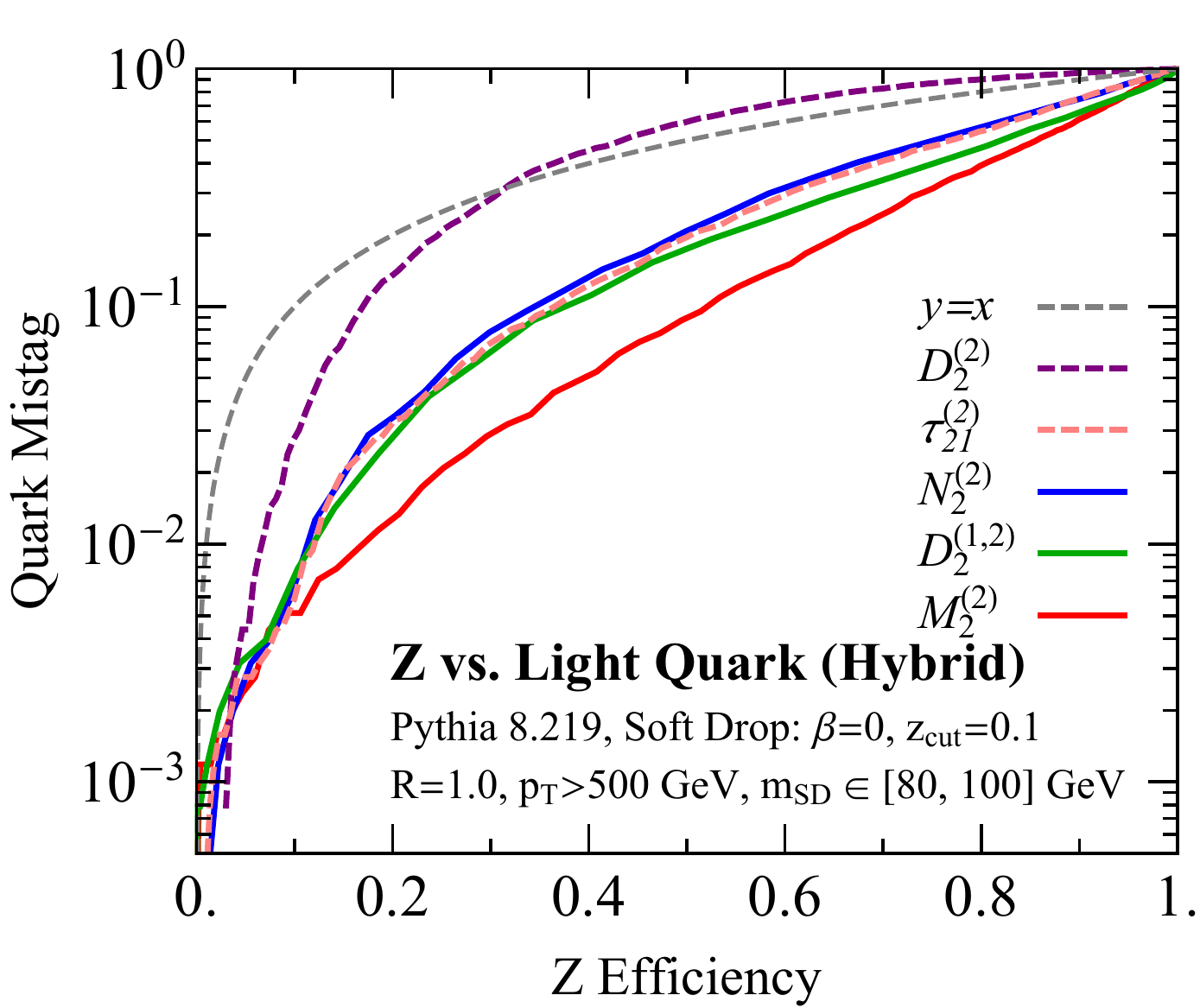}
}\qquad
\subfloat[]{\label{fig:hybrid_obs_b}
\includegraphics[width=6.5cm]{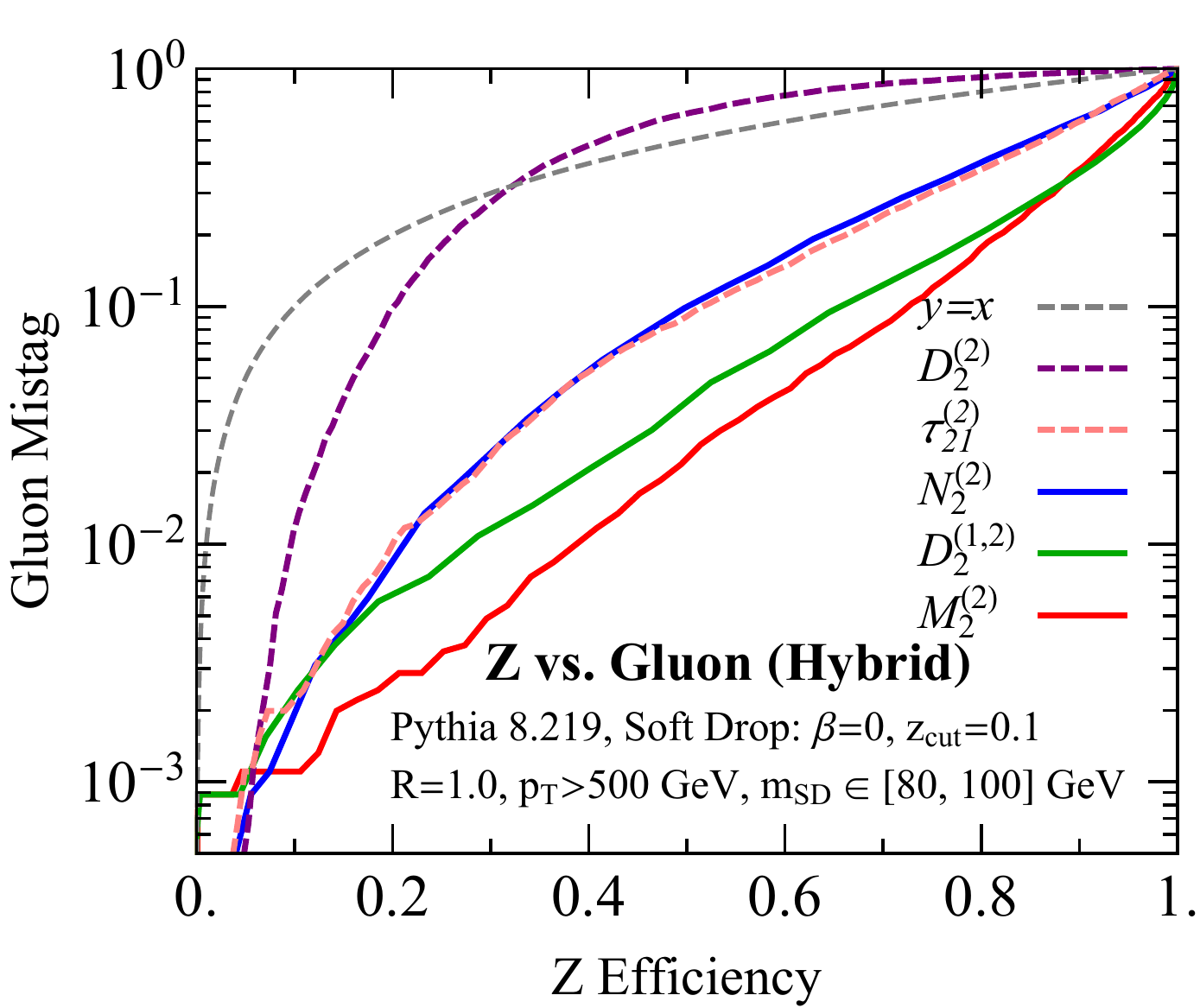}
}
\end{center}
\caption{Same as \Figs{fig:nD2_ps_quark_pythia_SD}{fig:nD2_ps_gluon_pythia_SD} but using a hybrid strategy where a cut is placed on the groomed jet mass, but the discriminants are ungroomed.
}
\label{fig:hybrid_obs}
\end{figure}

In \Fig{fig:hybrid_obs}, we show the ROC curves for boosted $Z$ discrimination, showing light quark and gluon backgrounds separately; this should be contrasted with \Fig{fig:nD2_ROC_2prong_quark}. The behavior of these hybrid observables can be understood using the power-counting analysis of \Sec{sec:twoprong_MC_stable}, where we analyzed the stability of the observables as a function of $m_J$ and $p_{TJ}$. For signal jets, a cut on the groomed mass has little effect due to the color singlet nature of the $Z$ boson, and therefore the hybrid observables should have a similar behavior to the ungroomed observables.   For background QCD jets, however, applying a groomed mass cut in the same mass window enforces a higher effective cut on the ungroomed mass.  This, in turn, enters the scaling relations for the background distributions given in \Sec{sec:twoprong_MC_stable}:
\begin{align}
\label{eq:app_2prong_scaling}
M^{(2), \text{peak}}_2\sim \frac{m_{J}^2}{p_{TJ}^2}\,, \qquad N^{(2), \text{max,peak}}_2 \sim \text{const}\,, \qquad     D^{(2), \text{max}}_2\sim \frac{p_{TJ}^2}{m_J^2}\,.
\end{align}
For $\Mobsnobeta{2}$, and similarly for $\Dobs{2}{1,2}$, using a groomed mass cut has the interesting effect of pushing the ungroomed background distribution to higher values, thereby improving discrimination power. For $\Nobsnobeta{2}$, the distribution is parametrically unmodified, and therefore similar discrimination power is expected for the ungroomed and hybrid observables. For $\Dobs{2}{2}$, larger effective mass values push the distribution to lower values, thereby worsening discrimination power.  These power-counting predictions are seen clearly in \Fig{fig:hybrid_obs}.

The above behavior is perhaps counterintuitive, especially the poor performance of $\Dobs{2}{2}$ and the good performance of $\Mobsnobeta{2}$, but it follows straightforwardly from the power-counting analysis.  That said, the quantitative discrimination power depends crucially on the choice of mass window, and one must keep in mind that this study is based on a relatively narrow soft-dropped mass cut around $m_Z$.  Further studies are therefore warranted to test whether discrimination performance can indeed be improved by simultaneously using information before and after grooming.

\section{Supplemental Quark/Gluon Plots}\label{app:add_qg}

\begin{figure}
\begin{center}
\subfloat[]{\label{fig:2prong_obs_gluon_diffbeta_a_app}
\includegraphics[width=6.5cm]{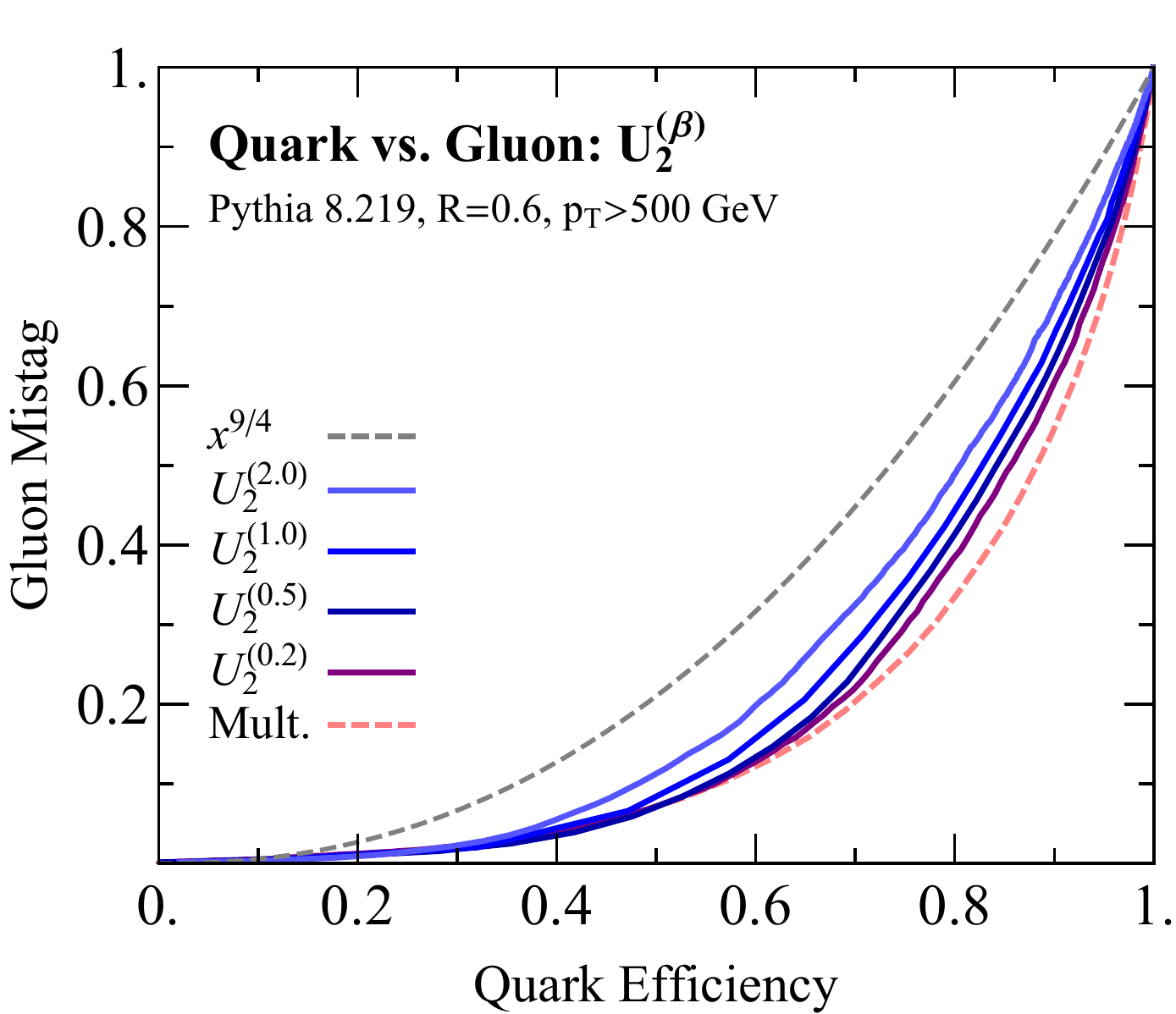}
}\qquad
\subfloat[]{\label{fig:2prong_obs_gluon_diffbeta_b_app}
\includegraphics[width=6.5cm]{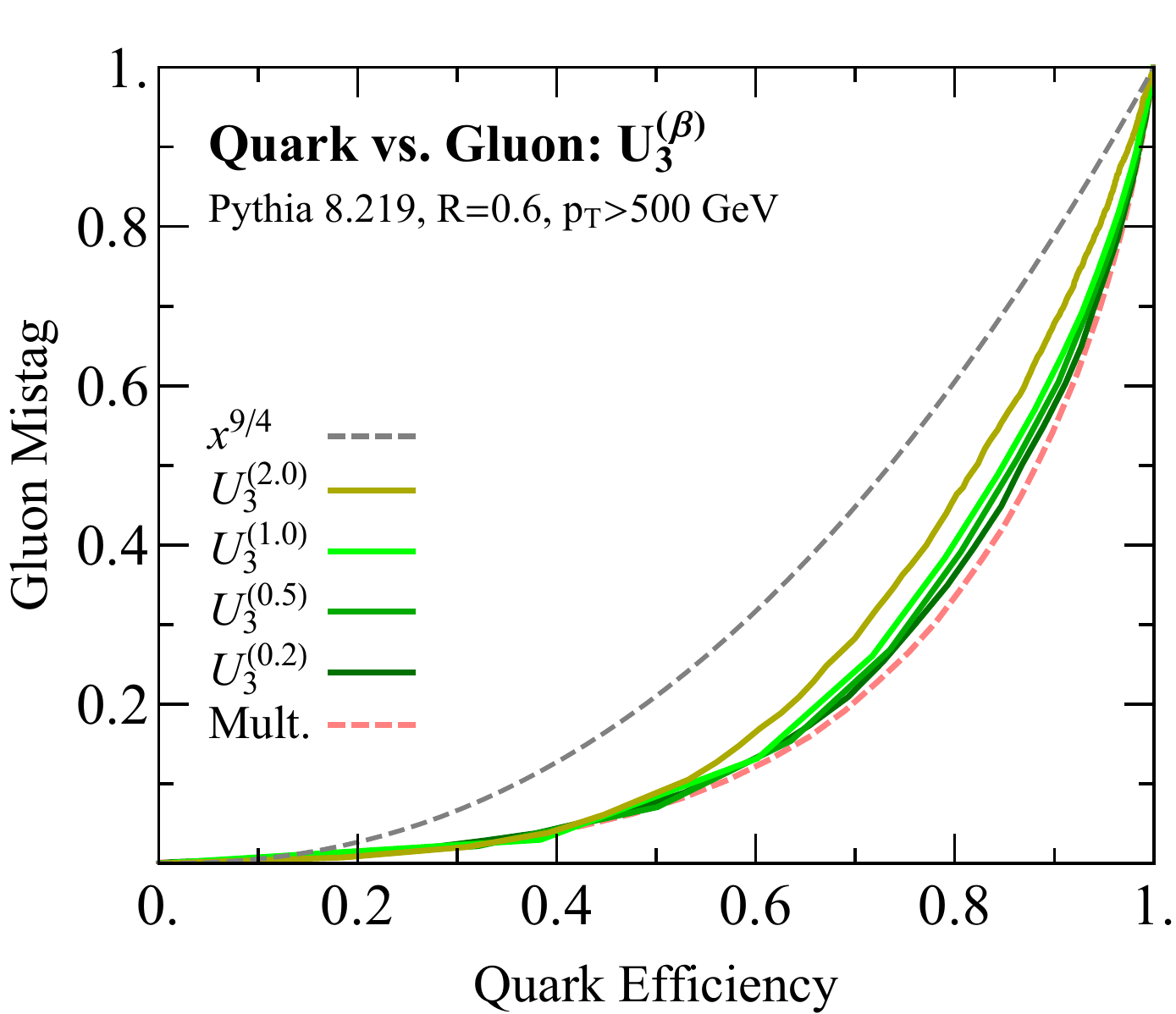}
}
\end{center}
\caption{ROC curves as the angular exponent $\beta$ is varied for (a) $\Uobsnobeta{2}$ and (b) $\Uobsnobeta{3}$.  For $\Uobsnobeta{3}$, a more stable ROC curve is observed throughout the entire distribution.  
}
\label{fig:2prong_obs_gluon_diffbeta_app}
\end{figure}

\begin{figure}
\begin{center}
\subfloat[]{\label{fig:2prong_obs_gluon_diffbeta_groom_a}
\includegraphics[width=6.5cm]{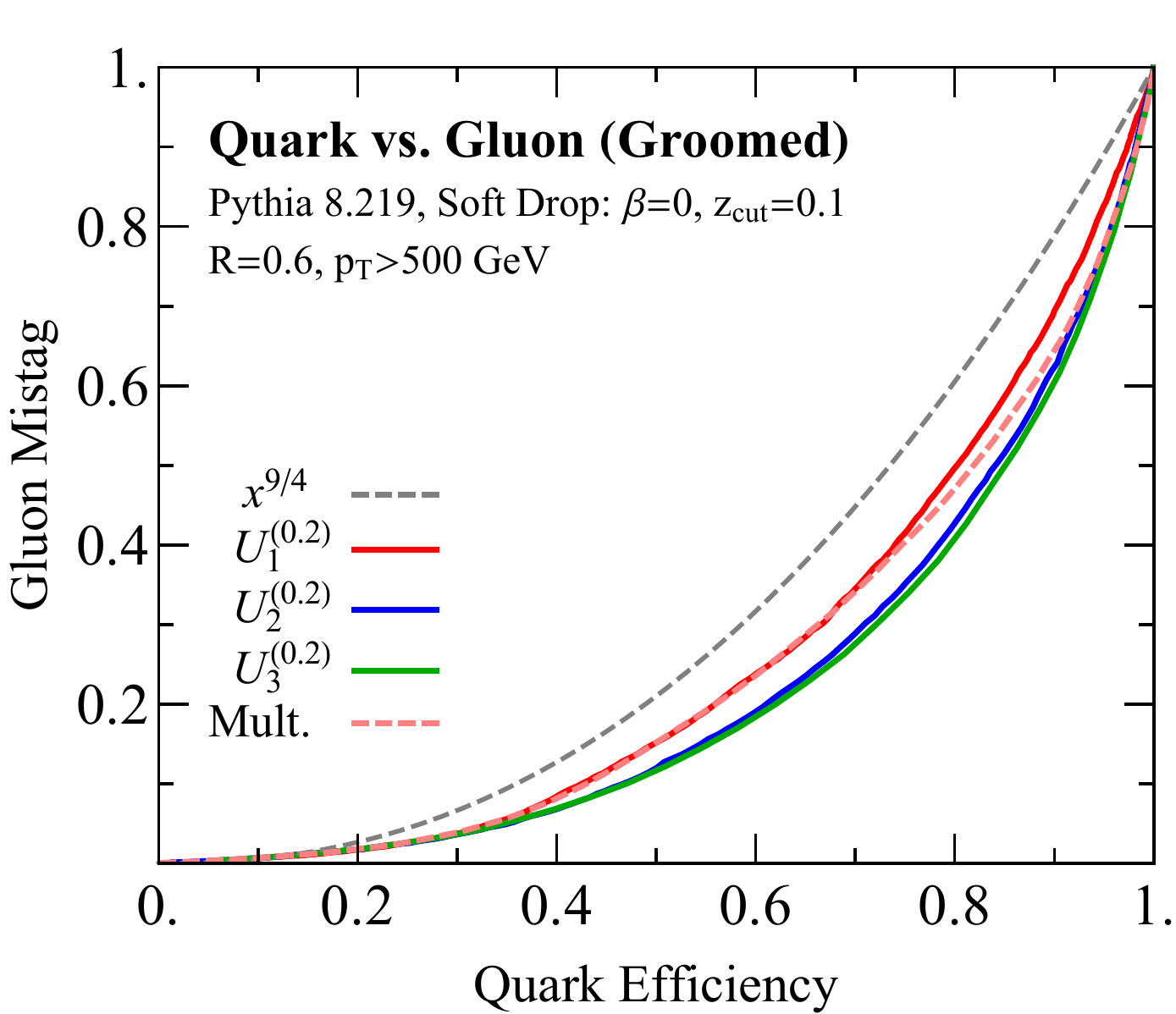}
}\qquad
\subfloat[]{\label{fig:2prong_obs_gluon_diffbeta_groom_b}
\includegraphics[width=6.5cm]{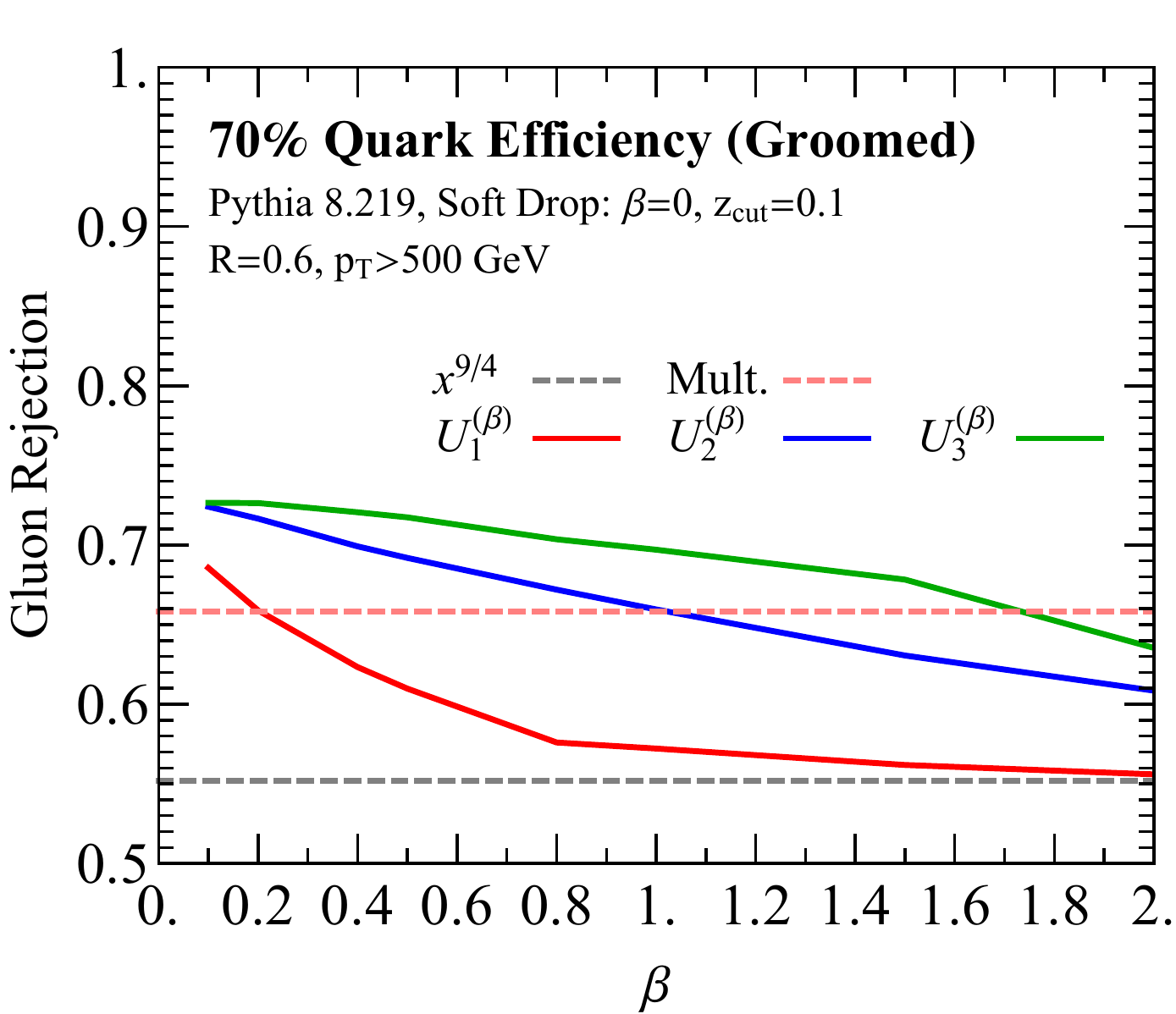}
}
\end{center}
\caption{Same as \Fig{fig:2prong_obs_gluon_diffbeta}, but after grooming.  The improved performance of $\Uobsnobeta{2}$ and $\Uobsnobeta{3}$ relative to $\Uobsnobeta{1}$ is robust to removing soft radiation.  
}
\label{fig:2prong_obs_gluon_diffbeta_groom}
\end{figure}

In \Fig{fig:2prong_obs_gluon_diffbeta_a}, we emphasized the stability of $U_i$ for $i=2,3$ as a function of the angular exponent $\beta$.  In \Fig{fig:2prong_obs_gluon_diffbeta_app}, we show the full ROC curves for both $\Uobsnobeta{2}$ and $\Uobsnobeta{3}$ as a function of the angular exponent $\beta$.  Neither observable asymptotes to the Casimir scaling prediction, even at high efficiencies or high $\beta$ values.  Furthermore, the $\Uobsnobeta{3}$ distributions exhibit stability as a function of $\beta$ throughout the whole ROC curve. This would be interesting to verify in an analytic calculation.

In \Fig{fig:2prong_obs_gluon_diffbeta_groom_a}, we show the ROC curves for $\Uobsnobeta{2}$ and $\Uobsnobeta{3}$ after grooming for $\beta = 0.2$, showing that the $U_i$ series continues to perform better for larger values of $i$.  In \Fig{fig:2prong_obs_gluon_diffbeta_groom_b}, we show the performance as a function of $\beta$, demonstrating the stability of $\Uobsnobeta{3}$, even after grooming. 

\bibliography{ECFVariant}

\providecommand{\href}[2]{#2}\begingroup\raggedright\begin{thebibliography}{100}

\bibitem{Chatrchyan:2012sn}
{\bf CMS} Collaboration, S.~Chatrchyan et~al., {\it {Search for a Higgs boson
  in the decay channel $H$ to ZZ(*) to $q$ qbar $\ell^-$ l+ in $pp$ collisions
  at $\sqrt{s}=7$ TeV}},  {\em JHEP} {\bf 1204} (2012) 036,
  [\href{http://arxiv.org/abs/1202.1416}{{\tt arXiv:1202.1416}}].

\bibitem{CMS:2013cda}
{\bf CMS} Collaboration, {\it {Search for a Standard Model-like Higgs boson
  decaying into WW to l nu qqbar in pp collisions at sqrt s = 8 TeV}},  Tech.
  Rep. CMS-PAS-HIG-13-008, 2013.

\bibitem{Aad:2015cua}
{\bf ATLAS} Collaboration, G.~Aad et~al., {\it {Measurement of jet charge in
  dijet events from $\sqrt{s}$=8  TeV pp collisions with the ATLAS
  detector}},  {\em Phys. Rev.} {\bf D93} (2016), no.~5 052003,
  [\href{http://arxiv.org/abs/1509.05190}{{\tt arXiv:1509.05190}}].

\bibitem{Aad:2015lxa}
{\bf ATLAS} Collaboration, G.~Aad et~al., {\it {Measurement of colour flow with
  the jet pull angle in $t\bar{t}$ events using the ATLAS detector at
  $\sqrt{s}=8$ TeV}},  {\em Phys. Lett.} {\bf B750} (2015) 475--493,
  [\href{http://arxiv.org/abs/1506.05629}{{\tt arXiv:1506.05629}}].

\bibitem{ATLAS-CONF-2015-035}
{\bf ATLAS} Collaboration, {\it {Performance of jet substructure techniques in
  early $\sqrt{s}=13$ TeV $pp$ collisions with the ATLAS detector}},  Tech.
  Rep. ATLAS-CONF-2015-035, 2015.

\bibitem{Aad:2015rpa}
{\bf ATLAS} Collaboration, G.~Aad et~al., {\it {Identification of boosted,
  hadronically decaying W bosons and comparisons with ATLAS data taken at
  $\sqrt{s} = 8$ TeV}},  {\em Eur. Phys. J.} {\bf C76} (2016), no.~3 154,
  [\href{http://arxiv.org/abs/1510.05821}{{\tt arXiv:1510.05821}}].

\bibitem{Aad:2015hna}
{\bf ATLAS} Collaboration, G.~Aad et~al., {\it {Measurement of the differential
  cross-section of highly boosted top quarks as a function of their transverse
  momentum in $\sqrt{s}$ = 8 TeV proton-proton collisions using the ATLAS
  detector}},  {\em Phys. Rev.} {\bf D93} (2016), no.~3 032009,
  [\href{http://arxiv.org/abs/1510.03818}{{\tt arXiv:1510.03818}}].

\bibitem{ATLAS-CONF-2016-002}
{\bf ATLAS} Collaboration, {\it {Studies of $b$-tagging performance and jet
  substructure in a high $p_{T}$ $g\rightarrow b\bar{b}$ rich sample of
  large-$R$ jets from $pp$ collisions at $\sqrt{s}=8$ TeV with the ATLAS
  detector}},  Tech. Rep. ATLAS-CONF-2016-002, CERN, 2016.

\bibitem{ATLAS-CONF-2016-039}
{\bf ATLAS} Collaboration, {\it {Boosted Higgs ($\rightarrow b\bar{b}$) Boson
  Identification with the ATLAS Detector at $\sqrt{s} = 13$ TeV}},  Tech. Rep.
  ATLAS-CONF-2016-039, CERN, Geneva, Aug, 2016.

\bibitem{ATLAS-CONF-2016-034}
{\bf ATLAS} Collaboration, {\it {Discrimination of Light Quark and Gluon Jets
  in $pp$ collisions at $\sqrt{s} = 8$ TeV with the ATLAS Detector}},  Tech.
  Rep. ATLAS-CONF-2016-034, CERN, Geneva, Jul, 2016.

\bibitem{CMS-PAS-TOP-16-013}
{\bf CMS} Collaboration, {\it {Measurement of the ${\rm t}{\rm \bar{t}}$
  production cross section at 13 TeV in the all-jets final state}},  Tech. Rep.
  CMS-PAS-TOP-16-013, CERN, Geneva, 2016.

\bibitem{CMS-PAS-HIG-16-004}
{\bf CMS} Collaboration, {\it {Search for $\mathrm{t\overline{t}H}$ production
  in the $\mathrm{H}\rightarrow \mathrm{b\overline{b}}$ decay channel with
  $\sqrt{s}=13~\mathrm{TeV}$ pp collisions at the CMS experiment}},  Tech. Rep.
  CMS-PAS-HIG-16-004, CERN, Geneva, 2016.

\bibitem{CMS:2011bqa}
{\bf CMS} Collaboration, {\it {Search for BSM ttbar Production in the Boosted
  All-Hadronic Final State}},  Tech. Rep. CMS-PAS-EXO-11-006, 2011.

\bibitem{Fleischmann:2013woa}
{\bf ATLAS, CMS} Collaboration, S.~Fleischmann, {\it {Boosted top quark
  techniques and searches for $t\bar{t}$ resonances at the LHC}},  {\em
  J.Phys.Conf.Ser.} {\bf 452} (2013), no.~1 012034.

\bibitem{Pilot:2013bla}
{\bf ATLAS, CMS} Collaboration, J.~Pilot, {\it {Boosted Top Quarks, Top Pair
  Resonances, and Top Partner Searches at the LHC}},  {\em EPJ Web Conf.} {\bf
  60} (2013) 09003.

\bibitem{TheATLAScollaboration:2013qia}
{\bf ATLAS} Collaboration, {\it {Performance of boosted top quark
  identification in 2012 ATLAS data}},  Tech. Rep. ATLAS-CONF-2013-084,
  ATLAS-COM-CONF-2013-074, 2013.

\bibitem{Chatrchyan:2012ku}
{\bf CMS} Collaboration, S.~Chatrchyan et~al., {\it {Search for Anomalous
  $t\bar{t}$ Production in the Highly-Boosted All-Hadronic Final State}},  {\em
  JHEP} {\bf 1209} (2012) 029, [\href{http://arxiv.org/abs/1204.2488}{{\tt
  arXiv:1204.2488}}].

\bibitem{CMS-PAS-B2G-14-001}
{\bf CMS Collaboration} Collaboration, {\it {Search for pair-produced
  vector-like quarks of charge -1/3 decaying to bH using boosted Higgs
  jet-tagging in pp collisions at sqrt(s) = 8 TeV}},  Tech. Rep.
  CMS-PAS-B2G-14-001, CERN, Geneva, 2014.

\bibitem{CMS-PAS-B2G-14-002}
{\bf CMS} Collaboration, {\it {Search for top-Higgs resonances in all-hadronic
  final states using jet substructure methods}},  Tech. Rep.
  CMS-PAS-B2G-14-002, CERN, Geneva, 2014.

\bibitem{Khachatryan:2015axa}
{\bf CMS} Collaboration, V.~Khachatryan et~al., {\it {Search for vector-like T
  quarks decaying to top quarks and Higgs bosons in the all-hadronic channel
  using jet substructure}},  {\em JHEP} {\bf 06} (2015) 080,
  [\href{http://arxiv.org/abs/1503.01952}{{\tt arXiv:1503.01952}}].

\bibitem{Khachatryan:2015bma}
{\bf CMS} Collaboration, V.~Khachatryan et~al., {\it {Search for a massive
  resonance decaying into a Higgs boson and a W or Z boson in hadronic final
  states in proton-proton collisions at $ \sqrt{s}=8 $ TeV}},  {\em JHEP} {\bf
  02} (2016) 145, [\href{http://arxiv.org/abs/1506.01443}{{\tt
  arXiv:1506.01443}}].

\bibitem{Aad:2015owa}
{\bf ATLAS} Collaboration, G.~Aad et~al., {\it {Search for high-mass diboson
  resonances with boson-tagged jets in proton-proton collisions at $ \sqrt{s}=8
  $ TeV with the ATLAS detector}},  {\em JHEP} {\bf 12} (2015) 055,
  [\href{http://arxiv.org/abs/1506.00962}{{\tt arXiv:1506.00962}}].

\bibitem{Aaboud:2016okv}
{\bf ATLAS} Collaboration, M.~Aaboud et~al., {\it {Searches for heavy diboson
  resonances in $pp$ collisions at $\sqrt{s}=13$ TeV with the ATLAS detector}},
   \href{http://arxiv.org/abs/1606.04833}{{\tt arXiv:1606.04833}}.

\bibitem{Aaboud:2016trl}
{\bf ATLAS} Collaboration, M.~Aaboud et~al., {\it {Search for heavy resonances
  decaying to a $Z$ boson and a photon in $pp$ collisions at $\sqrt{s}=13$ TeV
  with the ATLAS detector}},  \href{http://arxiv.org/abs/1607.06363}{{\tt
  arXiv:1607.06363}}.

\bibitem{Aaboud:2016qgg}
{\bf ATLAS} Collaboration, M.~Aaboud et~al., {\it {Search for dark matter
  produced in association with a hadronically decaying vector boson in pp
  collisions at $\sqrt(s)$=13 TeV with the ATLAS detector}},
  \href{http://arxiv.org/abs/1608.02372}{{\tt arXiv:1608.02372}}.

\bibitem{ATLAS-CONF-2016-055}
{\bf ATLAS} Collaboration, {\it {Search for resonances with boson-tagged jets
  in 15.5 fb$^{-1}$ of $pp$ collisions at $\sqrt{s} = 13$ TeV collected with
  the ATLAS detector}},  Tech. Rep. ATLAS-CONF-2016-055, CERN, Geneva, Aug,
  2016.

\bibitem{ATLAS-CONF-2015-071}
{\bf ATLAS} Collaboration, {\it {Search for diboson resonances in the llqq
  final state in pp collisions at $\sqrt{s}$ = 13 TeV with the ATLAS
  detector}},  Tech. Rep. ATLAS-CONF-2015-071, CERN, Geneva, Dec, 2015.

\bibitem{ATLAS-CONF-2015-068}
{\bf ATLAS} Collaboration, {\it {Search for diboson resonances in the $\nu\nu
  qq$ final state in $pp$ collisions at $\sqrt{s}=$13 TeV with the ATLAS
  detector}},  Tech. Rep. ATLAS-CONF-2015-068, CERN, Geneva, Dec, 2015.

\bibitem{CMS-PAS-EXO-16-037}
{\bf CMS} Collaboration, {\it {Search for dark matter in final states with an
  energetic jet, or a hadronically decaying W or Z boson using
  $12.9~\mathrm{fb}^{-1}$ of data at $\sqrt{s} = 13~\mathrm{TeV}$}},  Tech.
  Rep. CMS-PAS-EXO-16-037, CERN, Geneva, 2016.

\bibitem{CMS-PAS-EXO-16-040}
{\bf CMS} Collaboration, {\it {Search for new physics in a boosted hadronic
  monotop final state using $12.9~\mathrm{fb}^{-1}$ of
  $\sqrt{s}=13~\mathrm{TeV}$ data}},  Tech. Rep. CMS-PAS-EXO-16-040, CERN,
  Geneva, 2016.

\bibitem{Khachatryan:2016mdm}
{\bf CMS} Collaboration, V.~Khachatryan et~al., {\it {Search for dark matter in
  proton-proton collisions at 8 TeV with missing transverse momentum and vector
  boson tagged jets}},  {\em Submitted to: JHEP} (2016)
  [\href{http://arxiv.org/abs/1607.05764}{{\tt arXiv:1607.05764}}].

\bibitem{CMS-PAS-HIG-16-016}
{\bf CMS} Collaboration, {\it {Searches for invisible Higgs boson decays with
  the CMS detector.}},  Tech. Rep. CMS-PAS-HIG-16-016, CERN, Geneva, 2016.

\bibitem{CMS-PAS-B2G-15-003}
{\bf CMS} Collaboration, {\it {Search for top quark-antiquark resonances in the
  all-hadronic final state at sqrt(s)=13 TeV}},  Tech. Rep. CMS-PAS-B2G-15-003,
  CERN, Geneva, 2016.

\bibitem{CMS-PAS-EXO-16-017}
{\bf CMS} Collaboration, {\it {Search for dark matter in association with a
  boosted top quark in the all hadronic final state}},  Tech. Rep.
  CMS-PAS-EXO-16-017, CERN, Geneva, 2016.

\bibitem{Abdesselam:2010pt}
A.~Abdesselam, E.~B. Kuutmann, U.~Bitenc, G.~Brooijmans, J.~Butterworth,
  et~al., {\it {Boosted objects: A Probe of beyond the Standard Model
  physics}},  {\em Eur.Phys.J.} {\bf C71} (2011) 1661,
  [\href{http://arxiv.org/abs/1012.5412}{{\tt arXiv:1012.5412}}].

\bibitem{Altheimer:2012mn}
A.~Altheimer, S.~Arora, L.~Asquith, G.~Brooijmans, J.~Butterworth, et~al., {\it
  {Jet Substructure at the Tevatron and LHC: New results, new tools, new
  benchmarks}},  {\em J.Phys.} {\bf G39} (2012) 063001,
  [\href{http://arxiv.org/abs/1201.0008}{{\tt arXiv:1201.0008}}].

\bibitem{Altheimer:2013yza}
A.~Altheimer, A.~Arce, L.~Asquith, J.~Backus~Mayes, E.~Bergeaas~Kuutmann,
  et~al., {\it {Boosted objects and jet substructure at the LHC. Report of
  BOOST2012, held at IFIC Valencia, 23rd-27th of July 2012}},  {\em
  Eur.Phys.J.} {\bf C74} (2014) 2792,
  [\href{http://arxiv.org/abs/1311.2708}{{\tt arXiv:1311.2708}}].

\bibitem{Adams:2015hiv}
D.~Adams et~al., {\it {Towards an Understanding of the Correlations in Jet
  Substructure}},  {\em Eur. Phys. J.} {\bf C75} (2015), no.~9 409,
  [\href{http://arxiv.org/abs/1504.00679}{{\tt arXiv:1504.00679}}].

\bibitem{Feige:2012vc}
I.~Feige, M.~D. Schwartz, I.~W. Stewart, and J.~Thaler, {\it {Precision Jet
  Substructure from Boosted Event Shapes}},  {\em Phys.Rev.Lett.} {\bf 109}
  (2012) 092001, [\href{http://arxiv.org/abs/1204.3898}{{\tt
  arXiv:1204.3898}}].

\bibitem{Field:2012rw}
M.~Field, G.~Gur-Ari, D.~A. Kosower, L.~Mannelli, and G.~Perez, {\it
  {Three-Prong Distribution of Massive Narrow QCD Jets}},  {\em Phys.Rev.} {\bf
  D87} (2013), no.~9 094013, [\href{http://arxiv.org/abs/1212.2106}{{\tt
  arXiv:1212.2106}}].

\bibitem{Dasgupta:2013ihk}
M.~Dasgupta, A.~Fregoso, S.~Marzani, and G.~P. Salam, {\it {Towards an
  understanding of jet substructure}},  {\em JHEP} {\bf 1309} (2013) 029,
  [\href{http://arxiv.org/abs/1307.0007}{{\tt arXiv:1307.0007}}].

\bibitem{Dasgupta:2013via}
M.~Dasgupta, A.~Fregoso, S.~Marzani, and A.~Powling, {\it {Jet substructure
  with analytical methods}},  {\em Eur.Phys.J.} {\bf C73} (2013), no.~11 2623,
  [\href{http://arxiv.org/abs/1307.0013}{{\tt arXiv:1307.0013}}].

\bibitem{Larkoski:2014pca}
A.~J. Larkoski, J.~Thaler, and W.~J. Waalewijn, {\it {Gaining (Mutual)
  Information about Quark/Gluon Discrimination}},  {\em JHEP} {\bf 1411} (2014)
  129, [\href{http://arxiv.org/abs/1408.3122}{{\tt arXiv:1408.3122}}].

\bibitem{Dasgupta:2015yua}
M.~Dasgupta, A.~Powling, and A.~Siodmok, {\it {On jet substructure methods for
  signal jets}},  {\em JHEP} {\bf 08} (2015) 079,
  [\href{http://arxiv.org/abs/1503.01088}{{\tt arXiv:1503.01088}}].

\bibitem{Seymour:1997kj}
M.~Seymour, {\it {Jet shapes in hadron collisions: Higher orders, resummation
  and hadronization}},  {\em Nucl.Phys.} {\bf B513} (1998) 269--300,
  [\href{http://arxiv.org/abs/hep-ph/9707338}{{\tt hep-ph/9707338}}].

\bibitem{Li:2011hy}
H.-n. Li, Z.~Li, and C.-P. Yuan, {\it {QCD resummation for jet substructures}},
   {\em Phys.Rev.Lett.} {\bf 107} (2011) 152001,
  [\href{http://arxiv.org/abs/1107.4535}{{\tt arXiv:1107.4535}}].

\bibitem{Larkoski:2012eh}
A.~J. Larkoski, {\it {QCD Analysis of the Scale-Invariance of Jets}},  {\em
  Phys.Rev.} {\bf D86} (2012) 054004,
  [\href{http://arxiv.org/abs/1207.1437}{{\tt arXiv:1207.1437}}].

\bibitem{Jankowiak:2012na}
M.~Jankowiak and A.~J. Larkoski, {\it {Angular Scaling in Jets}},  {\em JHEP}
  {\bf 1204} (2012) 039, [\href{http://arxiv.org/abs/1201.2688}{{\tt
  arXiv:1201.2688}}].

\bibitem{Chien:2014nsa}
Y.-T. Chien and I.~Vitev, {\it {Jet Shape Resummation Using Soft-Collinear
  Effective Theory}},  {\em JHEP} {\bf 1412} (2014) 061,
  [\href{http://arxiv.org/abs/1405.4293}{{\tt arXiv:1405.4293}}].

\bibitem{Chien:2014zna}
Y.-T. Chien, {\it {Resummation of Jet Shapes and Extracting Properties of the
  Quark-Gluon Plasma}},  {\em Int.J.Mod.Phys.Conf.Ser.} {\bf 37} (2015)
  1560047, [\href{http://arxiv.org/abs/1411.0741}{{\tt arXiv:1411.0741}}].

\bibitem{Isaacson:2015fra}
J.~Isaacson, H.-n. Li, Z.~Li, and C.~P. Yuan, {\it {Factorization for
  substructures of boosted Higgs jets}},
  \href{http://arxiv.org/abs/1505.06368}{{\tt arXiv:1505.06368}}.

\bibitem{Krohn:2012fg}
D.~Krohn, M.~D. Schwartz, T.~Lin, and W.~J. Waalewijn, {\it {Jet Charge at the
  LHC}},  {\em Phys.Rev.Lett.} {\bf 110} (2013), no.~21 212001,
  [\href{http://arxiv.org/abs/1209.2421}{{\tt arXiv:1209.2421}}].

\bibitem{Waalewijn:2012sv}
W.~J. Waalewijn, {\it {Calculating the Charge of a Jet}},  {\em Phys.Rev.} {\bf
  D86} (2012) 094030, [\href{http://arxiv.org/abs/1209.3019}{{\tt
  arXiv:1209.3019}}].

\bibitem{Larkoski:2014tva}
A.~J. Larkoski, I.~Moult, and D.~Neill, {\it {Toward Multi-Differential Cross
  Sections: Measuring Two Angularities on a Single Jet}},  {\em JHEP} {\bf
  1409} (2014) 046, [\href{http://arxiv.org/abs/1401.4458}{{\tt
  arXiv:1401.4458}}].

\bibitem{Procura:2014cba}
M.~Procura, W.~J. Waalewijn, and L.~Zeune, {\it {Resummation of
  Double-Differential Cross Sections and Fully-Unintegrated Parton Distribution
  Functions}},  {\em JHEP} {\bf 1502} (2015) 117,
  [\href{http://arxiv.org/abs/1410.6483}{{\tt arXiv:1410.6483}}].

\bibitem{Bertolini:2015pka}
D.~Bertolini, J.~Thaler, and J.~R. Walsh, {\it {The First Calculation of
  Fractional Jets}},  {\em JHEP} {\bf 1505} (2015) 008,
  [\href{http://arxiv.org/abs/1501.01965}{{\tt arXiv:1501.01965}}].

\bibitem{Bhattacherjee:2015psa}
B.~Bhattacherjee, S.~Mukhopadhyay, M.~M. Nojiri, Y.~Sakaki, and B.~R. Webber,
  {\it {Associated jet and subjet rates in light-quark and gluon jet
  discrimination}},  {\em JHEP} {\bf 1504} (2015) 131,
  [\href{http://arxiv.org/abs/1501.04794}{{\tt arXiv:1501.04794}}].

\bibitem{Larkoski:2015kga}
A.~J. Larkoski, I.~Moult, and D.~Neill, {\it {Analytic Boosted Boson
  Discrimination}},  {\em JHEP} {\bf 05} (2016) 117,
  [\href{http://arxiv.org/abs/1507.03018}{{\tt arXiv:1507.03018}}].

\bibitem{Dasgupta:2015lxh}
M.~Dasgupta, L.~Schunk, and G.~Soyez, {\it {Jet shapes for boosted jet
  two-prong decays from first-principles}},  {\em JHEP} {\bf 04} (2016) 166,
  [\href{http://arxiv.org/abs/1512.00516}{{\tt arXiv:1512.00516}}].

\bibitem{Frye:2016okc}
C.~Frye, A.~J. Larkoski, M.~D. Schwartz, and K.~Yan, {\it {Precision physics
  with pile-up insensitive observables}},
  \href{http://arxiv.org/abs/1603.06375}{{\tt arXiv:1603.06375}}.

\bibitem{Frye:2016aiz}
C.~Frye, A.~J. Larkoski, M.~D. Schwartz, and K.~Yan, {\it {Factorization for
  groomed jet substructure beyond the next-to-leading logarithm}},  {\em JHEP}
  {\bf 07} (2016) 064, [\href{http://arxiv.org/abs/1603.09338}{{\tt
  arXiv:1603.09338}}].

\bibitem{Kang:2016ehg}
Z.-B. Kang, F.~Ringer, and I.~Vitev, {\it {Jet substructure using
  semi-inclusive jet functions within SCET}},
  \href{http://arxiv.org/abs/1606.07063}{{\tt arXiv:1606.07063}}.

\bibitem{Hornig:2016ahz}
A.~Hornig, Y.~Makris, and T.~Mehen, {\it {Jet Shapes in Dijet Events at the LHC
  in SCET}},  {\em JHEP} {\bf 04} (2016) 097,
  [\href{http://arxiv.org/abs/1601.01319}{{\tt arXiv:1601.01319}}].

\bibitem{Walsh:2011fz}
J.~R. Walsh and S.~Zuberi, {\it {Factorization Constraints on Jet
  Substructure}},  \href{http://arxiv.org/abs/1110.5333}{{\tt
  arXiv:1110.5333}}.

\bibitem{Larkoski:2014gra}
A.~J. Larkoski, I.~Moult, and D.~Neill, {\it {Power Counting to Better Jet
  Observables}},  {\em JHEP} {\bf 1412} (2014) 009,
  [\href{http://arxiv.org/abs/1409.6298}{{\tt arXiv:1409.6298}}].

\bibitem{Larkoski:2014zma}
A.~J. Larkoski, I.~Moult, and D.~Neill, {\it {Building a Better Boosted Top
  Tagger}},  {\em Phys.Rev.} {\bf D91} (2015), no.~3 034035,
  [\href{http://arxiv.org/abs/1411.0665}{{\tt arXiv:1411.0665}}].

\bibitem{Cogan:2014oua}
J.~Cogan, M.~Kagan, E.~Strauss, and A.~Schwarztman, {\it {Jet-Images: Computer
  Vision Inspired Techniques for Jet Tagging}},  {\em JHEP} {\bf 02} (2015)
  118, [\href{http://arxiv.org/abs/1407.5675}{{\tt arXiv:1407.5675}}].

\bibitem{deOliveira:2015xxd}
L.~de~Oliveira, M.~Kagan, L.~Mackey, B.~Nachman, and A.~Schwartzman, {\it
  {Jet-images --- deep learning edition}},  {\em JHEP} {\bf 07} (2016) 069,
  [\href{http://arxiv.org/abs/1511.05190}{{\tt arXiv:1511.05190}}].

\bibitem{Almeida:2015jua}
L.~G. Almeida, M.~Backovi{\'c}, M.~Cliche, S.~J. Lee, and M.~Perelstein, {\it
  {Playing Tag with ANN: Boosted Top Identification with Pattern Recognition}},
   {\em JHEP} {\bf 07} (2015) 086, [\href{http://arxiv.org/abs/1501.05968}{{\tt
  arXiv:1501.05968}}].

\bibitem{Baldi:2016fql}
P.~Baldi, K.~Bauer, C.~Eng, P.~Sadowski, and D.~Whiteson, {\it {Jet
  Substructure Classification in High-Energy Physics with Deep Neural
  Networks}},  {\em Phys. Rev.} {\bf D93} (2016), no.~9 094034,
  [\href{http://arxiv.org/abs/1603.09349}{{\tt arXiv:1603.09349}}].

\bibitem{Guest:2016iqz}
D.~Guest, J.~Collado, P.~Baldi, S.-C. Hsu, G.~Urban, and D.~Whiteson, {\it {Jet
  Flavor Classification in High-Energy Physics with Deep Neural Networks}},
  \href{http://arxiv.org/abs/1607.08633}{{\tt arXiv:1607.08633}}.

\bibitem{Conway:2016caq}
J.~S. Conway, R.~Bhaskar, R.~D. Erbacher, and J.~Pilot, {\it {Identification of
  High-Momentum Top Quarks, Higgs Bosons, and W and Z Bosons Using Boosted
  Event Shapes}},  \href{http://arxiv.org/abs/1606.06859}{{\tt
  arXiv:1606.06859}}.

\bibitem{Barnard:2016qma}
J.~Barnard, E.~N. Dawe, M.~J. Dolan, and N.~Rajcic, {\it {Parton Shower
  Uncertainties in Jet Substructure Analyses with Deep Neural Networks}},
  \href{http://arxiv.org/abs/1609.00607}{{\tt arXiv:1609.00607}}.

\bibitem{Larkoski:2013eya}
A.~J. Larkoski, G.~P. Salam, and J.~Thaler, {\it {Energy Correlation Functions
  for Jet Substructure}},  {\em JHEP} {\bf 1306} (2013) 108,
  [\href{http://arxiv.org/abs/1305.0007}{{\tt arXiv:1305.0007}}].

\bibitem{Butterworth:2008iy}
J.~M. Butterworth, A.~R. Davison, M.~Rubin, and G.~P. Salam, {\it {Jet
  substructure as a new Higgs search channel at the LHC}},  {\em
  Phys.Rev.Lett.} {\bf 100} (2008) 242001,
  [\href{http://arxiv.org/abs/0802.2470}{{\tt arXiv:0802.2470}}].

\bibitem{Ellis:2009su}
S.~D. Ellis, C.~K. Vermilion, and J.~R. Walsh, {\it {Techniques for improved
  heavy particle searches with jet substructure}},  {\em Phys.Rev.} {\bf D80}
  (2009) 051501, [\href{http://arxiv.org/abs/0903.5081}{{\tt
  arXiv:0903.5081}}].

\bibitem{Ellis:2009me}
S.~D. Ellis, C.~K. Vermilion, and J.~R. Walsh, {\it {Recombination Algorithms
  and Jet Substructure: Pruning as a Tool for Heavy Particle Searches}},  {\em
  Phys.Rev.} {\bf D81} (2010) 094023,
  [\href{http://arxiv.org/abs/0912.0033}{{\tt arXiv:0912.0033}}].

\bibitem{Krohn:2009th}
D.~Krohn, J.~Thaler, and L.-T. Wang, {\it {Jet Trimming}},  {\em JHEP} {\bf
  1002} (2010) 084, [\href{http://arxiv.org/abs/0912.1342}{{\tt
  arXiv:0912.1342}}].

\bibitem{Dolen:2016kst}
J.~Dolen, P.~Harris, S.~Marzani, S.~Rappoccio, and N.~Tran, {\it {Thinking
  outside the ROCs: Designing Decorrelated Taggers (DDT) for jet
  substructure}},  {\em JHEP} {\bf 05} (2016) 156,
  [\href{http://arxiv.org/abs/1603.00027}{{\tt arXiv:1603.00027}}].

\bibitem{ATL-PHYS-PUB-2015-033}
{\it {Identification of boosted, hadronically-decaying $W$ and $Z$ bosons in
  $\sqrt{s} = 13$ TeV Monte Carlo Simulations for ATLAS}},  Tech. Rep.
  ATL-PHYS-PUB-2015-033, CERN, Geneva, Aug, 2015.

\bibitem{CMS-PAS-EXO-16-030}
{\bf CMS} Collaboration, {\it {Search for light vector resonances decaying to
  quarks at $\sqrt{s}=13~\mathrm{TeV}$}},  Tech. Rep. CMS-PAS-EXO-16-030, CERN,
  Geneva, 2016.

\bibitem{Thaler:2010tr}
J.~Thaler and K.~Van~Tilburg, {\it {Identifying Boosted Objects with
  N-subjettiness}},  {\em JHEP} {\bf 1103} (2011) 015,
  [\href{http://arxiv.org/abs/1011.2268}{{\tt arXiv:1011.2268}}].

\bibitem{Thaler:2011gf}
J.~Thaler and K.~Van~Tilburg, {\it {Maximizing Boosted Top Identification by
  Minimizing N-subjettiness}},  {\em JHEP} {\bf 1202} (2012) 093,
  [\href{http://arxiv.org/abs/1108.2701}{{\tt arXiv:1108.2701}}].

\bibitem{Cacciari:2011ma}
M.~Cacciari, G.~P. Salam, and G.~Soyez, {\it {FastJet User Manual}},  {\em
  Eur.Phys.J.} {\bf C72} (2012) 1896,
  [\href{http://arxiv.org/abs/1111.6097}{{\tt arXiv:1111.6097}}].

\bibitem{fjcontrib}
``Fastjet contrib.'' \url{http://fastjet.hepforge.org/contrib/}.

\bibitem{Larkoski:2014wba}
A.~J. Larkoski, S.~Marzani, G.~Soyez, and J.~Thaler, {\it {Soft Drop}},  {\em
  JHEP} {\bf 1405} (2014) 146, [\href{http://arxiv.org/abs/1402.2657}{{\tt
  arXiv:1402.2657}}].

\bibitem{Catani:1992jc}
S.~Catani, G.~Turnock, and B.~Webber, {\it {Jet broadening measures in $e^{+}
  e^{-}$ annihilation}},  {\em Phys.Lett.} {\bf B295} (1992) 269--276.

\bibitem{Dokshitzer:1998kz}
Y.~L. Dokshitzer, A.~Lucenti, G.~Marchesini, and G.~Salam, {\it {On the QCD
  analysis of jet broadening}},  {\em JHEP} {\bf 9801} (1998) 011,
  [\href{http://arxiv.org/abs/hep-ph/9801324}{{\tt hep-ph/9801324}}].

\bibitem{Banfi:2004yd}
A.~Banfi, G.~P. Salam, and G.~Zanderighi, {\it {Principles of general
  final-state resummation and automated implementation}},  {\em JHEP} {\bf
  0503} (2005) 073, [\href{http://arxiv.org/abs/hep-ph/0407286}{{\tt
  hep-ph/0407286}}].

\bibitem{Larkoski:2014uqa}
A.~J. Larkoski, D.~Neill, and J.~Thaler, {\it {Jet Shapes with the Broadening
  Axis}},  {\em JHEP} {\bf 1404} (2014) 017,
  [\href{http://arxiv.org/abs/1401.2158}{{\tt arXiv:1401.2158}}].

\bibitem{Stewart:2010tn}
I.~W. Stewart, F.~J. Tackmann, and W.~J. Waalewijn, {\it {N-Jettiness: An
  Inclusive Event Shape to Veto Jets}},  {\em Phys.Rev.Lett.} {\bf 105} (2010)
  092002, [\href{http://arxiv.org/abs/1004.2489}{{\tt arXiv:1004.2489}}].

\bibitem{Stewart:2015waa}
I.~W. Stewart, F.~J. Tackmann, J.~Thaler, C.~K. Vermilion, and T.~F. Wilkason,
  {\it {XCone: N-jettiness as an Exclusive Cone Jet Algorithm}},  {\em JHEP}
  {\bf 11} (2015) 072, [\href{http://arxiv.org/abs/1508.01516}{{\tt
  arXiv:1508.01516}}].

\bibitem{Thaler:2015xaa}
J.~Thaler and T.~F. Wilkason, {\it {Resolving Boosted Jets with XCone}},  {\em
  JHEP} {\bf 12} (2015) 051, [\href{http://arxiv.org/abs/1508.01518}{{\tt
  arXiv:1508.01518}}].

\bibitem{Larkoski:2015uaa}
A.~J. Larkoski and I.~Moult, {\it {The Singular Behavior of Jet Substructure
  Observables}},  {\em Phys. Rev.} {\bf D93} (2016) 014017,
  [\href{http://arxiv.org/abs/1510.08459}{{\tt arXiv:1510.08459}}].

\bibitem{Farhi:1977sg}
E.~Farhi, {\it {A QCD Test for Jets}},  {\em Phys.Rev.Lett.} {\bf 39} (1977)
  1587--1588.

\bibitem{Cacciari:2007fd}
M.~Cacciari and G.~P. Salam, {\it {Pileup subtraction using jet areas}},  {\em
  Phys. Lett.} {\bf B659} (2008) 119--126,
  [\href{http://arxiv.org/abs/0707.1378}{{\tt arXiv:0707.1378}}].

\bibitem{Alon:2011xb}
R.~Alon, E.~Duchovni, G.~Perez, A.~P. Pranko, and P.~K. Sinervo, {\it {A
  Data-driven method of pile-up correction for the substructure of massive
  jets}},  {\em Phys. Rev.} {\bf D84} (2011) 114025,
  [\href{http://arxiv.org/abs/1101.3002}{{\tt arXiv:1101.3002}}].

\bibitem{Soyez:2012hv}
G.~Soyez, G.~P. Salam, J.~Kim, S.~Dutta, and M.~Cacciari, {\it {Pileup
  subtraction for jet shapes}},  {\em Phys.Rev.Lett.} {\bf 110} (2013), no.~16
  162001, [\href{http://arxiv.org/abs/1211.2811}{{\tt arXiv:1211.2811}}].

\bibitem{Tseng:2013dva}
J.~Tseng and H.~Evans, {\it {Sequential recombination algorithm for jet
  clustering and background subtraction}},  {\em Phys. Rev.} {\bf D88} (2013)
  014044, [\href{http://arxiv.org/abs/1304.1025}{{\tt arXiv:1304.1025}}].

\bibitem{Krohn:2013lba}
D.~Krohn, M.~D. Schwartz, M.~Low, and L.-T. Wang, {\it {Jet Cleansing: Pileup
  Removal at High Luminosity}},  {\em Phys. Rev.} {\bf D90} (2014), no.~6
  065020, [\href{http://arxiv.org/abs/1309.4777}{{\tt arXiv:1309.4777}}].

\bibitem{Cacciari:2014gra}
M.~Cacciari, G.~P. Salam, and G.~Soyez, {\it {SoftKiller, a particle-level
  pileup removal method}},  {\em Eur. Phys. J.} {\bf C75} (2015), no.~2 59,
  [\href{http://arxiv.org/abs/1407.0408}{{\tt arXiv:1407.0408}}].

\bibitem{Bertolini:2014bba}
D.~Bertolini, P.~Harris, M.~Low, and N.~Tran, {\it {Pileup Per Particle
  Identification}},  {\em JHEP} {\bf 10} (2014) 059,
  [\href{http://arxiv.org/abs/1407.6013}{{\tt arXiv:1407.6013}}].

\bibitem{Dasgupta:2001sh}
M.~Dasgupta and G.~Salam, {\it {Resummation of nonglobal QCD observables}},
  {\em Phys.Lett.} {\bf B512} (2001) 323--330,
  [\href{http://arxiv.org/abs/hep-ph/0104277}{{\tt hep-ph/0104277}}].

\bibitem{Cacciari:2008gp}
M.~Cacciari, G.~P. Salam, and G.~Soyez, {\it {The Anti-k(t) jet clustering
  algorithm}},  {\em JHEP} {\bf 0804} (2008) 063,
  [\href{http://arxiv.org/abs/0802.1189}{{\tt arXiv:0802.1189}}].

\bibitem{Dokshitzer:1997in}
Y.~L. Dokshitzer, G.~Leder, S.~Moretti, and B.~Webber, {\it {Better jet
  clustering algorithms}},  {\em JHEP} {\bf 9708} (1997) 001,
  [\href{http://arxiv.org/abs/hep-ph/9707323}{{\tt hep-ph/9707323}}].

\bibitem{Wobisch:1998wt}
M.~Wobisch and T.~Wengler, {\it {Hadronization corrections to jet
  cross-sections in deep inelastic scattering}},
  \href{http://arxiv.org/abs/hep-ph/9907280}{{\tt hep-ph/9907280}}.

\bibitem{Wobisch:2000dk}
M.~Wobisch, {\it {Measurement and QCD analysis of jet cross-sections in deep
  inelastic positron proton collisions at $\sqrt{s} = 300$~GeV}},  2000.

\bibitem{Bauer:2000ew}
C.~W. Bauer, S.~Fleming, and M.~E. Luke, {\it {Summing Sudakov logarithms in B
  -> X(s gamma) in effective field theory}},  {\em Phys.Rev.} {\bf D63} (2000)
  014006, [\href{http://arxiv.org/abs/hep-ph/0005275}{{\tt hep-ph/0005275}}].

\bibitem{Bauer:2000yr}
C.~W. Bauer, S.~Fleming, D.~Pirjol, and I.~W. Stewart, {\it {An Effective field
  theory for collinear and soft gluons: Heavy to light decays}},  {\em
  Phys.Rev.} {\bf D63} (2001) 114020,
  [\href{http://arxiv.org/abs/hep-ph/0011336}{{\tt hep-ph/0011336}}].

\bibitem{Bauer:2001ct}
C.~W. Bauer and I.~W. Stewart, {\it {Invariant operators in collinear effective
  theory}},  {\em Phys.Lett.} {\bf B516} (2001) 134--142,
  [\href{http://arxiv.org/abs/hep-ph/0107001}{{\tt hep-ph/0107001}}].

\bibitem{Bauer:2001yt}
C.~W. Bauer, D.~Pirjol, and I.~W. Stewart, {\it {Soft collinear factorization
  in effective field theory}},  {\em Phys.Rev.} {\bf D65} (2002) 054022,
  [\href{http://arxiv.org/abs/hep-ph/0109045}{{\tt hep-ph/0109045}}].

\bibitem{Manohar:2006nz}
A.~V. Manohar and I.~W. Stewart, {\it {The Zero-Bin and Mode Factorization in
  Quantum Field Theory}},  {\em Phys.Rev.} {\bf D76} (2007) 074002,
  [\href{http://arxiv.org/abs/hep-ph/0605001}{{\tt hep-ph/0605001}}].

\bibitem{Dasgupta:2014yra}
M.~Dasgupta, F.~Dreyer, G.~P. Salam, and G.~Soyez, {\it {Small-radius jets to
  all orders in QCD}},  {\em JHEP} {\bf 1504} (2015) 039,
  [\href{http://arxiv.org/abs/1411.5182}{{\tt arXiv:1411.5182}}].

\bibitem{Chien:2015cka}
Y.-T. Chien, A.~Hornig, and C.~Lee, {\it {Soft-collinear mode for jet cross
  sections in soft collinear effective theory}},  {\em Phys. Rev.} {\bf D93}
  (2016), no.~1 014033, [\href{http://arxiv.org/abs/1509.04287}{{\tt
  arXiv:1509.04287}}].

\bibitem{Kolodrubetz:2016dzb}
D.~W. Kolodrubetz, P.~Pietrulewicz, I.~W. Stewart, F.~J. Tackmann, and W.~J.
  Waalewijn, {\it {Factorization for Jet Radius Logarithms in Jet Mass Spectra
  at the LHC}},  \href{http://arxiv.org/abs/1605.08038}{{\tt
  arXiv:1605.08038}}.

\bibitem{Kang:2016mcy}
Z.-B. Kang, F.~Ringer, and I.~Vitev, {\it {The semi-inclusive jet function in
  SCET and small radius resummation for inclusive jet production}},
  \href{http://arxiv.org/abs/1606.06732}{{\tt arXiv:1606.06732}}.

\bibitem{Bauer:2011uc}
C.~W. Bauer, F.~J. Tackmann, J.~R. Walsh, and S.~Zuberi, {\it {Factorization
  and Resummation for Dijet Invariant Mass Spectra}},  {\em Phys.Rev.} {\bf
  D85} (2012) 074006, [\href{http://arxiv.org/abs/1106.6047}{{\tt
  arXiv:1106.6047}}].

\bibitem{Larkoski:2015zka}
A.~J. Larkoski, I.~Moult, and D.~Neill, {\it {Non-Global Logarithms,
  Factorization, and the Soft Substructure of Jets}},  {\em JHEP} {\bf 09}
  (2015) 143, [\href{http://arxiv.org/abs/1501.04596}{{\tt arXiv:1501.04596}}].

\bibitem{Pietrulewicz:2016nwo}
P.~Pietrulewicz, F.~J. Tackmann, and W.~J. Waalewijn, {\it {Factorization and
  Resummation for Generic Hierarchies between Jets}},  {\em JHEP} {\bf 08}
  (2016) 002, [\href{http://arxiv.org/abs/1601.05088}{{\tt arXiv:1601.05088}}].

\bibitem{Larkoski:2013paa}
A.~J. Larkoski and J.~Thaler, {\it {Unsafe but Calculable: Ratios of
  Angularities in Perturbative QCD}},  {\em JHEP} {\bf 1309} (2013) 137,
  [\href{http://arxiv.org/abs/1307.1699}{{\tt arXiv:1307.1699}}].

\bibitem{Larkoski:2015lea}
A.~J. Larkoski, S.~Marzani, and J.~Thaler, {\it {Sudakov Safety in Perturbative
  QCD}},  {\em Phys.Rev.} {\bf D91} (2015), no.~11 111501,
  [\href{http://arxiv.org/abs/1502.01719}{{\tt arXiv:1502.01719}}].

\bibitem{Tkachov:1995kk}
F.~V. Tkachov, {\it {Measuring multi - jet structure of hadronic energy flow or
  What is a jet?}},  {\em Int. J. Mod. Phys.} {\bf A12} (1997) 5411--5529,
  [\href{http://arxiv.org/abs/hep-ph/9601308}{{\tt hep-ph/9601308}}].

\bibitem{Sveshnikov:1995vi}
N.~Sveshnikov and F.~Tkachov, {\it {Jets and quantum field theory}},  {\em
  Phys.Lett.} {\bf B382} (1996) 403--408,
  [\href{http://arxiv.org/abs/hep-ph/9512370}{{\tt hep-ph/9512370}}].

\bibitem{Cherzor:1997ak}
P.~S. Cherzor and N.~A. Sveshnikov, {\it {Jet observables and energy momentum
  tensor}},  in {\em {Quantum field theory and high-energy physics.
  Proceedings, Workshop, QFTHEP'97, Samara, Russia, September 4-10, 1997}},
  pp.~402--407, 1997.
\newblock \href{http://arxiv.org/abs/hep-ph/9710349}{{\tt hep-ph/9710349}}.

\bibitem{Tkachov:1999py}
F.~V. Tkachov, {\it {A Theory of jet definition}},  {\em Int. J. Mod. Phys.}
  {\bf A17} (2002) 2783--2884, [\href{http://arxiv.org/abs/hep-ph/9901444}{{\tt
  hep-ph/9901444}}].

\bibitem{Fox:1978vu}
G.~C. Fox and S.~Wolfram, {\it {Observables for the Analysis of Event Shapes in
  e+ e- Annihilation and Other Processes}},  {\em Phys. Rev. Lett.} {\bf 41}
  (1978) 1581.

\bibitem{Fox:1978vw}
G.~C. Fox and S.~Wolfram, {\it {Event Shapes in e+ e- Annihilation}},  {\em
  Nucl. Phys.} {\bf B149} (1979) 413. [Erratum: Nucl. Phys.B157,543(1979)].

\bibitem{GurAri:2011vx}
G.~Gur-Ari, M.~Papucci, and G.~Perez, {\it {Classification of Energy Flow
  Observables in Narrow Jets}},  \href{http://arxiv.org/abs/1101.2905}{{\tt
  arXiv:1101.2905}}.

\bibitem{Kaplan:2008ie}
D.~E. Kaplan, K.~Rehermann, M.~D. Schwartz, and B.~Tweedie, {\it {Top Tagging:
  A Method for Identifying Boosted Hadronically Decaying Top Quarks}},  {\em
  Phys.Rev.Lett.} {\bf 101} (2008) 142001,
  [\href{http://arxiv.org/abs/0806.0848}{{\tt arXiv:0806.0848}}].

\bibitem{Thaler:2008ju}
J.~Thaler and L.-T. Wang, {\it {Strategies to Identify Boosted Tops}},  {\em
  JHEP} {\bf 0807} (2008) 092, [\href{http://arxiv.org/abs/0806.0023}{{\tt
  arXiv:0806.0023}}].

\bibitem{Almeida:2008yp}
L.~G. Almeida, S.~J. Lee, G.~Perez, G.~F. Sterman, I.~Sung, et~al., {\it
  {Substructure of high-$p_T$ Jets at the LHC}},  {\em Phys.Rev.} {\bf D79}
  (2009) 074017, [\href{http://arxiv.org/abs/0807.0234}{{\tt
  arXiv:0807.0234}}].

\bibitem{Almeida:2008tp}
L.~G. Almeida, S.~J. Lee, G.~Perez, I.~Sung, and J.~Virzi, {\it {Top Jets at
  the LHC}},  {\em Phys.Rev.} {\bf D79} (2009) 074012,
  [\href{http://arxiv.org/abs/0810.0934}{{\tt arXiv:0810.0934}}].

\bibitem{Plehn:2009rk}
T.~Plehn, G.~P. Salam, and M.~Spannowsky, {\it {Fat Jets for a Light Higgs}},
  {\em Phys.Rev.Lett.} {\bf 104} (2010) 111801,
  [\href{http://arxiv.org/abs/0910.5472}{{\tt arXiv:0910.5472}}].

\bibitem{Plehn:2010st}
T.~Plehn, M.~Spannowsky, M.~Takeuchi, and D.~Zerwas, {\it {Stop Reconstruction
  with Tagged Tops}},  {\em JHEP} {\bf 1010} (2010) 078,
  [\href{http://arxiv.org/abs/1006.2833}{{\tt arXiv:1006.2833}}].

\bibitem{Almeida:2010pa}
L.~G. Almeida, S.~J. Lee, G.~Perez, G.~Sterman, and I.~Sung, {\it {Template
  Overlap Method for Massive Jets}},  {\em Phys.Rev.} {\bf D82} (2010) 054034,
  [\href{http://arxiv.org/abs/1006.2035}{{\tt arXiv:1006.2035}}].

\bibitem{Jankowiak:2011qa}
M.~Jankowiak and A.~J. Larkoski, {\it {Jet Substructure Without Trees}},  {\em
  JHEP} {\bf 1106} (2011) 057, [\href{http://arxiv.org/abs/1104.1646}{{\tt
  arXiv:1104.1646}}].

\bibitem{Soper:2012pb}
D.~E. Soper and M.~Spannowsky, {\it {Finding top quarks with shower
  deconstruction}},  {\em Phys.Rev.} {\bf D87} (2013), no.~5 054012,
  [\href{http://arxiv.org/abs/1211.3140}{{\tt arXiv:1211.3140}}].

\bibitem{Anders:2013oga}
C.~Anders, C.~Bernaciak, G.~Kasieczka, T.~Plehn, and T.~Schell, {\it
  {Benchmarking an even better top tagger algorithm}},  {\em Phys. Rev.} {\bf
  D89} (2014), no.~7 074047, [\href{http://arxiv.org/abs/1312.1504}{{\tt
  arXiv:1312.1504}}].

\bibitem{Freytsis:2014hpa}
M.~Freytsis, T.~Volansky, and J.~R. Walsh, {\it {Tagging Partially
  Reconstructed Objects with Jet Substructure}},
  \href{http://arxiv.org/abs/1412.7540}{{\tt arXiv:1412.7540}}.

\bibitem{Larkoski:2015yqa}
A.~J. Larkoski, F.~Maltoni, and M.~Selvaggi, {\it {Tracking down hyper-boosted
  top quarks}},  {\em JHEP} {\bf 1506} (2015) 032,
  [\href{http://arxiv.org/abs/1503.03347}{{\tt arXiv:1503.03347}}].

\bibitem{Kasieczka:2015jma}
G.~Kasieczka, T.~Plehn, T.~Schell, T.~Strebler, and G.~P. Salam, {\it
  {Resonance Searches with an Updated Top Tagger}},  {\em JHEP} {\bf 06} (2015)
  203, [\href{http://arxiv.org/abs/1503.05921}{{\tt arXiv:1503.05921}}].

\bibitem{Lapsien:2016zor}
T.~Lapsien, R.~Kogler, and J.~Haller, {\it {A new tagger for hadronically
  decaying heavy particles at the LHC}},
  \href{http://arxiv.org/abs/1606.04961}{{\tt arXiv:1606.04961}}.

\bibitem{ATLAS-CONF-2012-100}
{\bf ATLAS} Collaboration, {\it {Identification and Tagging of Double b-hadron
  jets with the ATLAS Detector}},  Tech. Rep. ATLAS-CONF-2012-100, CERN,
  Geneva, Jul, 2012.

\bibitem{CMS:2013vea}
{\bf CMS} Collaboration, {\it {Performance of b tagging at sqrt(s)=8 TeV in
  multijet, ttbar and boosted topology events}},  Tech. Rep.
  CMS-PAS-BTV-13-001, 2013.

\bibitem{ATL-PHYS-PUB-2014-014}
{\bf ATLAS} Collaboration, {\it {b-tagging in dense environments}},  Tech. Rep.
  ATL-PHYS-PUB-2014-014, CERN, Geneva, Aug, 2014.

\bibitem{ATL-PHYS-PUB-2015-035}
{\bf ATLAS} Collaboration, {\it {Expected Performance of Boosted Higgs
  ($\rightarrow b\bar{b}$) Boson Identification with the ATLAS Detector at
  $\sqrt{s} = 13$ TeV}},  Tech. Rep. ATL-PHYS-PUB-2015-035, CERN, Geneva, Aug,
  2015.

\bibitem{CMS-PAS-BTV-15-001}
{\bf CMS} Collaboration, {\it {Identification of b quark jets at the CMS
  Experiment in the LHC Run 2}},  Tech. Rep. CMS-PAS-BTV-15-001, CERN, Geneva,
  2016.

\bibitem{CMS-PAS-BTV-15-002}
{\bf CMS} Collaboration, {\it {Identification of double-b quark jets in boosted
  event topologies}},  Tech. Rep. CMS-PAS-BTV-15-002, CERN, Geneva, 2016.

\bibitem{ATLAS-CONF-2016-001}
{\bf ATLAS} Collaboration, {\it {Calibration of ATLAS $b$-tagging algorithms in
  dense jet environments}},  Tech. Rep. ATLAS-CONF-2016-001, CERN, Geneva, Feb,
  2016.

\bibitem{Soper:2011cr}
D.~E. Soper and M.~Spannowsky, {\it {Finding physics signals with shower
  deconstruction}},  {\em Phys. Rev.} {\bf D84} (2011) 074002,
  [\href{http://arxiv.org/abs/1102.3480}{{\tt arXiv:1102.3480}}].

\bibitem{Soper:2014rya}
D.~E. Soper and M.~Spannowsky, {\it {Finding physics signals with event
  deconstruction}},  {\em Phys. Rev.} {\bf D89} (2014), no.~9 094005,
  [\href{http://arxiv.org/abs/1402.1189}{{\tt arXiv:1402.1189}}].

\bibitem{Aad:2016pux}
{\bf ATLAS} Collaboration, G.~Aad et~al., {\it {Identification of high
  transverse momentum top quarks in $pp$ collisions at $\sqrt{s}$ = 8 TeV with
  the ATLAS detector}},  {\em JHEP} {\bf 06} (2016) 093,
  [\href{http://arxiv.org/abs/1603.03127}{{\tt arXiv:1603.03127}}].

\bibitem{CMS-PAS-JME-15-002}
{\bf CMS} Collaboration, {\it {Top Tagging with New Approaches}},  Tech. Rep.
  CMS-PAS-JME-15-002, CERN, Geneva, 2016.

\bibitem{gregory_talk}
G.~Soyez, {\it {A QCD description of jet shapes for boosted jets}},  {\em BOOST
  Conference} (2016).

\bibitem{gregory_paper}
G.~P. Salam, L.~Schunk, and G.~Soyez, {\it {Towards a better use of
  N-subjettiness}},  {\em forthcoming} (2016).

\bibitem{Alwall:2014hca}
J.~Alwall, R.~Frederix, S.~Frixione, V.~Hirschi, F.~Maltoni, et~al., {\it {The
  automated computation of tree-level and next-to-leading order differential
  cross sections, and their matching to parton shower simulations}},  {\em
  JHEP} {\bf 1407} (2014) 079, [\href{http://arxiv.org/abs/1405.0301}{{\tt
  arXiv:1405.0301}}].

\bibitem{Sjostrand:2006za}
T.~Sjostrand, S.~Mrenna, and P.~Z. Skands, {\it {PYTHIA 6.4 Physics and
  Manual}},  {\em JHEP} {\bf 0605} (2006) 026,
  [\href{http://arxiv.org/abs/hep-ph/0603175}{{\tt hep-ph/0603175}}].

\bibitem{Sjostrand:2007gs}
T.~Sjostrand, S.~Mrenna, and P.~Z. Skands, {\it {A Brief Introduction to PYTHIA
  8.1}},  {\em Comput.Phys.Commun.} {\bf 178} (2008) 852--867,
  [\href{http://arxiv.org/abs/0710.3820}{{\tt arXiv:0710.3820}}].

\bibitem{Larkoski:2014bia}
A.~J. Larkoski and J.~Thaler, {\it {Aspects of jets at 100 TeV}},  {\em
  Phys.Rev.} {\bf D90} (2014), no.~3 034010,
  [\href{http://arxiv.org/abs/1406.7011}{{\tt arXiv:1406.7011}}].

\bibitem{ATL-PHYS-PUB-2015-053}
{\bf ATLAS} Collaboration, {\it {Boosted hadronic top identification at ATLAS
  for early 13 TeV data}},  Tech. Rep. ATL-PHYS-PUB-2015-053, CERN, Geneva,
  Dec, 2015.

\bibitem{ATLAS-CONF-2016-082}
{\bf ATLAS} Collaboration, {\it {Searches for heavy ZZ and ZW resonances in the
  llqq and vvqq final states in pp collisions at sqrt(s) = 13 TeV with the
  ATLAS detector}},  Tech. Rep. ATLAS-CONF-2016-082, CERN, Geneva, Aug, 2016.

\bibitem{ATLAS-CONF-2016-083}
{\bf ATLAS} Collaboration, {\it {A Search for Resonances Decaying to a $W$ or
  $Z$ Boson and a Higgs Boson in the $q\bar{q}^{(\prime)}b\bar{b}$ Final
  State}},  Tech. Rep. ATLAS-CONF-2016-083, CERN, Geneva, Aug, 2016.

\bibitem{CMS:2016pfl}
{\bf CMS} Collaboration, {\it {Search for new resonances decaying to
  $\mathrm{WW}/\mathrm{WZ} \to \ell\nu \mathrm{qq}$}},  Tech. Rep.
  CMS-PAS-B2G-16-020, CERN, 2016.

\bibitem{CMS:2016mvc}
{\bf CMS} Collaboration, {\it {Search for high-mass resonances in the
  $\mathrm{Z(q\overline{q})}\gamma$ final state at $\sqrt{s}=8~\mathrm{TeV}$}},
   Tech. Rep. CMS-PAS-EXO-16-025, CERN, 2016.

\bibitem{CMS:2016wev}
{\bf CMS} Collaboration, {\it {Combination of diboson resonance searches at 8
  and 13 TeV}},  Tech. Rep. CMS-PAS-B2G-16-007, CERN, 2016.

\bibitem{Goncalves:2015yua}
D.~Gon{\c c}alves, F.~Krauss, and M.~Spannowsky, {\it {Augmenting the diboson
  excess for the LHC Run II}},  {\em Phys. Rev.} {\bf D92} (2015), no.~5
  053010, [\href{http://arxiv.org/abs/1508.04162}{{\tt arXiv:1508.04162}}].

\bibitem{Martin:2016jdw}
A.~Martin and T.~S. Roy, {\it {Cautionary tale of mismeasured tails from q/g
  bias}},  {\em Phys. Rev.} {\bf D94} (2016), no.~1 014003,
  [\href{http://arxiv.org/abs/1604.05728}{{\tt arXiv:1604.05728}}].

\bibitem{Larkoski:2015npa}
A.~J. Larkoski and I.~Moult, {\it {Nonglobal correlations in collider
  physics}},  {\em Phys. Rev.} {\bf D93} (2016), no.~1 014012,
  [\href{http://arxiv.org/abs/1510.05657}{{\tt arXiv:1510.05657}}].

\bibitem{Giele:2007di}
W.~T. Giele, D.~A. Kosower, and P.~Z. Skands, {\it {A simple shower and
  matching algorithm}},  {\em Phys.Rev.} {\bf D78} (2008) 014026,
  [\href{http://arxiv.org/abs/0707.3652}{{\tt arXiv:0707.3652}}].

\bibitem{Giele:2011cb}
W.~Giele, D.~Kosower, and P.~Skands, {\it {Higher-Order Corrections to Timelike
  Jets}},  {\em Phys.Rev.} {\bf D84} (2011) 054003,
  [\href{http://arxiv.org/abs/1102.2126}{{\tt arXiv:1102.2126}}].

\bibitem{GehrmannDeRidder:2011dm}
A.~Gehrmann-De~Ridder, M.~Ritzmann, and P.~Z. Skands, {\it {Timelike
  Dipole-Antenna Showers with Massive Fermions}},  {\em Phys.Rev.} {\bf D85}
  (2012) 014013, [\href{http://arxiv.org/abs/1108.6172}{{\tt
  arXiv:1108.6172}}].

\bibitem{Ritzmann:2012ca}
M.~Ritzmann, D.~Kosower, and P.~Skands, {\it {Antenna Showers with Hadronic
  Initial States}},  {\em Phys.Lett.} {\bf B718} (2013) 1345--1350,
  [\href{http://arxiv.org/abs/1210.6345}{{\tt arXiv:1210.6345}}].

\bibitem{Hartgring:2013jma}
L.~Hartgring, E.~Laenen, and P.~Skands, {\it {Antenna Showers with One-Loop
  Matrix Elements}},  {\em JHEP} {\bf 1310} (2013) 127,
  [\href{http://arxiv.org/abs/1303.4974}{{\tt arXiv:1303.4974}}].

\bibitem{Larkoski:2013yi}
A.~J. Larkoski, J.~J. Lopez-Villarejo, and P.~Skands, {\it {Helicity-Dependent
  Showers and Matching with VINCIA}},  {\em Phys.Rev.} {\bf D87} (2013), no.~5
  054033, [\href{http://arxiv.org/abs/1301.0933}{{\tt arXiv:1301.0933}}].

\bibitem{Fischer:2016vfv}
N.~Fischer, S.~Prestel, M.~Ritzmann, and P.~Skands, {\it {Vincia for Hadron
  Colliders}},  \href{http://arxiv.org/abs/1605.06142}{{\tt arXiv:1605.06142}}.

\bibitem{ATLAS-CONF-2015-073}
{\bf ATLAS} Collaboration, {\it {Search for resonances with boson-tagged jets
  in 3.2 fb−1 of p p collisions at √ s = 13 TeV collected with the ATLAS
  detector}},  Tech. Rep. ATLAS-CONF-2015-073, CERN, Geneva, Dec, 2015.

\bibitem{ATLAS-CONF-2016-016}
{\bf ATLAS} Collaboration, {\it {Search for ZZ resonances in the $\ell\ell qq$
  final state in pp collisions at $\sqrt{s}$ = 13 TeV with the ATLAS
  detector}},  Tech. Rep. ATLAS-CONF-2016-016, CERN, Geneva, Mar, 2016.

\bibitem{CMS-PAS-JME-14-002}
{\bf CMS} Collaboration, {\it {V Tagging Observables and Correlations}},  Tech.
  Rep. CMS-PAS-JME-14-002, CERN, Geneva, 2014.

\bibitem{Gallicchio:2011xc}
J.~Gallicchio and M.~D. Schwartz, {\it {Pure Samples of Quark and Gluon Jets at
  the LHC}},  {\em JHEP} {\bf 1110} (2011) 103,
  [\href{http://arxiv.org/abs/1104.1175}{{\tt arXiv:1104.1175}}].

\bibitem{Gallicchio:2011xq}
J.~Gallicchio and M.~D. Schwartz, {\it {Quark and Gluon Tagging at the LHC}},
  {\em Phys.Rev.Lett.} {\bf 107} (2011) 172001,
  [\href{http://arxiv.org/abs/1106.3076}{{\tt arXiv:1106.3076}}].

\bibitem{Gallicchio:2012ez}
J.~Gallicchio and M.~D. Schwartz, {\it {Quark and Gluon Jet Substructure}},
  {\em JHEP} {\bf 1304} (2013) 090, [\href{http://arxiv.org/abs/1211.7038}{{\tt
  arXiv:1211.7038}}].

\bibitem{Badger:2016bpw}
J.~R. Andersen et~al., {\it {Les Houches 2015: Physics at TeV Colliders
  Standard Model Working Group Report}},  in {\em {9th Les Houches Workshop on
  Physics at TeV Colliders (PhysTeV 2015) Les Houches, France, June 1-19,
  2015}}, 2016.
\newblock \href{http://arxiv.org/abs/1605.04692}{{\tt arXiv:1605.04692}}.

\bibitem{CMS:2013wea}
{\bf CMS} Collaboration, {\it {Pileup Jet Identification}},  Tech. Rep.
  CMS-PAS-JME-13-005, 2013.

\bibitem{CMS:2013kfa}
{\bf CMS} Collaboration, {\it {Performance of quark/gluon discrimination in 8
  TeV pp data}},  Tech. Rep. CMS-PAS-JME-13-002, 2013.

\bibitem{Aad:2014gea}
{\bf ATLAS} Collaboration, G.~Aad et~al., {\it {Light-quark and gluon jet
  discrimination in pp collisions at $\sqrt{s}$ = 7 TeV with the ATLAS
  detector}},  \href{http://arxiv.org/abs/1405.6583}{{\tt arXiv:1405.6583}}.

\bibitem{Berger:2003iw}
C.~F. Berger, T.~Kucs, and G.~F. Sterman, {\it {Event shape / energy flow
  correlations}},  {\em Phys.Rev.} {\bf D68} (2003) 014012,
  [\href{http://arxiv.org/abs/hep-ph/0303051}{{\tt hep-ph/0303051}}].

\bibitem{Brodsky:1976mg}
S.~J. Brodsky and J.~F. Gunion, {\it {Hadron Multiplicity in Color Gauge Theory
  Models}},  {\em Phys. Rev. Lett.} {\bf 37} (1976) 402--405.

\bibitem{Konishi:1978yx}
K.~Konishi, A.~Ukawa, and G.~Veneziano, {\it {A Simple Algorithm for QCD
  Jets}},  {\em Phys. Lett.} {\bf B78} (1978) 243--248.

\bibitem{Mueller:1983cq}
A.~H. Mueller, {\it {Square Root of alpha (Q**2) Corrections to Particle
  Multiplicity Ratios in Gluon and Quark Jets}},  {\em Nucl. Phys.} {\bf B241}
  (1984) 141--154.

\bibitem{Malaza:1984vv}
E.~D. Malaza and B.~R. Webber, {\it {QCD CORRECTIONS TO JET MULTIPLICITY
  MOMENTS}},  {\em Phys. Lett.} {\bf B149} (1984) 501--503.

\bibitem{Gaffney:1984yd}
J.~B. Gaffney and A.~H. Mueller, {\it {Alpha (Q**2) Corrections to Particle
  Multiplicity Ratios in Gluon and Quark Jets}},  {\em Nucl. Phys.} {\bf B250}
  (1985) 109--142.

\bibitem{Malaza:1985jd}
E.~D. Malaza and B.~R. Webber, {\it {Multiplicity Distributions in Quark and
  Gluon Jets}},  {\em Nucl. Phys.} {\bf B267} (1986) 702--713.

\bibitem{Catani:1991pm}
S.~Catani, Y.~L. Dokshitzer, F.~Fiorani, and B.~R. Webber, {\it {Average number
  of jets in e+ e- annihilation}},  {\em Nucl. Phys.} {\bf B377} (1992)
  445--460.

\bibitem{Catani:1992tm}
S.~Catani, B.~R. Webber, Y.~L. Dokshitzer, and F.~Fiorani, {\it {Average
  multiplicities in two and three jet e+ e- annihilation events}},  {\em Nucl.
  Phys.} {\bf B383} (1992) 419--441.

\bibitem{Dremin:1993vq}
I.~M. Dremin and V.~A. Nechitailo, {\it {Moments of multiplicity distributions
  in higher order perturbative QCD}},  {\em JETP Lett.} {\bf 58} (1993)
  881--885.

\bibitem{Dremin:1994bj}
I.~M. Dremin and V.~A. Nechitailo, {\it {Average multiplicities in gluon and
  quark jets in higher order perturbative QCD}},  {\em Mod. Phys. Lett.} {\bf
  A9} (1994) 1471--1478, [\href{http://arxiv.org/abs/hep-ex/9406002}{{\tt
  hep-ex/9406002}}].

\bibitem{Capella:1999ms}
A.~Capella, I.~M. Dremin, J.~W. Gary, V.~A. Nechitailo, and J.~Tran Thanh~Van,
  {\it {Evolution of average multiplicities of quark and gluon jets}},  {\em
  Phys. Rev.} {\bf D61} (2000) 074009,
  [\href{http://arxiv.org/abs/hep-ph/9910226}{{\tt hep-ph/9910226}}].

\bibitem{Bolzoni:2012ii}
P.~Bolzoni, B.~A. Kniehl, and A.~V. Kotikov, {\it {Gluon and quark jet
  multiplicities at N$^3$LO+NNLL}},  {\em Phys. Rev. Lett.} {\bf 109} (2012)
  242002, [\href{http://arxiv.org/abs/1209.5914}{{\tt arXiv:1209.5914}}].

\bibitem{Aad:2016oit}
{\bf ATLAS} Collaboration, G.~Aad et~al., {\it {Measurement of the
  charged-particle multiplicity inside jets from $\sqrt{s}=8$ TeV $pp$
  collisions with the ATLAS detector}},  {\em Eur. Phys. J.} {\bf C76} (2016),
  no.~6 322, [\href{http://arxiv.org/abs/1602.00988}{{\tt arXiv:1602.00988}}].

\bibitem{FerreiradeLima:2016gcz}
D.~Ferreira~de Lima, P.~Petrov, D.~Soper, and M.~Spannowsky, {\it {Quark-Gluon
  tagging with Shower Deconstruction: Unearthing dark matter and Higgs
  couplings}},  \href{http://arxiv.org/abs/1607.06031}{{\tt arXiv:1607.06031}}.

\bibitem{Abdallah:2003xz}
{\bf DELPHI} Collaboration, J.~Abdallah et~al., {\it {A Study of the energy
  evolution of event shape distributions and their means with the DELPHI
  detector at LEP}},  {\em Eur. Phys. J.} {\bf C29} (2003) 285--312,
  [\href{http://arxiv.org/abs/hep-ex/0307048}{{\tt hep-ex/0307048}}].

\bibitem{Heister:2003aj}
{\bf ALEPH} Collaboration, A.~Heister et~al., {\it {Studies of QCD at e+ e-
  centre-of-mass energies between 91-GeV and 209-GeV}},  {\em Eur. Phys. J.}
  {\bf C35} (2004) 457--486.

\bibitem{Achard:2004sv}
{\bf L3} Collaboration, P.~Achard et~al., {\it {Studies of hadronic event
  structure in $e^{+} e^{-}$ annihilation from 30-GeV to 209-GeV with the L3
  detector}},  {\em Phys.Rept.} {\bf 399} (2004) 71--174,
  [\href{http://arxiv.org/abs/hep-ex/0406049}{{\tt hep-ex/0406049}}].

\bibitem{Abbiendi:2004qz}
{\bf OPAL} Collaboration, G.~Abbiendi et~al., {\it {Measurement of event shape
  distributions and moments in e+ e- ---> hadrons at 91-GeV - 209-GeV and a
  determination of alpha(s)}},  {\em Eur. Phys. J.} {\bf C40} (2005) 287--316,
  [\href{http://arxiv.org/abs/hep-ex/0503051}{{\tt hep-ex/0503051}}].

\bibitem{Korchemsky:1999kt}
G.~P. Korchemsky and G.~F. Sterman, {\it {Power corrections to event shapes and
  factorization}},  {\em Nucl.Phys.} {\bf B555} (1999) 335--351,
  [\href{http://arxiv.org/abs/hep-ph/9902341}{{\tt hep-ph/9902341}}].

\bibitem{Korchemsky:2000kp}
G.~Korchemsky and S.~Tafat, {\it {On power corrections to the event shape
  distributions in QCD}},  {\em JHEP} {\bf 0010} (2000) 010,
  [\href{http://arxiv.org/abs/hep-ph/0007005}{{\tt hep-ph/0007005}}].

\bibitem{Bosch:2004th}
S.~Bosch, B.~Lange, M.~Neubert, and G.~Paz, {\it {Factorization and shape
  function effects in inclusive B meson decays}},  {\em Nucl.Phys.} {\bf B699}
  (2004) 335--386, [\href{http://arxiv.org/abs/hep-ph/0402094}{{\tt
  hep-ph/0402094}}].

\bibitem{Hoang:2007vb}
A.~H. Hoang and I.~W. Stewart, {\it {Designing gapped soft functions for jet
  production}},  {\em Phys.Lett.} {\bf B660} (2008) 483--493,
  [\href{http://arxiv.org/abs/0709.3519}{{\tt arXiv:0709.3519}}].

\bibitem{Ligeti:2008ac}
Z.~Ligeti, I.~W. Stewart, and F.~J. Tackmann, {\it {Treating the b quark
  distribution function with reliable uncertainties}},  {\em Phys.Rev.} {\bf
  D78} (2008) 114014, [\href{http://arxiv.org/abs/0807.1926}{{\tt
  arXiv:0807.1926}}].

\bibitem{Boughezal:2015dva}
R.~Boughezal, C.~Focke, X.~Liu, and F.~Petriello, {\it {$W$-boson production in
  association with a jet at next-to-next-to-leading order in perturbative
  QCD}},  {\em Phys. Rev. Lett.} {\bf 115} (2015), no.~6 062002,
  [\href{http://arxiv.org/abs/1504.02131}{{\tt arXiv:1504.02131}}].

\bibitem{Gaunt:2015pea}
J.~Gaunt, M.~Stahlhofen, F.~J. Tackmann, and J.~R. Walsh, {\it {N-jettiness
  Subtractions for NNLO QCD Calculations}},  {\em JHEP} {\bf 09} (2015) 058,
  [\href{http://arxiv.org/abs/1505.04794}{{\tt arXiv:1505.04794}}].

\bibitem{Hook:2011cq}
A.~Hook, M.~Jankowiak, and J.~G. Wacker, {\it {Jet Dipolarity: Top Tagging with
  Color Flow}},  {\em JHEP} {\bf 04} (2012) 007,
  [\href{http://arxiv.org/abs/1102.1012}{{\tt arXiv:1102.1012}}].

\end{thebibliography}\endgroup

\end{document}